\documentclass[10pt,letterpaper]{article}
\usepackage{dvbench-release}
\newcommand{\DVBenchTaskCount}{95}

\newcommand{\DVBenchDomainCount}{9}
\newcommand{\DVBenchSlocMinimum}{142}
\newcommand{\DVBenchSlocMaximum}{22,722}

\newcommand{\DVBenchModuleMinimum}{1}
\newcommand{\DVBenchModuleMaximum}{64}

\hypersetup{
  pdftitle={DVBench: End-to-End Benchmark for Real-World UVM Design Verification Agents},
  pdfauthor={Jiaru Qian, Jiawei Li, Xiaolong Hu, Zheng Shen, LiMing Zhao, Xin Zhang, Siyuan Jiang, Yongqiang Wei, Dong Liu, Jianyang Li, Yihua Chen, Yangxu Liao, Yuxin Wu, Jing Liang, Bin Wen, Junjie Zhang, Yongsheng Han, Yihong Dong, Ge Li},
  pdfsubject={DVBench research report},
  pdfkeywords={design verification, UVM, LLM agents, benchmark}
}
\begin{document}
\thispagestyle{releasefirst}
\begin{tcolorbox}[releasecover]
\centering
{\fontsize{17.8}{21.5}\selectfont\bfseries\color{ReleaseNavy}
\textcolor{ReleaseBlue}{DVBench:} End-to-End Benchmark for Real-World
UVM Design Verification Agents\par}
\vspace{10pt}
\newcommand{\PaperAuthor}[2]{\mbox{#1\if\relax\detokenize{#2}\relax\else\textsuperscript{#2}\fi}}
{\fontsize{9.3}{12.3}\selectfont\bfseries
\PaperAuthor{Jiaru Qian}{1,2,8,*}\quad
\PaperAuthor{Jiawei Li}{2,*}\quad
\PaperAuthor{Xiaolong Hu}{2}\quad
\PaperAuthor{Zheng Shen}{2}\quad
\PaperAuthor{LiMing Zhao}{2}\quad
\PaperAuthor{Xin Zhang}{1,8}\\[1.5pt]
\PaperAuthor{Siyuan Jiang}{7}\quad
\PaperAuthor{Yongqiang Wei}{2}\quad
\PaperAuthor{Dong Liu}{2}\quad
\PaperAuthor{Jianyang Li}{2}\quad
\PaperAuthor{Yihua Chen}{2,4}\quad
\PaperAuthor{Yangxu Liao}{4}\\[1.5pt]
\PaperAuthor{Yuxin Wu}{5}\quad
\PaperAuthor{Jing Liang}{3}\quad
\PaperAuthor{Bin Wen}{3}\quad
\PaperAuthor{Junjie Zhang}{3}\quad
\PaperAuthor{Yongsheng Han}{3}\quad
\PaperAuthor{Yihong Dong}{8}\quad
\PaperAuthor{Ge Li}{1,\textdagger}\par}
\vspace{4pt}
{\fontsize{8}{10}\selectfont\color{ReleaseMuted}
\textsuperscript{*}Equal contribution\qquad
\textsuperscript{\textdagger}Corresponding author\par}

\vspace{12pt}
\makebox[\linewidth][c]{%
\raisebox{-0.5\height}{\includegraphics[width=88pt]{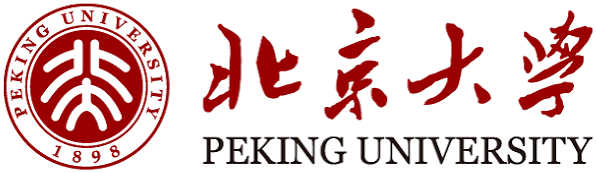}}\hspace{18pt}%
\raisebox{-0.5\height}{\includegraphics[width=97pt]{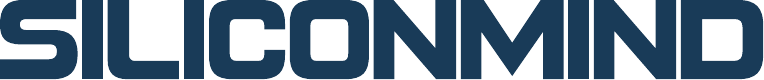}}\hspace{18pt}%
\raisebox{-0.5\height}{\includegraphics[width=82pt]{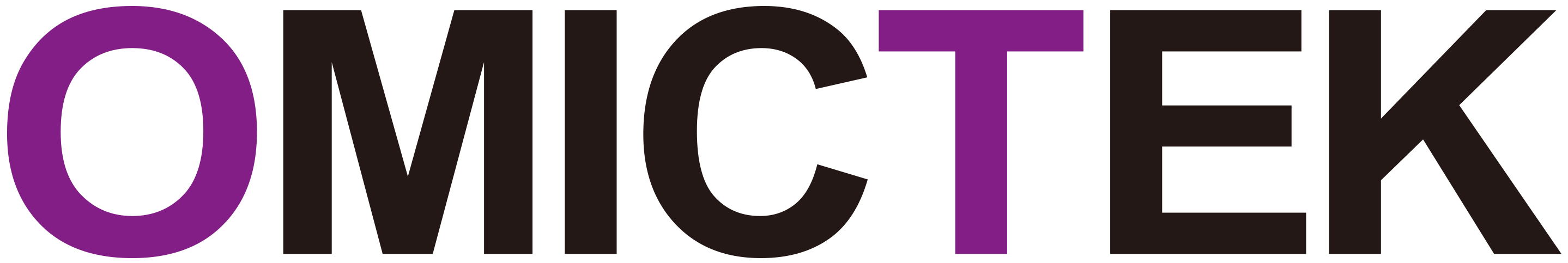}}\hspace{18pt}%
\raisebox{-0.5\height}{\includegraphics[width=88pt]{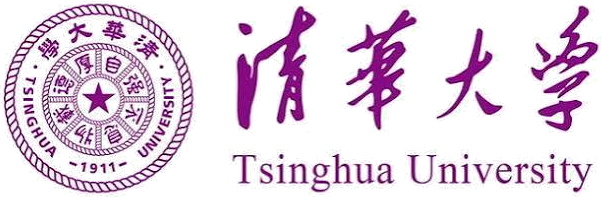}}}\par
\vspace{10pt}
\makebox[\linewidth][c]{%
\raisebox{-0.5\height}{\includegraphics[width=120pt]{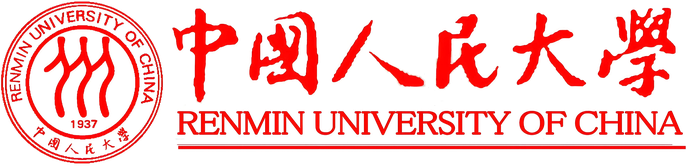}}\hspace{24pt}%
\raisebox{-0.5\height}{\includegraphics[width=90pt]{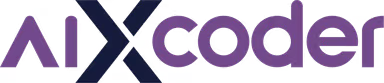}}\hspace{24pt}%
\raisebox{-0.5\height}{\includegraphics[width=117pt]{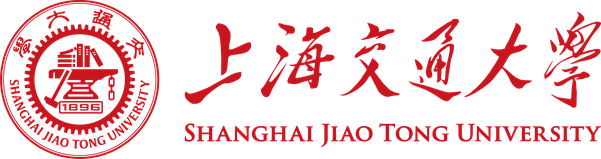}}}\par
\vspace{10pt}
{\fontsize{8.3}{11}\selectfont\color{ReleaseNavy}
\textsuperscript{1}School of Computer Science, Peking University\quad
\textsuperscript{2}SiliconMind\quad
\textsuperscript{3}Omnicom-tech\quad
\textsuperscript{4}Tsinghua University\\[1pt]
\textsuperscript{5}Renmin University of China\quad
\textsuperscript{6}Shanghai Jiao Tong University\quad
\textsuperscript{7}aiXcoder\\[1pt]
\textsuperscript{8}Key Laboratory of High Confidence Software Technologies, Peking University, Ministry of Education\quad \par}

\vspace{6pt}
{\sffamily\fontsize{8}{10}\selectfont\color{ReleaseBlue}
\href{mailto:qianjiaru26@stu.pku.edu.cn}{\faEnvelope\enspace qianjiaru26@stu.pku.edu.cn}\quad
\href{mailto:lige@pku.edu.cn}{\faEnvelope\enspace lige@pku.edu.cn}\\[3pt]
\href{https://github.com/JiaruQian/DVBench.git}{\faGithub\enspace github.com/JiaruQian/DVBench}\qquad
\href{https://huggingface.co/datasets/qianjiaru/DVBench}{\raisebox{-2pt}{\includegraphics[height=10pt]{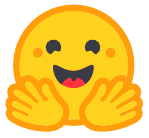}}\enspace Hugging Face: qianjiaru/DVBench}\par}
\vspace{8pt}
\par\raggedright
{\fontsize{9.3}{11.7}\selectfont
\textbf{\color{ReleaseBlue}Abstract.}\enspace
Design verification (DV) remains a major bottleneck in hardware development, with large-scale verification using the Universal Verification Methodology (UVM) being highly labor-intensive. Recent advances in large language models (LLMs) and agentic systems offer new opportunities for DV automation. However, existing benchmarks primarily
construct mutually decoupled subtasks, which deviates from the real-world workflow.  In this paper, we present DVBench, the first end-to-end benchmark for Real-World UVM design verification agents.
DVBench comprises 95 tasks across 9 functional domains and 5 difficulty levels, covering diverse verification scenarios and design scales. Each task requires agents to independently complete the full verification workflow. DVBench evaluates final testbenches via compilation, coverage, and bug detection, while assessing intermediate artifacts including feature lists, test plans, and reference models. We evaluate frontier LLMs (e.g. Claude Opus 5.5, GPT-6 Astra) and agent frameworks (e.g. Claude Code, Codex, OpenCode). Experimental results show that the best LLM-based agent (Codex \& GPT-6 Astra) can only achieve an overall success rate of 21.05\%, revealing substantial room for improvement in complex verification tasks.

\par}
\vspace{8pt}
{\fontsize{8}{10}\selectfont\color{ReleaseMuted}
\textbf{Keywords}\enspace Design verification\enspace /\enspace LLM agents\enspace /\enspace Benchmark \enspace /\enspace Universal Verification Methodology (UVM)\par }
\end{tcolorbox}
\vspace{5pt}
\begingroup
\centering
\includegraphics[width=\linewidth]{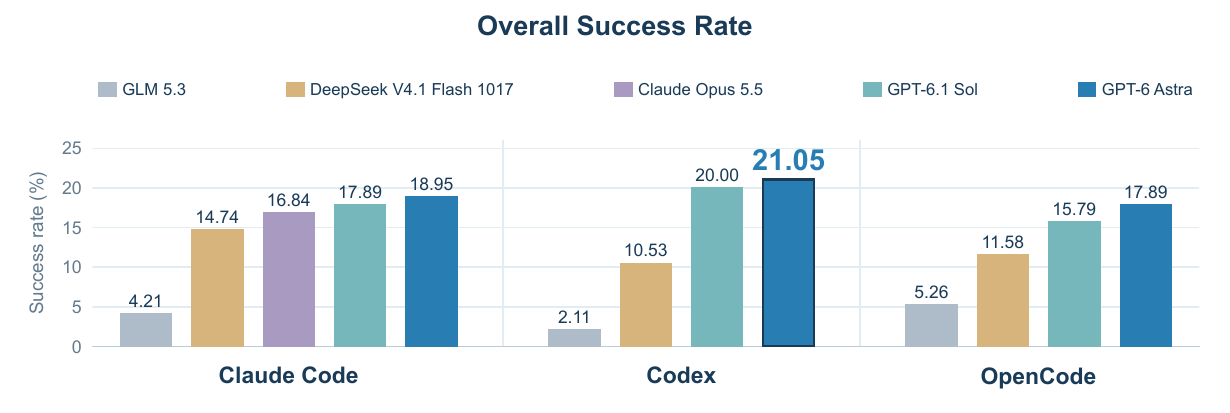}
\captionsetup{hypcap=false}
\captionof{figure}{\textbf{End-to-end UVM verification remains challenging.} Overall success across 13 agent--model configurations on the same 95 tasks, grouped by agent framework (Table~\ref{tab:main-results}). Colors denote the underlying models. The strongest configuration, Codex with GPT-6 Astra, achieves 21.05\% overall success rate.}
\label{fig:results-teaser}
\par\endgroup
\clearpage

\noindent\begin{minipage}{\linewidth}
\centering
\includegraphics[width=\linewidth]{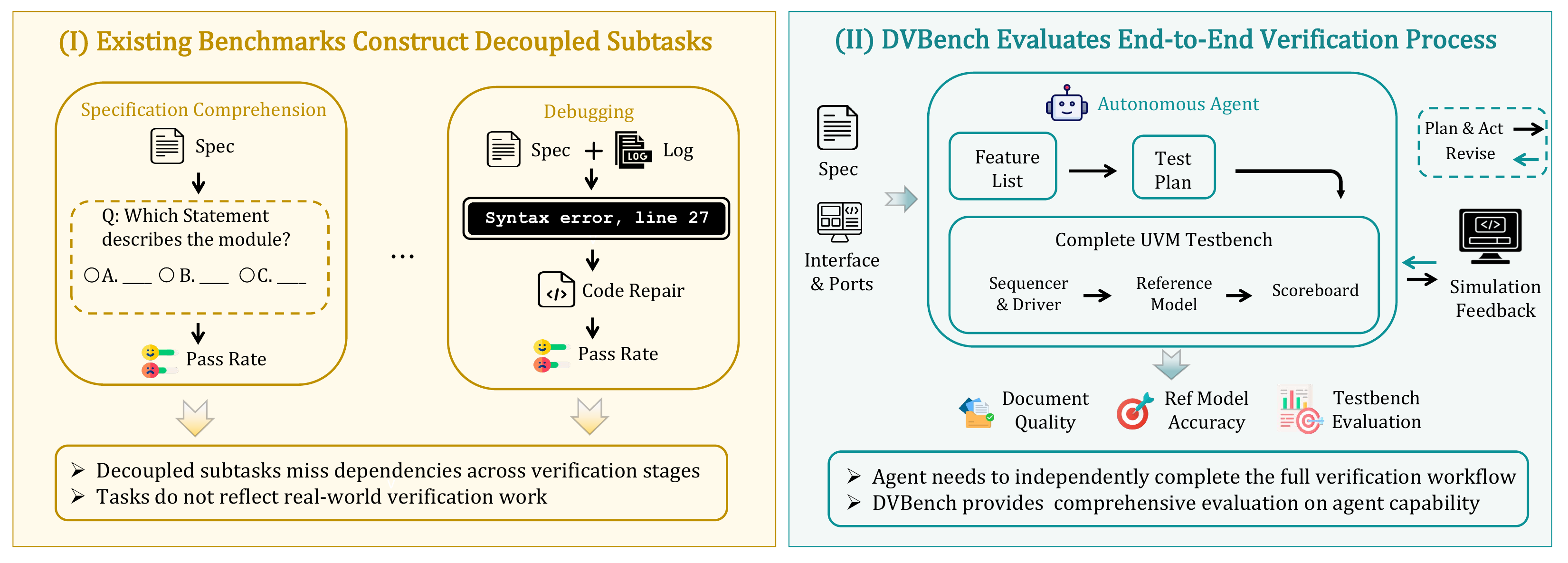}
\captionsetup{hypcap=false}
\captionof{figure}{\textbf{Motivation of DVBench.} (I) Existing benchmarks primarily construct mutually decoupled subtasks. Such setting overlooks the dependencies across verification stages and does not reflect the iterative nature of real-world verification. (II) DVBench provides comprehensive evaluation on the capability of the agent in real-world end-to-end verification process. The agent must independently complete the entire verification workflow.}
\label{fig: motivation}
\end{minipage}\par

\section{Introduction}
\label{sec:introduction}

Design verification (DV)~\cite{maropoulos2010design,lam2005hardware,mehta2018asic,pradhan2009practical,pinckney2025comprehensive,deng2000constraint} is one of the most demanding and time-consuming stages of chip development~\cite{dangwal2020agile,khailany2020accelerating,kenji2001current}. Before fabrication, engineers must establish that a design behaves as specified. The Universal Verification Methodology (UVM)~\cite{height2012practical,fiergolski2017simulation,harshitha2021introduction,qamar2020comprehensive,hosny2022unified} provides a standardized framework for building reusable verification environments. In this framework, engineers must derive a verification strategy from the specification and write checks that can expose defects. Execution feedback then guides further development and debugging. As designs grow, checks must account for longer transaction histories and interactions across components. Completing this iterative workflow demands substantial expertise and engineering time. 


Large language models (LLMs)~\cite{naveed2025comprehensive,minaee2024large,zhao2026survey,chang2024survey,shao2024survey} and tool-using agents~\cite{huang2024understanding,dong2025survey,masterman2024landscape} offer a path to automating this work. Agents can build UVM environments and refine them using simulation feedback~\cite{ye2025concept,ye2026uvmarvel,meng2026haven,lin2026autouvm,wang2026ucagent}. Heretofore, existing benchmarks~\cite{wan2026fixme,pinckney2025comprehensive,murthy2025verifllmbench,qiu2024autobench,kang2025fveval} primarily evaluate individual verification capabilities through mutually decoupled subtasks. Their task designs also largely target standalone LLMs rather than autonomous agents. As illustrated in Figure~\ref{fig: motivation}, such evaluations overlook the dependencies between verification stages. In practical workflows, decisions made during requirement analysis and verification planning directly shape subsequent modeling and testbench construction. Simulation results may further expose inconsistencies or uncovered behaviors, requiring earlier decisions to be revisited. Verification is therefore an iterative process rather than a sequence of independent tasks. Its intermediate artifacts and execution feedback jointly shape subsequent development. Autonomous UVM construction requires agents to reason across stage boundaries and adapt their actions based on evolving execution results. Assessing these capabilities calls for complete verification assignments that preserve inter-stage dependencies, together with fine-grained evaluation of intermediate artifacts and final verification environments.



To address these gaps, we present \textbf{DVBench}, the first end-to-end benchmark for evaluating agents in real-world UVM verification environment construction. DVBench comprises \textbf{\DVBenchTaskCount{} tasks} across \textbf{\DVBenchDomainCount{} functional domains} and \textbf{5 difficulty levels}, ranging from basic logic to integrated subsystems. Each task provides a  specification and the design's native interface, while keeping the implementation RTL hidden as a blackbox. We provide an EDA environment equipped with a Hardware Description Language (HDL) simulator to support compilation and simulation. Agents must independently complete the entire verification task.
This setup preserves dependencies across stages and evaluates the agent's ability to coordinate the full workflow.
Further, DVBench introduces \textbf{fine-grained assessment} grounded in practical verification requirements. It assesses the complete testbench via compilation, coverage and bug detection. Specifically, DVBench also evaluate the accuracy of the reference model within the testbench. Beyond the final testbench, DVBench also assesses the intermediate verification artifacts including feature lists and test plans for insight into the agent's reasoning and specification comprehension. Together, these assessments examine both the quality of intermediate verification artifacts and the effectiveness of the final testbench.

We evaluate combinations of frontier LLMs and agent frameworks, including GPT-6 Astra, GPT-6.1 Sol, Claude Opus 5.5, DeepSeek V4.1 Flash 1017 and GLM 5.3 on
Claude Code~\cite{anthropic_claude_code}, Codex~\cite{openai_codex}, and OpenCode~\cite{anomaly_opencode}. Experimental results show that Codex powered by GPT-6 Astra achieves the highest overall success rate of 21.05\%, while Claude Code with Claude Opus 5.5 achieves 16.84\%. Further analysis reveals that high coverage does not necessarily translate into effective bug detection. We also find that failures in end-to-end verification tasks often originate from incomplete feature analysis and test planning at the initial stages, whose shortcomings propagate to subsequent verification stages. End-to-end real-world UVM construction remains a challenging task for current frontier LLM-based agents.  In conclusion, our contributions are three-fold:
\begin{itemize}
    \item We propose the first real-world end-to-end UVM verification benchmark comprising \DVBenchTaskCount{} tasks across \DVBenchDomainCount{} functional domains and 5 difficulty levels, assessing agents' ability to independently construct complete verification environments.
    \item We conduct fine-grained assessment for the verification workflow. We evaluate the final testbench while separately assessing the accuracy of the reference model. We also assess intermediate artifacts to provide insight into agents' reasoning process.
    \item We conduct sufficient experiments on frontier LLM-based agents. Results reveal the strong association between early-stage document quality and overall success. Moreover, bug detection remains a key challenge in autonomous UVM construction.
\end{itemize}

\section{Related Work}
\label{sec:related-work}

\subsection{Benchmarks for Hardware Verification}
Existing benchmarks assess hardware verification at different task boundaries. FIXME~\cite{wan2026fixme} covers the verification workflow through five mutually decoupled task subsets. CVDP~\cite{pinckney2025comprehensive} includes both model-level and agentic RTL design and verification tasks. VerifLLMBench~\cite{murthy2025verifllmbench} evaluates UVM component generation within predefined harnesses. AutoBench's AutoEval~\cite{qiu2024autobench} evaluates generated testbenches through compilation, golden-design acceptance, and fault detection. Meanwhile, specialized evaluations address specification understanding and test planning~\cite{huang2026specread,kochar2026grpo}, assertion quality~\cite{wu2026assertllm2,nie2026hiersva,kang2025fveval}, and execution-aware coverage~\cite{zhang2026llm4cov}. Repository-level benchmarks also assess agents on hardware issue resolution~\cite{cui2026hwe,zou2026agentic}. DVBench evaluates autonomous construction of a complete UVM testbench while preserving dependencies across verification stages. Joint assessment of verification documents, the reference model, and overall testbench connects full-workflow capability to verification quality.

\subsection{Automated UVM Environment Construction}
UVM automation has progressed from assistance within existing environments to integrated testbench construction. UVLLM~\cite{hu2025uvllm} combines LLM-generated reference models with a pre-built UVM testbench to guide RTL repair. UVM$^2$~\cite{ye2025concept} coordinates specialized agents and reusable templates for planning, testbench generation, and coverage-driven refinement. UVMarvel~\cite{ye2026uvmarvel} constructs subsystem-level environments using a human-provided intermediate representation and a bus protocol library. HAVEN~\cite{meng2026haven} combines LLM-produced architectural plans with protocol templates and a sequence language to construct and refine UVM testbenches. These methods rely on specialized workflows and expert-designed resources. In this paper, DVBench evaluates general agents' ability to independently construct complete UVM environments.


\section{DVBench Construction}

\subsection{Overview}
In this section, we describe the construction process of DVBench. We first introduce the details of DVBench data collection and curation. We also clarify the structure and components of each task in DVBench. The overall process involved 19 experts over 3 months.

\label{sec:tasks}

\begin{figure}[t]
\centering
\includegraphics[width=0.72\linewidth]{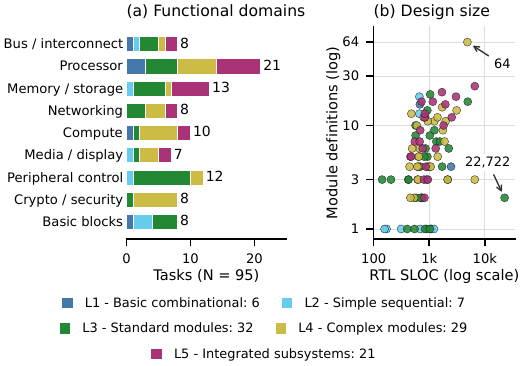}
\caption{DVBench diversity across \DVBenchTaskCount{} tasks. (a) Primary functional domains, stacked by difficulty level. (b) Per-task RTL SLOC and source module definitions, including supplied dependencies; both axes use logarithmic scales. Arrows mark the maximum on each axis. The shared legend names each difficulty level and reports its task count.}
\label{fig:benchmark-profile}
\end{figure}

\subsection{Data Collection and Curation}
To ensure task diversity across functional domains, we conduct large-scale mining of GitHub, collecting approximately 51k repositories containing synthesizable RTL. The repositories span a broad spectrum of hardware designs, ranging from coursework and personal projects to industrial-scale open-source projects and academic research prototypes. Representative sources include MegaBOOM, soDLA, an NVDLA-based deep learning accelerator.

We select high-complexity designs based on their IP categories and structural characteristics. For large repositories containing multiple IPs, agents identify independent IP units and subsystem-level modules. To establish reliable evaluation references, human experts complete the specifications, feature lists, and test plans for each selected design, and develop the corresponding UVM testbenches. A task is included in DVBench only if its expert-developed testbench compiles successfully and achieves over 90\% in all coverage metrics. Since a substantial portion of each task is newly developed by human experts, the risk of data contamination is substantially reduced.

Guided by real-world industrial verification practices, we classify tasks into five difficulty levels according to RTL complexity. L1 covers combinational behavior; L2 adds short state histories; L3 covers module protocols and boundaries; L4 targets complex algorithms and multistage transactions; and L5 captures coupled subsystem roles.
As illustrated in Figure~\ref{fig:benchmark-profile}, the resulting benchmark contains \DVBenchTaskCount{} tasks spanning \DVBenchDomainCount{} primary functional domains, with tasks represented across all five difficulty levels. The RTL bundles range from \DVBenchSlocMinimum{} to \DVBenchSlocMaximum{} SLOC and contain \DVBenchModuleMinimum{} to \DVBenchModuleMaximum{} module definitions. These counts include supplied dependencies, with paths deduplicated within each task.

\subsection{Task Construction Workflow}

\subsubsection{\textbf{Feature Lists and Test Plans}}
Feature Lists and Test Plans are the primary artifacts in real-world UVM testbench construction. Feature lists enumerate the observable behaviors and properties of the design, while test plans describe the scenarios, inputs, and expected outputs to verify those features. We adopt LLM-as-judge to evaluate the completeness and correctness of feature lists and test plans.
To construct reliable evaluation references, human experts write oracle feature lists, test plans, and rubrics for all \DVBenchTaskCount{} tasks.
For each task, human experts manually review and refine the oracle feature list, test plan, and corresponding evaluation rubrics. To validate the rubrics, experts prepare positive and negative examples and use them to test the LLM judge. Positive examples include correct descriptions and equivalent expressions, while negative examples contain omissions or incorrect verification requirements. Based on the judge's results, experts iteratively refine the rubrics and repeat the evaluation until the judge can reliably distinguish correct and incorrect submissions.

\subsubsection{\textbf{Reference Models}}
We construct hidden replay tests from golden RTL and its behavioral contract with fixed seeds $S=\{1,7,42\}$. Directed reset, recovery, boundary, and protocol scenarios complement seeded stimuli. Legal input histories and valid output observation windows define the comparison points.



\subsubsection{\textbf{Final Testbenches}}
We prepare a standardized VCS/UVM environment with fixed dependencies, compilation options, parameters, and entry points, together with Design Under Test (DUT) coverage instrumentation. A complete reference testbench, equipped with an independent input-history model, is first qualified against the golden RTL across three random seeds.

To assess defect detection, we design 10 RTL mutants specifically for each task. Each targets a specification requirement by linking a plausible implementation mistake to a legal trigger and an observable output difference. This requires reasoning about each design's protocol, state history, and numerical boundaries. For example, a FIFO mutant may suppress repeated data, requiring duplicate pushes to expose the error, while an AES-CTR mutant may skip counter initialization when a key is reused, requiring a second operation with a new IV. Experts manually review all mutants against the specification, repairing unsuitable cases and adding missing fault scenarios. Only mutants that compile under the native configuration and exhibit observable differences under legal inputs are retained. The final testbenches generated by agents are evaluated against the RTL mutants to assess their bug detection capabilities.

Appendix~\ref{app:mutants} makes this design process concrete through ten released mutant diffs. For each example, it connects the implementation change to a legal trigger, an observable output difference, and the checking obligation needed to distinguish it. The cases span bit boundaries, numeric interpretation, temporal sampling, transaction ownership, and cross-operation state, illustrating why task-specific mutation requires detailed hardware reasoning.

\subsection{Task Structure}
\subsubsection{\textbf{Public task}}
In each task, the agent receives an instruction document specifying the verification objectives and submission requirements. The agent is also provided with the design specification and the DUT's native interface, including its ports and parameters, as well as guidelines for artifact submission.

\subsubsection{\textbf{Hidden assessment assets}}
The golden RTL and its dependencies remain hidden throughout the task. Oracle feature lists, test plans, scoring rubrics, and mutant designs are also withheld from the agent. To support debugging without revealing these assets, DVBench provides a black-box simulation service through which the agent can interact with the golden design. The service returns bounded compilation diagnostics and simulation feedback, and exposes waveforms for public DUT interfaces and agent-owned signals. This allows the agent to refine its artifacts based on execution results without accessing the underlying RTL or hidden evaluation references.

\subsection{Agent Rollout Procedure}
Each task starts with an empty submission. The agent must independently produce a feature list and test plan in JSON and corresponding Markdown formats, together with a complete UVM testbench covering stimulus generation, monitoring, and checking. We provide a standard EDA environment to support compilation and simulation. During execution, the agent can use the debugging service to iteratively refine its submission, but receives no metric-related evaluation feedback. All submitted artifacts are assessed only after the final submission. This separation prevents agents from directly optimizing against benchmark scores and enables a more reliable evaluation of their ability to complete the entire verification workflow.
\section{DVBench Evaluation Protocol}
\label{sec:artifact-evaluation}

\subsection{Overview}

\begin{figure*}
    \centering
    \includegraphics[width=0.95\linewidth]{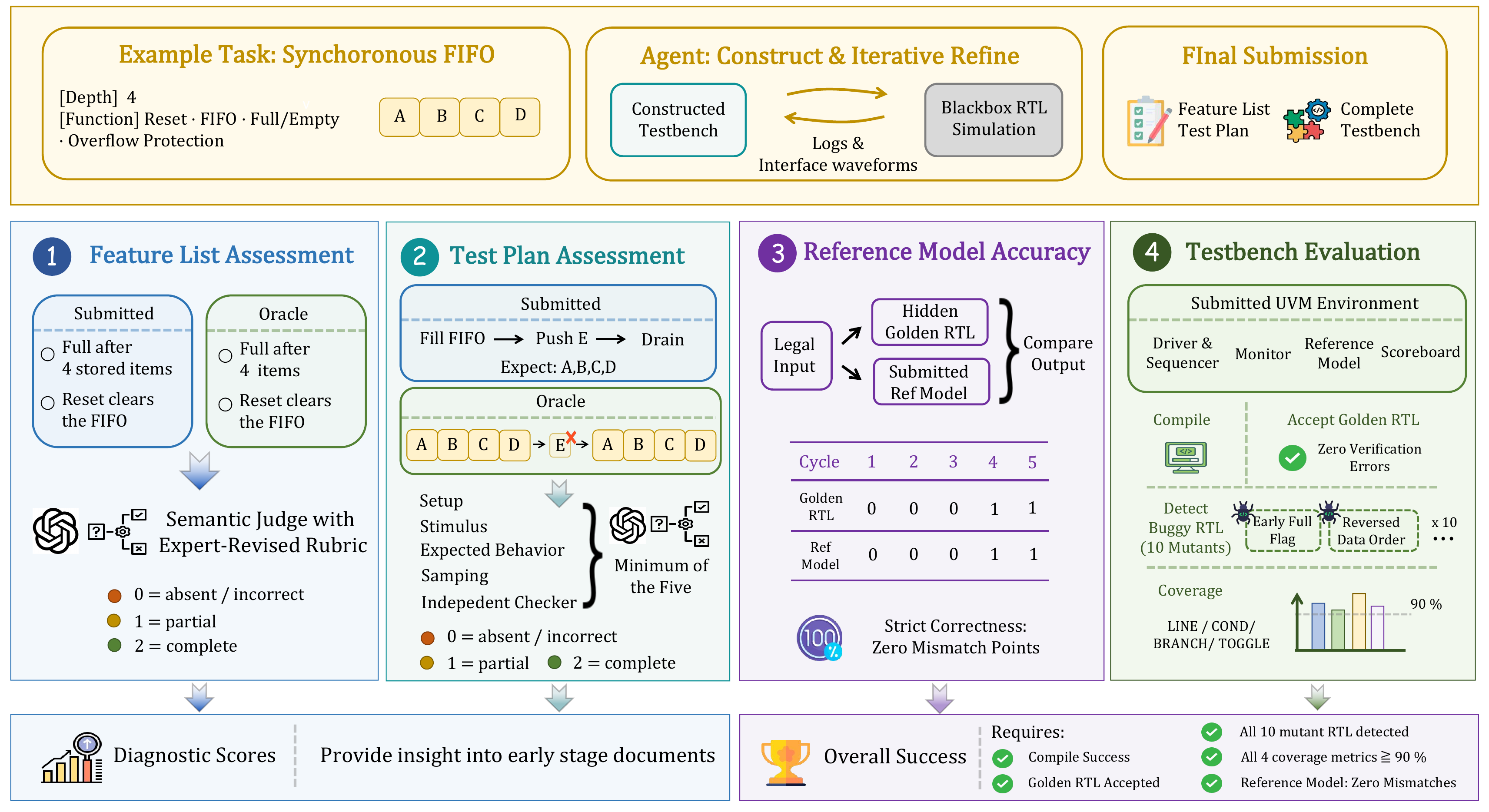}
    \caption{\textbf{DVBench evaluation protocol.} We take synchronous FIFO as an example. The final submission undergoes feature list assessment, test plan assessment, reference model accuracy checks, and complete testbench evaluation. Diagnostic scores provide  insights into agentic capabilities at earlier stages of the verification workflow. Overall success requires compilation, golden-RTL acceptance, mutant detection, sufficient DUT coverage, and strict reference model correctness. Detailed evaluation criteria are described from Section~\ref{sec:feature-list-assessment} to Section~\ref{sec:testbench-evaluation}.}
    \label{fig:artifact-evaluation-framework}
\end{figure*}

DVBench provides a comprehensive and fine-grained assessment of agents' design verification capabilities. For each final submission, DVBench first assesses the generated feature list (Section~\ref{sec:feature-list-assessment}) and test plan (Section~\ref{sec: testplan assessment}) to determine whether they accurately and comprehensively capture the functionality and verification requirements. Since the reference model is a critical component in UVM framework, its correctness is then assessed separately (Section~\ref{sec: reference model acc}) by checking whether its predictions strictly match the golden RTL outputs during simulation. Finally, DVBench
compiles the generated testbench and runs it against the golden RTL to verify its ability to correctly validate the golden design (Section~\ref{sec:testbench-evaluation}). The testbench is then evaluated against 10 buggy mutant designs to assess its effectiveness in detecting injected defects. DVBench considers the quality of artifacts produced at each stage to provide a comprehensive assessment of the agent's capabilities and determine whether a task achieves overall success. By combining end-to-end evaluation in a realistic design verification environment with fine-grained artifact assessment, DVBench provides a holistic view of agent performance.


\subsection{Feature List Assessment}
\label{sec:feature-list-assessment}
We assess whether the submitted feature list correctly describes the DUT's functionality and covers all required behaviors. Let $\mathcal{Q}$ denote the set of unique claims extracted from the submitted feature list, and $\mathcal{O}$ the set of required obligations in oracle feature list. A semantic judge compares the submitted claims against these obligations with rubric. Then the judge assigns each claim and obligation a score in $\{0,1,2\}$, representing absent or incorrect, partially supported, and fully supported, respectively. Specifically, $s_q$ measures how well a submitted claim is supported, while $g_o$ measures how completely a required obligation is covered.
Based on these scores, precision, recall, and F1 are calculated as follows:
\begin{equation}
    \begin{gathered}
P=\frac{\sum_{q\in\mathcal{Q}}s_q/2}{|\mathcal{Q}|},\qquad
R=\frac{\sum_{o\in\mathcal{O}}g_o/2}{|\mathcal{O}|},\qquad
\text{F1}=\frac{2PR}{P+R}.
\end{gathered}
\label{eq:semantic-scores}
\end{equation}
Precision reflects the correctness of the submitted claims, whereas recall measures how much of the required functionality is covered. F1 is set to zero when $P+R=0$. The assessment allows equivalent names, split or merged descriptions, and facts expressed jointly across multiple claims. However, contradictory claims about the same condition are treated as errors and affect the scores of the corresponding obligations. The submitted JSON file serves as the primary source for assessment, while the Markdown version must remain consistent with it. Deterministic checks are also applied to validate references and score aggregation.

\subsection{Test Plan Assessment}
\label{sec: testplan assessment}
We assess whether the test plan provides a concrete way to exercise and independently verify each required behavior. For each obligation, the assessment considers five aspects: whether the test establishes the necessary setup and input history, provides appropriate stimulus, specifies the expected behavior, samples the DUT at the correct time, and includes a checker capable of detecting incorrect behavior. The obligation receives the lowest score among these five aspects:
\begin{equation}
 g_o^{\mathrm{plan}}=\min\{g_o^{\mathrm{setup}},g_o^{\mathrm{stim}},
 g_o^{\mathrm{expect}},g_o^{\mathrm{sample}},g_o^{\mathrm{check}}\}.
\label{eq:plan-grade}
\end{equation}
An obligation may be covered by multiple tests, which can jointly provide the required evidence. We apply the same precision, recall, and F1 calculation in Equation~\ref{eq:semantic-scores} to the plan, using its deduplicated behavioral and methodological claims and the fixed set of test obligations. The coverage score $g_o$ is replaced by $g_o^{\mathrm{plan}}$.
Importantly, plan assessment is based on the submitted feature list and test plan themselves. Missing features do not reduce the number of obligations used for scoring. A test that specifies the correct expected behavior is not sufficient on its own. Since it must also provide appropriate stimulus and an checker to verify the DUT's actual behavior.

\subsection{Reference Model Accuracy}
\label{sec: reference model acc}
As a critical component of the testbench, the reference model must accurately predict the DUT's behavior for the verification process to succeed. We therefore evaluate its accuracy separately by comparing its predictions against the outputs of the golden RTL. Specifically, the golden RTL and the submitted reference model are driven by the same input sequences, and their outputs are compared at valid observation points. The reference model makes predictions based only on the input sequences, reset signals, and input history, without relying on the DUT outputs. A reference model is considered strictly correct only if it produces zero mismatches across all required seeds.

\subsection{Testbench Evaluation}
\label{sec:testbench-evaluation}
We prepare a standardized VCS/UVM setup with fixed dependencies, compilation options, parameters, and entry points, and equip it with DUT coverage instrumentation. Under this setup, each submitted testbench is assessed in terms of its coverage, compilation, acceptance of the golden RTL, and ability to detect injected design defects.

\subsubsection{\textbf{DUT coverage}}
We measure DUT coverage using four metrics: line (LINE), condition (COND), branch (BRANCH), and toggle (TOGGLE) coverage. Coverage databases from all required seeds are merged before computing the covered-to-total ratio for each metric. Overall success requires at least 90\% in all four metrics.

\subsubsection{\textbf{Compilation}}
The submission must compile with the fixed DUT/UVM configuration.

\subsubsection{\textbf{Golden-RTL acceptance}}
The generated testbench must terminate normally with zero verification errors on the golden RTL across all required seeds.

\subsubsection{\textbf{Mutant detection}}
We run the generated testbench against ten buggy RTL mutants, each containing an injected design defect, and assess whether its checks expose these bugs. The mutant kill rate is the fraction of the ten mutants successfully detected. Compilation failures and timeouts do not count as detections. Overall success requires detecting all ten bugs.



\subsection{Overall Success}
An agent achieves overall success only if its submission satisfies all evaluation criteria. The testbench should compile successfully, correctly verify the golden RTL, detect all injected bugs across the mutant designs, and achieve at least 90\% in each of the four coverage metrics. In addition, the reference model must produce strictly correct predictions of the golden RTL outputs. The assessment of feature list and test plan does not contribute to the overall success criterion. Instead, they provide additional insights into the capability of agents at earlier stages of the verification workflow, helping identify weaknesses that may not be apparent from final testbench performance alone. In conclusion, the comprehensive evaluation enables both a holistic measurement of an agent's ability to complete the full verification workflow and a more detailed analysis of its capabilities at different stages.

\section{Experiments}
\label{sec:experiments}

\begin{table*}[!t]
\centering
\caption{Main results on 95 tasks. All values are percentages; higher is better. Bold and underlined values denote the best and second-best distinct values across all configurations in each column, including ties.}
\label{tab:main-results}
\begingroup
\setlength{\tabcolsep}{0.5pt}
\renewcommand{\arraystretch}{2.1}
\fontsize{8.5}{12}\selectfont
\definecolor{overallshade}{HTML}{E9F5FD}
\newcommand{\resulthead}[1]{{\fontsize{8.5}{10}\selectfont\makecell[c]{#1}}}
\newcommand{\resultgroup}[2]{\multicolumn{#1}{c}{\raisebox{6pt}{\textbf{#2}}}}
\newcolumntype{R}[1]{>{\centering\arraybackslash}m{#1}}
\begin{tabular*}{\textwidth}{@{\extracolsep{\fill}}m{45pt}>{\begin{minipage}[c][22pt][c]{70pt}\raggedright}l<{\end{minipage}}|>{\columncolor{overallshade}}R{30pt}|R{32pt}R{26pt}R{28pt}|R{32pt}R{28pt}R{30pt}|R{26pt}R{26pt}R{29pt}R{29pt}|R{29pt}R{26pt}@{}}
\toprule
\multicolumn{1}{c}{\multirow{2}{*}[2pt]{\textbf{Agent}}} & \multicolumn{1}{c}{\multirow{2}{*}[2pt]{\textbf{LLM}}} & \multicolumn{1}{c}{\multirow{2}{*}[2pt]{\makecell[c]{\textbf{Overall}\\\textbf{success}}}} & \resultgroup{3}{Task-level criteria} & \resultgroup{3}{Testbench} & \resultgroup{4}{Mean DUT coverage} & \resultgroup{2}{Mean F1} \\[-12pt]
\cmidrule(lr){4-6}\cmidrule(lr){7-9}\cmidrule(lr){10-13}\cmidrule(l){14-15}
 & \multicolumn{1}{c}{} & \multicolumn{1}{c}{} & \resulthead{Kill all\\10\\mutants} & \resulthead{Ref.\\model\\exact} & \resulthead{All cov.\\$\geq90\%$} & \resulthead{Compile} & \resulthead{Golden\\pass} & \resulthead{Mutant\\kill\\(mean)} & \resulthead{Line} & \resulthead{Cond.} & \resulthead{Branch} & \resulthead{Toggle} & \resulthead{Feature \\ List} & \resulthead{Test \\ Plan} \\
\midrule
\multirow{5}{45pt}{Claude Code} & GLM 5.3 & 4.21 & 8.42 & 42.11 & 14.74 & 71.58 & 41.05 & 32.00 & 59.77 & 54.57 & 54.90 & 49.46 & 48.52 & 14.90 \\
 & DeepSeek V4.1 Flash 1017 & 14.74 & 46.32 & 58.95 & 29.47 & \textbf{100.00} & \underline{96.84} & 83.47 & 91.34 & 85.46 & 86.67 & 80.73 & 71.61 & 30.01 \\
 & Claude Opus 5.5 & 16.84 & 43.16 & \textbf{71.58} & \underline{35.79} & \underline{98.95} & 93.68 & 80.63 & 90.82 & 85.72 & 86.91 & 83.37 & 78.50 & 42.95 \\
 & GPT-6.1 Sol & 17.89 & 47.37 & 68.42 & \textbf{36.84} & \underline{98.95} & 91.58 & 81.58 & \underline{91.49} & 86.04 & 87.39 & 82.92 & 94.11 & 71.43 \\
 & GPT-6 Astra & 18.95 & 51.58 & \underline{70.53} & \underline{35.79} & \textbf{100.00} & 93.68 & 84.32 & \textbf{92.72} & \textbf{87.06} & \textbf{88.70} & \textbf{83.60} & 92.49 & 70.37 \\
\midrule
\multirow{4}{45pt}{OpenCode} & GLM 5.3 & 5.26 & 8.42 & 15.79 & 10.53 & 14.74 & 12.63 & 11.47 & 13.35 & 13.06 & 13.12 & 12.65 & 15.34 & 5.01 \\
 & DeepSeek V4.1 Flash 1017 & 11.58 & 41.05 & 60.00 & 31.58 & \underline{98.95} & \textbf{97.89} & 84.11 & 90.62 & 84.59 & 86.02 & 80.37 & 82.62 & 36.00 \\
 & GPT-6.1 Sol & 15.79 & 41.05 & 68.42 & \textbf{36.84} & 97.89 & 90.53 & 79.16 & 90.41 & 85.28 & 86.49 & 82.75 & \textbf{95.17} & 77.71 \\
 & GPT-6 Astra & 17.89 & 51.58 & \underline{70.53} & \underline{35.79} & \underline{98.95} & 93.68 & 84.95 & 90.66 & 85.37 & 86.66 & 82.98 & 94.08 & \underline{77.91} \\
\midrule
\multirow{4}{45pt}{Codex} & GLM 5.3 & 2.11 & 7.37 & 24.21 & 13.68 & 83.16 & 36.84 & 23.79 & 45.07 & 38.96 & 41.12 & 34.07 & 49.18 & 20.00 \\
 & DeepSeek V4.1 Flash 1017 & 10.53 & 29.47 & 58.95 & 28.42 & \textbf{100.00} & \underline{96.84} & 78.11 & 89.07 & 81.85 & 84.07 & 75.58 & 77.88 & 34.49 \\
 & GPT-6.1 Sol & \underline{20.00} & \underline{52.63} & \underline{70.53} & \underline{35.79} & \textbf{100.00} & 94.74 & \underline{85.68} & 86.34 & 81.08 & 82.92 & 78.74 & \underline{94.40} & \textbf{80.31} \\
 & GPT-6 Astra & \textbf{21.05} & \textbf{55.79} & \underline{70.53} & \textbf{36.84} & \textbf{100.00} & 94.74 & \textbf{85.79} & 91.48 & \underline{86.30} & \underline{87.40} & \underline{83.52} & 93.95 & 76.95 \\
\bottomrule
\end{tabular*}
\endgroup
\end{table*}

\begin{figure}[!t]
\centering
\includegraphics[width=0.72\linewidth]{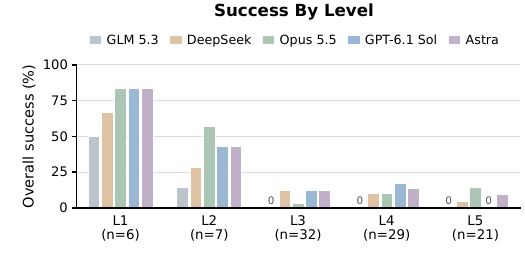}
\caption{Overall success by difficulty for five models using Claude Code. Each group follows the legend order; $n$ denotes the number of tasks.}
\label{fig:success-by-level}
\end{figure}

Our experiments establish the value of DVBench from four complementary perspectives: measuring complete verification capability, identifying the hardware reasoning that determines verification effectiveness, characterizing the performance--resource tradeoff, and assessing benchmark-specific data contamination. We organize the analysis around four research questions:
\textbf{RQ1:} Can agents independently complete the full UVM verification workflow, and what capabilities do they demonstrate?
\textbf{RQ2:} What are the major difficulties and sources of failure across this workflow?
\textbf{RQ3:} How do agents compare in efficiency and resource cost?
\textbf{RQ4:} Is there evidence of data contamination?
Together, these experiments connect the end-to-end outcome to the quality of the artifacts and decisions that produce it.

\subsection{Implementation Details}
\label{sec:implementation-details}
We run all agent frameworks in isolated server containers with access to the published task files, file-editing tools, and an HDL simulation service. Agents can iteratively inspect compilation and simulation reports and revise their submissions. The interaction budget is 300 agent turns, with a delivery reminder at turn 250. Request-level timeout or inactivity limits are set to 3,600\,s each turn. Per-response output budgets are set to 64,000 tokens. Evaluation uses seeds $\{1,7,42\}$ for sampling. 

\subsection{Main Results (RQ1)}
\label{sec:main-results}

We evaluate Claude Opus 5.5, GPT-6 Astra, GPT-6.1 Sol, GLM 5.3, and DeepSeek V4.1 Flash 1017 through 13 agent--model configurations on the same 95 tasks. Table~\ref{tab:main-results} reports both overall success and the component measurements. Overall success requires compilation, golden-RTL acceptance, detection of all ten mutants, exact reference-model predictions, and at least 90\% coverage in each of the four coverage metrics. The component measurements explain which capabilities a configuration demonstrates; their conjunction establishes whether those capabilities come together in a complete verification solution.

\noindent\textbf{End-to-end verification exposes a capability frontier beyond executable code.}
Codex with GPT-6 Astra achieves the highest overall success rate of 21.05\%, completing 20 tasks, followed by Codex with GPT-6.1 Sol at 20.00\% (19 tasks) and Claude Code with GPT-6 Astra at 18.95\% (18 tasks). Each successful task represents a complete verification environment whose stimulus generation, checking logic, reference model, and coverage jointly satisfy the evaluation requirements. DVBench therefore captures a concrete and demanding form of autonomous engineering: constructing a system that accepts correct hardware, rejects faulty hardware, and predicts behavior accurately across independently supplied histories.
The contrast between executable artifacts and end-to-end success is particularly revealing. Codex with GPT-6 Astra compiles on all 95 tasks and passes the golden RTL on 90, while 20 satisfy the complete verification contract. This separation makes DVBench informative even when compilation performance reaches saturation. A compilation benchmark would assign the configuration a perfect score; a golden-execution benchmark would report 94.74\%. DVBench distinguishes these achievements from the broader responsibility of verification, where a testbench must actively expose defects and support trustworthy checking across the design's behavioral space. The 21.05\% result thus establishes a clear frontier for autonomous verification and a substantial development target beyond code generation.

Figure~\ref{fig:success-by-level} further demonstrates that the benchmark spans meaningful verification demands. With Claude Code, GPT-6 Astra completes tasks at every difficulty level, including two L5 tasks, while the distribution of success varies across models and levels. The difficulty breakdown places these outcomes in context: agents must handle both local functional behavior and the longer histories, state transitions, and protocol interactions present in more involved designs. The complete success matrix in Appendix~\ref{app:results} makes these capabilities visible at task granularity.

\noindent\textbf{Runnable testbenches do not guarantee effective verification.}
DVBench's component measurements distinguish three complementary responsibilities: exercising the design, detecting incorrect behavior, and predicting correct behavior. Coverage measures how broadly stimulus reaches the implementation; mutation testing measures whether the stimulus and checker together distinguish plausible defects; independent reference-model replay measures whether the agent's behavioral understanding generalizes beyond its own generated tests. These responsibilities interact, but each supplies evidence that the others cannot replace.
For Codex with GPT-6 Astra, mean mutant detection reaches 85.79\%, while complete detection of all ten mutants is achieved on 55.79\% of tasks. The distinction is operationally meaningful: detecting many faults demonstrates broad checking capability, whereas detecting the entire set requires closing the remaining scenario-specific gaps. Similarly, 36.84\% of tasks meet all four coverage thresholds. Examining those 35 tasks directly, 25 detect all ten mutants, 29 have exact reference models, and 20 satisfy the full contract. Thus, even within the same high-coverage subset, fault detection and behavioral prediction add substantial discriminating power. Coverage establishes that the environment exercises the design; the additional checks establish what the environment can reliably conclude from that execution.

Independent replay is especially valuable because an agent develops its stimulus and reference model together. A shared simplification can make those artifacts agree on the agent's chosen scenarios without faithfully representing the specification. Among the 90 golden-passing tasks for Codex with GPT-6 Astra, 26 do not achieve exact reference-model predictions under independent evaluation. DVBench exposes this distinction by assessing the model on externally supplied histories. Its evaluation therefore tests both the verification environment's behavior on correct hardware and the correctness of the behavioral oracle that supports that environment. This complementary design is central to measuring verification competence rather than artifact plausibility.

\noindent\textbf{Verification documents provide the reasoning foundation for implementation.}
Feature lists and test plans make the transition from specification understanding to executable verification explicit. A feature list decomposes the required behavior into verifiable obligations; a test plan turns those obligations into stimulus conditions, observation points, and checking rules. Their value lies in carrying the specification's structure into the implementation, especially for behaviors that require carefully chosen histories or timing relationships.

The results show a strong association between document quality and end-to-end success. Across the 13 configuration means, Pearson correlations with overall success are 0.856 for feature-list F1 and 0.887 for test-plan F1. GPT-based configurations consistently achieve feature-list F1 above 92\% and test-plan F1 between 70.37\% and 80.31\%, alongside the strongest overall results within each framework. The relationship supports the original workflow rationale: a well-structured understanding of what to verify and how to verify it provides a sound basis for the subsequent construction process.

The distinction between feature lists and test plans also explains why both artifacts deserve evaluation. Recognizing a FIFO's full/empty behavior, for example, establishes a verification obligation; a concrete plan must additionally specify how to reach the boundary, which simultaneous operations to apply, and which cycle's outputs to check. Appendix~\ref{app:fifo} illustrates this progression with complete reference feature and test-plan records, while Appendix~\ref{app:advanced} extends it to width conversion and AXI interactions. By assessing documents and executable artifacts together, DVBench reveals how effectively an agent translates behavioral knowledge into discriminating tests. Document scores characterize the quality of that foundation, and execution metrics establish how successfully it is realized.

\noindent\textbf{Agent frameworks and LLM capabilities jointly determine verification performance.}
Claude Code achieves the highest mean overall success across its evaluated configurations. The same ordering holds when comparing the four models available in all three frameworks: Claude Code completes 53 of 380 task--configuration pairs (13.95\%), compared with 51 (13.42\%) for Codex and 48 (12.63\%) for OpenCode. This matched comparison highlights Claude Code's effectiveness across different underlying models. At the model level, the GPT family achieves the highest overall success within each framework, demonstrating strong end-to-end verification capability across execution environments.
Individual pairings reveal another useful dimension. Both GPT models achieve their best overall success in Codex: 19 tasks for GPT-6.1 Sol and 20 for GPT-6 Astra. DeepSeek V4.1 Flash 1017 completes 14 tasks with Claude Code, ten with Codex, and 11 with OpenCode. These results show why the agent--model configuration is the relevant unit of evaluation. The framework supplies the environment for tool use, feedback incorporation, and iterative revision; the model supplies the reasoning that interprets specifications and directs those actions. DVBench evaluates their combined ability to carry an engineering assignment through to a validated result, while the artifact-level measurements expose the capability profile behind each pairing.

\subsection{Workflow Difficulties and Failure Analysis (RQ2)}
\label{sec:failure-analysis}

\begin{figure}[htbp]
\centering
\includegraphics[width=0.65\linewidth]{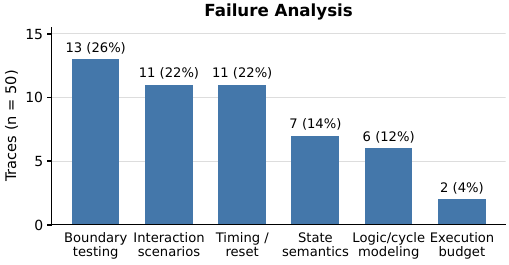}
\caption{Primary failure categories in 50 selected unsuccessful traces from Codex with GPT-6 Astra from an earlier evaluation snapshot. Labels show counts and percentages.}
\label{fig:failure-analysis}
\end{figure}

To identify the reasoning that determines verification effectiveness, we analyze 50 selected unsuccessful task traces from Codex with GPT-6 Astra retained from an earlier evaluation snapshot, assigning one primary diagnosis per trace. The analysis draws on the retained testbenches, reference models, evaluation diagnostics, and mutant differences. Figure~\ref{fig:failure-analysis} groups the cases into insufficient testing (70\%), reference-model defects (26\%), and simulation-budget issues (4\%). The resulting picture is actionable: effective verification depends on constructing observations that separate correct hardware from plausible alternatives and on modeling precisely what those observations mean.

\noindent\textbf{Boundary testing requires discriminating values, not just diverse values.}
Boundary scenarios account for 13 cases (26\% of the sample). Their common theme is the need to target the decision that determines correct behavior. A test can include many valid and invalid inputs yet leave a particular comparison or boundary condition indistinguishable from a faulty implementation. The useful question is whether the chosen values separate the intended predicate from a plausible incorrect one.

For example, in the network-filter task \texttt{dv-016}, negative EtherType tests do not isolate a change in the least significant bit of the accepted value \texttt{0x0800}. Likewise, ARP miss scenarios do not include targets that share the same low 16 bits while differing in the high 16 bits. Such paired inputs directly distinguish full-field comparisons from partial-field comparisons. Their value comes from the controlled relationship between inputs, rather than the size of the input set. DVBench's mutation assessment makes this distinction concrete: surviving mutants identify the behavioral boundary that the existing tests have not separated.

\noindent\textbf{Transaction interactions require controlled changes across histories.}
Interaction scenarios account for 11 cases (22\%). Stateful hardware often depends on whether a field changes, remains stable, or is accepted at a particular handshake. Consequently, independent transaction coverage does not automatically establish that the relationships between transactions have been exercised.

The AXI task \texttt{dv-080} illustrates this issue. Write data varies with the beat index and read-response codes change between beats, leaving scenarios such as equal write data with different write strobes, or equal response codes with different read data, untested. Those controlled combinations are precisely what can expose stale-field or caching behavior. In \texttt{dv-095}, address and write-data payloads remain held after their respective handshakes until the write response. Under that stimulus, an implementation that incorrectly reads live payload signals can appear equivalent to one that correctly captures them at acceptance. Legally changing the payload after the accepted handshake distinguishes the two behaviors. These cases identify an important verification skill: reasoning about which signals must vary independently, and when that variation becomes observable.

\noindent\textbf{Timing is part of the functionality being verified.}
Timing and reset scenarios contribute another 11 cases (22\%). In synchronous hardware, correctness depends on both the value and the time at which it is sampled or updated. A stimulus sequence can exercise the right logical operation while making an incorrect sampling rule observationally equivalent to the intended one.

For the FIFO task \texttt{dv-009}, input perturbations occur during the falling or low clock phase, with data remaining stable after the write edge. This pattern masks a mutant that incorrectly permits transparent writes during the high phase. A legal between-edge perturbation, respecting the interface's timing requirements, would separate edge-triggered capture from high-phase transparency. In the UART task \texttt{dv-019}, an input held for 64 clocks makes current-cycle and previous-cycle sampling produce the same value. These examples show why adding more transactions of the same temporal pattern may not increase fault discrimination. Effective stimulus must explicitly create the timing contrast that makes an incorrect implementation observable.

\noindent\textbf{Reference models must capture observable hardware semantics.}
Reference-model defects account for 13 cases (26\%): seven concern initialization or four-state semantics, and six concern functional or cycle behavior. This grouping highlights the breadth of a hardware oracle's responsibility. It must represent startup state, reset effects, masked updates, and cycle-level transitions in addition to the eventual arithmetic or logical result. Four-state values and resolved signals are also part of the interface semantics whenever they affect observable behavior.

In \texttt{dv-017}, the candidate model uses an all-ones default SRAM write mask during idle behavior, which does not match the task's masked-write profile. The same submission detects all ten mutants, making the independent model assessment particularly informative: strong fault detection and exact behavioral prediction capture different achievements. The initialization cases further draw attention to distinctions such as an undriven or released signal versus a driven logical value. These are concrete modeling obligations, and DVBench's independent replay turns them into measurable outcomes. The evaluation thus connects a diagnosis to the artifact that needs improvement, rather than reducing every unsuccessful submission to an undifferentiated task score.

\noindent\textbf{Scenario design must also account for simulation cost.}
The remaining two cases (4\%) involve sequences exceeding the simulation budget. In \texttt{dv-079}, one test phase spans more than 94 million clock cycles, followed by additional million-cycle scenarios, and all three golden-evaluation seeds time out. This case emphasizes the value of directed scenario construction: a verification plan should reach the state that distinguishes the target behavior with an execution length appropriate to the simulation environment. Taken together, the analysis identifies three concrete development priorities for verification agents: discriminating stimulus, temporally faithful models, and economical access to the states that matter. DVBench supplies evidence at each of these levels, turning the benchmark into a diagnostic basis for improving autonomous verification.

\subsection{Token Usage and Runtime (RQ3)}
\label{sec:resource-usage}
\begin{table}[htbp]
\centering
\caption{Mean token usage and elapsed time per task over 95 tasks.}
\label{tab:resource-usage}
\begingroup
\setlength{\tabcolsep}{2pt}
\renewcommand{\arraystretch}{1.05}
\fontsize{9}{11}\selectfont
\begin{tabular*}{0.88\columnwidth}{@{\extracolsep{\fill}}llcc@{}}
\toprule
Agent & LLM & \makecell[c]{Mean tokens\\($10^6$)} & \makecell[c]{Mean time\\(min)} \\
\midrule
\multirow{5}{*}{Claude Code} & GLM 5.3 & 17.27 & 258.55 \\
 & DeepSeek V4.1 Flash 1017 & 22.51 & 37.96 \\
 & Claude Opus 5.5 & 2.37 & 32.83 \\
 & GPT-6.1 Sol & 1.36 & 21.75 \\
 & GPT-6 Astra & 1.08 & 12.43 \\
\midrule
\multirow{4}{*}{OpenCode} & GLM 5.3 & 1.46 & 566.61 \\
 & DeepSeek V4.1 Flash 1017 & 11.03 & 32.63 \\
 & GPT-6.1 Sol & 2.19 & 26.64 \\
 & GPT-6 Astra & 1.42 & 19.06 \\
\midrule
\multirow{4}{*}{Codex} & GLM 5.3 & 10.19 & 99.18 \\
 & DeepSeek V4.1 Flash 1017 & 14.77 & 25.07 \\
 & GPT-6.1 Sol & 2.94 & 30.96 \\
 & GPT-6 Astra & 1.48 & 16.48 \\
\bottomrule
\end{tabular*}
\endgroup
\end{table}

\noindent\textbf{Stronger verification performance can coincide with substantial resource savings.}
Table~\ref{tab:resource-usage} reports arithmetic means over the same 95 tasks for each configuration. GPT-6 Astra achieves higher overall success than GPT-6.1 Sol and shorter elapsed time in all three frameworks, alongside lower token usage in Claude Code and Codex. Within Codex, it completes one additional task while reducing mean token usage from 2.94 to 1.48 million (49.6\%) and mean runtime from 30.96 to 16.48 minutes (46.8\%). The improvement therefore appears simultaneously in the verification outcome and in the resources required to produce it.
The pattern extends across execution environments. With Claude Code, Astra reduces mean token usage by 20.1\% and runtime by 42.8\% relative to Sol while completing one additional task. With OpenCode, it completes two additional tasks and reduces runtime by 28.5\%; the recorded token mean is a lower bound because an interrupted request has incomplete usage accounting. The success and runtime improvements across all three frameworks are important for agent development: progress on the full verification contract can come from more effective use of the available interaction budget. DVBench makes that progress visible by measuring completion quality alongside the resources consumed across tasks.

\noindent\textbf{Different pairings offer useful performance--resource tradeoffs.}
Claude Code with GPT-6 Astra achieves the lowest mean token usage and runtime among all evaluated configurations, at 1.08 million tokens and 12.43 minutes per task, while completing 18 tasks. Codex with the same model reaches the highest overall success, completing 20 tasks at 1.48 million tokens and 16.48 minutes. These configurations represent distinct operating points: the former emphasizes lower average resource consumption with strong completion performance, while the latter delivers the largest number of fully validated solutions. Reporting both makes the benchmark useful for selecting an agent configuration according to deployment priorities, including turnaround time, resource availability, and the value of an additional completed verification assignment.

\noindent\textbf{Efficiency is meaningful when tied to validated outcomes.}
Token usage and runtime capture different aspects of an agent system. Token totals sum input and output tokens, including cached input, while elapsed time also reflects tool execution and orchestration. Their joint interpretation is therefore more informative than either measure in isolation. For example, OpenCode with GLM 5.3 records at least 1.46 million tokens per task and 566.61 minutes of mean elapsed time, completing five tasks. Its recorded token volume alone does not capture the turnaround required to deliver those outcomes. Relating these quantities to overall success distinguishes low recorded interaction volume from efficient delivery of a working verification environment. By reporting resources over all 95 tasks, DVBench characterizes the cost of taking on the full assignment portfolio, providing a practical complement to its capability measurements.

\subsection{Data Contamination (RQ4)}
\label{sec:data-contamination}

\noindent\textbf{Expert-authored specifications provide a distinct target for contamination assessment.}
DVBench requires agents to reason from task-specific behavioral contracts. We therefore assess whether models can recover the wording of those contracts using a TS-Guessing-style~\cite{deng2024investigating} probe. For 60 selected tasks, we mask two expert-authored specification spans per task and ask each model to reconstruct the missing text. The same 120 masked spans are used for each of four models, yielding 480 responses. Each prompt requests one best guess in an independent session. Answers are extracted from the returned responses and compared with the original spans using case-sensitive exact matching after trimming outer whitespace. This directly probes recovery of the benchmark's specification text. We report the results in Table~\ref{tab:contamination}.

\begin{table}[htbp]
\centering
\caption{Exact specification recovery on 60 tasks, with 120 answers per model. Recovered counts are measured after answer extraction. Lower recovery indicates lower benchmark-specific contamination risk under this probe.}
\label{tab:contamination}
\small
\begin{tabular}{lrr}
\toprule
Model & Recovered & Recovery rate \\
\midrule
GPT-6.1 Sol & 6/120 & 5.0\% \\
Claude Opus 5.5 & 3/120 & 2.5\% \\
GPT-6 Astra & 4/120 & 3.3\% \\
DeepSeek V4.1 Flash 1017 & 0/120 & 0.0\% \\
\bottomrule
\end{tabular}
\end{table}

\noindent\textbf{Consistently low recovery supports low contamination risk.}
Table~\ref{tab:contamination} shows exact recovery rates between 0\% and 5.0\% across all four models. GPT-6.1 Sol recovers six spans, Claude Opus 5.5 recovers three, GPT-6 Astra recovers four, and DeepSeek V4.1 Flash 1017 recovers none. The consistency across models indicates that benchmark-specific specification knowledge is not readily accessible under the tested prompts, supporting a low risk of substantial data contamination in the expert-authored specifications.

This result complements the main evaluation in a useful way. DVBench's successful agents must translate the supplied specification into feature decompositions, executable scenarios, and accurate behavioral predictions. The strongest overall configuration, Codex with GPT-6 Astra, combines its 21.05\% end-to-end success with a 3.3\% exact specification-recovery rate in the separate probe. These measurements address complementary questions: the main experiment establishes what agents can construct from the task contract, and the contamination probe assesses access to its original wording. Together, they support DVBench's role as an evaluation of specification-driven verification capability, with expert-authored contracts providing a concrete foundation for the assessment.

\subsection{Case Study: From Agent Decisions to Verification Evidence}
\label{sec:case-study}

DVBench makes long-horizon verification inspectable: an agent's resource trajectory can be connected to the concrete evidence it establishes. We select three complementary runs. The FIFO task (\texttt{dv-009}, Codex with GPT-6 Astra) shows a short refinement loop; the instruction buffer (\texttt{dv-038}) and fetch target queue (\texttt{dv-061}), both using Claude Code with DeepSeek V4.1 Flash 1017, show long-horizon success and failure, respectively. The latter two share an agent--model configuration but reach different final outcomes after 113 and 253 model decisions. Together, they reveal what sustained interaction accomplishes, and which engineering distinctions determine whether that interaction closes the verification task.

\begin{figure}[p]
\centering
\includegraphics[width=\linewidth]{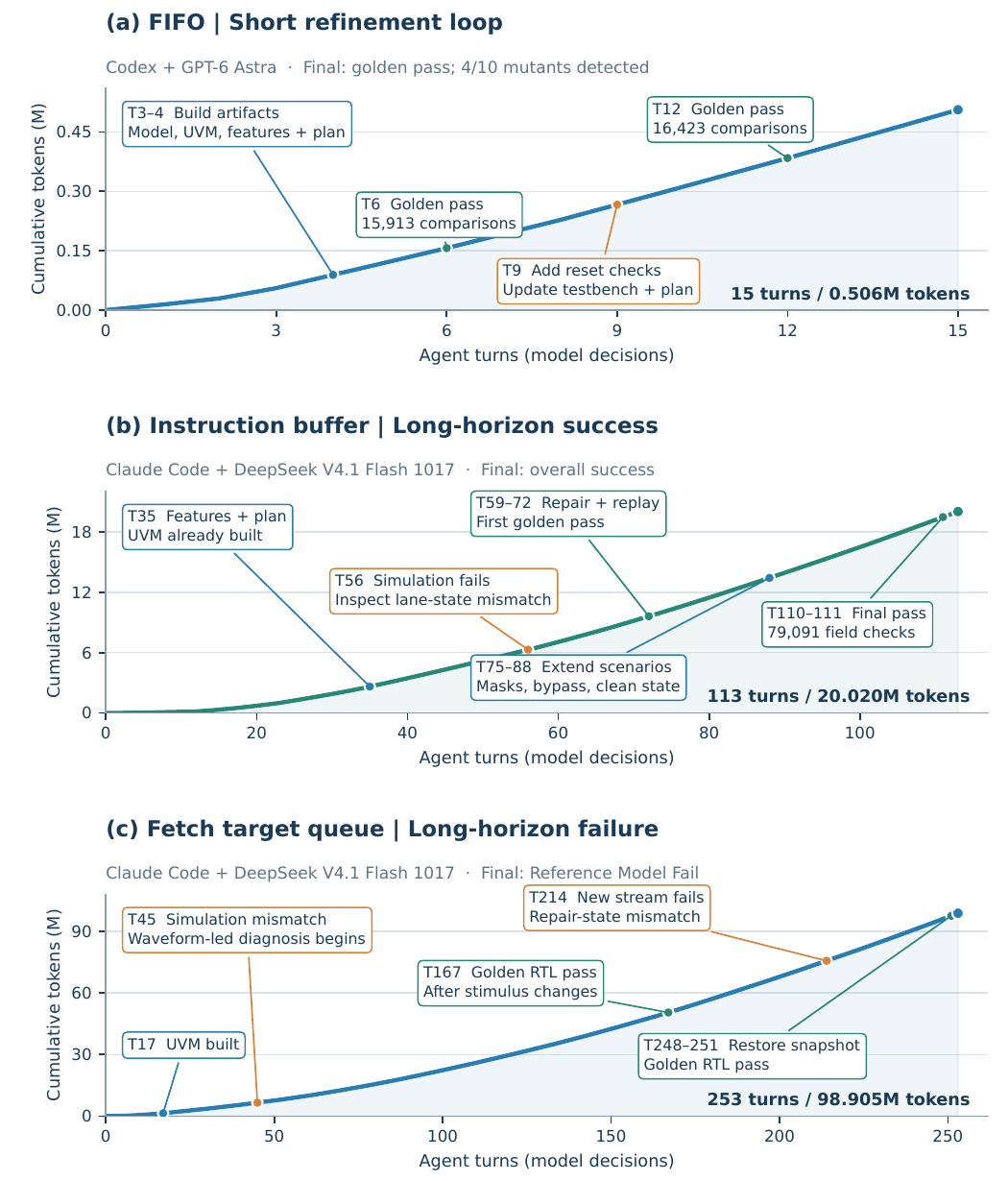}
\caption{\textbf{Three trajectories distinguish short refinement, long-horizon convergence, and long-horizon non-convergence.} Curves join measured cumulative token counts; markers identify engineering milestones. All panels use millions of tokens (M), with independent axis ranges. Golden RTL passes denote successful execution of the agent's testbench against golden RTL during development, using seed 1; final outcomes in subtitles come from frozen-submission assessment. The fetch target queue achieves a Golden RTL pass after restoring a snapshot, but its reference model fails independent evaluation. Hidden outcomes are not feedback available at the plotted turns.}
\label{fig:case-trajectories}
\end{figure}

Figure~\ref{fig:case-trajectories} plots cumulative input-plus-output tokens against completed model decisions, with tool actions and observed feedback attached to the corresponding turns. Cached input is included, consistently with Table~\ref{tab:resource-usage}; cumulative usage counts repeated context processing rather than unique generated text. Each turn is one completed model response, matched to an actual usage receipt, and may issue multiple tool calls. This common definition avoids conflating framework-specific CLI counters with model decisions. The panels use independent axis ranges to retain the short FIFO trace alongside the two substantially longer runs.

\noindent\textbf{A productive refinement loop can preserve a stimulus blind spot.}
In the FIFO trajectory, the agent creates the model and UVM sources at turn 3, completes the paired feature/test-plan documents at turn 4, and receives a golden pass at turn 6. The simulation log reports 15,913 comparisons. After inspecting the logs, the agent adds reset-held clock-edge checks at turn 9 and updates the test plan to match the implementation. A second golden pass arrives at turn 12, with 16,423 observations and zero UVM errors or fatals. Final delivery at turn 15 consumes 506,347 tokens. This is a substantive engineering loop: executable construction, inspection of real feedback, and a targeted extension of the verification scenario set.
 
DVBench evaluates more than this successful interaction with golden RTL. The FIFO driver changes inputs at the falling edge and performs its extra data perturbation during the low clock phase. After the rising edge, it samples the outputs while the write data remains stable. A mutant that incorrectly allows transparent writes during the high phase is therefore not distinguished by these perturbations. Increasing the number of comparisons, or adding another reset scenario, does not establish this particular timing contrast. Final assessment records golden acceptance, exact independent-model predictions, and satisfaction of all four coverage thresholds, alongside detection of four of ten mutants. The case demonstrates DVBench's diagnostic value: the trace shows that the agent can build and refine a coherent environment, while mutation assessment identifies the behavioral distinctions that this environment has yet to establish.

\noindent\textbf{Long-horizon success accumulates semantic corrections and stronger scenarios.}
The instruction buffer requires the agent to coordinate eight banks, paired instruction lanes, pending queues, readiness, flushes, and retained payloads. It builds the model and UVM environment before completing the feature and test-plan documents at turn 35. Following interface and compilation fixes, simulation feedback at turn 56 exposes mismatches in lane validity and payloads. The agent inspects the earliest discrepant observations, repairs the lane-retirement mask at turn 59, and develops an offline replay of the interface waveform. Crucially, the replay distinguishes pre-edge eligibility from updated state: retiring an active pair must not automatically permit a pending pair to load on that same edge. With the replay aligned to that ordering, turn 69 reports zero waveform mismatches, and turn 72 observes the first golden pass.

The subsequent work expands the verification argument. Turns 75--88 add explicit empty-queue bypass checks, flush/enqueue coincidences, zero-mask payload scenarios, and clean-state preparation for directed sequences. These additions make individual semantic obligations observable under controlled histories. The final debug pass at turn 110 is followed by a log and waveform audit at turn 111: 1,673 sampled cycles produce 79,091 field checks with zero mismatches. The run ends at turn 113 with 20,019,893 cumulative tokens. Frozen assessment confirms overall success, exact independent-model replay, detection of every evaluated mutant, and line/condition/branch/toggle coverage of 100/97.72/100/99.64\%. The distinctive achievement is the alignment of state-transition reasoning, independent prediction, and scenario construction. Extended interaction turns local diagnostics into a verification environment that satisfies the benchmark's final checks.

\noindent\textbf{Maintaining global model consistency remains a challenge during long-horizon debugging.}
The fetch target queue illustrates the difficulty of resolving local mismatches while maintaining correct reference-model behavior across interacting features. Within the same agent--model configuration as the successful instruction-buffer run, the agent builds its UVM package by turn 17 and inspects simulation mismatches spanning queue indices, history fields, and branch-predictor updates at turn 45. Much of the subsequent interaction concentrates on repair-mode timing, update suppression, stop PCs, and pointer capture. These behaviors share state and depend on input history, so a correction that explains one observed mismatch must also remain consistent with transitions elsewhere in the model. The trajectory shows that sustained work on a local discrepancy does not by itself establish that broader consistency.

A Golden RTL pass at turn 167 follows stimulus changes that remove redirect/repair traffic from one extended segment. Here, a Golden RTL pass means that the agent's testbench completes execution against golden RTL without reported UVM errors or fatals; it does not denote independent reference-model acceptance. Later iterations revisit reset during repair and obtain further passes, yet replacing the random stream at turn 212 reopens the problem: simulation at turn 214 reports 13 UVM errors involving predictor-update validity and repair-update flags. Additional model changes and PC diversification produce further failing checks at turns 234 and 239. The agent eventually restores an earlier snapshot at turn 248, disables branch-mispredict traffic in the post-reset window at turn 250, and obtains another Golden RTL pass at turn 251. After 253 turns and 98,905,394 cumulative tokens, the frozen reference model still fails independent evaluation, with 1,888 differing points out of 24,561.

The recurring mismatches under changed stimulus reveal a persistent challenge for long-horizon agents: resolving local defects while maintaining globally consistent reference-model behavior across interacting features. In this run, repeated attention to repair-state discrepancies produces locally passing configurations without establishing correctness across the broader set of input histories. The contrast with the instruction-buffer case is concrete: that run combines semantic corrections with additional distinguishing scenarios, whereas the fetch-target-queue run repeatedly revisits repair behavior and ultimately returns to an earlier snapshot. DVBench makes this difference measurable through independent model evaluation, exposing whether extended debugging preserves the correctness of the model as a whole. Appendix~\ref{app:uvm} exposes the FIFO implementation and a complementary width-converter environment; Appendix~\ref{app:mutants} shows how targeted mutants encode the semantic distinctions that a successful environment must establish.

\section{Conclusion}
We propose DVBench, an end-to-end benchmark for real-world UVM verification environment construction, comprising 95 tasks across 9 functional domains and 5 difficulty levels. By jointly assessing verification documents, reference models, and executable testbenches, DVBench connects verification intent with demonstrated checking effectiveness. We conduct experiments across 13 agent–model configurations. Results reveal a substantial gap between runnable testbenches and complete verification success, with the strongest configuration achieving 21.05\%.  DVBench provides a foundation for evaluating and improving agents that independently translate specifications into effective verification environments.

\par\bigskip
\begingroup
\fontsize{9}{11}\selectfont
\bibliographystyle{IEEEtran}
\bibliography{references}
\endgroup
\clearpage
\appendix
\begin{AppendixCard}{A guide to the supplementary materials}
\textbf{Appendix~\ref{app:results}:} overall success and resource usage, success counts by evaluation criterion, difficulty/domain results, and task-level outcomes.\par\smallskip
\textbf{Appendix~\ref{app:instructions}:} task inputs, submission schema, the complete instantiated instruction for a representative task, and feedback/version details.\par\smallskip
\textbf{Appendix~\ref{app:fifo}:} a complete four-feature/four-test FIFO reference example with requirement traceability.\par\smallskip
\textbf{Appendix~\ref{app:advanced}:} selected width-conversion and AXI examples that expose parameter, timing, and transaction-history obligations.\par\smallskip
\textbf{Appendix~\ref{app:uvm}:} generated UVM architectures, frozen source excerpts, and their connection to the recorded agent trajectories.\par\smallskip
\textbf{Appendix~\ref{app:mutants}:} ten real mutant diffs with legal triggers, observation windows, and design insights.\par\smallskip
\textbf{Appendix~\ref{app:catalog}:} the 95-task catalog. \textbf{Appendix~\ref{app:measurements}:} all 1,235 per-task measurement records, grouped by configuration.
\end{AppendixCard}

\section{Detailed Experimental Results}
\label{app:results}

\subsection{Overall Success and Resource Usage}
We evaluate 13 agent--model configurations on 95 tasks each. Codex with GPT-6 Astra achieves the highest overall success, completing 20 tasks (21.05\%). Claude Code with GPT-6 Astra completes 18 tasks (18.95\%) with the lowest mean resource usage: 1.08 million tokens and 12.43 minutes per task. Token and runtime averages include all 95 tasks for each configuration.

\begingroup
\fontsize{8.8}{10.8}\selectfont
\setlength{\tabcolsep}{3pt}\renewcommand{\arraystretch}{1.04}
\setlength{\LTleft}{0pt}\setlength{\LTright}{0pt}
\Needspace{10\baselineskip}
\captionsetup{hypcap=false}
\captionof{table}{Overall success and mean resource usage across 95 tasks.}\label{tab:app-configs}
\addtocounter{table}{-1}
\begin{longtable}{>{\raggedright\arraybackslash}p{\dimexpr 0.0550000\textwidth-2\tabcolsep\relax}>{\raggedright\arraybackslash}p{\dimexpr 0.1500000\textwidth-2\tabcolsep\relax}>{\raggedright\arraybackslash}p{\dimexpr 0.2850000\textwidth-2\tabcolsep\relax}>{\raggedleft\arraybackslash}p{\dimexpr 0.0900000\textwidth-2\tabcolsep\relax}>{\raggedleft\arraybackslash}p{\dimexpr 0.1100000\textwidth-2\tabcolsep\relax}>{\raggedleft\arraybackslash}p{\dimexpr 0.1550000\textwidth-2\tabcolsep\relax}>{\raggedleft\arraybackslash}p{\dimexpr 0.1550000\textwidth-2\tabcolsep\relax}}
\toprule
\rowcolor{ReleasePale}\textbf{ID} & \textbf{Framework} & \textbf{Model} & \textbf{Solved} & \textbf{Rate (\%)} & \textbf{Tokens (M)} & \textbf{Time (min)} \\
\midrule\endfirsthead
\multicolumn{7}{l}{\small\color{ReleaseMuted}Table~\ref{tab:app-configs} (continued)}\\
\toprule
\rowcolor{ReleasePale}\textbf{ID} & \textbf{Framework} & \textbf{Model} & \textbf{Solved} & \textbf{Rate (\%)} & \textbf{Tokens (M)} & \textbf{Time (min)} \\
\midrule\endhead
\midrule
\multicolumn{7}{r}{\footnotesize\color{ReleaseMuted}Continued on next page}\\\endfoot
\bottomrule\endlastfoot
C01 & Claude Code & GLM 5.3 & 4/95 & 4.21 & 17.27 & 258.55 \\
C02 & Claude Code & DeepSeek V4.1 Flash 1017 & 14/95 & 14.74 & 22.51 & 37.96 \\
C03 & Claude Code & Claude Opus 5.5 & 16/95 & 16.84 & 2.37 & 32.83 \\
C04 & Claude Code & GPT-6.1 Sol & 17/95 & 17.89 & 1.36 & 21.75 \\
C05 & Claude Code & GPT-6 Astra & 18/95 & 18.95 & 1.08 & 12.43 \\
C06 & Codex & GLM 5.3 & 2/95 & 2.11 & 10.19 & 99.18 \\
C07 & Codex & DeepSeek V4.1 Flash 1017 & 10/95 & 10.53 & 14.77 & 25.07 \\
C08 & Codex & GPT-6.1 Sol & 19/95 & 20.00 & 2.94 & 30.96 \\
C09 & Codex & GPT-6 Astra & 20/95 & 21.05 & 1.48 & 16.48 \\
C10 & OpenCode & GLM 5.3 & 5/95 & 5.26 & 1.46 & 566.61 \\
C11 & OpenCode & DeepSeek V4.1 Flash 1017 & 11/95 & 11.58 & 11.03 & 32.63 \\
C12 & OpenCode & GPT-6.1 Sol & 15/95 & 15.79 & 2.19 & 26.64 \\
C13 & OpenCode & GPT-6 Astra & 17/95 & 17.89 & 1.42 & 19.06 \\
\end{longtable}\endgroup

\Needspace{13\baselineskip}
\subsection{Success Counts by Evaluation Criterion}
Each column counts tasks satisfying one evaluation criterion: compilation, golden-RTL acceptance, detection of all ten evaluated mutants, exact reference-model predictions, or at least 90\% coverage in each of the four metrics. Overall success requires all criteria simultaneously. For Codex with GPT-6 Astra, 95 tasks compile, 90 pass golden RTL, 53 detect all ten mutants, 67 have exact reference models, and 35 meet all coverage thresholds; 20 satisfy the complete set. These counts show how executable testbenches, fault detection, behavioral prediction, and coverage contribute distinct evidence of verification capability.

\begingroup
\fontsize{9}{11}\selectfont
\setlength{\tabcolsep}{3pt}\renewcommand{\arraystretch}{1.04}
\setlength{\LTleft}{0pt}\setlength{\LTright}{0pt}
\Needspace{10\baselineskip}
\captionsetup{hypcap=false}
\captionof{table}{Tasks satisfying each evaluation criterion, out of 95 per configuration. Each criterion is counted independently; Overall requires all criteria.}\label{tab:app-gates}
\addtocounter{table}{-1}
\begin{longtable}{>{\raggedright\arraybackslash}p{\dimexpr 0.0700000\textwidth-2\tabcolsep\relax}>{\raggedleft\arraybackslash}p{\dimexpr 0.1400000\textwidth-2\tabcolsep\relax}>{\raggedleft\arraybackslash}p{\dimexpr 0.1400000\textwidth-2\tabcolsep\relax}>{\raggedleft\arraybackslash}p{\dimexpr 0.1400000\textwidth-2\tabcolsep\relax}>{\raggedleft\arraybackslash}p{\dimexpr 0.1900000\textwidth-2\tabcolsep\relax}>{\raggedleft\arraybackslash}p{\dimexpr 0.1600000\textwidth-2\tabcolsep\relax}>{\raggedleft\arraybackslash}p{\dimexpr 0.1600000\textwidth-2\tabcolsep\relax}}
\toprule
\rowcolor{ReleasePale}\textbf{ID} & \textbf{Compile} & \textbf{Golden} & \textbf{Kill all} & \textbf{Exact model} & \textbf{Cov. $\geq$90} & \textbf{Overall} \\
\midrule\endfirsthead
\multicolumn{7}{l}{\small\color{ReleaseMuted}Table~\ref{tab:app-gates} (continued)}\\
\toprule
\rowcolor{ReleasePale}\textbf{ID} & \textbf{Compile} & \textbf{Golden} & \textbf{Kill all} & \textbf{Exact model} & \textbf{Cov. $\geq$90} & \textbf{Overall} \\
\midrule\endhead
\midrule
\multicolumn{7}{r}{\footnotesize\color{ReleaseMuted}Continued on next page}\\\endfoot
\bottomrule\endlastfoot
C01 & 68 & 39 & 8 & 40 & 14 & 4 \\
C02 & 95 & 92 & 44 & 56 & 28 & 14 \\
C03 & 94 & 89 & 41 & 68 & 34 & 16 \\
C04 & 94 & 87 & 45 & 65 & 35 & 17 \\
C05 & 95 & 89 & 49 & 67 & 34 & 18 \\
C06 & 79 & 35 & 7 & 23 & 13 & 2 \\
C07 & 95 & 92 & 28 & 56 & 27 & 10 \\
C08 & 95 & 90 & 50 & 67 & 34 & 19 \\
C09 & 95 & 90 & 53 & 67 & 35 & 20 \\
C10 & 14 & 12 & 8 & 15 & 10 & 5 \\
C11 & 94 & 93 & 39 & 57 & 30 & 11 \\
C12 & 93 & 86 & 39 & 65 & 35 & 15 \\
C13 & 94 & 89 & 49 & 67 & 34 & 17 \\
\end{longtable}\endgroup

\Needspace{27\baselineskip}
\subsection{Success by Difficulty and Functional Domain}
The five difficulty levels contain 6, 7, 32, 29, and 21 tasks, respectively. The strongest configurations solve five of six L1 tasks. Codex with GPT-6.1 Sol solves five of seven L2 tasks, while Claude Code with Claude Opus 5.5 completes three L5 tasks. Both GPT-6 Astra and DeepSeek V4.1 Flash 1017 achieve success at every difficulty level when paired with Claude Code.
\begingroup
\fontsize{9}{11}\selectfont
\setlength{\tabcolsep}{3pt}\renewcommand{\arraystretch}{1.04}
\setlength{\LTleft}{0pt}\setlength{\LTright}{0pt}
\Needspace{10\baselineskip}
\captionsetup{hypcap=false}
\captionof{table}{Overall success by difficulty level, shown as successful tasks / total tasks.}\label{tab:app-levels}
\addtocounter{table}{-1}
\begin{longtable}{>{\raggedright\arraybackslash}p{\dimexpr 0.1500000\textwidth-2\tabcolsep\relax}>{\raggedleft\arraybackslash}p{\dimexpr 0.1700000\textwidth-2\tabcolsep\relax}>{\raggedleft\arraybackslash}p{\dimexpr 0.1700000\textwidth-2\tabcolsep\relax}>{\raggedleft\arraybackslash}p{\dimexpr 0.1700000\textwidth-2\tabcolsep\relax}>{\raggedleft\arraybackslash}p{\dimexpr 0.1700000\textwidth-2\tabcolsep\relax}>{\raggedleft\arraybackslash}p{\dimexpr 0.1700000\textwidth-2\tabcolsep\relax}}
\toprule
\rowcolor{ReleasePale}\textbf{ID} & \textbf{L1} & \textbf{L2} & \textbf{L3} & \textbf{L4} & \textbf{L5} \\
\midrule\endfirsthead
\multicolumn{6}{l}{\small\color{ReleaseMuted}Table~\ref{tab:app-levels} (continued)}\\
\toprule
\rowcolor{ReleasePale}\textbf{ID} & \textbf{L1} & \textbf{L2} & \textbf{L3} & \textbf{L4} & \textbf{L5} \\
\midrule\endhead
\midrule
\multicolumn{6}{r}{\footnotesize\color{ReleaseMuted}Continued on next page}\\\endfoot
\bottomrule\endlastfoot
C01 & 3/6 & 1/7 & 0/32 & 0/29 & 0/21 \\
C02 & 4/6 & 2/7 & 4/32 & 3/29 & 1/21 \\
C03 & 5/6 & 4/7 & 1/32 & 3/29 & 3/21 \\
C04 & 5/6 & 3/7 & 4/32 & 5/29 & 0/21 \\
C05 & 5/6 & 3/7 & 4/32 & 4/29 & 2/21 \\
C06 & 1/6 & 1/7 & 0/32 & 0/29 & 0/21 \\
C07 & 4/6 & 2/7 & 3/32 & 0/29 & 1/21 \\
C08 & 5/6 & 5/7 & 4/32 & 3/29 & 2/21 \\
C09 & 5/6 & 4/7 & 4/32 & 5/29 & 2/21 \\
C10 & 4/6 & 1/7 & 0/32 & 0/29 & 0/21 \\
C11 & 5/6 & 3/7 & 3/32 & 0/29 & 0/21 \\
C12 & 5/6 & 3/7 & 3/32 & 3/29 & 1/21 \\
C13 & 5/6 & 4/7 & 3/32 & 3/29 & 2/21 \\
\end{longtable}\endgroup

\Needspace{20\baselineskip}
Performance also varies across the nine functional domains. Several configurations complete four of eight arithmetic tasks and five of 24 digital-design tasks. Both GPT models in Codex complete three of 13 memory tasks. Processor tasks yield up to four successful solutions among 16 designs, demonstrating end-to-end verification capability beyond standalone logic blocks.
\begingroup
\fontsize{8}{10}\selectfont
\setlength{\tabcolsep}{3pt}\renewcommand{\arraystretch}{1.04}
\setlength{\LTleft}{0pt}\setlength{\LTright}{0pt}
\Needspace{10\baselineskip}
\captionsetup{hypcap=false}
\captionof{table}{Successful tasks by functional domain. N is the total number of tasks in each domain.}\label{tab:app-domains}
\addtocounter{table}{-1}
\begin{longtable}{>{\raggedright\arraybackslash}p{\dimexpr 0.2000000\textwidth-2\tabcolsep\relax}>{\raggedleft\arraybackslash}p{\dimexpr 0.0400000\textwidth-2\tabcolsep\relax}>{\raggedleft\arraybackslash}p{\dimexpr 0.0584615\textwidth-2\tabcolsep\relax}>{\raggedleft\arraybackslash}p{\dimexpr 0.0584615\textwidth-2\tabcolsep\relax}>{\raggedleft\arraybackslash}p{\dimexpr 0.0584615\textwidth-2\tabcolsep\relax}>{\raggedleft\arraybackslash}p{\dimexpr 0.0584615\textwidth-2\tabcolsep\relax}>{\raggedleft\arraybackslash}p{\dimexpr 0.0584615\textwidth-2\tabcolsep\relax}>{\raggedleft\arraybackslash}p{\dimexpr 0.0584615\textwidth-2\tabcolsep\relax}>{\raggedleft\arraybackslash}p{\dimexpr 0.0584615\textwidth-2\tabcolsep\relax}>{\raggedleft\arraybackslash}p{\dimexpr 0.0584615\textwidth-2\tabcolsep\relax}>{\raggedleft\arraybackslash}p{\dimexpr 0.0584615\textwidth-2\tabcolsep\relax}>{\raggedleft\arraybackslash}p{\dimexpr 0.0584615\textwidth-2\tabcolsep\relax}>{\raggedleft\arraybackslash}p{\dimexpr 0.0584615\textwidth-2\tabcolsep\relax}>{\raggedleft\arraybackslash}p{\dimexpr 0.0584615\textwidth-2\tabcolsep\relax}>{\raggedleft\arraybackslash}p{\dimexpr 0.0584615\textwidth-2\tabcolsep\relax}}
\toprule
\rowcolor{ReleasePale}\textbf{Domain} & \textbf{N} & \textbf{C01} & \textbf{C02} & \textbf{C03} & \textbf{C04} & \textbf{C05} & \textbf{C06} & \textbf{C07} & \textbf{C08} & \textbf{C09} & \textbf{C10} & \textbf{C11} & \textbf{C12} & \textbf{C13} \\
\midrule\endfirsthead
\multicolumn{15}{l}{\small\color{ReleaseMuted}Table~\ref{tab:app-domains} (continued)}\\
\toprule
\rowcolor{ReleasePale}\textbf{Domain} & \textbf{N} & \textbf{C01} & \textbf{C02} & \textbf{C03} & \textbf{C04} & \textbf{C05} & \textbf{C06} & \textbf{C07} & \textbf{C08} & \textbf{C09} & \textbf{C10} & \textbf{C11} & \textbf{C12} & \textbf{C13} \\
\midrule\endhead
\midrule
\multicolumn{15}{r}{\footnotesize\color{ReleaseMuted}Continued on next page}\\\endfoot
\bottomrule\endlastfoot
Arithmetic & 8 & 2 & 3 & 4 & 4 & 4 & 1 & 3 & 4 & 4 & 3 & 4 & 4 & 4 \\
Memory & 13 & 0 & 2 & 2 & 1 & 1 & 0 & 1 & 3 & 3 & 0 & 1 & 1 & 2 \\
Digital design & 24 & 1 & 3 & 3 & 5 & 5 & 0 & 4 & 5 & 5 & 1 & 3 & 4 & 5 \\
Interconnect & 8 & 1 & 1 & 1 & 1 & 1 & 1 & 1 & 1 & 1 & 1 & 1 & 1 & 1 \\
Control & 7 & 0 & 0 & 1 & 1 & 2 & 0 & 0 & 1 & 1 & 0 & 1 & 1 & 2 \\
Communication & 5 & 0 & 1 & 0 & 0 & 1 & 0 & 0 & 1 & 1 & 0 & 0 & 0 & 0 \\
Processor & 16 & 0 & 3 & 3 & 4 & 3 & 0 & 1 & 3 & 4 & 0 & 1 & 4 & 2 \\
Cryptography & 10 & 0 & 1 & 2 & 1 & 1 & 0 & 0 & 1 & 1 & 0 & 0 & 0 & 1 \\
Signal processing & 4 & 0 & 0 & 0 & 0 & 0 & 0 & 0 & 0 & 0 & 0 & 0 & 0 & 0 \\
\end{longtable}\endgroup

\subsection{Task-Level Success Matrix}
The matrix presents the overall outcomes for all 95 tasks across 13 configurations. Rows identify individual tasks and their difficulty levels; columns C01--C13 identify the agent--model configurations. A blue 1 indicates that all success criteria are met, and 0 indicates that at least one criterion is unmet.
\begingroup
\fontsize{8.4}{10.4}\selectfont
\setlength{\tabcolsep}{3pt}\renewcommand{\arraystretch}{1.04}
\setlength{\LTleft}{0pt}\setlength{\LTright}{0pt}
\Needspace{10\baselineskip}
\captionsetup{hypcap=false}
\captionof{table}{Overall success for each task and configuration. Blue 1: all criteria met; 0: at least one criterion unmet.}\label{tab:app-matrix}
\addtocounter{table}{-1}
\begin{longtable}{>{\raggedright\arraybackslash}p{\dimexpr 0.1000000\textwidth-2\tabcolsep\relax}>{\raggedright\arraybackslash}p{\dimexpr 0.0550000\textwidth-2\tabcolsep\relax}>{\raggedleft\arraybackslash}p{\dimexpr 0.0650000\textwidth-2\tabcolsep\relax}>{\raggedleft\arraybackslash}p{\dimexpr 0.0650000\textwidth-2\tabcolsep\relax}>{\raggedleft\arraybackslash}p{\dimexpr 0.0650000\textwidth-2\tabcolsep\relax}>{\raggedleft\arraybackslash}p{\dimexpr 0.0650000\textwidth-2\tabcolsep\relax}>{\raggedleft\arraybackslash}p{\dimexpr 0.0650000\textwidth-2\tabcolsep\relax}>{\raggedleft\arraybackslash}p{\dimexpr 0.0650000\textwidth-2\tabcolsep\relax}>{\raggedleft\arraybackslash}p{\dimexpr 0.0650000\textwidth-2\tabcolsep\relax}>{\raggedleft\arraybackslash}p{\dimexpr 0.0650000\textwidth-2\tabcolsep\relax}>{\raggedleft\arraybackslash}p{\dimexpr 0.0650000\textwidth-2\tabcolsep\relax}>{\raggedleft\arraybackslash}p{\dimexpr 0.0650000\textwidth-2\tabcolsep\relax}>{\raggedleft\arraybackslash}p{\dimexpr 0.0650000\textwidth-2\tabcolsep\relax}>{\raggedleft\arraybackslash}p{\dimexpr 0.0650000\textwidth-2\tabcolsep\relax}>{\raggedleft\arraybackslash}p{\dimexpr 0.0650000\textwidth-2\tabcolsep\relax}}
\toprule
\rowcolor{ReleasePale}\textbf{Task} & \textbf{Lv.} & \textbf{C01} & \textbf{C02} & \textbf{C03} & \textbf{C04} & \textbf{C05} & \textbf{C06} & \textbf{C07} & \textbf{C08} & \textbf{C09} & \textbf{C10} & \textbf{C11} & \textbf{C12} & \textbf{C13} \\
\midrule\endfirsthead
\multicolumn{15}{l}{\small\color{ReleaseMuted}Table~\ref{tab:app-matrix} (continued)}\\
\toprule
\rowcolor{ReleasePale}\textbf{Task} & \textbf{Lv.} & \textbf{C01} & \textbf{C02} & \textbf{C03} & \textbf{C04} & \textbf{C05} & \textbf{C06} & \textbf{C07} & \textbf{C08} & \textbf{C09} & \textbf{C10} & \textbf{C11} & \textbf{C12} & \textbf{C13} \\
\midrule\endhead
\midrule
\multicolumn{15}{r}{\footnotesize\color{ReleaseMuted}Continued on next page}\\\endfoot
\bottomrule\endlastfoot
dv-001 & L1 & \cellcolor{ReleasePale}\textbf{\color{ReleaseBlue}1} & \cellcolor{ReleasePale}\textbf{\color{ReleaseBlue}1} & \cellcolor{ReleasePale}\textbf{\color{ReleaseBlue}1} & \cellcolor{ReleasePale}\textbf{\color{ReleaseBlue}1} & \cellcolor{ReleasePale}\textbf{\color{ReleaseBlue}1} & \cellcolor{ReleasePale}\textbf{\color{ReleaseBlue}1} & \cellcolor{ReleasePale}\textbf{\color{ReleaseBlue}1} & \cellcolor{ReleasePale}\textbf{\color{ReleaseBlue}1} & \cellcolor{ReleasePale}\textbf{\color{ReleaseBlue}1} & \cellcolor{ReleasePale}\textbf{\color{ReleaseBlue}1} & \cellcolor{ReleasePale}\textbf{\color{ReleaseBlue}1} & \cellcolor{ReleasePale}\textbf{\color{ReleaseBlue}1} & \cellcolor{ReleasePale}\textbf{\color{ReleaseBlue}1} \\
dv-002 & L1 & \textcolor{ReleaseMuted}{0} & \cellcolor{ReleasePale}\textbf{\color{ReleaseBlue}1} & \cellcolor{ReleasePale}\textbf{\color{ReleaseBlue}1} & \cellcolor{ReleasePale}\textbf{\color{ReleaseBlue}1} & \cellcolor{ReleasePale}\textbf{\color{ReleaseBlue}1} & \textcolor{ReleaseMuted}{0} & \cellcolor{ReleasePale}\textbf{\color{ReleaseBlue}1} & \cellcolor{ReleasePale}\textbf{\color{ReleaseBlue}1} & \cellcolor{ReleasePale}\textbf{\color{ReleaseBlue}1} & \cellcolor{ReleasePale}\textbf{\color{ReleaseBlue}1} & \cellcolor{ReleasePale}\textbf{\color{ReleaseBlue}1} & \cellcolor{ReleasePale}\textbf{\color{ReleaseBlue}1} & \cellcolor{ReleasePale}\textbf{\color{ReleaseBlue}1} \\
dv-003 & L1 & \cellcolor{ReleasePale}\textbf{\color{ReleaseBlue}1} & \textcolor{ReleaseMuted}{0} & \cellcolor{ReleasePale}\textbf{\color{ReleaseBlue}1} & \cellcolor{ReleasePale}\textbf{\color{ReleaseBlue}1} & \cellcolor{ReleasePale}\textbf{\color{ReleaseBlue}1} & \textcolor{ReleaseMuted}{0} & \cellcolor{ReleasePale}\textbf{\color{ReleaseBlue}1} & \cellcolor{ReleasePale}\textbf{\color{ReleaseBlue}1} & \cellcolor{ReleasePale}\textbf{\color{ReleaseBlue}1} & \cellcolor{ReleasePale}\textbf{\color{ReleaseBlue}1} & \cellcolor{ReleasePale}\textbf{\color{ReleaseBlue}1} & \cellcolor{ReleasePale}\textbf{\color{ReleaseBlue}1} & \cellcolor{ReleasePale}\textbf{\color{ReleaseBlue}1} \\
dv-004 & L1 & \textcolor{ReleaseMuted}{0} & \textcolor{ReleaseMuted}{0} & \textcolor{ReleaseMuted}{0} & \textcolor{ReleaseMuted}{0} & \textcolor{ReleaseMuted}{0} & \textcolor{ReleaseMuted}{0} & \textcolor{ReleaseMuted}{0} & \textcolor{ReleaseMuted}{0} & \textcolor{ReleaseMuted}{0} & \textcolor{ReleaseMuted}{0} & \textcolor{ReleaseMuted}{0} & \textcolor{ReleaseMuted}{0} & \textcolor{ReleaseMuted}{0} \\
dv-005 & L1 & \cellcolor{ReleasePale}\textbf{\color{ReleaseBlue}1} & \cellcolor{ReleasePale}\textbf{\color{ReleaseBlue}1} & \cellcolor{ReleasePale}\textbf{\color{ReleaseBlue}1} & \cellcolor{ReleasePale}\textbf{\color{ReleaseBlue}1} & \cellcolor{ReleasePale}\textbf{\color{ReleaseBlue}1} & \textcolor{ReleaseMuted}{0} & \cellcolor{ReleasePale}\textbf{\color{ReleaseBlue}1} & \cellcolor{ReleasePale}\textbf{\color{ReleaseBlue}1} & \cellcolor{ReleasePale}\textbf{\color{ReleaseBlue}1} & \cellcolor{ReleasePale}\textbf{\color{ReleaseBlue}1} & \cellcolor{ReleasePale}\textbf{\color{ReleaseBlue}1} & \cellcolor{ReleasePale}\textbf{\color{ReleaseBlue}1} & \cellcolor{ReleasePale}\textbf{\color{ReleaseBlue}1} \\
dv-006 & L1 & \textcolor{ReleaseMuted}{0} & \cellcolor{ReleasePale}\textbf{\color{ReleaseBlue}1} & \cellcolor{ReleasePale}\textbf{\color{ReleaseBlue}1} & \cellcolor{ReleasePale}\textbf{\color{ReleaseBlue}1} & \cellcolor{ReleasePale}\textbf{\color{ReleaseBlue}1} & \textcolor{ReleaseMuted}{0} & \textcolor{ReleaseMuted}{0} & \cellcolor{ReleasePale}\textbf{\color{ReleaseBlue}1} & \cellcolor{ReleasePale}\textbf{\color{ReleaseBlue}1} & \textcolor{ReleaseMuted}{0} & \cellcolor{ReleasePale}\textbf{\color{ReleaseBlue}1} & \cellcolor{ReleasePale}\textbf{\color{ReleaseBlue}1} & \cellcolor{ReleasePale}\textbf{\color{ReleaseBlue}1} \\
dv-007 & L2 & \textcolor{ReleaseMuted}{0} & \textcolor{ReleaseMuted}{0} & \textcolor{ReleaseMuted}{0} & \textcolor{ReleaseMuted}{0} & \textcolor{ReleaseMuted}{0} & \textcolor{ReleaseMuted}{0} & \textcolor{ReleaseMuted}{0} & \cellcolor{ReleasePale}\textbf{\color{ReleaseBlue}1} & \textcolor{ReleaseMuted}{0} & \textcolor{ReleaseMuted}{0} & \textcolor{ReleaseMuted}{0} & \textcolor{ReleaseMuted}{0} & \textcolor{ReleaseMuted}{0} \\
dv-008 & L2 & \textcolor{ReleaseMuted}{0} & \textcolor{ReleaseMuted}{0} & \cellcolor{ReleasePale}\textbf{\color{ReleaseBlue}1} & \textcolor{ReleaseMuted}{0} & \textcolor{ReleaseMuted}{0} & \textcolor{ReleaseMuted}{0} & \textcolor{ReleaseMuted}{0} & \cellcolor{ReleasePale}\textbf{\color{ReleaseBlue}1} & \cellcolor{ReleasePale}\textbf{\color{ReleaseBlue}1} & \textcolor{ReleaseMuted}{0} & \textcolor{ReleaseMuted}{0} & \textcolor{ReleaseMuted}{0} & \cellcolor{ReleasePale}\textbf{\color{ReleaseBlue}1} \\
dv-009 & L2 & \textcolor{ReleaseMuted}{0} & \textcolor{ReleaseMuted}{0} & \textcolor{ReleaseMuted}{0} & \textcolor{ReleaseMuted}{0} & \textcolor{ReleaseMuted}{0} & \textcolor{ReleaseMuted}{0} & \textcolor{ReleaseMuted}{0} & \textcolor{ReleaseMuted}{0} & \textcolor{ReleaseMuted}{0} & \textcolor{ReleaseMuted}{0} & \textcolor{ReleaseMuted}{0} & \textcolor{ReleaseMuted}{0} & \textcolor{ReleaseMuted}{0} \\
dv-010 & L2 & \cellcolor{ReleasePale}\textbf{\color{ReleaseBlue}1} & \cellcolor{ReleasePale}\textbf{\color{ReleaseBlue}1} & \cellcolor{ReleasePale}\textbf{\color{ReleaseBlue}1} & \cellcolor{ReleasePale}\textbf{\color{ReleaseBlue}1} & \cellcolor{ReleasePale}\textbf{\color{ReleaseBlue}1} & \cellcolor{ReleasePale}\textbf{\color{ReleaseBlue}1} & \cellcolor{ReleasePale}\textbf{\color{ReleaseBlue}1} & \cellcolor{ReleasePale}\textbf{\color{ReleaseBlue}1} & \cellcolor{ReleasePale}\textbf{\color{ReleaseBlue}1} & \cellcolor{ReleasePale}\textbf{\color{ReleaseBlue}1} & \cellcolor{ReleasePale}\textbf{\color{ReleaseBlue}1} & \cellcolor{ReleasePale}\textbf{\color{ReleaseBlue}1} & \cellcolor{ReleasePale}\textbf{\color{ReleaseBlue}1} \\
dv-011 & L2 & \textcolor{ReleaseMuted}{0} & \textcolor{ReleaseMuted}{0} & \textcolor{ReleaseMuted}{0} & \textcolor{ReleaseMuted}{0} & \textcolor{ReleaseMuted}{0} & \textcolor{ReleaseMuted}{0} & \textcolor{ReleaseMuted}{0} & \textcolor{ReleaseMuted}{0} & \textcolor{ReleaseMuted}{0} & \textcolor{ReleaseMuted}{0} & \textcolor{ReleaseMuted}{0} & \textcolor{ReleaseMuted}{0} & \textcolor{ReleaseMuted}{0} \\
dv-012 & L2 & \textcolor{ReleaseMuted}{0} & \cellcolor{ReleasePale}\textbf{\color{ReleaseBlue}1} & \cellcolor{ReleasePale}\textbf{\color{ReleaseBlue}1} & \cellcolor{ReleasePale}\textbf{\color{ReleaseBlue}1} & \cellcolor{ReleasePale}\textbf{\color{ReleaseBlue}1} & \textcolor{ReleaseMuted}{0} & \cellcolor{ReleasePale}\textbf{\color{ReleaseBlue}1} & \cellcolor{ReleasePale}\textbf{\color{ReleaseBlue}1} & \cellcolor{ReleasePale}\textbf{\color{ReleaseBlue}1} & \textcolor{ReleaseMuted}{0} & \cellcolor{ReleasePale}\textbf{\color{ReleaseBlue}1} & \cellcolor{ReleasePale}\textbf{\color{ReleaseBlue}1} & \cellcolor{ReleasePale}\textbf{\color{ReleaseBlue}1} \\
dv-013 & L2 & \textcolor{ReleaseMuted}{0} & \textcolor{ReleaseMuted}{0} & \cellcolor{ReleasePale}\textbf{\color{ReleaseBlue}1} & \cellcolor{ReleasePale}\textbf{\color{ReleaseBlue}1} & \cellcolor{ReleasePale}\textbf{\color{ReleaseBlue}1} & \textcolor{ReleaseMuted}{0} & \textcolor{ReleaseMuted}{0} & \cellcolor{ReleasePale}\textbf{\color{ReleaseBlue}1} & \cellcolor{ReleasePale}\textbf{\color{ReleaseBlue}1} & \textcolor{ReleaseMuted}{0} & \cellcolor{ReleasePale}\textbf{\color{ReleaseBlue}1} & \cellcolor{ReleasePale}\textbf{\color{ReleaseBlue}1} & \cellcolor{ReleasePale}\textbf{\color{ReleaseBlue}1} \\
dv-014 & L3 & \textcolor{ReleaseMuted}{0} & \textcolor{ReleaseMuted}{0} & \textcolor{ReleaseMuted}{0} & \textcolor{ReleaseMuted}{0} & \textcolor{ReleaseMuted}{0} & \textcolor{ReleaseMuted}{0} & \textcolor{ReleaseMuted}{0} & \textcolor{ReleaseMuted}{0} & \textcolor{ReleaseMuted}{0} & \textcolor{ReleaseMuted}{0} & \textcolor{ReleaseMuted}{0} & \textcolor{ReleaseMuted}{0} & \textcolor{ReleaseMuted}{0} \\
dv-015 & L3 & \textcolor{ReleaseMuted}{0} & \textcolor{ReleaseMuted}{0} & \textcolor{ReleaseMuted}{0} & \textcolor{ReleaseMuted}{0} & \textcolor{ReleaseMuted}{0} & \textcolor{ReleaseMuted}{0} & \textcolor{ReleaseMuted}{0} & \textcolor{ReleaseMuted}{0} & \textcolor{ReleaseMuted}{0} & \textcolor{ReleaseMuted}{0} & \textcolor{ReleaseMuted}{0} & \textcolor{ReleaseMuted}{0} & \textcolor{ReleaseMuted}{0} \\
dv-016 & L3 & \textcolor{ReleaseMuted}{0} & \textcolor{ReleaseMuted}{0} & \textcolor{ReleaseMuted}{0} & \textcolor{ReleaseMuted}{0} & \textcolor{ReleaseMuted}{0} & \textcolor{ReleaseMuted}{0} & \textcolor{ReleaseMuted}{0} & \textcolor{ReleaseMuted}{0} & \textcolor{ReleaseMuted}{0} & \textcolor{ReleaseMuted}{0} & \textcolor{ReleaseMuted}{0} & \textcolor{ReleaseMuted}{0} & \textcolor{ReleaseMuted}{0} \\
dv-017 & L3 & \textcolor{ReleaseMuted}{0} & \textcolor{ReleaseMuted}{0} & \textcolor{ReleaseMuted}{0} & \textcolor{ReleaseMuted}{0} & \textcolor{ReleaseMuted}{0} & \textcolor{ReleaseMuted}{0} & \textcolor{ReleaseMuted}{0} & \textcolor{ReleaseMuted}{0} & \textcolor{ReleaseMuted}{0} & \textcolor{ReleaseMuted}{0} & \textcolor{ReleaseMuted}{0} & \textcolor{ReleaseMuted}{0} & \textcolor{ReleaseMuted}{0} \\
dv-018 & L3 & \textcolor{ReleaseMuted}{0} & \cellcolor{ReleasePale}\textbf{\color{ReleaseBlue}1} & \cellcolor{ReleasePale}\textbf{\color{ReleaseBlue}1} & \cellcolor{ReleasePale}\textbf{\color{ReleaseBlue}1} & \cellcolor{ReleasePale}\textbf{\color{ReleaseBlue}1} & \textcolor{ReleaseMuted}{0} & \cellcolor{ReleasePale}\textbf{\color{ReleaseBlue}1} & \cellcolor{ReleasePale}\textbf{\color{ReleaseBlue}1} & \cellcolor{ReleasePale}\textbf{\color{ReleaseBlue}1} & \textcolor{ReleaseMuted}{0} & \cellcolor{ReleasePale}\textbf{\color{ReleaseBlue}1} & \cellcolor{ReleasePale}\textbf{\color{ReleaseBlue}1} & \cellcolor{ReleasePale}\textbf{\color{ReleaseBlue}1} \\
dv-019 & L3 & \textcolor{ReleaseMuted}{0} & \textcolor{ReleaseMuted}{0} & \textcolor{ReleaseMuted}{0} & \textcolor{ReleaseMuted}{0} & \textcolor{ReleaseMuted}{0} & \textcolor{ReleaseMuted}{0} & \textcolor{ReleaseMuted}{0} & \textcolor{ReleaseMuted}{0} & \textcolor{ReleaseMuted}{0} & \textcolor{ReleaseMuted}{0} & \textcolor{ReleaseMuted}{0} & \textcolor{ReleaseMuted}{0} & \textcolor{ReleaseMuted}{0} \\
dv-020 & L3 & \textcolor{ReleaseMuted}{0} & \textcolor{ReleaseMuted}{0} & \textcolor{ReleaseMuted}{0} & \textcolor{ReleaseMuted}{0} & \textcolor{ReleaseMuted}{0} & \textcolor{ReleaseMuted}{0} & \textcolor{ReleaseMuted}{0} & \textcolor{ReleaseMuted}{0} & \cellcolor{ReleasePale}\textbf{\color{ReleaseBlue}1} & \textcolor{ReleaseMuted}{0} & \textcolor{ReleaseMuted}{0} & \textcolor{ReleaseMuted}{0} & \textcolor{ReleaseMuted}{0} \\
dv-021 & L3 & \textcolor{ReleaseMuted}{0} & \textcolor{ReleaseMuted}{0} & \textcolor{ReleaseMuted}{0} & \textcolor{ReleaseMuted}{0} & \cellcolor{ReleasePale}\textbf{\color{ReleaseBlue}1} & \textcolor{ReleaseMuted}{0} & \textcolor{ReleaseMuted}{0} & \textcolor{ReleaseMuted}{0} & \textcolor{ReleaseMuted}{0} & \textcolor{ReleaseMuted}{0} & \textcolor{ReleaseMuted}{0} & \textcolor{ReleaseMuted}{0} & \cellcolor{ReleasePale}\textbf{\color{ReleaseBlue}1} \\
dv-022 & L3 & \textcolor{ReleaseMuted}{0} & \textcolor{ReleaseMuted}{0} & \textcolor{ReleaseMuted}{0} & \textcolor{ReleaseMuted}{0} & \textcolor{ReleaseMuted}{0} & \textcolor{ReleaseMuted}{0} & \textcolor{ReleaseMuted}{0} & \textcolor{ReleaseMuted}{0} & \textcolor{ReleaseMuted}{0} & \textcolor{ReleaseMuted}{0} & \textcolor{ReleaseMuted}{0} & \textcolor{ReleaseMuted}{0} & \textcolor{ReleaseMuted}{0} \\
dv-023 & L3 & \textcolor{ReleaseMuted}{0} & \cellcolor{ReleasePale}\textbf{\color{ReleaseBlue}1} & \textcolor{ReleaseMuted}{0} & \cellcolor{ReleasePale}\textbf{\color{ReleaseBlue}1} & \cellcolor{ReleasePale}\textbf{\color{ReleaseBlue}1} & \textcolor{ReleaseMuted}{0} & \cellcolor{ReleasePale}\textbf{\color{ReleaseBlue}1} & \cellcolor{ReleasePale}\textbf{\color{ReleaseBlue}1} & \cellcolor{ReleasePale}\textbf{\color{ReleaseBlue}1} & \textcolor{ReleaseMuted}{0} & \cellcolor{ReleasePale}\textbf{\color{ReleaseBlue}1} & \cellcolor{ReleasePale}\textbf{\color{ReleaseBlue}1} & \cellcolor{ReleasePale}\textbf{\color{ReleaseBlue}1} \\
dv-024 & L3 & \textcolor{ReleaseMuted}{0} & \textcolor{ReleaseMuted}{0} & \textcolor{ReleaseMuted}{0} & \textcolor{ReleaseMuted}{0} & \textcolor{ReleaseMuted}{0} & \textcolor{ReleaseMuted}{0} & \textcolor{ReleaseMuted}{0} & \textcolor{ReleaseMuted}{0} & \textcolor{ReleaseMuted}{0} & \textcolor{ReleaseMuted}{0} & \textcolor{ReleaseMuted}{0} & \textcolor{ReleaseMuted}{0} & \textcolor{ReleaseMuted}{0} \\
dv-025 & L3 & \textcolor{ReleaseMuted}{0} & \textcolor{ReleaseMuted}{0} & \textcolor{ReleaseMuted}{0} & \textcolor{ReleaseMuted}{0} & \textcolor{ReleaseMuted}{0} & \textcolor{ReleaseMuted}{0} & \textcolor{ReleaseMuted}{0} & \textcolor{ReleaseMuted}{0} & \textcolor{ReleaseMuted}{0} & \textcolor{ReleaseMuted}{0} & \textcolor{ReleaseMuted}{0} & \textcolor{ReleaseMuted}{0} & \textcolor{ReleaseMuted}{0} \\
dv-026 & L3 & \textcolor{ReleaseMuted}{0} & \textcolor{ReleaseMuted}{0} & \textcolor{ReleaseMuted}{0} & \textcolor{ReleaseMuted}{0} & \textcolor{ReleaseMuted}{0} & \textcolor{ReleaseMuted}{0} & \textcolor{ReleaseMuted}{0} & \textcolor{ReleaseMuted}{0} & \textcolor{ReleaseMuted}{0} & \textcolor{ReleaseMuted}{0} & \textcolor{ReleaseMuted}{0} & \textcolor{ReleaseMuted}{0} & \textcolor{ReleaseMuted}{0} \\
dv-027 & L3 & \textcolor{ReleaseMuted}{0} & \textcolor{ReleaseMuted}{0} & \textcolor{ReleaseMuted}{0} & \textcolor{ReleaseMuted}{0} & \textcolor{ReleaseMuted}{0} & \textcolor{ReleaseMuted}{0} & \textcolor{ReleaseMuted}{0} & \textcolor{ReleaseMuted}{0} & \textcolor{ReleaseMuted}{0} & \textcolor{ReleaseMuted}{0} & \textcolor{ReleaseMuted}{0} & \textcolor{ReleaseMuted}{0} & \textcolor{ReleaseMuted}{0} \\
dv-028 & L3 & \textcolor{ReleaseMuted}{0} & \cellcolor{ReleasePale}\textbf{\color{ReleaseBlue}1} & \textcolor{ReleaseMuted}{0} & \cellcolor{ReleasePale}\textbf{\color{ReleaseBlue}1} & \cellcolor{ReleasePale}\textbf{\color{ReleaseBlue}1} & \textcolor{ReleaseMuted}{0} & \cellcolor{ReleasePale}\textbf{\color{ReleaseBlue}1} & \cellcolor{ReleasePale}\textbf{\color{ReleaseBlue}1} & \cellcolor{ReleasePale}\textbf{\color{ReleaseBlue}1} & \textcolor{ReleaseMuted}{0} & \cellcolor{ReleasePale}\textbf{\color{ReleaseBlue}1} & \cellcolor{ReleasePale}\textbf{\color{ReleaseBlue}1} & \textcolor{ReleaseMuted}{0} \\
dv-029 & L3 & \textcolor{ReleaseMuted}{0} & \textcolor{ReleaseMuted}{0} & \textcolor{ReleaseMuted}{0} & \textcolor{ReleaseMuted}{0} & \textcolor{ReleaseMuted}{0} & \textcolor{ReleaseMuted}{0} & \textcolor{ReleaseMuted}{0} & \textcolor{ReleaseMuted}{0} & \textcolor{ReleaseMuted}{0} & \textcolor{ReleaseMuted}{0} & \textcolor{ReleaseMuted}{0} & \textcolor{ReleaseMuted}{0} & \textcolor{ReleaseMuted}{0} \\
dv-030 & L3 & \textcolor{ReleaseMuted}{0} & \textcolor{ReleaseMuted}{0} & \textcolor{ReleaseMuted}{0} & \textcolor{ReleaseMuted}{0} & \textcolor{ReleaseMuted}{0} & \textcolor{ReleaseMuted}{0} & \textcolor{ReleaseMuted}{0} & \textcolor{ReleaseMuted}{0} & \textcolor{ReleaseMuted}{0} & \textcolor{ReleaseMuted}{0} & \textcolor{ReleaseMuted}{0} & \textcolor{ReleaseMuted}{0} & \textcolor{ReleaseMuted}{0} \\
dv-031 & L3 & \textcolor{ReleaseMuted}{0} & \textcolor{ReleaseMuted}{0} & \textcolor{ReleaseMuted}{0} & \textcolor{ReleaseMuted}{0} & \textcolor{ReleaseMuted}{0} & \textcolor{ReleaseMuted}{0} & \textcolor{ReleaseMuted}{0} & \textcolor{ReleaseMuted}{0} & \textcolor{ReleaseMuted}{0} & \textcolor{ReleaseMuted}{0} & \textcolor{ReleaseMuted}{0} & \textcolor{ReleaseMuted}{0} & \textcolor{ReleaseMuted}{0} \\
dv-032 & L3 & \textcolor{ReleaseMuted}{0} & \textcolor{ReleaseMuted}{0} & \textcolor{ReleaseMuted}{0} & \textcolor{ReleaseMuted}{0} & \textcolor{ReleaseMuted}{0} & \textcolor{ReleaseMuted}{0} & \textcolor{ReleaseMuted}{0} & \textcolor{ReleaseMuted}{0} & \textcolor{ReleaseMuted}{0} & \textcolor{ReleaseMuted}{0} & \textcolor{ReleaseMuted}{0} & \textcolor{ReleaseMuted}{0} & \textcolor{ReleaseMuted}{0} \\
dv-033 & L3 & \textcolor{ReleaseMuted}{0} & \textcolor{ReleaseMuted}{0} & \textcolor{ReleaseMuted}{0} & \textcolor{ReleaseMuted}{0} & \textcolor{ReleaseMuted}{0} & \textcolor{ReleaseMuted}{0} & \textcolor{ReleaseMuted}{0} & \textcolor{ReleaseMuted}{0} & \textcolor{ReleaseMuted}{0} & \textcolor{ReleaseMuted}{0} & \textcolor{ReleaseMuted}{0} & \textcolor{ReleaseMuted}{0} & \textcolor{ReleaseMuted}{0} \\
dv-034 & L3 & \textcolor{ReleaseMuted}{0} & \textcolor{ReleaseMuted}{0} & \textcolor{ReleaseMuted}{0} & \textcolor{ReleaseMuted}{0} & \textcolor{ReleaseMuted}{0} & \textcolor{ReleaseMuted}{0} & \textcolor{ReleaseMuted}{0} & \textcolor{ReleaseMuted}{0} & \textcolor{ReleaseMuted}{0} & \textcolor{ReleaseMuted}{0} & \textcolor{ReleaseMuted}{0} & \textcolor{ReleaseMuted}{0} & \textcolor{ReleaseMuted}{0} \\
dv-035 & L3 & \textcolor{ReleaseMuted}{0} & \textcolor{ReleaseMuted}{0} & \textcolor{ReleaseMuted}{0} & \textcolor{ReleaseMuted}{0} & \textcolor{ReleaseMuted}{0} & \textcolor{ReleaseMuted}{0} & \textcolor{ReleaseMuted}{0} & \textcolor{ReleaseMuted}{0} & \textcolor{ReleaseMuted}{0} & \textcolor{ReleaseMuted}{0} & \textcolor{ReleaseMuted}{0} & \textcolor{ReleaseMuted}{0} & \textcolor{ReleaseMuted}{0} \\
dv-036 & L3 & \textcolor{ReleaseMuted}{0} & \textcolor{ReleaseMuted}{0} & \textcolor{ReleaseMuted}{0} & \textcolor{ReleaseMuted}{0} & \textcolor{ReleaseMuted}{0} & \textcolor{ReleaseMuted}{0} & \textcolor{ReleaseMuted}{0} & \textcolor{ReleaseMuted}{0} & \textcolor{ReleaseMuted}{0} & \textcolor{ReleaseMuted}{0} & \textcolor{ReleaseMuted}{0} & \textcolor{ReleaseMuted}{0} & \textcolor{ReleaseMuted}{0} \\
dv-037 & L3 & \textcolor{ReleaseMuted}{0} & \textcolor{ReleaseMuted}{0} & \textcolor{ReleaseMuted}{0} & \textcolor{ReleaseMuted}{0} & \textcolor{ReleaseMuted}{0} & \textcolor{ReleaseMuted}{0} & \textcolor{ReleaseMuted}{0} & \textcolor{ReleaseMuted}{0} & \textcolor{ReleaseMuted}{0} & \textcolor{ReleaseMuted}{0} & \textcolor{ReleaseMuted}{0} & \textcolor{ReleaseMuted}{0} & \textcolor{ReleaseMuted}{0} \\
dv-038 & L3 & \textcolor{ReleaseMuted}{0} & \cellcolor{ReleasePale}\textbf{\color{ReleaseBlue}1} & \textcolor{ReleaseMuted}{0} & \textcolor{ReleaseMuted}{0} & \textcolor{ReleaseMuted}{0} & \textcolor{ReleaseMuted}{0} & \textcolor{ReleaseMuted}{0} & \textcolor{ReleaseMuted}{0} & \textcolor{ReleaseMuted}{0} & \textcolor{ReleaseMuted}{0} & \textcolor{ReleaseMuted}{0} & \textcolor{ReleaseMuted}{0} & \textcolor{ReleaseMuted}{0} \\
dv-039 & L3 & \textcolor{ReleaseMuted}{0} & \textcolor{ReleaseMuted}{0} & \textcolor{ReleaseMuted}{0} & \textcolor{ReleaseMuted}{0} & \textcolor{ReleaseMuted}{0} & \textcolor{ReleaseMuted}{0} & \textcolor{ReleaseMuted}{0} & \textcolor{ReleaseMuted}{0} & \textcolor{ReleaseMuted}{0} & \textcolor{ReleaseMuted}{0} & \textcolor{ReleaseMuted}{0} & \textcolor{ReleaseMuted}{0} & \textcolor{ReleaseMuted}{0} \\
dv-040 & L3 & \textcolor{ReleaseMuted}{0} & \textcolor{ReleaseMuted}{0} & \textcolor{ReleaseMuted}{0} & \textcolor{ReleaseMuted}{0} & \textcolor{ReleaseMuted}{0} & \textcolor{ReleaseMuted}{0} & \textcolor{ReleaseMuted}{0} & \textcolor{ReleaseMuted}{0} & \textcolor{ReleaseMuted}{0} & \textcolor{ReleaseMuted}{0} & \textcolor{ReleaseMuted}{0} & \textcolor{ReleaseMuted}{0} & \textcolor{ReleaseMuted}{0} \\
dv-041 & L3 & \textcolor{ReleaseMuted}{0} & \textcolor{ReleaseMuted}{0} & \textcolor{ReleaseMuted}{0} & \textcolor{ReleaseMuted}{0} & \textcolor{ReleaseMuted}{0} & \textcolor{ReleaseMuted}{0} & \textcolor{ReleaseMuted}{0} & \textcolor{ReleaseMuted}{0} & \textcolor{ReleaseMuted}{0} & \textcolor{ReleaseMuted}{0} & \textcolor{ReleaseMuted}{0} & \textcolor{ReleaseMuted}{0} & \textcolor{ReleaseMuted}{0} \\
dv-042 & L3 & \textcolor{ReleaseMuted}{0} & \textcolor{ReleaseMuted}{0} & \textcolor{ReleaseMuted}{0} & \textcolor{ReleaseMuted}{0} & \textcolor{ReleaseMuted}{0} & \textcolor{ReleaseMuted}{0} & \textcolor{ReleaseMuted}{0} & \textcolor{ReleaseMuted}{0} & \textcolor{ReleaseMuted}{0} & \textcolor{ReleaseMuted}{0} & \textcolor{ReleaseMuted}{0} & \textcolor{ReleaseMuted}{0} & \textcolor{ReleaseMuted}{0} \\
dv-043 & L3 & \textcolor{ReleaseMuted}{0} & \textcolor{ReleaseMuted}{0} & \textcolor{ReleaseMuted}{0} & \textcolor{ReleaseMuted}{0} & \textcolor{ReleaseMuted}{0} & \textcolor{ReleaseMuted}{0} & \textcolor{ReleaseMuted}{0} & \textcolor{ReleaseMuted}{0} & \textcolor{ReleaseMuted}{0} & \textcolor{ReleaseMuted}{0} & \textcolor{ReleaseMuted}{0} & \textcolor{ReleaseMuted}{0} & \textcolor{ReleaseMuted}{0} \\
dv-044 & L3 & \textcolor{ReleaseMuted}{0} & \textcolor{ReleaseMuted}{0} & \textcolor{ReleaseMuted}{0} & \textcolor{ReleaseMuted}{0} & \textcolor{ReleaseMuted}{0} & \textcolor{ReleaseMuted}{0} & \textcolor{ReleaseMuted}{0} & \textcolor{ReleaseMuted}{0} & \textcolor{ReleaseMuted}{0} & \textcolor{ReleaseMuted}{0} & \textcolor{ReleaseMuted}{0} & \textcolor{ReleaseMuted}{0} & \textcolor{ReleaseMuted}{0} \\
dv-045 & L3 & \textcolor{ReleaseMuted}{0} & \textcolor{ReleaseMuted}{0} & \textcolor{ReleaseMuted}{0} & \cellcolor{ReleasePale}\textbf{\color{ReleaseBlue}1} & \textcolor{ReleaseMuted}{0} & \textcolor{ReleaseMuted}{0} & \textcolor{ReleaseMuted}{0} & \cellcolor{ReleasePale}\textbf{\color{ReleaseBlue}1} & \textcolor{ReleaseMuted}{0} & \textcolor{ReleaseMuted}{0} & \textcolor{ReleaseMuted}{0} & \textcolor{ReleaseMuted}{0} & \textcolor{ReleaseMuted}{0} \\
dv-046 & L4 & \textcolor{ReleaseMuted}{0} & \textcolor{ReleaseMuted}{0} & \textcolor{ReleaseMuted}{0} & \textcolor{ReleaseMuted}{0} & \textcolor{ReleaseMuted}{0} & \textcolor{ReleaseMuted}{0} & \textcolor{ReleaseMuted}{0} & \textcolor{ReleaseMuted}{0} & \textcolor{ReleaseMuted}{0} & \textcolor{ReleaseMuted}{0} & \textcolor{ReleaseMuted}{0} & \textcolor{ReleaseMuted}{0} & \textcolor{ReleaseMuted}{0} \\
dv-047 & L4 & \textcolor{ReleaseMuted}{0} & \cellcolor{ReleasePale}\textbf{\color{ReleaseBlue}1} & \cellcolor{ReleasePale}\textbf{\color{ReleaseBlue}1} & \cellcolor{ReleasePale}\textbf{\color{ReleaseBlue}1} & \cellcolor{ReleasePale}\textbf{\color{ReleaseBlue}1} & \textcolor{ReleaseMuted}{0} & \textcolor{ReleaseMuted}{0} & \cellcolor{ReleasePale}\textbf{\color{ReleaseBlue}1} & \cellcolor{ReleasePale}\textbf{\color{ReleaseBlue}1} & \textcolor{ReleaseMuted}{0} & \textcolor{ReleaseMuted}{0} & \textcolor{ReleaseMuted}{0} & \cellcolor{ReleasePale}\textbf{\color{ReleaseBlue}1} \\
dv-048 & L4 & \textcolor{ReleaseMuted}{0} & \textcolor{ReleaseMuted}{0} & \textcolor{ReleaseMuted}{0} & \textcolor{ReleaseMuted}{0} & \textcolor{ReleaseMuted}{0} & \textcolor{ReleaseMuted}{0} & \textcolor{ReleaseMuted}{0} & \textcolor{ReleaseMuted}{0} & \textcolor{ReleaseMuted}{0} & \textcolor{ReleaseMuted}{0} & \textcolor{ReleaseMuted}{0} & \textcolor{ReleaseMuted}{0} & \textcolor{ReleaseMuted}{0} \\
dv-049 & L4 & \textcolor{ReleaseMuted}{0} & \cellcolor{ReleasePale}\textbf{\color{ReleaseBlue}1} & \textcolor{ReleaseMuted}{0} & \textcolor{ReleaseMuted}{0} & \cellcolor{ReleasePale}\textbf{\color{ReleaseBlue}1} & \textcolor{ReleaseMuted}{0} & \textcolor{ReleaseMuted}{0} & \cellcolor{ReleasePale}\textbf{\color{ReleaseBlue}1} & \cellcolor{ReleasePale}\textbf{\color{ReleaseBlue}1} & \textcolor{ReleaseMuted}{0} & \textcolor{ReleaseMuted}{0} & \textcolor{ReleaseMuted}{0} & \textcolor{ReleaseMuted}{0} \\
dv-050 & L4 & \textcolor{ReleaseMuted}{0} & \textcolor{ReleaseMuted}{0} & \textcolor{ReleaseMuted}{0} & \textcolor{ReleaseMuted}{0} & \textcolor{ReleaseMuted}{0} & \textcolor{ReleaseMuted}{0} & \textcolor{ReleaseMuted}{0} & \textcolor{ReleaseMuted}{0} & \textcolor{ReleaseMuted}{0} & \textcolor{ReleaseMuted}{0} & \textcolor{ReleaseMuted}{0} & \textcolor{ReleaseMuted}{0} & \textcolor{ReleaseMuted}{0} \\
dv-051 & L4 & \textcolor{ReleaseMuted}{0} & \textcolor{ReleaseMuted}{0} & \textcolor{ReleaseMuted}{0} & \cellcolor{ReleasePale}\textbf{\color{ReleaseBlue}1} & \textcolor{ReleaseMuted}{0} & \textcolor{ReleaseMuted}{0} & \textcolor{ReleaseMuted}{0} & \textcolor{ReleaseMuted}{0} & \cellcolor{ReleasePale}\textbf{\color{ReleaseBlue}1} & \textcolor{ReleaseMuted}{0} & \textcolor{ReleaseMuted}{0} & \textcolor{ReleaseMuted}{0} & \textcolor{ReleaseMuted}{0} \\
dv-052 & L4 & \textcolor{ReleaseMuted}{0} & \textcolor{ReleaseMuted}{0} & \textcolor{ReleaseMuted}{0} & \textcolor{ReleaseMuted}{0} & \textcolor{ReleaseMuted}{0} & \textcolor{ReleaseMuted}{0} & \textcolor{ReleaseMuted}{0} & \textcolor{ReleaseMuted}{0} & \textcolor{ReleaseMuted}{0} & \textcolor{ReleaseMuted}{0} & \textcolor{ReleaseMuted}{0} & \textcolor{ReleaseMuted}{0} & \textcolor{ReleaseMuted}{0} \\
dv-053 & L4 & \textcolor{ReleaseMuted}{0} & \textcolor{ReleaseMuted}{0} & \textcolor{ReleaseMuted}{0} & \textcolor{ReleaseMuted}{0} & \textcolor{ReleaseMuted}{0} & \textcolor{ReleaseMuted}{0} & \textcolor{ReleaseMuted}{0} & \textcolor{ReleaseMuted}{0} & \textcolor{ReleaseMuted}{0} & \textcolor{ReleaseMuted}{0} & \textcolor{ReleaseMuted}{0} & \textcolor{ReleaseMuted}{0} & \textcolor{ReleaseMuted}{0} \\
dv-054 & L4 & \textcolor{ReleaseMuted}{0} & \textcolor{ReleaseMuted}{0} & \textcolor{ReleaseMuted}{0} & \textcolor{ReleaseMuted}{0} & \textcolor{ReleaseMuted}{0} & \textcolor{ReleaseMuted}{0} & \textcolor{ReleaseMuted}{0} & \textcolor{ReleaseMuted}{0} & \textcolor{ReleaseMuted}{0} & \textcolor{ReleaseMuted}{0} & \textcolor{ReleaseMuted}{0} & \textcolor{ReleaseMuted}{0} & \textcolor{ReleaseMuted}{0} \\
dv-055 & L4 & \textcolor{ReleaseMuted}{0} & \textcolor{ReleaseMuted}{0} & \textcolor{ReleaseMuted}{0} & \textcolor{ReleaseMuted}{0} & \textcolor{ReleaseMuted}{0} & \textcolor{ReleaseMuted}{0} & \textcolor{ReleaseMuted}{0} & \textcolor{ReleaseMuted}{0} & \textcolor{ReleaseMuted}{0} & \textcolor{ReleaseMuted}{0} & \textcolor{ReleaseMuted}{0} & \textcolor{ReleaseMuted}{0} & \textcolor{ReleaseMuted}{0} \\
dv-056 & L4 & \textcolor{ReleaseMuted}{0} & \textcolor{ReleaseMuted}{0} & \textcolor{ReleaseMuted}{0} & \cellcolor{ReleasePale}\textbf{\color{ReleaseBlue}1} & \cellcolor{ReleasePale}\textbf{\color{ReleaseBlue}1} & \textcolor{ReleaseMuted}{0} & \textcolor{ReleaseMuted}{0} & \textcolor{ReleaseMuted}{0} & \cellcolor{ReleasePale}\textbf{\color{ReleaseBlue}1} & \textcolor{ReleaseMuted}{0} & \textcolor{ReleaseMuted}{0} & \cellcolor{ReleasePale}\textbf{\color{ReleaseBlue}1} & \cellcolor{ReleasePale}\textbf{\color{ReleaseBlue}1} \\
dv-057 & L4 & \textcolor{ReleaseMuted}{0} & \textcolor{ReleaseMuted}{0} & \textcolor{ReleaseMuted}{0} & \textcolor{ReleaseMuted}{0} & \textcolor{ReleaseMuted}{0} & \textcolor{ReleaseMuted}{0} & \textcolor{ReleaseMuted}{0} & \textcolor{ReleaseMuted}{0} & \textcolor{ReleaseMuted}{0} & \textcolor{ReleaseMuted}{0} & \textcolor{ReleaseMuted}{0} & \textcolor{ReleaseMuted}{0} & \textcolor{ReleaseMuted}{0} \\
dv-058 & L4 & \textcolor{ReleaseMuted}{0} & \textcolor{ReleaseMuted}{0} & \textcolor{ReleaseMuted}{0} & \textcolor{ReleaseMuted}{0} & \textcolor{ReleaseMuted}{0} & \textcolor{ReleaseMuted}{0} & \textcolor{ReleaseMuted}{0} & \textcolor{ReleaseMuted}{0} & \textcolor{ReleaseMuted}{0} & \textcolor{ReleaseMuted}{0} & \textcolor{ReleaseMuted}{0} & \textcolor{ReleaseMuted}{0} & \textcolor{ReleaseMuted}{0} \\
dv-059 & L4 & \textcolor{ReleaseMuted}{0} & \textcolor{ReleaseMuted}{0} & \textcolor{ReleaseMuted}{0} & \textcolor{ReleaseMuted}{0} & \textcolor{ReleaseMuted}{0} & \textcolor{ReleaseMuted}{0} & \textcolor{ReleaseMuted}{0} & \textcolor{ReleaseMuted}{0} & \textcolor{ReleaseMuted}{0} & \textcolor{ReleaseMuted}{0} & \textcolor{ReleaseMuted}{0} & \textcolor{ReleaseMuted}{0} & \textcolor{ReleaseMuted}{0} \\
dv-060 & L4 & \textcolor{ReleaseMuted}{0} & \cellcolor{ReleasePale}\textbf{\color{ReleaseBlue}1} & \cellcolor{ReleasePale}\textbf{\color{ReleaseBlue}1} & \cellcolor{ReleasePale}\textbf{\color{ReleaseBlue}1} & \cellcolor{ReleasePale}\textbf{\color{ReleaseBlue}1} & \textcolor{ReleaseMuted}{0} & \textcolor{ReleaseMuted}{0} & \textcolor{ReleaseMuted}{0} & \textcolor{ReleaseMuted}{0} & \textcolor{ReleaseMuted}{0} & \textcolor{ReleaseMuted}{0} & \cellcolor{ReleasePale}\textbf{\color{ReleaseBlue}1} & \textcolor{ReleaseMuted}{0} \\
dv-061 & L4 & \textcolor{ReleaseMuted}{0} & \textcolor{ReleaseMuted}{0} & \cellcolor{ReleasePale}\textbf{\color{ReleaseBlue}1} & \cellcolor{ReleasePale}\textbf{\color{ReleaseBlue}1} & \textcolor{ReleaseMuted}{0} & \textcolor{ReleaseMuted}{0} & \textcolor{ReleaseMuted}{0} & \cellcolor{ReleasePale}\textbf{\color{ReleaseBlue}1} & \cellcolor{ReleasePale}\textbf{\color{ReleaseBlue}1} & \textcolor{ReleaseMuted}{0} & \textcolor{ReleaseMuted}{0} & \cellcolor{ReleasePale}\textbf{\color{ReleaseBlue}1} & \cellcolor{ReleasePale}\textbf{\color{ReleaseBlue}1} \\
dv-062 & L4 & \textcolor{ReleaseMuted}{0} & \textcolor{ReleaseMuted}{0} & \textcolor{ReleaseMuted}{0} & \textcolor{ReleaseMuted}{0} & \textcolor{ReleaseMuted}{0} & \textcolor{ReleaseMuted}{0} & \textcolor{ReleaseMuted}{0} & \textcolor{ReleaseMuted}{0} & \textcolor{ReleaseMuted}{0} & \textcolor{ReleaseMuted}{0} & \textcolor{ReleaseMuted}{0} & \textcolor{ReleaseMuted}{0} & \textcolor{ReleaseMuted}{0} \\
dv-063 & L4 & \textcolor{ReleaseMuted}{0} & \textcolor{ReleaseMuted}{0} & \textcolor{ReleaseMuted}{0} & \textcolor{ReleaseMuted}{0} & \textcolor{ReleaseMuted}{0} & \textcolor{ReleaseMuted}{0} & \textcolor{ReleaseMuted}{0} & \textcolor{ReleaseMuted}{0} & \textcolor{ReleaseMuted}{0} & \textcolor{ReleaseMuted}{0} & \textcolor{ReleaseMuted}{0} & \textcolor{ReleaseMuted}{0} & \textcolor{ReleaseMuted}{0} \\
dv-064 & L4 & \textcolor{ReleaseMuted}{0} & \textcolor{ReleaseMuted}{0} & \textcolor{ReleaseMuted}{0} & \textcolor{ReleaseMuted}{0} & \textcolor{ReleaseMuted}{0} & \textcolor{ReleaseMuted}{0} & \textcolor{ReleaseMuted}{0} & \textcolor{ReleaseMuted}{0} & \textcolor{ReleaseMuted}{0} & \textcolor{ReleaseMuted}{0} & \textcolor{ReleaseMuted}{0} & \textcolor{ReleaseMuted}{0} & \textcolor{ReleaseMuted}{0} \\
dv-065 & L4 & \textcolor{ReleaseMuted}{0} & \textcolor{ReleaseMuted}{0} & \textcolor{ReleaseMuted}{0} & \textcolor{ReleaseMuted}{0} & \textcolor{ReleaseMuted}{0} & \textcolor{ReleaseMuted}{0} & \textcolor{ReleaseMuted}{0} & \textcolor{ReleaseMuted}{0} & \textcolor{ReleaseMuted}{0} & \textcolor{ReleaseMuted}{0} & \textcolor{ReleaseMuted}{0} & \textcolor{ReleaseMuted}{0} & \textcolor{ReleaseMuted}{0} \\
dv-066 & L4 & \textcolor{ReleaseMuted}{0} & \textcolor{ReleaseMuted}{0} & \textcolor{ReleaseMuted}{0} & \textcolor{ReleaseMuted}{0} & \textcolor{ReleaseMuted}{0} & \textcolor{ReleaseMuted}{0} & \textcolor{ReleaseMuted}{0} & \textcolor{ReleaseMuted}{0} & \textcolor{ReleaseMuted}{0} & \textcolor{ReleaseMuted}{0} & \textcolor{ReleaseMuted}{0} & \textcolor{ReleaseMuted}{0} & \textcolor{ReleaseMuted}{0} \\
dv-067 & L4 & \textcolor{ReleaseMuted}{0} & \textcolor{ReleaseMuted}{0} & \textcolor{ReleaseMuted}{0} & \textcolor{ReleaseMuted}{0} & \textcolor{ReleaseMuted}{0} & \textcolor{ReleaseMuted}{0} & \textcolor{ReleaseMuted}{0} & \textcolor{ReleaseMuted}{0} & \textcolor{ReleaseMuted}{0} & \textcolor{ReleaseMuted}{0} & \textcolor{ReleaseMuted}{0} & \textcolor{ReleaseMuted}{0} & \textcolor{ReleaseMuted}{0} \\
dv-068 & L4 & \textcolor{ReleaseMuted}{0} & \textcolor{ReleaseMuted}{0} & \textcolor{ReleaseMuted}{0} & \textcolor{ReleaseMuted}{0} & \textcolor{ReleaseMuted}{0} & \textcolor{ReleaseMuted}{0} & \textcolor{ReleaseMuted}{0} & \textcolor{ReleaseMuted}{0} & \textcolor{ReleaseMuted}{0} & \textcolor{ReleaseMuted}{0} & \textcolor{ReleaseMuted}{0} & \textcolor{ReleaseMuted}{0} & \textcolor{ReleaseMuted}{0} \\
dv-069 & L4 & \textcolor{ReleaseMuted}{0} & \textcolor{ReleaseMuted}{0} & \textcolor{ReleaseMuted}{0} & \textcolor{ReleaseMuted}{0} & \textcolor{ReleaseMuted}{0} & \textcolor{ReleaseMuted}{0} & \textcolor{ReleaseMuted}{0} & \textcolor{ReleaseMuted}{0} & \textcolor{ReleaseMuted}{0} & \textcolor{ReleaseMuted}{0} & \textcolor{ReleaseMuted}{0} & \textcolor{ReleaseMuted}{0} & \textcolor{ReleaseMuted}{0} \\
dv-070 & L4 & \textcolor{ReleaseMuted}{0} & \textcolor{ReleaseMuted}{0} & \textcolor{ReleaseMuted}{0} & \textcolor{ReleaseMuted}{0} & \textcolor{ReleaseMuted}{0} & \textcolor{ReleaseMuted}{0} & \textcolor{ReleaseMuted}{0} & \textcolor{ReleaseMuted}{0} & \textcolor{ReleaseMuted}{0} & \textcolor{ReleaseMuted}{0} & \textcolor{ReleaseMuted}{0} & \textcolor{ReleaseMuted}{0} & \textcolor{ReleaseMuted}{0} \\
dv-071 & L4 & \textcolor{ReleaseMuted}{0} & \textcolor{ReleaseMuted}{0} & \textcolor{ReleaseMuted}{0} & \textcolor{ReleaseMuted}{0} & \textcolor{ReleaseMuted}{0} & \textcolor{ReleaseMuted}{0} & \textcolor{ReleaseMuted}{0} & \textcolor{ReleaseMuted}{0} & \textcolor{ReleaseMuted}{0} & \textcolor{ReleaseMuted}{0} & \textcolor{ReleaseMuted}{0} & \textcolor{ReleaseMuted}{0} & \textcolor{ReleaseMuted}{0} \\
dv-072 & L4 & \textcolor{ReleaseMuted}{0} & \textcolor{ReleaseMuted}{0} & \textcolor{ReleaseMuted}{0} & \textcolor{ReleaseMuted}{0} & \textcolor{ReleaseMuted}{0} & \textcolor{ReleaseMuted}{0} & \textcolor{ReleaseMuted}{0} & \textcolor{ReleaseMuted}{0} & \textcolor{ReleaseMuted}{0} & \textcolor{ReleaseMuted}{0} & \textcolor{ReleaseMuted}{0} & \textcolor{ReleaseMuted}{0} & \textcolor{ReleaseMuted}{0} \\
dv-073 & L4 & \textcolor{ReleaseMuted}{0} & \textcolor{ReleaseMuted}{0} & \textcolor{ReleaseMuted}{0} & \textcolor{ReleaseMuted}{0} & \textcolor{ReleaseMuted}{0} & \textcolor{ReleaseMuted}{0} & \textcolor{ReleaseMuted}{0} & \textcolor{ReleaseMuted}{0} & \textcolor{ReleaseMuted}{0} & \textcolor{ReleaseMuted}{0} & \textcolor{ReleaseMuted}{0} & \textcolor{ReleaseMuted}{0} & \textcolor{ReleaseMuted}{0} \\
dv-074 & L4 & \textcolor{ReleaseMuted}{0} & \textcolor{ReleaseMuted}{0} & \textcolor{ReleaseMuted}{0} & \textcolor{ReleaseMuted}{0} & \textcolor{ReleaseMuted}{0} & \textcolor{ReleaseMuted}{0} & \textcolor{ReleaseMuted}{0} & \textcolor{ReleaseMuted}{0} & \textcolor{ReleaseMuted}{0} & \textcolor{ReleaseMuted}{0} & \textcolor{ReleaseMuted}{0} & \textcolor{ReleaseMuted}{0} & \textcolor{ReleaseMuted}{0} \\
dv-075 & L5 & \textcolor{ReleaseMuted}{0} & \textcolor{ReleaseMuted}{0} & \textcolor{ReleaseMuted}{0} & \textcolor{ReleaseMuted}{0} & \textcolor{ReleaseMuted}{0} & \textcolor{ReleaseMuted}{0} & \textcolor{ReleaseMuted}{0} & \textcolor{ReleaseMuted}{0} & \textcolor{ReleaseMuted}{0} & \textcolor{ReleaseMuted}{0} & \textcolor{ReleaseMuted}{0} & \textcolor{ReleaseMuted}{0} & \textcolor{ReleaseMuted}{0} \\
dv-076 & L5 & \textcolor{ReleaseMuted}{0} & \textcolor{ReleaseMuted}{0} & \textcolor{ReleaseMuted}{0} & \textcolor{ReleaseMuted}{0} & \textcolor{ReleaseMuted}{0} & \textcolor{ReleaseMuted}{0} & \textcolor{ReleaseMuted}{0} & \textcolor{ReleaseMuted}{0} & \textcolor{ReleaseMuted}{0} & \textcolor{ReleaseMuted}{0} & \textcolor{ReleaseMuted}{0} & \textcolor{ReleaseMuted}{0} & \textcolor{ReleaseMuted}{0} \\
dv-077 & L5 & \textcolor{ReleaseMuted}{0} & \textcolor{ReleaseMuted}{0} & \textcolor{ReleaseMuted}{0} & \textcolor{ReleaseMuted}{0} & \textcolor{ReleaseMuted}{0} & \textcolor{ReleaseMuted}{0} & \textcolor{ReleaseMuted}{0} & \textcolor{ReleaseMuted}{0} & \textcolor{ReleaseMuted}{0} & \textcolor{ReleaseMuted}{0} & \textcolor{ReleaseMuted}{0} & \textcolor{ReleaseMuted}{0} & \textcolor{ReleaseMuted}{0} \\
dv-078 & L5 & \textcolor{ReleaseMuted}{0} & \textcolor{ReleaseMuted}{0} & \textcolor{ReleaseMuted}{0} & \textcolor{ReleaseMuted}{0} & \textcolor{ReleaseMuted}{0} & \textcolor{ReleaseMuted}{0} & \textcolor{ReleaseMuted}{0} & \textcolor{ReleaseMuted}{0} & \textcolor{ReleaseMuted}{0} & \textcolor{ReleaseMuted}{0} & \textcolor{ReleaseMuted}{0} & \textcolor{ReleaseMuted}{0} & \textcolor{ReleaseMuted}{0} \\
dv-079 & L5 & \textcolor{ReleaseMuted}{0} & \textcolor{ReleaseMuted}{0} & \textcolor{ReleaseMuted}{0} & \textcolor{ReleaseMuted}{0} & \textcolor{ReleaseMuted}{0} & \textcolor{ReleaseMuted}{0} & \textcolor{ReleaseMuted}{0} & \textcolor{ReleaseMuted}{0} & \textcolor{ReleaseMuted}{0} & \textcolor{ReleaseMuted}{0} & \textcolor{ReleaseMuted}{0} & \textcolor{ReleaseMuted}{0} & \textcolor{ReleaseMuted}{0} \\
dv-080 & L5 & \textcolor{ReleaseMuted}{0} & \textcolor{ReleaseMuted}{0} & \textcolor{ReleaseMuted}{0} & \textcolor{ReleaseMuted}{0} & \textcolor{ReleaseMuted}{0} & \textcolor{ReleaseMuted}{0} & \textcolor{ReleaseMuted}{0} & \textcolor{ReleaseMuted}{0} & \textcolor{ReleaseMuted}{0} & \textcolor{ReleaseMuted}{0} & \textcolor{ReleaseMuted}{0} & \textcolor{ReleaseMuted}{0} & \textcolor{ReleaseMuted}{0} \\
dv-081 & L5 & \textcolor{ReleaseMuted}{0} & \textcolor{ReleaseMuted}{0} & \textcolor{ReleaseMuted}{0} & \textcolor{ReleaseMuted}{0} & \textcolor{ReleaseMuted}{0} & \textcolor{ReleaseMuted}{0} & \textcolor{ReleaseMuted}{0} & \textcolor{ReleaseMuted}{0} & \textcolor{ReleaseMuted}{0} & \textcolor{ReleaseMuted}{0} & \textcolor{ReleaseMuted}{0} & \textcolor{ReleaseMuted}{0} & \textcolor{ReleaseMuted}{0} \\
dv-082 & L5 & \textcolor{ReleaseMuted}{0} & \cellcolor{ReleasePale}\textbf{\color{ReleaseBlue}1} & \cellcolor{ReleasePale}\textbf{\color{ReleaseBlue}1} & \textcolor{ReleaseMuted}{0} & \cellcolor{ReleasePale}\textbf{\color{ReleaseBlue}1} & \textcolor{ReleaseMuted}{0} & \textcolor{ReleaseMuted}{0} & \cellcolor{ReleasePale}\textbf{\color{ReleaseBlue}1} & \cellcolor{ReleasePale}\textbf{\color{ReleaseBlue}1} & \textcolor{ReleaseMuted}{0} & \textcolor{ReleaseMuted}{0} & \cellcolor{ReleasePale}\textbf{\color{ReleaseBlue}1} & \cellcolor{ReleasePale}\textbf{\color{ReleaseBlue}1} \\
dv-083 & L5 & \textcolor{ReleaseMuted}{0} & \textcolor{ReleaseMuted}{0} & \cellcolor{ReleasePale}\textbf{\color{ReleaseBlue}1} & \textcolor{ReleaseMuted}{0} & \cellcolor{ReleasePale}\textbf{\color{ReleaseBlue}1} & \textcolor{ReleaseMuted}{0} & \cellcolor{ReleasePale}\textbf{\color{ReleaseBlue}1} & \cellcolor{ReleasePale}\textbf{\color{ReleaseBlue}1} & \cellcolor{ReleasePale}\textbf{\color{ReleaseBlue}1} & \textcolor{ReleaseMuted}{0} & \textcolor{ReleaseMuted}{0} & \textcolor{ReleaseMuted}{0} & \cellcolor{ReleasePale}\textbf{\color{ReleaseBlue}1} \\
dv-084 & L5 & \textcolor{ReleaseMuted}{0} & \textcolor{ReleaseMuted}{0} & \textcolor{ReleaseMuted}{0} & \textcolor{ReleaseMuted}{0} & \textcolor{ReleaseMuted}{0} & \textcolor{ReleaseMuted}{0} & \textcolor{ReleaseMuted}{0} & \textcolor{ReleaseMuted}{0} & \textcolor{ReleaseMuted}{0} & \textcolor{ReleaseMuted}{0} & \textcolor{ReleaseMuted}{0} & \textcolor{ReleaseMuted}{0} & \textcolor{ReleaseMuted}{0} \\
dv-085 & L5 & \textcolor{ReleaseMuted}{0} & \textcolor{ReleaseMuted}{0} & \textcolor{ReleaseMuted}{0} & \textcolor{ReleaseMuted}{0} & \textcolor{ReleaseMuted}{0} & \textcolor{ReleaseMuted}{0} & \textcolor{ReleaseMuted}{0} & \textcolor{ReleaseMuted}{0} & \textcolor{ReleaseMuted}{0} & \textcolor{ReleaseMuted}{0} & \textcolor{ReleaseMuted}{0} & \textcolor{ReleaseMuted}{0} & \textcolor{ReleaseMuted}{0} \\
dv-086 & L5 & \textcolor{ReleaseMuted}{0} & \textcolor{ReleaseMuted}{0} & \textcolor{ReleaseMuted}{0} & \textcolor{ReleaseMuted}{0} & \textcolor{ReleaseMuted}{0} & \textcolor{ReleaseMuted}{0} & \textcolor{ReleaseMuted}{0} & \textcolor{ReleaseMuted}{0} & \textcolor{ReleaseMuted}{0} & \textcolor{ReleaseMuted}{0} & \textcolor{ReleaseMuted}{0} & \textcolor{ReleaseMuted}{0} & \textcolor{ReleaseMuted}{0} \\
dv-087 & L5 & \textcolor{ReleaseMuted}{0} & \textcolor{ReleaseMuted}{0} & \textcolor{ReleaseMuted}{0} & \textcolor{ReleaseMuted}{0} & \textcolor{ReleaseMuted}{0} & \textcolor{ReleaseMuted}{0} & \textcolor{ReleaseMuted}{0} & \textcolor{ReleaseMuted}{0} & \textcolor{ReleaseMuted}{0} & \textcolor{ReleaseMuted}{0} & \textcolor{ReleaseMuted}{0} & \textcolor{ReleaseMuted}{0} & \textcolor{ReleaseMuted}{0} \\
dv-088 & L5 & \textcolor{ReleaseMuted}{0} & \textcolor{ReleaseMuted}{0} & \textcolor{ReleaseMuted}{0} & \textcolor{ReleaseMuted}{0} & \textcolor{ReleaseMuted}{0} & \textcolor{ReleaseMuted}{0} & \textcolor{ReleaseMuted}{0} & \textcolor{ReleaseMuted}{0} & \textcolor{ReleaseMuted}{0} & \textcolor{ReleaseMuted}{0} & \textcolor{ReleaseMuted}{0} & \textcolor{ReleaseMuted}{0} & \textcolor{ReleaseMuted}{0} \\
dv-089 & L5 & \textcolor{ReleaseMuted}{0} & \textcolor{ReleaseMuted}{0} & \textcolor{ReleaseMuted}{0} & \textcolor{ReleaseMuted}{0} & \textcolor{ReleaseMuted}{0} & \textcolor{ReleaseMuted}{0} & \textcolor{ReleaseMuted}{0} & \textcolor{ReleaseMuted}{0} & \textcolor{ReleaseMuted}{0} & \textcolor{ReleaseMuted}{0} & \textcolor{ReleaseMuted}{0} & \textcolor{ReleaseMuted}{0} & \textcolor{ReleaseMuted}{0} \\
dv-090 & L5 & \textcolor{ReleaseMuted}{0} & \textcolor{ReleaseMuted}{0} & \textcolor{ReleaseMuted}{0} & \textcolor{ReleaseMuted}{0} & \textcolor{ReleaseMuted}{0} & \textcolor{ReleaseMuted}{0} & \textcolor{ReleaseMuted}{0} & \textcolor{ReleaseMuted}{0} & \textcolor{ReleaseMuted}{0} & \textcolor{ReleaseMuted}{0} & \textcolor{ReleaseMuted}{0} & \textcolor{ReleaseMuted}{0} & \textcolor{ReleaseMuted}{0} \\
dv-091 & L5 & \textcolor{ReleaseMuted}{0} & \textcolor{ReleaseMuted}{0} & \cellcolor{ReleasePale}\textbf{\color{ReleaseBlue}1} & \textcolor{ReleaseMuted}{0} & \textcolor{ReleaseMuted}{0} & \textcolor{ReleaseMuted}{0} & \textcolor{ReleaseMuted}{0} & \textcolor{ReleaseMuted}{0} & \textcolor{ReleaseMuted}{0} & \textcolor{ReleaseMuted}{0} & \textcolor{ReleaseMuted}{0} & \textcolor{ReleaseMuted}{0} & \textcolor{ReleaseMuted}{0} \\
dv-092 & L5 & \textcolor{ReleaseMuted}{0} & \textcolor{ReleaseMuted}{0} & \textcolor{ReleaseMuted}{0} & \textcolor{ReleaseMuted}{0} & \textcolor{ReleaseMuted}{0} & \textcolor{ReleaseMuted}{0} & \textcolor{ReleaseMuted}{0} & \textcolor{ReleaseMuted}{0} & \textcolor{ReleaseMuted}{0} & \textcolor{ReleaseMuted}{0} & \textcolor{ReleaseMuted}{0} & \textcolor{ReleaseMuted}{0} & \textcolor{ReleaseMuted}{0} \\
dv-093 & L5 & \textcolor{ReleaseMuted}{0} & \textcolor{ReleaseMuted}{0} & \textcolor{ReleaseMuted}{0} & \textcolor{ReleaseMuted}{0} & \textcolor{ReleaseMuted}{0} & \textcolor{ReleaseMuted}{0} & \textcolor{ReleaseMuted}{0} & \textcolor{ReleaseMuted}{0} & \textcolor{ReleaseMuted}{0} & \textcolor{ReleaseMuted}{0} & \textcolor{ReleaseMuted}{0} & \textcolor{ReleaseMuted}{0} & \textcolor{ReleaseMuted}{0} \\
dv-094 & L5 & \textcolor{ReleaseMuted}{0} & \textcolor{ReleaseMuted}{0} & \textcolor{ReleaseMuted}{0} & \textcolor{ReleaseMuted}{0} & \textcolor{ReleaseMuted}{0} & \textcolor{ReleaseMuted}{0} & \textcolor{ReleaseMuted}{0} & \textcolor{ReleaseMuted}{0} & \textcolor{ReleaseMuted}{0} & \textcolor{ReleaseMuted}{0} & \textcolor{ReleaseMuted}{0} & \textcolor{ReleaseMuted}{0} & \textcolor{ReleaseMuted}{0} \\
dv-095 & L5 & \textcolor{ReleaseMuted}{0} & \textcolor{ReleaseMuted}{0} & \textcolor{ReleaseMuted}{0} & \textcolor{ReleaseMuted}{0} & \textcolor{ReleaseMuted}{0} & \textcolor{ReleaseMuted}{0} & \textcolor{ReleaseMuted}{0} & \textcolor{ReleaseMuted}{0} & \textcolor{ReleaseMuted}{0} & \textcolor{ReleaseMuted}{0} & \textcolor{ReleaseMuted}{0} & \textcolor{ReleaseMuted}{0} & \textcolor{ReleaseMuted}{0} \\
\end{longtable}\endgroup

\clearpage
\section{Task Instructions and Submission Contract}
\label{app:instructions}
\subsection{The public task package}
An assignment provides a behavioral specification, the native interface and fixed parameter profile, a submission contract, and a golden-design debugging interface. The agent constructs its own verification artifacts without seeing implementation RTL or maintenance-side reference answers. The examples below describe the released public package; later sections reproduce explicitly labeled reference documents to explain what a concrete verification plan looks like.

\begin{AppendixCard}{Public inputs and required outputs}
\textbf{Inputs:} \texttt{instruction.md}, \texttt{public/spec.md}, \texttt{public/ports.md}, \texttt{public/submission\_format.md}, \texttt{submission\_contract.json}, and the public feedback/budget descriptions.\par\smallskip
\textbf{Documents:} \texttt{flow/features.json} and \texttt{flow/features.md}; \texttt{flow/testplan.json} and \texttt{flow/testplan.md}.\par\smallskip
\textbf{Executable artifacts:} a self-contained \texttt{model/model.sv}, the complete UVM environment, and dependency-ordered \texttt{sources.f}. The native model is reused for independent model assessment and scoreboard prediction.\par\smallskip
\textbf{Entry points:} UVM 1.2, top \texttt{dvbench\_top}, test \texttt{dvbench\_test}, and independent module \texttt{dvbench\_model}.
\end{AppendixCard}

\subsection{Document fields and traceability}
Feature documents state \emph{what} is observable; plans state \emph{how} to establish, stimulate, sample, and check it. JSON is the primary evidence carrier. Markdown presents the same IDs, references, and facts, and cannot repair a fact missing from JSON. IDs are links, not a prescribed ontology or a target number of entries. Equivalent splitting or merging is permitted by the assessment protocol.

\begin{AppendixCard}{Feature-list schema}
The root contains a nonempty \texttt{features} array. Each feature has nonempty strings \texttt{feature\_id}, \texttt{module}, \texttt{definition}, \texttt{timing}, and \texttt{boundary}; nonempty arrays \texttt{requirement\_ids} and \texttt{interface}; and optionally \texttt{parent\_id}. Feature IDs are unique, requirement IDs refer to the public specification, interface names are native ports, and parent references must be valid and acyclic.
\end{AppendixCard}
\begin{AppendixCard}{Test-plan schema}
The root contains a nonempty \texttt{tests} array. Each test has nonempty strings \texttt{test\_id}, \texttt{stimulus}, \texttt{expected}, \texttt{checker}, \texttt{implementation\_id}, and \texttt{source}, plus nonempty arrays \texttt{feature\_ids} and \texttt{requirement\_ids}. The semantic fields jointly state legal preparation, the actual drive domain, independent expected behavior, valid sampling, comparison, and mismatch reporting. Implementation fields link the plan to submitted checks.
\end{AppendixCard}

\clearpage
\subsection{JSON format examples supplied with the task}
The public submission-format document uses the fictional \texttt{lamp\_latch} module to demonstrate the schema without supplying an answer to the target design. The following paired records preserve its original field values and are only reformatted as indented JSON. The feature describes capture/hold behavior; the test first establishes a known state, then exercises capture and hold with an independent checker. These are format examples, distinct from the real-task reference examples in Appendices~\ref{app:fifo} and~\ref{app:advanced}.
\par\noindent\begin{minipage}{\linewidth}
\begin{lstlisting}[style=appendixcode]
{
  "features": [
    {
      "feature_id": "F-L",
      "requirement_ids": [
        "R-example-1"
      ],
      "module": "lamp_latch",
      "definition": "When enable is 1, capture the lamp input d on the rising edge of clk; otherwise retain q.",
      "interface": [
        "clk",
        "enable",
        "d",
        "q"
      ],
      "timing": "Sample on the rising edge; q is valid after the edge.",
      "boundary": "Hold when enable=0; inputs are two-state."
    }
  ]
}
\end{lstlisting}
\end{minipage}\par
\noindent\begin{minipage}{\linewidth}
\begin{lstlisting}[style=appendixcode]
{
  "tests": [
    {
      "test_id": "T-L",
      "feature_ids": [
        "F-L"
      ],
      "requirement_ids": [
        "R-example-1"
      ],
      "stimulus": "First set enable=1,d=0 on a falling edge and pass through a rising edge to establish q=0; then set d=1 on a falling edge and pass through a rising edge; finally set enable=0,d=0 and pass through another rising edge.",
      "expected": "q after the three edges is 0, 1, and 1, respectively.",
      "checker": "Build an independent capture/hold model from the drive history. After each edge, when stable, compare actual q with the prediction using case equality; report UVM_ERROR on a mismatch. Confirm that all three comparisons are triggered.",
      "implementation_id": "V-L",
      "source": "lamp_pkg.sv:lamp_scoreboard.write"
    }
  ]
}
\end{lstlisting}
\end{minipage}\par

\subsection{Complete instantiated instruction: CVA6 FIFO}
The following is the complete body of the released \texttt{dv-009} instruction, with its Markdown heading and inline-code typography converted for print. It is an instantiated task template, not a reconstructed prompt from an agent trajectory. Its opening paragraph supplies task-specific ports and parameters; the remaining sections describe the shared deliverables, independence rules, execution interface, and completion criteria. Embedded commands are text from the task package.
\clearpage
\begin{AppendixCard}{Task assignment and deliverables}
Read \texttt{public/spec.md}, \texttt{public/ports.md}, \texttt{public/submission\_format.md}, and \texttt{submission\_contract.json}. The native DUT top is \texttt{cva6\_fifo\_v3}. Use fixed DEPTH=8, FALL\_THROUGH=1, DATA\_WIDTH=32, dtype=logic[31:0], FPGA\_EN=0, and FPGA\_ALTERA=0; ADDR\_DEPTH=3 is derived. Preserve usage\_o [2:0] and data\_i/data\_o [31:0]. clk\_i updates state on rising edges; rst\_ni is asynchronous active-low and flush\_i is a synchronous clear input. RTL is hidden. Preserve the exact native port names, directions, widths, packed and unpacked index directions, and parameter profile described in the interface contract.\par\medskip
Submit \texttt{flow/features.md}, \texttt{flow/features.json}, \texttt{flow/testplan.md}, \texttt{flow/testplan.json}, self-contained \texttt{model/model.sv}, complete UVM source code, and \texttt{sources.f}. Features describe what must be verified; the testplan explains the legal preparation and stimulus, independent expected behavior, valid sampling, and failure checks. JSON must contain all facts; Markdown must present the same IDs, references, and meaning and cannot supply facts missing from JSON. Follow the field types, references, and unrelated paired examples in the submission format. Reasonable feature and test grouping is allowed; names and entry counts do not establish correctness.\par\medskip
The feature JSON root is an object with a nonempty \texttt{features} array. Each entry has nonempty strings \texttt{feature\_id}, \texttt{module}, \texttt{definition}, \texttt{timing}, and \texttt{boundary}, and nonempty arrays of distinct strings \texttt{requirement\_ids} and \texttt{interface}. Use unique feature IDs, public R-ID references, and native port names. Optional \texttt{parent\_id} references an existing feature or is null, without cycles. The plan JSON root is an object with a nonempty \texttt{tests} array. Each entry has nonempty strings \texttt{test\_id}, \texttt{stimulus}, \texttt{expected}, \texttt{checker}, \texttt{implementation\_id}, and \texttt{source}, and nonempty arrays of distinct strings \texttt{feature\_ids} and \texttt{requirement\_ids}. Use unique test IDs and valid references to your features and public requirements; \texttt{source} and \texttt{implementation\_id} locate real submitted checks. State preparation, actual stimulus domain, expected relationships, sampling, and mismatch reporting concretely in the existing fields. Contradictory Markdown/JSON facts under the same conditions and window are errors for the affected obligations.\par\medskip
Use UVM 1.2 with top \texttt{dvbench\_top} and test \texttt{dvbench\_test}. Implement the stimulus, driver, monitor, environment, independent predictor, scoreboard, checks, and coverage needed for the complete environment. Connect the independent expected values and DUT observations to actual/expected scoreboard comparisons. Report failures with \texttt{UVM\_ERROR} or \texttt{UVM\_FATAL}. After legal stimulus and checking finish, in-flight observations drain, and ERROR/FATAL counts are zero, emit \texttt{DVBench\_COMPLETED} exactly once. Completion and coverage do not replace effective checking.\par\medskip
Interact with the DUT through its native ports. Do not access internal hierarchy, use force/release or DPI, submit a DUT copy, or probe hidden files, the network, or the operating system. \texttt{sources.f} must list one submission-relative \texttt{.sv} or \texttt{.v} path per line in dependency order and include \texttt{model/model.sv}; it must contain no compiler options, absolute paths, parent traversal, or wildcards. Follow the submission format for permitted includes; keep submitted HDL dependencies in compilation order. Preserve the clock, reset, bidirectional-net, and indexing rules in the native interface contract.\par\medskip
\end{AppendixCard}
\begin{AppendixCard}{Mandatory Independent Model Entry Point}
\texttt{model/model.sv} must define the self-contained SystemVerilog module \texttt{dvbench\_model} with the DUT's native ports and contracted parameter profile, including the exact directions, widths, and index directions. Use this identical file for independent model execution and prediction in the UVM scoreboard. It must work without UVM or other submitted source files. Predict only from reset, stimulus and its history. Do not instantiate the DUT, copy its implementation, use DUT actual outputs as the expected result, or replay hidden expected outputs. Wire expected outputs separately from actual observations. Classes, packages, functions, protocol checks, or a disconnected model alone do not satisfy the module and scoreboard contract.\par\medskip
\end{AppendixCard}
\begin{AppendixCard}{Debugging}
Read \texttt{public/debug\_feedback.md}. Submit a snapshot with \texttt{bash /public/environment/run.sh simulate --submission /submission --output /tmp/feedback-01.json}. Keep feedback outside the submission directory. The service uses the golden design and seed 1; mutation results are available only after the submission is frozen.\par\medskip
\end{AppendixCard}
\begin{AppendixCard}{Execution budget and completion}
You have at most \textbf{300 cumulative agent decision turns}, with a delivery reminder at turn \textbf{250}. The budget persists across compaction and recovery of the same run. There is no whole-task wall-time or cumulative token limit. Compilation, simulation, and individual model requests have separate limits; read \texttt{public/deployment\_budget.json}.\par\medskip
Read failed feedback and its listed compiler logs, simulation logs, and public-interface waveforms. Repair the submission and rerun while budget remains. A golden PASS is a debugging milestone: complete every required document, the independent model, and the UVM environment, and address known failures before finishing. Save the complete submission and report its actual verification status. Do not use subagents.\par\medskip
Use the framework's native file-editing tools. In Codex and OpenCode, \texttt{apply\_patch} is a tool call, not a shell command; Claude Code uses \texttt{Write} and \texttt{Edit}. Use the framework's shell tool for commands. If required tools are unavailable, report an execution-environment failure instead of pretending that a tool call succeeded.\par\medskip
Run \texttt{bash /public/environment/run.sh simulate --submission /submission --output /tmp/feedback-01.json}, choosing a new output filename for each attempt. Read \texttt{public/debug\_feedback.md} for artifact retrieval. Keep feedback outside \texttt{/submission}. Use the same self-contained \texttt{model/model.sv} for independent model evaluation and scoreboard predictions.
\par\medskip
\end{AppendixCard}

\subsection{Feedback and configuration versions}
During development, feedback uses the golden design and seed 1. The published debugging contract allows 50 feedback calls, a 900-second compilation limit, and a separate 600-second golden-simulation limit. Final assessment uses seeds 1, 7, and 42. The returned index exposes candidate compilation/simulation logs and public-interface waveforms; downloading an artifact does not start another simulation. Hidden mutation results are withheld until submission freeze.

\begin{lstlisting}[style=appendixcode]
bash /public/environment/run.sh simulate 
  --submission /submission --output /tmp/feedback-01.json
bash /public/environment/run.sh artifact 
  --attempt 1 --name compile.log --output /tmp/compile-01.log
\end{lstlisting}
The commands are shown wrapped for print; each invocation is issued as one shell command. Feedback is stored outside the submission tree. A golden pass is a debugging observation, not proof that independent prediction, coverage, or mutation criteria have been met.

The paper's experiment setting in Section~\ref{sec:implementation-details} uses a 64,000-token per-response output budget. The public deployment file captured with these examples is dated November 17, 2026 and permits 128,000 output tokens per response. Both specify 300 cumulative decision turns, a reminder at turn 250, a 3,600-second request timeout, and no whole-task wall-time or cumulative token cap. The later public deployment limit is documented separately and must not be substituted for the recorded experiment setting.

\clearpage
\section{Worked Example: FIFO Features and Test Plans}
\label{app:fifo}
\subsection{Why this example is useful}
Task \texttt{dv-009}, the L2 CVA6 FIFO, is small enough to show the entire feature list and test plan while retaining behavior that a generic queue model can miss. It has depth 8, 32-bit data, fall-through enabled, asynchronous active-low reset, and synchronous flush. Its three-bit usage output wraps to zero at full capacity; zero usage alone therefore cannot establish emptiness. Empty-queue push can make data visible before a clock edge, and simultaneous push/pop can consume that item without storing it.

The four features and four tests below come from the maintenance-side \texttt{solution/reference/flow} files. They are \textbf{reference artifacts}, not generated submissions from any configuration in the results tables. Their inclusion illustrates document structure and traceability; it does not assert a fresh simulation or a perfect assessment score. All field values are retained, with only layout and typography changed. The paired original JSON and Markdown files are included with the source materials.

\begin{center}
\begin{tikzpicture}[node distance=10pt, every node/.style={font=\sffamily\small,align=center}, block/.style={draw=ReleaseRule,fill=ReleasePale,rounded corners=2pt,minimum height=36pt,text width=96pt}, >=Stealth]
\node[block] (requirement) {Public requirement\\\textbf{R-dv-009-004}};
\node[block,right=of requirement] (feature) {Observable feature\\\textbf{F-4}};
\node[block,right=of feature] (plan) {Stimulus and checker\\\textbf{T-4}};
\node[block,right=of plan] (implementation) {Implementation link\\\textbf{V-4}};
\draw[->,ReleaseBlue] (requirement)--(feature);
\draw[->,ReleaseBlue] (feature)--(plan);
\draw[->,ReleaseBlue] (plan)--(implementation);
\end{tikzpicture}
\end{center}
This chain links a public obligation to a behavioral claim, then to a concrete test and source location. The independent model supplies expectations from stimulus history; native DUT observations supply actual values. The source location is a document reference, not evidence that a check was dynamically reached.

\subsection{Complete reference feature list and test plan}
\begin{AppendixCard}[breakable=false]{dv-009 / F-1 / Reference feature}
\textbf{\color{ReleaseNavy}Requirements:} R-dv-009-001\par\smallskip
\textbf{\color{ReleaseNavy}Module:} cva6\_fifo\_v3\par\smallskip
\textbf{\color{ReleaseNavy}Definition:} rst\_ni low asynchronously clears the queue; it remains empty after deassertion. flush\_i clears it on the rising edge of clk\_i, taking priority over push/pop; flush does not take effect immediately between edges. After clearing, usage=0 and full=0; empty is still governed by the combinational fall-through rule for the current push.\par\smallskip
\textbf{\color{ReleaseNavy}Interface:} clk\_i, rst\_ni, flush\_i, push\_i, pop\_i, usage\_o, full\_o, empty\_o\par\smallskip
\textbf{\color{ReleaseNavy}Timing:} After reset assertion and before the next clock; before the flush edge and after NBA.\par\smallskip
\textbf{\color{ReleaseNavy}Boundary:} Fixed DEPTH=8, FALL\_THROUGH=1, DATA\_WIDTH=32, dtype=logic[31:0], FPGA\_EN=0, FPGA\_ALTERA=0; ADDR\_DEPTH=3 is a derived value. Inputs are two-state, and non-reset inputs avoid rising edges. push\_i=1 is prohibited when full\_o=1, and pop\_i=1 is prohibited when empty\_o=1; simultaneous push/pop on an empty queue is legal due to fall-through. data\_o is not scored when empty\_o=1. testmode\_i has no observable effect. The original source did not provide a spec; the following is the observable contract of the fixed RTL.\par\smallskip
\end{AppendixCard}
\begin{AppendixCard}[breakable=false]{dv-009 / T-1 / Reference test plan}
\textbf{\color{ReleaseNavy}Feature links:} F-1\par\smallskip
\textbf{\color{ReleaseNavy}Requirements:} R-dv-009-001\par\smallskip
\textbf{\color{ReleaseNavy}Stimulus and preparation:} Legally fill the queue to a nonempty state, then stop clk\_i with push/pop both 0. Assert reset between edges; idle after deassertion; when not empty, raise flush between edges and then pass through a rising edge; apply flush simultaneously with legal push/pop.\par\smallskip
\textbf{\color{ReleaseNavy}Expected behavior:} rst\_ni low asynchronously clears the queue; it remains empty after deassertion. flush\_i clears it on the rising edge of clk\_i, taking priority over push/pop; flush does not take effect immediately between edges. After clearing, usage=0 and full=0; empty is still governed by the combinational fall-through rule for the current push.\par\smallskip
\textbf{\color{ReleaseNavy}Checker and sampling:} Predict using an independent input-history model, without reading actual DUT outputs. After reset assertion and before the next clock; before the flush edge and after NBA. Use bitwise four-state comparison of actual and expected, report UVM\_ERROR on mismatch, record the comparison count, and trigger this scenario's check at least once.\par\smallskip
\textbf{\color{ReleaseNavy}Implementation:} V-1\par\smallskip
\textbf{\color{ReleaseNavy}Source:} reference\_pkg.sv:native\_scoreboard.write\par\smallskip
\end{AppendixCard}
\begin{AppendixCard}[breakable=false]{dv-009 / F-2 / Reference feature}
\textbf{\color{ReleaseNavy}Requirements:} R-dv-009-002\par\smallskip
\textbf{\color{ReleaseNavy}Module:} cva6\_fifo\_v3\par\smallskip
\textbf{\color{ReleaseNavy}Definition:} The queue capacity is 8, and full\_o is 1 when the stored entry count is 8; usage\_o is the low 3 bits of the stored entry count and is 0 when full, so emptiness cannot be inferred from usage=0. A push when not full enqueues, a pop when not empty dequeues, and legal simultaneous push/pop preserves the stored entry count.\par\smallskip
\textbf{\color{ReleaseNavy}Interface:} clk\_i, push\_i, pop\_i, full\_o, empty\_o, usage\_o, data\_i\par\smallskip
\textbf{\color{ReleaseNavy}Timing:} After combinational state changes and after the rising-edge NBA.\par\smallskip
\textbf{\color{ReleaseNavy}Boundary:} Fixed DEPTH=8, FALL\_THROUGH=1, DATA\_WIDTH=32, dtype=logic[31:0], FPGA\_EN=0, FPGA\_ALTERA=0; ADDR\_DEPTH=3 is a derived value. Inputs are two-state, and non-reset inputs avoid rising edges. push\_i=1 is prohibited when full\_o=1, and pop\_i=1 is prohibited when empty\_o=1; simultaneous push/pop on an empty queue is legal due to fall-through. data\_o is not scored when empty\_o=1. testmode\_i has no observable effect. The original source did not provide a spec; the following is the observable contract of the fixed RTL.\par\smallskip
\end{AppendixCard}
\begin{AppendixCard}[breakable=false]{dv-009 / T-2 / Reference test plan}
\textbf{\color{ReleaseNavy}Feature links:} F-2\par\smallskip
\textbf{\color{ReleaseNavy}Requirements:} R-dv-009-002\par\smallskip
\textbf{\color{ReleaseNavy}Stimulus and preparation:} Reset to an empty queue, with all operations obeying the legal full/empty domain. Fill one entry at a time from 1 to 8 entries and drain one entry at a time; perform simultaneous push/pop at 1 and 7 entries, wrap around repeatedly, and insert idle periods.\par\smallskip
\textbf{\color{ReleaseNavy}Expected behavior:} The queue capacity is 8, and full\_o is 1 when the stored entry count is 8; usage\_o is the low 3 bits of the stored entry count and is 0 when full, so emptiness cannot be inferred from usage=0. A push when not full enqueues, a pop when not empty dequeues, and legal simultaneous push/pop preserves the stored entry count.\par\smallskip
\textbf{\color{ReleaseNavy}Checker and sampling:} Predict using an independent input-history model, without reading actual DUT outputs. After combinational state changes and after the rising-edge NBA. Use bitwise four-state comparison of actual and expected, report UVM\_ERROR on mismatch, record the comparison count, and trigger this scenario's check at least once.\par\smallskip
\textbf{\color{ReleaseNavy}Implementation:} V-2\par\smallskip
\textbf{\color{ReleaseNavy}Source:} reference\_pkg.sv:native\_scoreboard.write\par\smallskip
\end{AppendixCard}
\begin{AppendixCard}[breakable=false]{dv-009 / F-3 / Reference feature}
\textbf{\color{ReleaseNavy}Requirements:} R-dv-009-003\par\smallskip
\textbf{\color{ReleaseNavy}Module:} cva6\_fifo\_v3\par\smallskip
\textbf{\color{ReleaseNavy}Definition:} When not empty, data\_o combinationally presents the oldest data not yet dequeued; push accepts data\_i on the rising edge of clk\_i, and pop removes the queue head on the rising edge. Without pop, the queue head is retained; repeated wraparound and simultaneous push/pop do not change FIFO order. Arbitrary changes to testmode\_i do not change these behaviors.\par\smallskip
\textbf{\color{ReleaseNavy}Interface:} clk\_i, push\_i, pop\_i, data\_i, data\_o, empty\_o, testmode\_i\par\smallskip
\textbf{\color{ReleaseNavy}Timing:} Stable before and after the edge when not empty; data with empty=1 is not checked.\par\smallskip
\textbf{\color{ReleaseNavy}Boundary:} Fixed DEPTH=8, FALL\_THROUGH=1, DATA\_WIDTH=32, dtype=logic[31:0], FPGA\_EN=0, FPGA\_ALTERA=0; ADDR\_DEPTH=3 is a derived value. Inputs are two-state, and non-reset inputs avoid rising edges. push\_i=1 is prohibited when full\_o=1, and pop\_i=1 is prohibited when empty\_o=1; simultaneous push/pop on an empty queue is legal due to fall-through. data\_o is not scored when empty\_o=1. testmode\_i has no observable effect. The original source did not provide a spec; the following is the observable contract of the fixed RTL.\par\smallskip
\end{AppendixCard}
\begin{AppendixCard}[breakable=false]{dv-009 / T-3 / Reference test plan}
\textbf{\color{ReleaseNavy}Feature links:} F-3\par\smallskip
\textbf{\color{ReleaseNavy}Requirements:} R-dv-009-003\par\smallskip
\textbf{\color{ReleaseNavy}Stimulus and preparation:} After reset, enqueue different 32-bit data values in sequence; the input mirror comes only from the driving history. Different data, 0/all 1s/random high-bit patterns, filling/draining, multiple wraparounds, interleaved same-cycle enqueue/dequeue and holds, and toggling testmode.\par\smallskip
\textbf{\color{ReleaseNavy}Expected behavior:} When not empty, data\_o combinationally presents the oldest data not yet dequeued; push accepts data\_i on the rising edge of clk\_i, and pop removes the queue head on the rising edge. Without pop, the queue head is retained; repeated wraparound and simultaneous push/pop do not change FIFO order. Arbitrary changes to testmode\_i do not change these behaviors.\par\smallskip
\textbf{\color{ReleaseNavy}Checker and sampling:} Predict using an independent input-history model, without reading actual DUT outputs. Stable before and after the edge when not empty; data with empty=1 is not checked. Use bitwise four-state comparison of actual and expected, report UVM\_ERROR on mismatch, record the comparison count, and trigger this scenario's check at least once.\par\smallskip
\textbf{\color{ReleaseNavy}Implementation:} V-3\par\smallskip
\textbf{\color{ReleaseNavy}Source:} reference\_pkg.sv:native\_scoreboard.write\par\smallskip
\end{AppendixCard}
\begin{AppendixCard}[breakable=false]{dv-009 / F-4 / Reference feature}
\textbf{\color{ReleaseNavy}Requirements:} R-dv-009-004\par\smallskip
\textbf{\color{ReleaseNavy}Module:} cva6\_fifo\_v3\par\smallskip
\textbf{\color{ReleaseNavy}Definition:} When storage is empty and push\_i=1, empty\_o immediately becomes 0 without a clock, and data\_o combinationally equals data\_i. In this state, a rising edge with pop\_i=0 retains the data as 1 entry; if pop\_i=1, the entry is consumed through fall-through, and storage still contains 0 entries. Without push, an empty queue has empty\_o=1.\par\smallskip
\textbf{\color{ReleaseNavy}Interface:} clk\_i, data\_i, push\_i, pop\_i, data\_o, empty\_o, usage\_o\par\smallskip
\textbf{\color{ReleaseNavy}Timing:} When the combinational inputs are stable, do not wait for a clock; then check the stored entry count after the rising edge and the state after withdrawing push.\par\smallskip
\textbf{\color{ReleaseNavy}Boundary:} Fixed DEPTH=8, FALL\_THROUGH=1, DATA\_WIDTH=32, dtype=logic[31:0], FPGA\_EN=0, FPGA\_ALTERA=0; ADDR\_DEPTH=3 is a derived value. Inputs are two-state, and non-reset inputs avoid rising edges. push\_i=1 is prohibited when full\_o=1, and pop\_i=1 is prohibited when empty\_o=1; simultaneous push/pop on an empty queue is legal due to fall-through. data\_o is not scored when empty\_o=1. testmode\_i has no observable effect. The original source did not provide a spec; the following is the observable contract of the fixed RTL.\par\smallskip
\end{AppendixCard}
\begin{AppendixCard}[breakable=false]{dv-009 / T-4 / Reference test plan}
\textbf{\color{ReleaseNavy}Feature links:} F-4\par\smallskip
\textbf{\color{ReleaseNavy}Requirements:} R-dv-009-004\par\smallskip
\textbf{\color{ReleaseNavy}Stimulus and preparation:} Establish an empty queue using reset or flush, and hold clk\_i low. Change push/data between edges to check fall-through, crossing the edge with pop0 and pop1 separately; withdraw push after fall-through, and repeat with different data.\par\smallskip
\textbf{\color{ReleaseNavy}Expected behavior:} When storage is empty and push\_i=1, empty\_o immediately becomes 0 without a clock, and data\_o combinationally equals data\_i. In this state, a rising edge with pop\_i=0 retains the data as 1 entry; if pop\_i=1, the entry is consumed through fall-through, and storage still contains 0 entries. Without push, an empty queue has empty\_o=1.\par\smallskip
\textbf{\color{ReleaseNavy}Checker and sampling:} Predict using an independent input-history model, without reading actual DUT outputs. When the combinational inputs are stable, do not wait for a clock; then check the stored entry count after the rising edge and the state after withdrawing push. Use bitwise four-state comparison of actual and expected, report UVM\_ERROR on mismatch, record the comparison count, and trigger this scenario's check at least once.\par\smallskip
\textbf{\color{ReleaseNavy}Implementation:} V-4\par\smallskip
\textbf{\color{ReleaseNavy}Source:} reference\_pkg.sv:native\_scoreboard.write\par\smallskip
\end{AppendixCard}

\subsection{Reading the example as a verification argument}
For reset, a nonempty initial state distinguishes clearing behavior from an already-empty queue, and the between-edge observation distinguishes asynchronous reset from a synchronous implementation. For capacity, reaching eight stored entries exposes the usage wraparound. For ordering, repeated pointer wraparound and distinct data values make reordering observable. For fall-through, the plan deliberately observes before the clock edge and exercises both retention and direct consumption. These are distinct verification obligations, even when one sequence or scoreboard implements several of them.

The checker also records whether comparisons occurred. A test that completes without exercising its claimed condition offers no evidence for that condition. Similarly, predicting from DUT outputs would remove the independence needed to detect a faulty design. These examples motivate the separate setup, stimulus, expectation, sampling, and checking dimensions used in Section~\ref{sec:artifact-evaluation}.

\clearpage
\section{Examples of Parameter and Transaction Complexity}
\label{app:advanced}
\subsection{Width conversion: three parameter profiles}
Task \texttt{dv-018} is an L3 width converter with 4/8, 8/4, and 8/8 input/output-width profiles. The first two contain state; the equal-width case is combinational. The reference records selected here are F03/T03 and F05/T05, two of the five feature/test pairs. They expose two common modeling shortcuts: sampling only when output valid is asserted, and applying one ready/valid interpretation to all parameter profiles.

In the 4/8 case, a first segment can update defined output data while output valid remains low. In the 8/4 case, consecutive segments are obtained from their respective current input values, so a model that latches one input word and splits it unconditionally would implement a different contract. Backpressure must also be profile-specific: it holds registered data but does not freeze the combinational pass-through profile. The reproduced fields describe the source contract; they are not generalized requirements for all width converters.
\begin{AppendixCard}[breakable=false]{dv-018 / F03 / Reference feature}
\textbf{\color{ReleaseNavy}Requirements:} dv-018/R03\par\smallskip
\textbf{\color{ReleaseNavy}Module:} width\_convert\par\smallskip
\textbf{\color{ReleaseNavy}Definition:} 4/8 writes two successively processed half-words into rd\_data[7:4] and [3:0]; each updated segment remains defined when rd\_vld=0. The second segment sets rd\_vld, and a newly completed word takes priority over consumption on the same edge. 8/4 alternately takes the high half-word and low half-word from wr\_data at that edge and registers the segment into rd\_data; each processing edge sets rd\_vld, otherwise it holds if the old rd\_vld is asserted and ready is not asserted, and clears to 0 in all other cases. rd\_data holds on an edge with no processing. For 8/8, rd\_data combinationally equals wr\_data, adding no latency.\par\smallskip
\textbf{\color{ReleaseNavy}Interface:} clock, wr\_data, wr\_vld, rd\_data, rd\_vld\par\smallskip
\textbf{\color{ReleaseNavy}Timing:} For registered branches, compare after signals settle following the rising edge of clock; compare asynchronous reset between edges; compare the pass-through branch and the combinational relationship of wr\_ready after inputs settle and before the edge.\par\smallskip
\textbf{\color{ReleaseNavy}Boundary:} 4/8 writes two successively processed half-words into rd\_data[7:4] and [3:0]; each updated segment remains defined when rd\_vld=0. The second segment sets rd\_vld, and a newly completed word takes priority over consumption on the same edge. 8/4 alternately takes the high half-word and low half-word from wr\_data at that edge and registers the segment into rd\_data; each processing edge sets rd\_vld, otherwise it holds if the old rd\_vld is asserted and ready is not asserted, and clears to 0 in all other cases. rd\_data holds on an edge with no processing. For 8/8, rd\_data combinationally equals wr\_data, adding no latency.\par\smallskip
\end{AppendixCard}
\begin{AppendixCard}[breakable=false]{dv-018 / T03 / Reference test plan}
\textbf{\color{ReleaseNavy}Feature links:} F03\par\smallskip
\textbf{\color{ReleaseNavy}Requirements:} dv-018/R03\par\smallskip
\textbf{\color{ReleaseNavy}Stimulus and preparation:} For each profile, input 00, ff, different high and low half-words, bit-by-bit values, and random values; for 4/8, assemble words continuously and check the defined data during the invalid period after the first segment; for 8/4, change wr\_data on each edge to confirm that the high and low segments both come from their respective current inputs; for 8/8, change data between edges and observe combinational propagation.\par\smallskip
\textbf{\color{ReleaseNavy}Expected behavior:} 4/8 writes two successively processed half-words into rd\_data[7:4] and [3:0]; each updated segment remains defined when rd\_vld=0. The second segment sets rd\_vld, and a newly completed word takes priority over consumption on the same edge. 8/4 alternately takes the high half-word and low half-word from wr\_data at that edge and registers the segment into rd\_data; each processing edge sets rd\_vld, otherwise it holds if the old rd\_vld is asserted and ready is not asserted, and clears to 0 in all other cases. rd\_data holds on an edge with no processing. For 8/8, rd\_data combinationally equals wr\_data, adding no latency.\par\smallskip
\textbf{\color{ReleaseNavy}Checker and sampling:} Independently calculate allocation/packet sequences or segment arithmetic from input history, without reading actual DUT outputs to generate expected; for registered branches, compare after signals settle following the rising edge of clock; compare asynchronous reset between edges; compare the pass-through branch and the combinational relationship of wr\_ready after inputs settle and before the edge. Compare each valid port using four-state !==; report UVM\_ERROR for X/Z or mismatches; drain after legal stimulus ends, and report an error if the comparison count is 0.\par\smallskip
\textbf{\color{ReleaseNavy}Implementation:} CHK03\par\smallskip
\textbf{\color{ReleaseNavy}Source:} tb/package.sv:dvbench\_scoreboard.write\par\smallskip
\end{AppendixCard}
\begin{AppendixCard}[breakable=false]{dv-018 / F05 / Reference feature}
\textbf{\color{ReleaseNavy}Requirements:} dv-018/R05\par\smallskip
\textbf{\color{ReleaseNavy}Module:} width\_convert\par\smallskip
\textbf{\color{ReleaseNavy}Definition:} 4/8 and 8/4 do not process input when rd\_ready=0, and rd\_data, rd\_vld, and rd\_last hold; however, wr\_ready for 8/4 still changes according to rule R02. After ready recovers, processing uses the current edge's inputs; input acceptance must not be inferred from DUT outputs. Observe registered branches after the native edge and NBA settling, and asynchronous reset between edges; 8/8 can be observed once inputs are stable, and even rd\_ready=0 does not freeze pass-through data/valid/last.\par\smallskip
\textbf{\color{ReleaseNavy}Interface:} clock, rd\_ready, wr\_ready, rd\_data, rd\_last, rd\_vld\par\smallskip
\textbf{\color{ReleaseNavy}Timing:} For registered branches, compare after signals settle following the rising edge of clock; compare asynchronous reset between edges; compare the pass-through branch and the combinational relationship of wr\_ready after inputs settle and before the edge.\par\smallskip
\textbf{\color{ReleaseNavy}Boundary:} 4/8 and 8/4 do not process input when rd\_ready=0, and rd\_data, rd\_vld, and rd\_last hold; however, wr\_ready for 8/4 still changes according to rule R02. After ready recovers, processing uses the current edge's inputs; input acceptance must not be inferred from DUT outputs. Observe registered branches after the native edge and NBA settling, and asynchronous reset between edges; 8/8 can be observed once inputs are stable, and even rd\_ready=0 does not freeze pass-through data/valid/last.\par\smallskip
\end{AppendixCard}
\begin{AppendixCard}[breakable=false]{dv-018 / T05 / Reference test plan}
\textbf{\color{ReleaseNavy}Feature links:} F05\par\smallskip
\textbf{\color{ReleaseNavy}Requirements:} dv-018/R05\par\smallskip
\textbf{\color{ReleaseNavy}Stimulus and preparation:} After a valid output in each registered profile, drive ready low and hold it for multiple cycles, changing inputs during this time and covering wr\_vld=0/1, then restore ready for continuous processing; for 8/8, change data/valid/last while ready is low to confirm that pass-through continues.\par\smallskip
\textbf{\color{ReleaseNavy}Expected behavior:} 4/8 and 8/4 do not process input when rd\_ready=0, and rd\_data, rd\_vld, and rd\_last hold; however, wr\_ready for 8/4 still changes according to rule R02. After ready recovers, processing uses the current edge's inputs; input acceptance must not be inferred from DUT outputs. Observe registered branches after the native edge and NBA settling, and asynchronous reset between edges; 8/8 can be observed once inputs are stable, and even rd\_ready=0 does not freeze pass-through data/valid/last.\par\smallskip
\textbf{\color{ReleaseNavy}Checker and sampling:} Independently calculate allocation/packet sequences or segment arithmetic from input history, without reading actual DUT outputs to generate expected; for registered branches, compare after signals settle following the rising edge of clock; compare asynchronous reset between edges; compare the pass-through branch and the combinational relationship of wr\_ready after inputs settle and before the edge. Compare each valid port using four-state !==; report UVM\_ERROR for X/Z or mismatches; drain after legal stimulus ends, and report an error if the comparison count is 0.\par\smallskip
\textbf{\color{ReleaseNavy}Implementation:} CHK05\par\smallskip
\textbf{\color{ReleaseNavy}Source:} tb/package.sv:dvbench\_scoreboard.write\par\smallskip
\end{AppendixCard}

\subsection{AXI interconnect: serialization and history-dependent control}
Task \texttt{dv-080} is an L5 4-by-4 AXI interconnect with fixed default parameters. The selected records F-axi-03/T-axi-03 and F-axi-11/T-axi-11 are two of twelve feature/test pairs. The first makes arbitration release observable by withholding a final response while other addresses remain pending. The second compares control signals across successive target selections and reset, including a historical target-index effect specified by the task.

This design serializes one transaction globally. An address handshake alone does not release ownership: a write waits for the upstream B response handshake and a read waits for the upstream final R beat. The second example also demonstrates why a nominally protocol-correct model may still fail exact replay: the task specifies cycle-level public control behavior, including which historical target is projected immediately after decoding. These are properties of the fixed released design, not universal AXI rules. As in the source, planned semantic objectives and observed HDL execution coverage are distinct.
\begin{AppendixCard}[breakable=false]{dv-080 / F-axi-03 / Reference feature}
\textbf{\color{ReleaseNavy}Requirements:} R-dv-080-003\par\smallskip
\textbf{\color{ReleaseNavy}Module:} axi\_interconnect\par\smallskip
\textbf{\color{ReleaseNavy}Definition:} Only one transaction is serviced globally; among the 8 requests, index 2*n is AW for s[n], and 2*n+1 is AR. After reset, round-robin priority starts at the lowest index; after a selection, the next starting point is the selected index+1, wrapping around. The grant remains asserted until the corresponding s-side B handshake or Rlast handshake releases it. AW/AR completion must not release it early; simultaneous reads and writes and different request ports are serialized according to this rule.\par\smallskip
\textbf{\color{ReleaseNavy}Interface:} clk, m\_axi\_arready, m\_axi\_arvalid, m\_axi\_awready, m\_axi\_awvalid, m\_axi\_bready, m\_axi\_bvalid, m\_axi\_rready, m\_axi\_rvalid, m\_axi\_wready, m\_axi\_wvalid, rst, s\_axi\_arready, s\_axi\_arvalid, s\_axi\_awready, s\_axi\_awvalid, s\_axi\_bready, s\_axi\_bvalid, s\_axi\_rready, s\_axi\_rvalid, s\_axi\_wready, s\_axi\_wvalid\par\smallskip
\textbf{\color{ReleaseNavy}Timing:} The exact timing and validity domain of this clause follow definition; do not assume additional pipelining or out-of-order behavior.\par\smallskip
\textbf{\color{ReleaseNavy}Boundary:} Use the fixed public default parameters and the domain defined by this clause; do not add requirements from other AXI versions.\par\smallskip
\end{AppendixCard}
\begin{AppendixCard}[breakable=false]{dv-080 / T-axi-03 / Reference test plan}
\textbf{\color{ReleaseNavy}Feature links:} F-axi-03\par\smallskip
\textbf{\color{ReleaseNavy}Requirements:} R-dv-080-003\par\smallskip
\textbf{\color{ReleaseNavy}Stimulus and preparation:} Have all 4 s ports hold AW and AR requests simultaneously, marking the 8 requests with different addresses and IDs; complete data and responses according to the selected transaction, check the round-robin order, and block the final R beat or B to verify that other addresses cannot get ahead; cover wraparound with two consecutive rounds. Drive on falling edges and hold payload until its valid/ready handshake; sample stably before each rising edge and sample after waiting 1ns after the edge; handshakes are determined by the old ready and valid. Always compare controls, compare valid payloads, and do not score before the first synchronous reset.\par\smallskip
\textbf{\color{ReleaseNavy}Expected behavior:} Only one transaction is serviced globally; among the 8 requests, index 2*n is AW for s[n], and 2*n+1 is AR. After reset, round-robin priority starts at the lowest index; after a selection, the next starting point is the selected index+1, wrapping around. The grant remains asserted until the corresponding s-side B handshake or Rlast handshake releases it. AW/AR completion must not release it early; simultaneous reads and writes and different request ports are serialized according to this rule.\par\smallskip
\textbf{\color{ReleaseNavy}Checker and sampling:} Predict using an independent transaction and beat-queue model based on input history; the monitor reads actual native ports, and the scoreboard performs field-by-field four-state comparisons by request/beat sequence number. Report valid differences independently as errors, and report valid X/Z or data/routing/order/beat-count mismatches as UVM\_ERROR; at completion, check that all transactions have drained and the comparison count is nonzero; do not use DUT outputs as expected values.\par\smallskip
\textbf{\color{ReleaseNavy}Implementation:} axi\_reference\_sequence\par\smallskip
\textbf{\color{ReleaseNavy}Source:} tb.sv; model/model.sv; semantic coverage objectives and current HDL execution coverage are accepted separately; do not falsely claim dynamic completion in text.\par\smallskip
\end{AppendixCard}
\begin{AppendixCard}[breakable=false]{dv-080 / F-axi-11 / Reference feature}
\textbf{\color{ReleaseNavy}Requirements:} R-dv-080-011\par\smallskip
\textbf{\color{ReleaseNavy}Module:} axi\_interconnect\par\smallskip
\textbf{\color{ReleaseNavy}Definition:} In the first cycle after a legal read address handshake, the onehot position of m\_rready uses the target index saved from the previous decoding, with an initial value of 0 at the start of simulation; synchronous reset does not clear the saved target index, and the reset edge still commits the index obtained from combinational decoding before reset. The subsequent first beat projects according to this historical index; from the next cycle onward, use the current target index. Validity is determined by the read buffer availability condition at that time. Legal endpoints do not send R before the corresponding AR handshake, so this ready projection cannot create data from nothing; this clause only specifies the exact timing of public control outputs.\par\smallskip
\textbf{\color{ReleaseNavy}Interface:} clk, m\_axi\_araddr, m\_axi\_arburst, m\_axi\_arcache, m\_axi\_arid, m\_axi\_arlen, m\_axi\_arlock, m\_axi\_arprot, m\_axi\_arqos, m\_axi\_arready, m\_axi\_arregion, m\_axi\_arsize, m\_axi\_aruser, m\_axi\_arvalid, m\_axi\_awaddr, m\_axi\_awburst, m\_axi\_awprot, m\_axi\_awready, m\_axi\_awregion, m\_axi\_awuser, m\_axi\_bready, m\_axi\_bresp, m\_axi\_buser, m\_axi\_rdata, m\_axi\_rid, m\_axi\_rlast, m\_axi\_rready, m\_axi\_rresp, m\_axi\_ruser, m\_axi\_rvalid, m\_axi\_wready, m\_axi\_wstrb, m\_axi\_wuser, rst, s\_axi\_araddr, s\_axi\_arburst, s\_axi\_arcache, s\_axi\_arid, s\_axi\_arlen, s\_axi\_arlock, s\_axi\_arprot, s\_axi\_arqos, s\_axi\_arready, s\_axi\_arsize, s\_axi\_aruser, s\_axi\_arvalid, s\_axi\_awaddr, s\_axi\_awburst, s\_axi\_awprot, s\_axi\_awready, s\_axi\_awuser, s\_axi\_bready, s\_axi\_bresp, s\_axi\_buser, s\_axi\_rdata, s\_axi\_rid, s\_axi\_rlast, s\_axi\_rready, s\_axi\_rresp, s\_axi\_ruser, s\_axi\_rvalid, s\_axi\_wready, s\_axi\_wstrb, s\_axi\_wuser\par\smallskip
\textbf{\color{ReleaseNavy}Timing:} The exact timing and validity domain of this clause follow definition; do not assume additional pipelining or out-of-order behavior.\par\smallskip
\textbf{\color{ReleaseNavy}Boundary:} Use the fixed public default parameters and the domain defined by this clause; do not add requirements from other AXI versions.\par\smallskip
\end{AppendixCard}
\begin{AppendixCard}[breakable=false]{dv-080 / T-axi-11 / Reference test plan}
\textbf{\color{ReleaseNavy}Feature links:} F-axi-11\par\smallskip
\textbf{\color{ReleaseNavy}Requirements:} R-dv-080-011\par\smallskip
\textbf{\color{ReleaseNavy}Stimulus and preparation:} Alternate targets 0/1/3/2 in successive reads, keeping m\_rvalid at 0 before AR; compare edge by edge that ready uses the old target in the first cycle after decoding and the new target in the next cycle, and send R only after the target AR completes. Make the first read target after the initial reset nonzero to check the initial value of 0; reset again after a nonzero target history to check index retention, and also cover being in the decode stage before the reset edge. Drive on the falling edge and hold payload until its valid/ready handshake; sample stable values before each rising edge and sample after waiting 1ns after the edge, with handshakes determined by the old ready and valid. Always compare controls, compare valid payload, and do not score before the initial synchronous reset.\par\smallskip
\textbf{\color{ReleaseNavy}Expected behavior:} In the first cycle after a legal read address handshake, the onehot position of m\_rready uses the target index saved from the previous decoding, with an initial value of 0 at the start of simulation; synchronous reset does not clear the saved target index, and the reset edge still commits the index obtained from combinational decoding before reset. The subsequent first beat projects according to this historical index; from the next cycle onward, use the current target index. Validity is determined by the read buffer availability condition at that time. Legal endpoints do not send R before the corresponding AR handshake, so this ready projection cannot create data from nothing; this clause only specifies the exact timing of public control outputs.\par\smallskip
\textbf{\color{ReleaseNavy}Checker and sampling:} Predict using an independent transaction and beat-queue model based on input history; the monitor reads actual native ports, and the scoreboard performs field-by-field four-state comparisons by request/beat sequence number. Report valid differences independently as errors, and report valid X/Z or data/routing/order/beat-count mismatches as UVM\_ERROR; at completion, check that all transactions have drained and the comparison count is nonzero; do not use DUT outputs as expected values.\par\smallskip
\textbf{\color{ReleaseNavy}Implementation:} axi\_reference\_sequence\par\smallskip
\textbf{\color{ReleaseNavy}Source:} tb.sv; model/model.sv; semantic coverage objectives and current HDL execution coverage are accepted separately; do not falsely claim dynamic completion in text.\par\smallskip
\end{AppendixCard}

\subsection{What the three cases cover}
The FIFO case emphasizes short state histories and combinational observation windows; the converter case adds parameter-dependent semantics and partial updates; the interconnect case adds cross-channel ownership, backpressure, and transaction history. They were chosen for explanatory complementarity and document availability, not as a random sample or as evidence of average agent behavior. For each case, the complete original specification and paired reference documents are preserved, even when only selected records are printed.

\clearpage
\clearpage
\section{Inside the Generated UVM Environments}
\label{app:uvm}

The reference documents in Appendices~\ref{app:fifo} and~\ref{app:advanced} describe verification intent. This appendix shows executable artifacts from the FIFO run in Section~\ref{sec:case-study} and a complementary width-converter run (\texttt{dv-018}). Both are Codex with GPT-6 Astra submissions, distinct from the maintenance-side reference solutions. The width converter passes independent model replay and detects all ten mutants across its three parameter profiles, illustrating parameter-aware checking alongside the FIFO's timing-oriented stimulus. Excerpts reproduce the frozen source files, with line wrapping for print. Complete submissions, source hashes, per-turn token counts, and milestone evidence for these runs and the two long-horizon cases accompany the paper in \texttt{data/case-study/}.

\subsection{FIFO: connecting stimulus, prediction, and checking}
The FIFO submission uses a UVM sequence/sequencer/driver path and an event-driven monitor. A shared interface drives both the DUT and a separate \texttt{dvbench\_model}. The model receives inputs and their history; its outputs populate expected-value fields. The monitor captures actual and predicted observations, and an analysis connection delivers them to the scoreboard. The scoreboard compares flags and occupancy on every observation and compares data when the expected FIFO is nonempty. It also records behavioral coverage and checks that enough observations occurred.

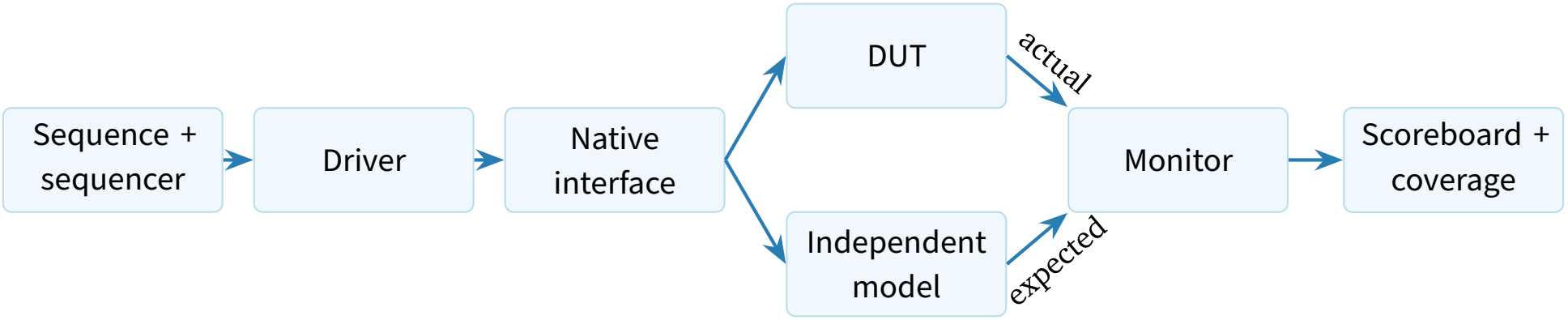
\begin{figure}[htbp]
\centering
\begin{tikzpicture}[x=1cm,y=1cm,>=Stealth,
  block/.style={draw=ReleaseRule,fill=ReleasePale,rounded corners=2pt,align=center,font=\sffamily\fontsize{8}{10}\selectfont,text width=1.55cm,minimum height=0.85cm},
  link/.style={->,draw=ReleaseBlue,thick}]
\node[block] (seq) at (0,0) {Sequence +\\sequencer};
\node[block] (driver) at (2.05,0) {Driver};
\node[block] (bus) at (4.1,0) {Native\\interface};
\node[block] (dut) at (6.4,0.85) {DUT};
\node[block] (model) at (6.4,-0.85) {Independent\\model};
\node[block] (monitor) at (8.7,0) {Monitor};
\node[block] (scoreboard) at (10.95,0) {Scoreboard +\\coverage};
\draw[link] (seq)--(driver);
\draw[link] (driver)--(bus);
\draw[link] (bus.east)--(dut.west);
\draw[link] (bus.east)--(model.west);
\draw[link] (dut.east)--node[above,sloped,font=\scriptsize]{actual}(monitor.north west);
\draw[link] (model.east)--node[below,sloped,font=\scriptsize]{expected}(monitor.south west);
\draw[link] (monitor)--(scoreboard);
\end{tikzpicture}
\caption{The generated FIFO verification architecture. DUT and model receive the same inputs. The model predicts independently of DUT outputs; the monitor packages observations for the scoreboard. The diagram summarizes the frozen submission.}
\label{fig:uvm-architecture}
\end{figure}

\begin{AppendixCard}{Frozen FIFO submission: file responsibilities}
\texttt{sources.f}: dependency-ordered compilation manifest.\par
\texttt{model/model.sv}: standalone native-port prediction module.\par
\texttt{tb/fifo\_if.sv}: inputs, actual/expected outputs, observation event, and phase tag.\par
\texttt{tb/fifo\_pkg.sv}: transactions, driver, monitor, scoreboard, sequence, environment, and test.\par
\texttt{tb/top.sv}: DUT/model instantiation, interface registration, and simulation watchdog.\par
\texttt{flow/}: six features and six test plans, with matching Markdown documents.
\end{AppendixCard}

\noindent\textbf{Environment assembly.} This wiring instantiates the components and connects stimulus delivery and observation flow, making the dependencies between verification artifacts executable.
\begin{lstlisting}[style=appendixcode,basicstyle=\ttfamily\fontsize{8}{9}\selectfont,language=Verilog,firstline=302,lastline=320]
  class fifo_env extends uvm_env;
    `uvm_component_utils(fifo_env)
    fifo_driver driver;
    fifo_monitor monitor;
    fifo_scoreboard scoreboard;
    uvm_sequencer #(fifo_item) sequencer;
    function new(string name, uvm_component parent); super.new(name,parent); endfunction
    function void build_phase(uvm_phase phase);
      super.build_phase(phase);
      driver=fifo_driver::type_id::create("driver",this);
      monitor=fifo_monitor::type_id::create("monitor",this);
      scoreboard=fifo_scoreboard::type_id::create("scoreboard",this);
      sequencer=uvm_sequencer#(fifo_item)::type_id::create("sequencer",this);
    endfunction
    function void connect_phase(uvm_phase phase);
      driver.seq_item_port.connect(sequencer.seq_item_export);
      monitor.observed.connect(scoreboard.observed);
    endfunction
  endclass
\end{lstlisting}

\clearpage
\subsection{Where the timing decisions live}
The driver determines which temporal distinctions the environment observes. Stimulus is applied at the falling edge; reset assertions, releases, and extra data perturbations receive distinct sample tags. The rising-edge sample follows DUT and model updates. This schedule makes the environment inspectable at the level of simulation events, rather than just component names.
\begin{lstlisting}[style=appendixcode,language=Verilog,firstline=39,lastline=76]
    task run_phase(uvm_phase phase);
      fifo_item req_item;
      forever begin
        seq_item_port.get_next_item(req_item);
        @(negedge vif.clk_i);
        vif.scenario = req_item.scenario;
        vif.rst_ni = 1;
        vif.push_i = req_item.push;
        vif.pop_i = req_item.pop;
        vif.flush_i = req_item.flush;
        vif.testmode_i = req_item.testmode;
        vif.data_i = req_item.data;
        #1;
        sample(0);
        if (req_item.reset_pulse) begin
          #1; vif.rst_ni = 0;
          #1; sample(2);
          if (req_item.reset_hold) begin
            repeat (2) begin
              @(posedge vif.clk_i);
              #1; sample(2);
            end
            @(negedge vif.clk_i);
          end else #1;
          vif.rst_ni = 1;
          #1; sample(3);
        end else begin
          #2;
          vif.testmode_i = ~req_item.testmode;
          if (req_item.perturb) vif.data_i = req_item.alternate;
          #1; sample(1);
        end
        @(posedge vif.clk_i);
        #1; sample(4);
        #1;
        seq_item_port.item_done();
      end
    endtask
\end{lstlisting}
Adding \texttt{reset\_hold} extends observations across active clock edges while reset is asserted, increasing golden comparisons. Ordinary data perturbation still precedes the rising edge, so it does not distinguish high-phase transparent writes from edge-triggered storage. A sound UVM component structure and extensive activity can therefore coexist with a missing temporal distinction. Mutation assessment exposes the required scenario.

\clearpage
\subsection{Scoreboards embody the observation contract}
The FIFO scoreboard compares flags and usage on every observation, but checks data when the expected FIFO is nonempty. Four-state case inequality preserves differences involving unknown values. Sample and data-check counters link claimed verification activity to actual observations.
\begin{lstlisting}[style=appendixcode,language=Verilog,firstline=130,lastline=144]
    function void write(fifo_observation obs);
      samples++;
      if (obs.full_actual !== obs.full_expected ||
          obs.empty_actual !== obs.empty_expected ||
          obs.usage_actual !== obs.usage_expected)
        `uvm_error("FLAGS",$sformatf("%s phase=%0d push/pop/flush=%b%b%b actual full/empty/usage=%b/%b/%0d expected=%b/%b/%0d",
          obs.scenario,obs.phase,obs.push,obs.pop,obs.flush,
          obs.full_actual,obs.empty_actual,obs.usage_actual,
          obs.full_expected,obs.empty_expected,obs.usage_expected))
      if (obs.empty_expected === 1'b0) begin
        data_checks++;
        if (obs.data_actual !== obs.data_expected)
          `uvm_error("DATA",$sformatf("%s phase=%0d actual=%08h expected=%08h",
            obs.scenario,obs.phase,obs.data_actual,obs.data_expected))
      end
\end{lstlisting}

\noindent\textbf{Width conversion requires a different checking rule.} The second environment reuses a testbench and parameterized model across three profiles. Its checker compares data, valid, last, and ready together, retaining data observations when valid is low. This matches the contract's defined partial-data updates.
\begin{lstlisting}[style=appendixcode,language=Verilog,firstline=209,lastline=219]
    function void check_outputs(observation actual, observation expected);
      comparisons++;
      if (actual.reset_active) reset_checks++;
      if ({actual.data,actual.valid,actual.last,actual.ready} !==
          {expected.data,expected.valid,expected.last,expected.ready})
        `uvm_error("OUTPUT", $sformatf("IW=%0d OW=%0d time=%0t reset=%0b inputs=%h/%b%b%b%b actual=%h/%b%b%b expected=%h/%b%b%b",
          IW,OW,actual.stamp,actual.reset_active,actual.input_data,
          actual.input_valid,actual.input_ready,actual.input_last,actual.input_align,
          actual.data,actual.valid,actual.last,actual.ready,
          expected.data,expected.valid,expected.last,expected.ready))
    endfunction
\end{lstlisting}

The model separately implements pass-through, assembly, and splitting. In the assembly branch, data-segment updates and output-valid updates have distinct conditions. A partial write can change a defined portion of \texttt{rd\_data} without asserting \texttt{rd\_vld}; the checker observes that behavior.
\begin{lstlisting}[style=appendixcode,language=Verilog,firstline=36,lastline=49]
          if (wr_vld && rd_ready) begin
            if (low_segment)
              rd_data[ISIZE-1:0] <= wr_data;
            else
              rd_data[OSIZE-1:ISIZE] <= wr_data;
            low_segment <= (wr_last || wr_align_last) ? 1'b0 : !low_segment;
          end
          if (wr_vld && rd_ready && low_segment) begin
            rd_vld <= 1;
            rd_last <= wr_last;
          end else if (rd_vld && rd_ready) begin
            rd_vld <= 0;
            rd_last <= 0;
          end
\end{lstlisting}

\clearpage
\section{Representative Mutant Designs and Discriminating Tests}
\label{app:mutants}

DVBench's mutants turn specification obligations into executable tests of verification effectiveness. A useful mutant connects a plausible implementation mistake, a legal input history that activates it, and an observable difference at a defined sampling point. Designing this connection requires understanding the DUT's parameters, protocol, state, and observation contract. The ten examples below span arithmetic, clock-domain crossing, filtering, memory, bus control, and cryptographic state, illustrating the reasoning behind the ten-mutant sets in Section~\ref{sec:tasks}.

Each task uses mutant identifiers m1--m10. The examples below show the corresponding golden and mutant code: minus lines are golden code, and plus lines are mutant code. Each example describes the changed behavior and an input sequence that exposes it.

\begin{AppendixCard}{From an implementation edit to a verification obligation}
\textbf{Semantic target:} identify the exact requirement changed by the edit.\par
\textbf{Activation:} choose legal values, parameters, and histories that make the implementations diverge.\par
\textbf{Propagation:} carry the divergence to a scored output without masking it through another condition.\par
\textbf{Observation:} specify the output, expected values, and sampling window that expose it.
\end{AppendixCard}

\begin{table}[htbp]
\centering\small
\caption{Representative mutants and the verification distinctions they require.}
\label{tab:mutant-examples}
\begin{tabular*}{\linewidth}{@{\extracolsep{\fill}}llll@{}}
\toprule
Task & Mutant & Semantic target & Discriminating scenario\\
\midrule
\texttt{dv-044} & m4 & Rounding bit & Discarded/retained bits differ\\
\texttt{dv-037} & m2 & Unsigned comparison & Opposite operand sign bits\\
\texttt{dv-008} & m1 & Gray encoding & Adjacent binary ones\\
\texttt{dv-016} & m7 & Any-match reduction & Two simultaneous matches\\
\texttt{dv-017} & m7 & Response ownership & Two ports, two returns\\
\texttt{dv-009} & m1 & FIFO multiplicity & Consecutive equal pushes\\
\texttt{dv-031} & m2 & Write-first memory & Observe the write cycle\\
\texttt{dv-094} & m1 & Asynchronous reset & Reset with no clock edge\\
\texttt{dv-095} & m1 & Accepted write address & Change address after handshake\\
\texttt{dv-063} & m1 & CTR context reload & Same key, different IV\\
\bottomrule
\end{tabular*}
\end{table}

\clearpage
\subsection{Numeric interpretation and encoding}
\subsubsection{Rounding at the discarded-bit boundary: \texttt{dv-044}/m4}
The multiplier's narrowing stage rounds using the highest discarded bit. Mutant m4 moves the index up by one, so the lowest retained bit controls the correction instead. The multiplication remains intact, isolating a conversion-stage obligation.
\begin{lstlisting}[style=appendixcode]
--- golden/axi_round.v
+++ m4/axi_round.v
@@ -42,7 +42,7 @@
 
       assign round_corr_trunc = 0;
       assign round_corr_rtz = (i_tdata[WIDTH_IN-1] & |i_tdata[WIDTH_IN-WIDTH_OUT-1:0]);
-      assign round_corr_nearest = i_tdata[WIDTH_IN-WIDTH_OUT-1];
+      assign round_corr_nearest = i_tdata[WIDTH_IN-WIDTH_OUT];
 
       // Detect special case of maximum positive number.
       // Rounding up is not possible in this case to avoid a wrap to a negative number.
\end{lstlisting}
In the released 16-by-16 signed profile with \texttt{WIDTH\_P=16}, \texttt{DROP\_TOP\_P=1}, and rounding enabled, 15 bits are discarded. Inputs $A=B=128$ produce \texttt{0x4000}: bit 14 is one and bit 15 is zero. Golden output is 1; the mutant produces 0. Check the corresponding valid output after the pipeline delay. Boundary-bit combinations are a more precise target than a broad selection of large and small operands.

\subsubsection{Unsigned window comparison: \texttt{dv-037}/m2}
The watchdog compares an unsigned counter with an unsigned threshold. Casting both to signed values preserves their bits but changes their ordering when the most significant bits differ.
\begin{lstlisting}[style=appendixcode]
--- golden/common/wd_timer/watchdog_timer.sv
+++ m2/common/wd_timer/watchdog_timer.sv
@@ -653,7 +653,7 @@
                 //--------------------------------------
                 // Window Violation
                 //--------------------------------------
-                if (window_en_wdt && (watchdog_counter > window_value_wdt)) begin
+                if (window_en_wdt && ($signed(watchdog_counter) > $signed(window_value_wdt))) begin
 
                     window_violation_wdt <= 1'b1;
 		    trace_event_id_wdt <= TRACE_WINDOW_VIOLATION;
\end{lstlisting}
Load \texttt{timeout=0x80001000} and \texttt{window=0x00000020}, enable the window, and perform the legal two-key refresh while the counter's high bit remains set. Golden logic reports a window violation; the mutant treats the counter as negative and permits reloading. Respect the cross-domain synchronization delay when reading status. Loading a high-bit value reaches the boundary directly, without waiting billions of cycles. The test obligation is the crossing of numeric interpretations, not simply exercising refresh.

\clearpage
\subsubsection{Gray-code operator replacement: \texttt{dv-008}/m1}
Replacing XOR with OR in Gray encoding changes the code whenever adjacent binary bits are both one, while leaving many sparse inputs unchanged.
\begin{lstlisting}[style=appendixcode]
--- golden/xpm_cdc_stolen.v
+++ m1/xpm_cdc_stolen.v
@@ -238,7 +238,7 @@
   assign async_path = src_gray_ff;
 
   assign synco_gray = dest_graysync_ff[DEST_SYNC_FF-1];
-  assign gray_enc = src_in_bin ^ (src_in_bin >> 1);
+  assign gray_enc = src_in_bin | (src_in_bin >> 1);
 
   // Convert gray code back to binary
   integer j;
\end{lstlisting}
In the width-four context, hold source values 0, 1, 2, and 3 long enough for source sampling and destination convergence. For binary \texttt{0011}, golden Gray code is \texttt{0010}; the mutant produces \texttt{0011}. The unchanged decoder returns 3 versus 2. Account for synchronizer latency and the enabled output register. This digitally observable fault tests the specified encoding/decoding relation.

\subsubsection{Any match versus odd parity: \texttt{dv-016}/m7}
The network filter accepts an address when at least one of eight entries matches. Replacing reduction OR with XOR changes the predicate from nonempty membership to odd match count.
\begin{lstlisting}[style=appendixcode]
--- golden/gbe_rx_filter.v
+++ m7/gbe_rx_filter.v
@@ -606,7 +606,7 @@
 assign  ipaddr_match_t[6]       =   (dst_ip==filter_ip_addr7)   ?   1'b1    :   1'b0;
 assign  ipaddr_match_t[7]       =   (dst_ip==filter_ip_addr8)   ?   1'b1    :   1'b0;
 
-assign  ip_addr_filter_ok_flag  =   |(ipaddr_match_t[7:0]);                                                     
+assign  ip_addr_filter_ok_flag  =   ^(ipaddr_match_t[7:0]);                                                     
 
 always@(posedge clk or posedge rst)
 begin
\end{lstlisting}
Set two entries to the destination IP and leave the others different. Send a complete, error-free non-UDP IPv4 frame that reaches address filtering. The match vector is \texttt{00000011}: golden logic reports \texttt{pkt\_type=2} at frame end, whereas the mutant reports 0. Unique table entries mask the defect. The example turns overlapping conditions into a verification obligation and ties activation to the frame-end observation point.

\clearpage
\subsection{Transaction multiplicity and delayed consequences}
\subsubsection{A response that leaves stale ownership: \texttt{dv-017}/m7}
The SRAM arbiter stores ownership of outstanding reads. A response must reach its port and remove its ownership entry. Replacing reduction OR with AND prevents a one-hot steering vector from acknowledging removal.
\begin{lstlisting}[style=appendixcode]
--- golden/prim_sram_arbiter.sv
+++ m7/prim_sram_arbiter.sv
@@ -122,7 +122,7 @@
   logic [N-1:0] steer;    // Steering sram_rvalid_i
   logic sram_ack;         // Ack for rvalid. |sram_rvalid_i
 
-  assign sram_ack = sram_rvalid_i & (|steer);
+  assign sram_ack = sram_rvalid_i & (&steer);
 
   // Request FIFO
   prim_fifo_sync #(
\end{lstlisting}
After reset and the specified startup cycle, issue reads from ports 0 and 1, stop issuing requests, and supply two responses. The first can reach port 0 correctly in both implementations. The second belongs to port 1 in golden logic but remains associated with port 0 in the mutant. Keep outstanding requests within the four-entry capacity. Checking returned data alone misses ownership errors on a broadcast data bus; the port-valid vector must also be checked. Evidence appears in the next transaction, after the faulty update.

\subsubsection{Equal data are still two transactions: \texttt{dv-009}/m1}
Every legal FIFO push must be preserved, including repeated values. Mutant m1 suppresses a push into a nonempty queue when there is no simultaneous pop and the value equals the previous item.
\begin{lstlisting}[style=appendixcode]
--- golden/cva6_fifo_v3.sv
+++ m1/cva6_fifo_v3.sv
@@ -68,7 +68,7 @@
     gate_clock = 1'b1;
 
     // push a new element to the queue
-    if (push_i && ~full_o) begin
+    if (push_i && ~full_o && (pop_i || status_cnt_q == 0 || data_i != mem_q[write_pointer_q == 0 ? FifoDepth-1 : write_pointer_q-1])) begin
         // push the data onto the queue
         mem_n[write_pointer_q] = data_i;
         // un-gate the clock, we want to write something
@@ -206,7 +206,7 @@
     end
 
     // push a new element to the queue
-    if (push_i && ~full_o) begin
+    if (push_i && ~full_o && (pop_i || status_cnt_q == 0 || data_i != mem_q[write_pointer_q == 0 ? FifoDepth-1 : write_pointer_q-1])) begin
       if (FPGA_EN) begin
         fifo_ram_we = 1'b1;
         fifo_ram_write_address = write_pointer_q;
\end{lstlisting}
Push \texttt{0x11223344} on two consecutive cycles with pop low, then deassert push. Golden usage is 2; mutant usage is 1. After one legal pop, golden still holds an item while the mutant is empty. Occupancy and empty status distinguish the implementations without using undefined empty data. Independent random 32-bit payloads produce adjacent equality with probability $2^{-32}$, making a directed repeat the appropriate trigger. Verification must preserve transaction multiplicity as well as value equality.

\clearpage
\subsection{Observation windows and accepted transactions}
\subsubsection{Write-first versus read-first memory: \texttt{dv-031}/m2}
Changing the port-A primitive configuration from write-first to read-first preserves stored data but changes the output during the write itself.
\begin{lstlisting}[style=appendixcode]
--- golden/BRAM_TDP_24w_256d_sim_netlist.v
+++ m2/BRAM_TDP_24w_256d_sim_netlist.v
@@ -780,7 +780,7 @@
     .SLEEP_ASYNC("FALSE"),
     .SRVAL_A(36'h000000000),
     .SRVAL_B(36'h000000000),
-    .WRITE_MODE_A("WRITE_FIRST"),
+    .WRITE_MODE_A("READ_FIRST"),
     .WRITE_MODE_B("WRITE_FIRST"),
     .WRITE_WIDTH_A(36),
     .WRITE_WIDTH_B(36)) 
\end{lstlisting}
After startup and stabilization, initialize address \texttt{0x12} to \texttt{0x111111}, then write \texttt{0x222222} there with reset inactive. At least 1\,ns after the write edge, golden \texttt{douta} is the new value; the mutant shows the old value. Avoid conflicting port-B access. A later read returns the new value in both, so eventual contents cannot establish write-cycle correctness. Primitive parameters encode functional obligations just as procedural RTL does.

\subsubsection{Reset without a clock edge: \texttt{dv-094}/m1}
The APB NTT wrapper specifies asynchronous active-low control reset. Removing the reset event leaves the branch present but changes when it executes.
\begin{lstlisting}[style=appendixcode]
--- golden/apb_ntt_apb_if.sv
+++ m1/apb_ntt_apb_if.sv
@@ -83,7 +83,7 @@
     end
   end
 
-  always_ff @(posedge i_ntt_pclk or negedge i_ntt_presetn) begin
+  always_ff @(posedge i_ntt_pclk) begin
     if (!i_ntt_presetn) begin
       o_ntt_start       <= 1'b0;
       o_ntt_mode        <= MODE_FWD_NTT;
\end{lstlisting}
Write \texttt{COEFF\_ADDR=0x37} through legal APB traffic, pause the clock, and apply a reset pulse with no rising edge. Before restarting the clock, observe the combinational read at \texttt{0x08}. Golden control state is zero; the mutant retains \texttt{0x37}. Committed coefficient storage retains its specified contents. Reset value and reset timing are separate obligations: a pulse spanning clock edges does not distinguish these implementations.

\clearpage
\subsubsection{Accepted address versus live address: \texttt{dv-095}/m1}
AXI4-Lite accepts address and data independently. Write effects must use the accepted address; this mutant instead decodes the live address bus at commit.
\begin{lstlisting}[style=appendixcode]
--- golden/csr.sv
+++ m1/csr.sv
@@ -127,7 +127,7 @@
 
             // ---- register write effects (bus) ----
             if (do_write) begin
-                case (awaddr_q)
+                case (s_axil_awaddr)
                     ADDR_CTRL: begin
                         if (wdata_q[1]) begin
                             soft_reset_reg <= 1'b1;   // abort: pulse reset...
\end{lstlisting}
Accept AW at \texttt{0x0C} (interrupt enable), deassert AWVALID, and legally change AWADDR to \texttt{0x00}. Deliver write data 1 afterwards. Golden logic enables the interrupt; the mutant interprets a control-register write. Readback at \texttt{0x0C} distinguishes the effects even if both return OKAY. Holding the address until response hides the fault by constraining the sender beyond protocol requirements. This is the temporal distinction discussed in Section~\ref{sec:failure-analysis}.

\clearpage
\subsection{Cross-operation cryptographic context}
\subsubsection{A new IV with an unchanged key: \texttt{dv-063}/m1}
AES-CTR must load the supplied IV on every accepted idle-state load. Mutant m1 remembers the prior key and reloads the counter only when the key changes, tying two independent context updates together.
\begin{lstlisting}[style=appendixcode]
--- golden/AesCTR.sv
+++ m1/AesCTR.sv
@@ -17,6 +17,17 @@
 	logic[127:0] cnt, nextCnt;
 	logic[127:0] regIBlock;
 	logic enRegIBlock;
+	logic loaded_key_valid;
+	logic [127:0] loaded_key;
+	always_ff @(posedge clk or posedge rst) begin
+		if (rst) begin
+			loaded_key_valid <= 1'b0;
+			loaded_key <= '0;
+		end else if (step == IDLE && load) begin
+			loaded_key_valid <= 1'b1;
+			loaded_key <= key;
+		end
+	end
 
 	// ==== state register ====
 	always_ff @(posedge clk or posedge rst)
@@ -64,7 +75,7 @@
 			case(step)
 				IDLE :
 					if (load) begin
-						nextCnt  = iv;
+						if (!loaded_key_valid || loaded_key != key) nextCnt = iv;
 						nextStep = KEY_EXPANSION;
 					end else if (start) begin
 						nextStep    = RUNNING;
\end{lstlisting}
Load key zero and IV zero, process one all-zero block, and wait until idle; the old counter is then 1. Load the same key with IV 2 and process another all-zero block. Golden ciphertext is $\mathrm{AES}_K(2)$; the mutant produces $\mathrm{AES}_K(1)$. AES is a permutation, so the outputs differ. Compare at the contract's completion point. Single-session vectors, or tests that change key and IV together, do not exercise the key-reuse obligation.

\noindent\textbf{What this collection demonstrates.} Small edits require precise and diverse verification arguments: carefully chosen bits, repeated values, overlapping matches, observations between clock edges, or a second transaction exposing corrupted history. DVBench makes those arguments executable. Coupling mutation assessment with golden acceptance requires a generated testbench to accept correct behavior and distinguish realistic deviations from it.

\clearpage
\section{Released Task Catalog}
\label{app:catalog}
The catalog gives the public identifier, released difficulty, primary domain, and design name for each task. It connects the worked examples and the numerical appendices without relying on historical directory names. The domain counts are 8 arithmetic, 13 memory, 24 digital-design, 8 interconnect, 7 control, 5 communication, 16 processor, 10 cryptography, and 4 signal-processing tasks.
\begingroup
\fontsize{8.8}{10.8}\selectfont
\setlength{\tabcolsep}{3pt}\renewcommand{\arraystretch}{1.04}
\setlength{\LTleft}{0pt}\setlength{\LTright}{0pt}
\Needspace{10\baselineskip}
\captionsetup{hypcap=false}
\captionof{table}{Released task catalog used to join the experiment records. Published task IDs, rather than legacy names or row positions, define the join.}\label{tab:app-catalog}
\addtocounter{table}{-1}
\begin{longtable}{>{\raggedright\arraybackslash}p{\dimexpr 0.1200000\textwidth-2\tabcolsep\relax}>{\raggedright\arraybackslash}p{\dimexpr 0.0900000\textwidth-2\tabcolsep\relax}>{\raggedright\arraybackslash}p{\dimexpr 0.2400000\textwidth-2\tabcolsep\relax}>{\raggedright\arraybackslash}p{\dimexpr 0.5500000\textwidth-2\tabcolsep\relax}}
\toprule
\rowcolor{ReleasePale}\textbf{Task} & \textbf{Level} & \textbf{Domain} & \textbf{Design} \\
\midrule\endfirsthead
\multicolumn{4}{l}{\small\color{ReleaseMuted}Table~\ref{tab:app-catalog} (continued)}\\
\toprule
\rowcolor{ReleasePale}\textbf{Task} & \textbf{Level} & \textbf{Domain} & \textbf{Design} \\
\midrule\endhead
\midrule
\multicolumn{4}{r}{\footnotesize\color{ReleaseMuted}Continued on next page}\\\endfoot
\bottomrule\endlastfoot
dv-001 & L1 & Arithmetic & four-bit-alu-display \\
dv-002 & L1 & Arithmetic & openmsp430-alu \\
dv-003 & L1 & Arithmetic & aluexeunit \\
dv-004 & L1 & Memory & tlfifofixer \\
dv-005 & L1 & Digital design & calculator \\
dv-006 & L1 & Arithmetic & ibex-alu \\
dv-007 & L2 & Memory & ram-2x336 \\
dv-008 & L2 & Memory & gray-cdc-and-buffers \\
dv-009 & L2 & Memory & cva6-fifo \\
dv-010 & L2 & Interconnect & hellapeekingarbiter \\
dv-011 & L2 & Digital design & vga \\
dv-012 & L2 & Digital design & s1196 \\
dv-013 & L2 & Control & timer \\
dv-014 & L3 & Digital design & ap3216c-sensor \\
dv-015 & L3 & Communication & i2c-controller \\
dv-016 & L3 & Digital design & gbe-rx-filter \\
dv-017 & L3 & Digital design & sram-arbiter \\
dv-018 & L3 & Memory & fifo-width-convert \\
dv-019 & L3 & Communication & ps2-keyboard-uart \\
dv-020 & L3 & Memory & watch-sdram-controller \\
dv-021 & L3 & Control & stopwatch-watch \\
dv-022 & L3 & Digital design & softmc-reorder-queue \\
dv-023 & L3 & Digital design & fifo \\
dv-024 & L3 & Memory & tlbuffer \\
dv-025 & L3 & Control & tlplic \\
dv-026 & L3 & Control & clint \\
dv-027 & L3 & Digital design & camera-reader \\
dv-028 & L3 & Processor & registerread \\
dv-029 & L3 & Memory & psx-memory-arbitrator \\
dv-030 & L3 & Control & digital-password-lock \\
dv-031 & L3 & Memory & bram-24x256 \\
dv-032 & L3 & Digital design & jtag-apb-debug \\
dv-033 & L3 & Digital design & zx48-user-io \\
dv-034 & L3 & Control & alarm-clock \\
dv-035 & L3 & Memory & rocev2-ip-fifo \\
dv-036 & L3 & Communication & mdio-controller \\
dv-037 & L3 & Control & watchdog-timer \\
dv-038 & L3 & Memory & ibuffer \\
dv-039 & L3 & Interconnect & nv-nvdla-cdp-rdma-cq \\
dv-040 & L3 & Digital design & nv-nvdla-glb \\
dv-041 & L3 & Digital design & boomiomshr \\
dv-042 & L3 & Processor & csrexe \\
dv-043 & L3 & Processor & rvc-decoder \\
dv-044 & L3 & Arithmetic & multiply \\
dv-045 & L3 & Digital design & s38584 \\
dv-046 & L4 & Arithmetic & fpnew-fma \\
dv-047 & L4 & Cryptography & decryption \\
dv-048 & L4 & Digital design & dpc-5x5 \\
dv-049 & L4 & Communication & xgmii-pcs-10g-tx \\
dv-050 & L4 & Cryptography & sha256-core \\
dv-051 & L4 & Processor & bimbranchpredictorbank \\
dv-052 & L4 & Cryptography & aes128-decrypt-pipeline \\
dv-053 & L4 & Processor & btbbranchpredictorbank \\
dv-054 & L4 & Cryptography & eth-rxethmac \\
dv-055 & L4 & Signal processing & fft-wrapper \\
dv-056 & L4 & Digital design & decoder \\
dv-057 & L4 & Signal processing & minicnn \\
dv-058 & L4 & Signal processing & video-filter-top \\
dv-059 & L4 & Cryptography & ascon-top \\
dv-060 & L4 & Processor & directory-controller-stage3 \\
dv-061 & L4 & Processor & fetchtargetqueue \\
dv-062 & L4 & Arithmetic & fp32-sigmoid-pipeline \\
dv-063 & L4 & Cryptography & aesctr \\
dv-064 & L4 & Interconnect & apb-timer-unit \\
dv-065 & L4 & Digital design & nv-nvdla-cdma-cvt \\
dv-066 & L4 & Processor & simt-stack \\
dv-067 & L4 & Processor & tagetable \\
dv-068 & L4 & Interconnect & tlfragmenter \\
dv-069 & L4 & Digital design & vmul-top \\
dv-070 & L4 & Digital design & elevator \\
dv-071 & L4 & Communication & ethernet-tx \\
dv-072 & L4 & Cryptography & des \\
dv-073 & L4 & Cryptography & sha3-mmio \\
dv-074 & L4 & Arithmetic & fpu-sqrt \\
dv-075 & L5 & Interconnect & axi-rdma \\
dv-076 & L5 & Processor & abd-riscv-soc \\
dv-077 & L5 & Cryptography & logic-hash \\
dv-078 & L5 & Processor & fpga-cpu \\
dv-079 & L5 & Digital design & pong \\
dv-080 & L5 & Interconnect & axi-4x4-interconnect \\
dv-081 & L5 & Processor & 16bit-cpu \\
dv-082 & L5 & Processor & rv32i-single-cycle-core \\
dv-083 & L5 & Digital design & lm32-dcache \\
dv-084 & L5 & Interconnect & yapp-router \\
dv-085 & L5 & Memory & cv32e40p-icache \\
dv-086 & L5 & Memory & instruction-cache \\
dv-087 & L5 & Digital design & vga \\
dv-088 & L5 & Digital design & nv-nvdla-bdma \\
dv-089 & L5 & Processor & will-cpu \\
dv-090 & L5 & Processor & mips \\
dv-091 & L5 & Cryptography & shared-mem \\
dv-092 & L5 & Processor & warp-scheduler \\
dv-093 & L5 & Digital design & nv-nvdla-csb-master \\
dv-094 & L5 & Interconnect & apb-ntt \\
dv-095 & L5 & Signal processing & systolic \\
\end{longtable}\endgroup

\clearpage
\section{Complete Per-Task Measurements}
\label{app:measurements}
The following tables retain every configuration/task pair. Coverage, document F1, reference point accuracy, and mutant kill are percentages; compile, golden, and overall fields are binary. Token and runtime columns use millions and minutes for legibility. The accompanying \texttt{task-metrics.csv} and JSON preserve full numeric precision, input/output token splits, feature/test-plan precision and recall, strict-model flags, legacy IDs, source sheet/row coordinates, and the recorded status notes. Success decisions are made before presentation rounding.

The source manifest records workbook, catalog, and example-document hashes. The accompanying builder checks extraction consistency; it does not rerun simulation or requalify mutation assets.
\subsection{C01: Claude Code + GLM 5.3}
\noindent\footnotesize Percentages: Ln/Cn/Br/Tg = line/condition/branch/toggle coverage; F1/P1 = feature/test-plan F1; Ref = reference-model point accuracy; Kill = mutant kill rate. C/G and O are binary compile/golden and overall results. Tok = input plus output tokens in millions; Min = elapsed minutes. Overall uses exact model correctness and all four coverage thresholds. \par\normalsize
\begingroup
\fontsize{8.2}{10.2}\selectfont
\setlength{\tabcolsep}{3pt}\renewcommand{\arraystretch}{1.04}
\setlength{\LTleft}{0pt}\setlength{\LTright}{0pt}
\Needspace{10\baselineskip}
\captionsetup{hypcap=false}
\captionof{table}{Per-task measurements for C01. The companion CSV retains unrounded values, precision/recall, exact-model flags, input/output token splits, and source-cell coordinates.}\label{tab:app-detail-C01}
\addtocounter{table}{-1}
\begin{longtable}{>{\raggedright\arraybackslash}p{\dimexpr 0.1050000\textwidth-2\tabcolsep\relax}>{\raggedleft\arraybackslash}p{\dimexpr 0.0745833\textwidth-2\tabcolsep\relax}>{\raggedleft\arraybackslash}p{\dimexpr 0.0745833\textwidth-2\tabcolsep\relax}>{\raggedleft\arraybackslash}p{\dimexpr 0.0745833\textwidth-2\tabcolsep\relax}>{\raggedleft\arraybackslash}p{\dimexpr 0.0745833\textwidth-2\tabcolsep\relax}>{\raggedleft\arraybackslash}p{\dimexpr 0.0745833\textwidth-2\tabcolsep\relax}>{\raggedleft\arraybackslash}p{\dimexpr 0.0745833\textwidth-2\tabcolsep\relax}>{\raggedleft\arraybackslash}p{\dimexpr 0.0745833\textwidth-2\tabcolsep\relax}>{\raggedleft\arraybackslash}p{\dimexpr 0.0745833\textwidth-2\tabcolsep\relax}>{\raggedleft\arraybackslash}p{\dimexpr 0.0745833\textwidth-2\tabcolsep\relax}>{\raggedleft\arraybackslash}p{\dimexpr 0.0745833\textwidth-2\tabcolsep\relax}>{\raggedleft\arraybackslash}p{\dimexpr 0.0745833\textwidth-2\tabcolsep\relax}>{\raggedleft\arraybackslash}p{\dimexpr 0.0745833\textwidth-2\tabcolsep\relax}}
\toprule
\rowcolor{ReleasePale}\textbf{Task} & \textbf{Ln} & \textbf{Cn} & \textbf{Br} & \textbf{Tg} & \textbf{F1} & \textbf{P1} & \textbf{Ref} & \textbf{C/G} & \textbf{Kill} & \textbf{O} & \textbf{Tok} & \textbf{Min} \\
\midrule\endfirsthead
\multicolumn{13}{l}{\small\color{ReleaseMuted}Table~\ref{tab:app-detail-C01} (continued)}\\
\toprule
\rowcolor{ReleasePale}\textbf{Task} & \textbf{Ln} & \textbf{Cn} & \textbf{Br} & \textbf{Tg} & \textbf{F1} & \textbf{P1} & \textbf{Ref} & \textbf{C/G} & \textbf{Kill} & \textbf{O} & \textbf{Tok} & \textbf{Min} \\
\midrule\endhead
\midrule
\multicolumn{13}{r}{\footnotesize\color{ReleaseMuted}Continued on next page}\\\endfoot
\bottomrule\endlastfoot
dv-001 & 100.0 & 93.0 & 100.0 & 96.9 & 96.0 & 96.7 & 100.0 & 1/1 & 100.0 & \textbf{\color{ReleaseBlue}1} & 0.984 & 14.77 \\
dv-002 & 100.0 & 84.9 & 73.1 & 95.9 & 87.0 & 15.0 & 100.0 & 1/0 & 0.0 & 0 & 18.877 & 102.72 \\
dv-003 & 100.0 & 100.0 & 100.0 & 100.0 & 67.6 & 38.8 & 100.0 & 1/1 & 100.0 & \textbf{\color{ReleaseBlue}1} & 21.158 & 128.54 \\
dv-004 & 50.5 & 26.3 & 13.5 & 0.3 & 100.0 & 0.0 & 100.0 & 1/0 & 0.0 & 0 & 25.912 & 196.85 \\
dv-005 & 100.0 & 99.3 & 100.0 & 98.9 & 91.9 & 44.7 & 100.0 & 1/1 & 100.0 & \textbf{\color{ReleaseBlue}1} & 2.043 & 43.65 \\
dv-006 & 0.0 & 0.0 & 0.0 & 0.0 & 0.0 & 0.0 & 99.7 & 1/0 & 0.0 & 0 & 20.887 & 619.16 \\
dv-007 & 100.0 & 100.0 & 100.0 & 100.0 & 76.7 & 21.8 & 100.0 & 1/0 & 0.0 & 0 & 4.505 & 140.04 \\
dv-008 & 80.8 & 80.1 & 77.1 & 86.1 & 0.0 & 0.0 & 0.0 & 1/0 & 0.0 & 0 & 25.656 & 461.27 \\
dv-009 & 59.3 & 50.0 & 47.8 & 85.9 & 0.0 & 0.0 & 44.6 & 1/0 & 0.0 & 0 & 8.057 & 276.58 \\
dv-010 & 100.0 & 100.0 & 100.0 & 99.3 & 100.0 & 0.0 & 100.0 & 1/1 & 100.0 & \textbf{\color{ReleaseBlue}1} & 8.792 & 155.69 \\
dv-011 & 98.4 & 93.5 & 91.3 & 52.9 & 100.0 & 83.9 & 100.0 & 1/1 & 100.0 & 0 & 0.372 & 3.71 \\
dv-012 & 6.2 & 1.4 & 50.0 & 0.2 & 0.0 & 0.0 & 100.0 & 1/0 & 0.0 & 0 & 21.804 & 141.47 \\
dv-013 & 0.0 & 0.0 & 0.0 & 0.0 & 92.8 & 0.0 & 100.0 & 1/1 & 90.0 & 0 & 8.923 & 195.88 \\
dv-014 & 0.0 & 0.0 & 0.0 & 0.0 & 0.0 & 0.0 & 0.0 & 0/0 & 0.0 & 0 & 10.402 & 230.44 \\
dv-015 & 100.0 & 100.0 & 97.3 & 89.4 & 45.4 & 28.3 & 100.0 & 1/1 & 90.0 & 0 & 54.466 & 631.10 \\
dv-016 & 94.5 & 79.0 & 86.7 & 45.6 & 0.0 & 0.0 & 80.4 & 1/1 & 30.0 & 0 & 26.808 & 284.81 \\
dv-017 & 96.8 & 77.8 & 88.4 & 90.3 & 92.3 & 81.9 & 89.9 & 1/1 & 90.0 & 0 & 20.266 & 174.52 \\
dv-018 & 100.0 & 100.0 & 100.0 & 94.8 & 0.0 & 0.0 & 100.0 & 0/0 & 0.0 & 0 & 14.810 & 202.88 \\
dv-019 & 74.6 & 75.8 & 54.5 & 98.5 & 0.0 & 0.0 & 98.4 & 1/0 & 0.0 & 0 & 22.785 & 295.39 \\
dv-020 & 97.1 & 93.2 & 85.5 & 95.1 & 0.0 & 0.0 & 70.8 & 1/0 & 0.0 & 0 & 22.636 & 291.64 \\
dv-021 & 68.9 & 42.0 & 58.8 & 54.8 & 0.0 & 0.0 & 93.3 & 1/0 & 0.0 & 0 & 12.573 & 198.40 \\
dv-022 & 0.0 & 0.0 & 0.0 & 0.0 & 0.0 & 0.0 & 4.3 & 0/0 & 0.0 & 0 & 25.726 & 389.65 \\
dv-023 & 100.0 & 100.0 & 100.0 & 100.0 & 0.0 & 0.0 & 100.0 & 1/1 & 70.0 & 0 & 4.693 & 242.60 \\
dv-024 & 42.9 & 57.2 & 26.6 & 32.0 & 38.1 & 0.0 & 99.8 & 1/1 & 20.0 & 0 & 26.202 & 222.90 \\
dv-025 & 40.1 & 30.8 & 22.6 & 0.1 & 0.0 & 0.0 & 100.0 & 1/0 & 0.0 & 0 & 26.581 & 190.11 \\
dv-026 & 42.1 & 48.8 & 24.9 & 8.1 & 91.1 & 53.0 & 100.0 & 1/1 & 60.0 & 0 & 12.204 & 207.08 \\
dv-027 & 0.0 & 0.0 & 0.0 & 0.0 & 0.0 & 0.0 & 0.0 & 0/0 & 0.0 & 0 & 7.163 & 651.49 \\
dv-028 & 100.0 & 98.0 & 100.0 & 98.7 & 21.0 & 0.0 & 100.0 & 1/1 & 90.0 & 0 & 20.593 & 204.16 \\
dv-029 & 91.7 & 88.6 & 80.7 & 19.5 & 77.0 & 10.3 & 3.5 & 1/0 & 0.0 & 0 & 25.519 & 377.31 \\
dv-030 & 0.0 & 0.0 & 0.0 & 0.0 & 0.0 & 0.0 & 5.5 & 1/0 & 0.0 & 0 & 13.139 & 837.38 \\
dv-031 & 0.0 & 0.0 & 0.0 & 0.0 & 0.0 & 0.0 & 0.0 & 0/0 & 0.0 & 0 & 0.168 & 206.50 \\
dv-032 & 84.6 & 54.4 & 76.3 & 29.5 & 0.0 & 0.0 & 1.8 & 1/0 & 0.0 & 0 & 19.064 & 335.90 \\
dv-033 & 77.4 & 67.7 & 64.6 & 38.1 & 0.0 & 0.0 & 0.0 & 1/0 & 0.0 & 0 & 22.744 & 384.59 \\
dv-034 & 78.3 & 92.9 & 72.3 & 39.3 & 31.9 & 0.0 & 100.0 & 0/0 & 0.0 & 0 & 4.958 & 181.81 \\
dv-035 & 0.0 & 0.0 & 0.0 & 0.0 & 0.0 & 0.0 & 36.7 & 0/0 & 0.0 & 0 & 1.628 & 163.47 \\
dv-036 & 94.4 & 83.4 & 86.3 & 70.3 & 61.7 & 0.0 & 34.2 & 1/1 & 50.0 & 0 & 18.588 & 574.52 \\
dv-037 & 89.6 & 77.9 & 84.3 & 28.6 & 0.0 & 0.0 & 16.2 & 1/0 & 0.0 & 0 & 26.322 & 378.68 \\
dv-038 & 100.0 & 97.7 & 100.0 & 99.6 & 95.0 & 0.0 & 100.0 & 1/1 & 90.0 & 0 & 13.623 & 171.90 \\
dv-039 & 99.6 & 86.3 & 99.6 & 36.7 & 84.8 & 13.2 & 100.0 & 1/1 & 80.0 & 0 & 16.120 & 204.48 \\
dv-040 & 100.0 & 93.9 & 99.2 & 68.3 & 90.9 & 13.2 & 100.0 & 1/1 & 70.0 & 0 & 13.485 & 201.49 \\
dv-041 & 99.3 & 97.7 & 85.0 & 99.7 & 88.5 & 0.0 & 99.8 & 1/1 & 90.0 & 0 & 23.227 & 231.69 \\
dv-042 & 0.0 & 0.0 & 0.0 & 0.0 & 0.0 & 0.0 & 22.7 & 1/0 & 0.0 & 0 & 4.950 & 154.80 \\
dv-043 & 85.6 & 95.1 & 75.9 & 100.0 & 8.1 & 59.2 & 100.0 & 1/1 & 100.0 & 0 & 0.644 & 12.01 \\
dv-044 & 0.0 & 0.0 & 0.0 & 0.0 & 0.0 & 0.0 & 56.3 & 1/0 & 0.0 & 0 & 10.456 & 422.83 \\
dv-045 & 0.0 & 0.0 & 0.0 & 0.0 & 0.0 & 0.0 & 0.0 & 0/0 & 0.0 & 0 & 3.980 & 19.28 \\
dv-046 & 0.0 & 0.0 & 0.0 & 0.0 & 84.6 & 0.0 & 0.0 & 0/0 & 0.0 & 0 & 26.724 & 1201.02 \\
dv-047 & 0.0 & 0.0 & 0.0 & 0.0 & 77.2 & 0.0 & 0.0 & 0/0 & 0.0 & 0 & 25.976 & 140.64 \\
dv-048 & 0.0 & 0.0 & 0.0 & 0.0 & 83.9 & 13.1 & 57.7 & 0/0 & 0.0 & 0 & 21.243 & 242.73 \\
dv-049 & 99.7 & 99.7 & 95.5 & 98.0 & 79.6 & 21.8 & 14.8 & 0/0 & 0.0 & 0 & 23.664 & 800.32 \\
dv-050 & 97.7 & 80.0 & 98.2 & 92.7 & 69.1 & 50.8 & 54.3 & 1/0 & 0.0 & 0 & 24.345 & 195.15 \\
dv-051 & 71.2 & 65.7 & 33.9 & 86.7 & 70.1 & 0.0 & 54.5 & 1/0 & 0.0 & 0 & 25.057 & 386.95 \\
dv-052 & 95.9 & 38.8 & 93.5 & 99.6 & 64.0 & 0.0 & 100.0 & 1/1 & 90.0 & 0 & 12.734 & 212.11 \\
dv-053 & 0.0 & 0.0 & 0.0 & 0.0 & 0.0 & 0.0 & 65.2 & 0/0 & 0.0 & 0 & 33.100 & 482.08 \\
dv-054 & 0.0 & 0.0 & 0.0 & 0.0 & 0.0 & 0.0 & 0.0 & 0/0 & 0.0 & 0 & 20.867 & 180.22 \\
dv-055 & 0.0 & 0.0 & 0.0 & 0.0 & 93.3 & 38.1 & 100.0 & 0/0 & 0.0 & 0 & 28.230 & 162.74 \\
dv-056 & 0.0 & 0.0 & 0.0 & 0.0 & 0.0 & 0.0 & 0.0 & 0/0 & 0.0 & 0 & 19.904 & 155.40 \\
dv-057 & 0.0 & 0.0 & 0.0 & 0.0 & 73.3 & 0.0 & 0.0 & 0/0 & 0.0 & 0 & 24.751 & 112.42 \\
dv-058 & 0.0 & 0.0 & 0.0 & 0.0 & 0.0 & 0.0 & 0.0 & 0/0 & 0.0 & 0 & 25.089 & 200.05 \\
dv-059 & 0.0 & 0.0 & 0.0 & 0.0 & 0.0 & 0.0 & 0.0 & 0/0 & 0.0 & 0 & 6.835 & 51.19 \\
dv-060 & 0.0 & 0.0 & 0.0 & 0.0 & 92.9 & 0.0 & 0.0 & 0/0 & 0.0 & 0 & 10.849 & 149.99 \\
dv-061 & 99.5 & 81.7 & 99.2 & 79.9 & 0.0 & 0.0 & 0.2 & 1/0 & 0.0 & 0 & 27.389 & 1114.26 \\
dv-062 & 0.0 & 0.0 & 0.0 & 0.0 & 0.0 & 0.0 & 93.9 & 0/0 & 0.0 & 0 & 18.605 & 119.50 \\
dv-063 & 63.5 & 100.0 & 62.9 & 96.8 & 38.0 & 0.0 & 63.1 & 1/0 & 0.0 & 0 & 23.374 & 224.29 \\
dv-064 & 96.5 & 89.5 & 93.6 & 72.2 & 69.9 & 0.0 & 97.4 & 1/1 & 90.0 & 0 & 16.960 & 157.59 \\
dv-065 & 98.4 & 74.2 & 79.6 & 89.3 & 80.9 & 10.5 & 100.0 & 1/1 & 70.0 & 0 & 23.964 & 340.65 \\
dv-066 & 95.8 & 95.3 & 93.2 & 19.3 & 100.0 & 0.0 & 100.0 & 1/1 & 80.0 & 0 & 21.281 & 198.39 \\
dv-067 & 80.2 & 87.1 & 81.2 & 90.3 & 41.9 & 0.0 & 99.6 & 1/1 & 50.0 & 0 & 19.510 & 358.81 \\
dv-068 & 0.0 & 0.0 & 0.0 & 0.0 & 96.3 & 0.0 & 63.9 & 0/0 & 0.0 & 0 & 24.092 & 165.15 \\
dv-069 & 100.0 & 80.7 & 100.0 & 82.6 & 58.0 & 0.0 & 100.0 & 1/1 & 100.0 & 0 & 17.907 & 179.99 \\
dv-070 & 73.0 & 46.5 & 52.2 & 29.8 & 88.1 & 52.6 & 72.9 & 1/1 & 40.0 & 0 & 1.947 & 23.46 \\
dv-071 & 99.6 & 62.1 & 85.1 & 87.9 & 89.7 & 93.3 & 100.0 & 1/1 & 90.0 & 0 & 1.418 & 18.36 \\
dv-072 & 100.0 & 100.0 & 100.0 & 99.9 & 100.0 & 80.8 & 100.0 & 1/1 & 80.0 & 0 & 1.669 & 14.52 \\
dv-073 & 96.8 & 84.2 & 92.2 & 89.7 & 100.0 & 94.7 & 100.0 & 1/1 & 70.0 & 0 & 19.663 & 149.04 \\
dv-074 & 100.0 & 100.0 & 100.0 & 92.6 & 99.1 & 97.1 & 100.0 & 1/0 & 0.0 & 0 & 24.417 & 255.06 \\
dv-075 & 98.1 & 83.9 & 88.5 & 75.4 & 0.0 & 0.0 & 0.0 & 1/0 & 0.0 & 0 & 17.853 & 199.31 \\
dv-076 & 72.8 & 77.8 & 55.4 & 76.5 & 95.7 & 0.0 & 100.0 & 1/1 & 90.0 & 0 & 13.834 & 80.76 \\
dv-077 & 0.0 & 0.0 & 0.0 & 0.0 & 89.8 & 0.0 & 0.0 & 0/0 & 0.0 & 0 & 13.491 & 524.08 \\
dv-078 & 71.3 & 76.6 & 63.3 & 45.0 & 0.0 & 0.0 & 0.0 & 1/1 & 90.0 & 0 & 0.445 & 45.81 \\
dv-079 & 89.6 & 68.4 & 77.6 & 68.9 & 54.3 & 21.2 & 100.0 & 1/1 & 30.0 & 0 & 25.560 & 54.28 \\
dv-080 & 0.0 & 0.0 & 0.0 & 0.0 & 70.5 & 19.5 & 94.3 & 0/0 & 0.0 & 0 & 25.704 & 67.27 \\
dv-081 & 43.9 & 44.6 & 41.8 & 39.3 & 90.8 & 0.0 & 100.0 & 1/0 & 0.0 & 0 & 19.481 & 90.31 \\
dv-082 & 96.2 & 97.0 & 92.9 & 99.4 & 55.5 & 24.7 & 99.7 & 1/1 & 100.0 & 0 & 18.313 & 226.72 \\
dv-083 & 0.0 & 0.0 & 0.0 & 0.0 & 0.0 & 0.0 & 99.9 & 0/0 & 0.0 & 0 & 14.136 & 216.49 \\
dv-084 & 95.0 & 76.5 & 83.2 & 61.5 & 69.3 & 33.6 & 16.1 & 1/0 & 0.0 & 0 & 17.445 & 271.22 \\
dv-085 & 0.0 & 0.0 & 0.0 & 0.0 & 77.6 & 0.0 & 97.2 & 0/0 & 0.0 & 0 & 22.164 & 249.85 \\
dv-086 & 96.1 & 81.1 & 91.0 & 96.5 & 52.0 & 0.0 & 100.0 & 1/0 & 0.0 & 0 & 21.685 & 237.75 \\
dv-087 & 0.0 & 0.0 & 0.0 & 0.0 & 96.3 & 33.5 & 100.0 & 0/0 & 0.0 & 0 & 21.470 & 542.13 \\
dv-088 & 88.5 & 67.8 & 72.9 & 25.2 & 45.8 & 0.0 & 0.0 & 1/0 & 0.0 & 0 & 23.868 & 462.23 \\
dv-089 & 79.3 & 50.5 & 64.8 & 58.6 & 52.2 & 0.0 & 7.7 & 1/0 & 0.0 & 0 & 18.569 & 55.53 \\
dv-090 & 36.6 & 60.9 & 38.1 & 65.6 & 73.0 & 78.4 & 100.0 & 1/1 & 80.0 & 0 & 9.438 & 28.96 \\
dv-091 & 97.3 & 100.0 & 95.6 & 99.4 & 0.0 & 0.0 & 100.0 & 1/1 & 70.0 & 0 & 11.911 & 251.95 \\
dv-092 & 96.4 & 78.5 & 77.6 & 83.9 & 81.8 & 0.0 & 100.0 & 1/1 & 80.0 & 0 & 5.685 & 98.72 \\
dv-093 & 99.1 & 87.8 & 98.6 & 85.9 & 75.4 & 30.7 & 98.8 & 1/1 & 90.0 & 0 & 28.774 & 595.73 \\
dv-094 & 98.3 & 98.5 & 94.4 & 61.1 & 51.4 & 39.1 & 100.0 & 1/1 & 80.0 & 0 & 19.324 & 327.85 \\
dv-095 & 99.4 & 82.1 & 97.9 & 96.3 & 90.8 & 21.8 & 100.0 & 1/1 & 50.0 & 0 & 25.630 & 187.33 \\
\end{longtable}\endgroup

\clearpage
\subsection{C02: Claude Code + DeepSeek V4.1 Flash 1017}
\noindent\footnotesize Percentages: Ln/Cn/Br/Tg = line/condition/branch/toggle coverage; F1/P1 = feature/test-plan F1; Ref = reference-model point accuracy; Kill = mutant kill rate. C/G and O are binary compile/golden and overall results. Tok = input plus output tokens in millions; Min = elapsed minutes. Overall uses exact model correctness and all four coverage thresholds. \par\normalsize
\begingroup
\fontsize{8.2}{10.2}\selectfont
\setlength{\tabcolsep}{3pt}\renewcommand{\arraystretch}{1.04}
\setlength{\LTleft}{0pt}\setlength{\LTright}{0pt}
\Needspace{10\baselineskip}
\captionsetup{hypcap=false}
\captionof{table}{Per-task measurements for C02. The companion CSV retains unrounded values, precision/recall, exact-model flags, input/output token splits, and source-cell coordinates.}\label{tab:app-detail-C02}
\addtocounter{table}{-1}
\begin{longtable}{>{\raggedright\arraybackslash}p{\dimexpr 0.1050000\textwidth-2\tabcolsep\relax}>{\raggedleft\arraybackslash}p{\dimexpr 0.0745833\textwidth-2\tabcolsep\relax}>{\raggedleft\arraybackslash}p{\dimexpr 0.0745833\textwidth-2\tabcolsep\relax}>{\raggedleft\arraybackslash}p{\dimexpr 0.0745833\textwidth-2\tabcolsep\relax}>{\raggedleft\arraybackslash}p{\dimexpr 0.0745833\textwidth-2\tabcolsep\relax}>{\raggedleft\arraybackslash}p{\dimexpr 0.0745833\textwidth-2\tabcolsep\relax}>{\raggedleft\arraybackslash}p{\dimexpr 0.0745833\textwidth-2\tabcolsep\relax}>{\raggedleft\arraybackslash}p{\dimexpr 0.0745833\textwidth-2\tabcolsep\relax}>{\raggedleft\arraybackslash}p{\dimexpr 0.0745833\textwidth-2\tabcolsep\relax}>{\raggedleft\arraybackslash}p{\dimexpr 0.0745833\textwidth-2\tabcolsep\relax}>{\raggedleft\arraybackslash}p{\dimexpr 0.0745833\textwidth-2\tabcolsep\relax}>{\raggedleft\arraybackslash}p{\dimexpr 0.0745833\textwidth-2\tabcolsep\relax}>{\raggedleft\arraybackslash}p{\dimexpr 0.0745833\textwidth-2\tabcolsep\relax}}
\toprule
\rowcolor{ReleasePale}\textbf{Task} & \textbf{Ln} & \textbf{Cn} & \textbf{Br} & \textbf{Tg} & \textbf{F1} & \textbf{P1} & \textbf{Ref} & \textbf{C/G} & \textbf{Kill} & \textbf{O} & \textbf{Tok} & \textbf{Min} \\
\midrule\endfirsthead
\multicolumn{13}{l}{\small\color{ReleaseMuted}Table~\ref{tab:app-detail-C02} (continued)}\\
\toprule
\rowcolor{ReleasePale}\textbf{Task} & \textbf{Ln} & \textbf{Cn} & \textbf{Br} & \textbf{Tg} & \textbf{F1} & \textbf{P1} & \textbf{Ref} & \textbf{C/G} & \textbf{Kill} & \textbf{O} & \textbf{Tok} & \textbf{Min} \\
\midrule\endhead
\midrule
\multicolumn{13}{r}{\footnotesize\color{ReleaseMuted}Continued on next page}\\\endfoot
\bottomrule\endlastfoot
dv-001 & 100.0 & 93.0 & 100.0 & 96.9 & 64.9 & 100.0 & 100.0 & 1/1 & 100.0 & \textbf{\color{ReleaseBlue}1} & 4.685 & 11.96 \\
dv-002 & 100.0 & 100.0 & 100.0 & 97.5 & 98.0 & 56.8 & 100.0 & 1/1 & 100.0 & \textbf{\color{ReleaseBlue}1} & 6.446 & 11.84 \\
dv-003 & 100.0 & 100.0 & 100.0 & 100.0 & 89.3 & 75.9 & 100.0 & 1/1 & 90.0 & 0 & 7.245 & 12.78 \\
dv-004 & 99.0 & 94.6 & 96.2 & 84.3 & 98.7 & 0.0 & 100.0 & 1/1 & 100.0 & 0 & 37.346 & 63.46 \\
dv-005 & 100.0 & 99.1 & 100.0 & 98.9 & 72.7 & 65.7 & 100.0 & 1/1 & 100.0 & \textbf{\color{ReleaseBlue}1} & 5.862 & 28.82 \\
dv-006 & 99.5 & 99.8 & 99.8 & 92.8 & 93.8 & 75.2 & 100.0 & 1/1 & 100.0 & \textbf{\color{ReleaseBlue}1} & 11.564 & 17.03 \\
dv-007 & 100.0 & 100.0 & 100.0 & 100.0 & 100.0 & 0.0 & 100.0 & 1/1 & 80.0 & 0 & 2.537 & 9.49 \\
dv-008 & 97.6 & 90.0 & 95.0 & 96.8 & 73.0 & 0.0 & 89.2 & 1/1 & 100.0 & 0 & 22.645 & 47.74 \\
dv-009 & 100.0 & 91.3 & 100.0 & 100.0 & 83.6 & 0.0 & 100.0 & 1/1 & 40.0 & 0 & 6.243 & 14.23 \\
dv-010 & 100.0 & 100.0 & 100.0 & 99.3 & 55.3 & 56.9 & 100.0 & 1/1 & 100.0 & \textbf{\color{ReleaseBlue}1} & 5.126 & 12.42 \\
dv-011 & 98.4 & 93.5 & 91.3 & 52.9 & 65.3 & 70.2 & 100.0 & 1/1 & 100.0 & 0 & 5.220 & 10.78 \\
dv-012 & 100.0 & 99.9 & 100.0 & 100.0 & 91.4 & 80.4 & 100.0 & 1/1 & 100.0 & \textbf{\color{ReleaseBlue}1} & 15.233 & 21.86 \\
dv-013 & 99.0 & 89.4 & 97.1 & 93.7 & 78.9 & 77.9 & 100.0 & 1/1 & 100.0 & 0 & 7.197 & 14.66 \\
dv-014 & 85.2 & 71.4 & 81.2 & 64.7 & 66.9 & 54.8 & 100.0 & 1/1 & 100.0 & 0 & 46.615 & 56.16 \\
dv-015 & 100.0 & 100.0 & 97.3 & 83.7 & 76.4 & 28.2 & 99.9 & 1/1 & 60.0 & 0 & 24.674 & 29.68 \\
dv-016 & 99.3 & 96.9 & 98.9 & 57.8 & 66.5 & 10.3 & 100.0 & 1/1 & 70.0 & 0 & 33.768 & 47.64 \\
dv-017 & 96.8 & 81.3 & 88.4 & 68.2 & 55.9 & 79.1 & 89.9 & 1/1 & 100.0 & 0 & 9.401 & 19.73 \\
dv-018 & 97.4 & 100.0 & 95.7 & 94.8 & 100.0 & 0.0 & 100.0 & 1/1 & 100.0 & \textbf{\color{ReleaseBlue}1} & 9.191 & 22.86 \\
dv-019 & 79.3 & 87.1 & 63.5 & 99.0 & 67.6 & 35.0 & 41.3 & 1/1 & 80.0 & 0 & 26.048 & 39.97 \\
dv-020 & 95.1 & 93.8 & 86.4 & 95.3 & 73.6 & 39.1 & 99.6 & 1/1 & 100.0 & 0 & 46.375 & 60.96 \\
dv-021 & 89.5 & 74.7 & 84.1 & 83.2 & 57.1 & 81.5 & 100.0 & 1/1 & 90.0 & 0 & 25.920 & 62.64 \\
dv-022 & 100.0 & 95.0 & 98.1 & 72.3 & 70.3 & 64.3 & 9.7 & 1/1 & 90.0 & 0 & 27.985 & 51.96 \\
dv-023 & 100.0 & 100.0 & 100.0 & 100.0 & 65.1 & 28.2 & 100.0 & 1/1 & 100.0 & \textbf{\color{ReleaseBlue}1} & 5.611 & 17.29 \\
dv-024 & 43.3 & 53.4 & 27.1 & 36.5 & 51.6 & 0.0 & 100.0 & 1/1 & 10.0 & 0 & 59.099 & 70.23 \\
dv-025 & 40.7 & 54.9 & 23.9 & 15.9 & 94.7 & 0.0 & 100.0 & 1/1 & 20.0 & 0 & 26.304 & 51.69 \\
dv-026 & 42.1 & 59.3 & 24.9 & 20.0 & 84.3 & 66.7 & 100.0 & 1/1 & 80.0 & 0 & 14.586 & 27.02 \\
dv-027 & 97.8 & 90.3 & 96.4 & 57.5 & 43.4 & 19.6 & 80.2 & 1/1 & 70.0 & 0 & 91.845 & 67.80 \\
dv-028 & 100.0 & 98.4 & 100.0 & 98.7 & 79.7 & 71.5 & 100.0 & 1/1 & 100.0 & \textbf{\color{ReleaseBlue}1} & 10.534 & 24.86 \\
dv-029 & 98.8 & 98.9 & 92.9 & 88.5 & 34.9 & 49.5 & 17.9 & 1/1 & 70.0 & 0 & 59.980 & 71.99 \\
dv-030 & 49.4 & 38.5 & 47.6 & 71.4 & 48.6 & 0.0 & 100.0 & 1/1 & 60.0 & 0 & 17.148 & 99.82 \\
dv-031 & 62.8 & 4.0 & 38.8 & 31.8 & 52.2 & 39.4 & 99.9 & 1/1 & 80.0 & 0 & 10.211 & 19.49 \\
dv-032 & 96.6 & 83.8 & 94.6 & 92.6 & 77.9 & 0.0 & 74.5 & 1/1 & 50.0 & 0 & 46.597 & 77.93 \\
dv-033 & 78.3 & 74.2 & 73.2 & 42.3 & 43.6 & 15.1 & 100.0 & 1/1 & 80.0 & 0 & 32.867 & 39.94 \\
dv-034 & 79.8 & 94.0 & 74.8 & 39.6 & 78.4 & 44.7 & 100.0 & 1/1 & 100.0 & 0 & 9.900 & 18.71 \\
dv-035 & 79.0 & 52.0 & 62.9 & 95.6 & 67.9 & 32.4 & 65.1 & 1/1 & 90.0 & 0 & 30.796 & 46.41 \\
dv-036 & 99.2 & 90.9 & 96.6 & 77.3 & 67.3 & 13.2 & 90.8 & 1/1 & 80.0 & 0 & 44.765 & 85.12 \\
dv-037 & 99.7 & 99.4 & 98.9 & 81.9 & 55.9 & 24.6 & 80.3 & 1/1 & 70.0 & 0 & 29.166 & 40.50 \\
dv-038 & 100.0 & 97.7 & 100.0 & 99.6 & 38.9 & 0.0 & 100.0 & 1/1 & 100.0 & \textbf{\color{ReleaseBlue}1} & 20.020 & 31.81 \\
dv-039 & 99.6 & 99.0 & 99.6 & 98.7 & 100.0 & 13.3 & 100.0 & 1/1 & 80.0 & 0 & 6.413 & 15.45 \\
dv-040 & 100.0 & 98.9 & 99.2 & 68.5 & 88.5 & 0.0 & 100.0 & 1/1 & 90.0 & 0 & 12.600 & 27.98 \\
dv-041 & 99.3 & 99.0 & 85.0 & 99.7 & 64.0 & 28.1 & 100.0 & 1/1 & 80.0 & 0 & 12.443 & 31.86 \\
dv-042 & 100.0 & 98.9 & 98.7 & 73.0 & 83.8 & 0.0 & 100.0 & 1/1 & 100.0 & 0 & 15.964 & 26.27 \\
dv-043 & 91.1 & 100.0 & 85.2 & 100.0 & 99.3 & 0.0 & 97.5 & 1/1 & 100.0 & 0 & 9.681 & 17.12 \\
dv-044 & 98.6 & 81.6 & 98.7 & 94.4 & 78.4 & 96.7 & 100.0 & 1/1 & 100.0 & 0 & 18.280 & 28.54 \\
dv-045 & 100.0 & 88.8 & 94.7 & 91.0 & 63.8 & 47.8 & 100.0 & 1/1 & 30.0 & 0 & 11.547 & 28.49 \\
dv-046 & 98.7 & 97.5 & 96.8 & 75.8 & 88.2 & 0.0 & 100.0 & 1/1 & 100.0 & 0 & 20.843 & 41.95 \\
dv-047 & 100.0 & 100.0 & 100.0 & 100.0 & 69.7 & 0.0 & 100.0 & 1/1 & 100.0 & \textbf{\color{ReleaseBlue}1} & 14.112 & 28.62 \\
dv-048 & 100.0 & 100.0 & 100.0 & 98.7 & 96.3 & 13.2 & 0.0 & 1/1 & 100.0 & 0 & 7.286 & 21.95 \\
dv-049 & 99.7 & 99.8 & 95.5 & 98.3 & 81.5 & 11.6 & 100.0 & 1/1 & 100.0 & \textbf{\color{ReleaseBlue}1} & 8.215 & 23.55 \\
dv-050 & 100.0 & 80.0 & 99.1 & 100.0 & 85.7 & 24.7 & 100.0 & 1/1 & 90.0 & 0 & 4.768 & 16.84 \\
dv-051 & 98.9 & 98.3 & 97.5 & 93.6 & 57.1 & 0.0 & 97.7 & 1/1 & 90.0 & 0 & 42.217 & 73.55 \\
dv-052 & 99.9 & 38.9 & 97.6 & 99.9 & 93.3 & 39.2 & 100.0 & 1/1 & 100.0 & 0 & 5.992 & 23.64 \\
dv-053 & 100.0 & 24.7 & 100.0 & 94.4 & 62.9 & 0.0 & 100.0 & 1/1 & 100.0 & 0 & 22.879 & 41.82 \\
dv-054 & 100.0 & 99.3 & 100.0 & 100.0 & 84.9 & 0.0 & 100.0 & 1/1 & 90.0 & 0 & 40.381 & 77.53 \\
dv-055 & 90.0 & 64.2 & 91.6 & 79.8 & 100.0 & 39.3 & 100.0 & 1/1 & 100.0 & 0 & 7.372 & 20.56 \\
dv-056 & 99.7 & 94.3 & 95.5 & 92.3 & 70.1 & 43.5 & 92.3 & 1/1 & 100.0 & 0 & 24.751 & 37.62 \\
dv-057 & 88.4 & 77.9 & 74.7 & 97.4 & 69.7 & 13.1 & 95.7 & 1/1 & 100.0 & 0 & 10.488 & 33.65 \\
dv-058 & 82.9 & 91.9 & 68.1 & 98.1 & 76.2 & 26.1 & 100.0 & 1/1 & 100.0 & 0 & 66.927 & 97.77 \\
dv-059 & 99.1 & 100.0 & 97.1 & 100.0 & 55.9 & 0.0 & 4.8 & 1/1 & 100.0 & 0 & 29.639 & 38.09 \\
dv-060 & 96.8 & 90.3 & 94.6 & 96.8 & 72.8 & 0.0 & 100.0 & 1/1 & 100.0 & \textbf{\color{ReleaseBlue}1} & 11.911 & 22.98 \\
dv-061 & 77.8 & 97.8 & 68.1 & 91.8 & 0.0 & 0.0 & 92.3 & 1/0 & 0.0 & 0 & 98.905 & 74.61 \\
dv-062 & 98.7 & 64.6 & 69.3 & 81.7 & 97.3 & 53.1 & 95.2 & 1/1 & 90.0 & 0 & 19.449 & 31.28 \\
dv-063 & 84.5 & 100.0 & 84.0 & 99.4 & 0.0 & 0.0 & 100.0 & 1/1 & 90.0 & 0 & 36.133 & 53.59 \\
dv-064 & 99.2 & 96.7 & 98.6 & 76.1 & 93.5 & 0.0 & 100.0 & 1/1 & 90.0 & 0 & 25.481 & 37.70 \\
dv-065 & 95.7 & 72.3 & 75.3 & 90.5 & 76.9 & 19.8 & 100.0 & 1/1 & 70.0 & 0 & 30.953 & 48.62 \\
dv-066 & 96.9 & 97.3 & 94.4 & 37.1 & 74.6 & 0.0 & 100.0 & 1/1 & 90.0 & 0 & 23.570 & 40.48 \\
dv-067 & 80.2 & 95.5 & 81.7 & 99.6 & 88.7 & 0.0 & 25.6 & 1/1 & 80.0 & 0 & 14.330 & 36.07 \\
dv-068 & 48.2 & 74.9 & 33.1 & 37.6 & 63.9 & 0.0 & 100.0 & 1/1 & 20.0 & 0 & 12.461 & 49.82 \\
dv-069 & 100.0 & 80.7 & 100.0 & 82.6 & 94.7 & 24.6 & 100.0 & 1/1 & 100.0 & 0 & 6.841 & 18.78 \\
dv-070 & 83.9 & 66.7 & 70.7 & 44.7 & 54.6 & 21.7 & 72.6 & 1/1 & 50.0 & 0 & 28.001 & 73.13 \\
dv-071 & 99.6 & 58.3 & 97.4 & 88.7 & 53.4 & 0.0 & 98.8 & 1/1 & 90.0 & 0 & 21.544 & 32.78 \\
dv-072 & 100.0 & 100.0 & 100.0 & 100.0 & 94.7 & 66.7 & 47.5 & 1/1 & 100.0 & 0 & 7.683 & 18.66 \\
dv-073 & 96.8 & 84.2 & 92.2 & 89.8 & 32.2 & 56.2 & 100.0 & 1/1 & 100.0 & 0 & 16.510 & 28.57 \\
dv-074 & 100.0 & 100.0 & 100.0 & 92.6 & 96.8 & 46.0 & 24.3 & 1/1 & 100.0 & 0 & 11.670 & 22.48 \\
dv-075 & 98.4 & 83.2 & 89.9 & 83.1 & 82.8 & 56.3 & 0.0 & 1/1 & 80.0 & 0 & 28.507 & 45.58 \\
dv-076 & 72.8 & 77.8 & 56.4 & 76.5 & 100.0 & 21.6 & 100.0 & 1/1 & 100.0 & 0 & 6.375 & 16.51 \\
dv-077 & 100.0 & 94.0 & 99.0 & 17.8 & 58.0 & 0.0 & 0.0 & 1/0 & 0.0 & 0 & 20.219 & 43.90 \\
dv-078 & 70.0 & 76.6 & 60.9 & 45.0 & 71.9 & 81.8 & 100.0 & 1/1 & 100.0 & 0 & 11.355 & 23.81 \\
dv-079 & 90.4 & 71.5 & 80.6 & 73.1 & 57.2 & 38.2 & 99.8 & 1/1 & 60.0 & 0 & 40.160 & 77.40 \\
dv-080 & 95.7 & 77.9 & 90.7 & 69.5 & 74.1 & 15.3 & 0.0 & 1/1 & 90.0 & 0 & 35.359 & 46.14 \\
dv-081 & 37.6 & 34.8 & 31.6 & 14.6 & 76.5 & 19.3 & 100.0 & 1/1 & 100.0 & 0 & 16.769 & 30.00 \\
dv-082 & 96.2 & 100.0 & 93.8 & 99.4 & 74.9 & 0.0 & 100.0 & 1/1 & 100.0 & \textbf{\color{ReleaseBlue}1} & 12.843 & 26.18 \\
dv-083 & 100.0 & 95.8 & 97.7 & 87.5 & 67.6 & 65.0 & 100.0 & 1/1 & 90.0 & 0 & 13.965 & 40.54 \\
dv-084 & 97.2 & 80.6 & 90.1 & 82.0 & 68.1 & 38.9 & 5.9 & 1/1 & 100.0 & 0 & 42.101 & 46.98 \\
dv-085 & 98.6 & 91.2 & 95.5 & 92.6 & 31.7 & 0.0 & 100.0 & 1/1 & 80.0 & 0 & 31.851 & 39.36 \\
dv-086 & 96.1 & 86.4 & 94.2 & 97.8 & 62.8 & 21.8 & 100.0 & 1/1 & 80.0 & 0 & 30.076 & 37.46 \\
dv-087 & 98.3 & 89.4 & 96.2 & 75.4 & 84.8 & 64.6 & 100.0 & 1/1 & 90.0 & 0 & 29.564 & 66.62 \\
dv-088 & 91.9 & 79.8 & 81.7 & 66.3 & 56.0 & 0.0 & 65.1 & 1/0 & 0.0 & 0 & 25.914 & 43.29 \\
dv-089 & 97.8 & 90.3 & 94.3 & 71.7 & 33.3 & 28.1 & 85.1 & 1/1 & 50.0 & 0 & 39.938 & 44.56 \\
dv-090 & 36.6 & 60.9 & 38.1 & 65.3 & 60.6 & 79.0 & 100.0 & 1/1 & 90.0 & 0 & 13.794 & 28.99 \\
dv-091 & 97.3 & 100.0 & 95.6 & 99.4 & 63.7 & 0.0 & 100.0 & 1/1 & 70.0 & 0 & 29.565 & 39.33 \\
dv-092 & 97.2 & 95.5 & 95.1 & 83.6 & 92.2 & 35.6 & 100.0 & 1/1 & 100.0 & 0 & 10.996 & 27.65 \\
dv-093 & 99.1 & 87.4 & 98.6 & 76.2 & 53.0 & 30.8 & 88.7 & 1/1 & 100.0 & 0 & 6.207 & 22.83 \\
dv-094 & 98.3 & 98.5 & 94.4 & 61.1 & 100.0 & 74.6 & 100.0 & 1/1 & 90.0 & 0 & 12.721 & 22.85 \\
dv-095 & 99.9 & 93.7 & 99.7 & 87.7 & 86.7 & 39.3 & 35.5 & 1/1 & 80.0 & 0 & 25.543 & 34.51 \\
\end{longtable}\endgroup

\clearpage
\subsection{C03: Claude Code + Claude Opus 5.5}
\noindent\footnotesize Percentages: Ln/Cn/Br/Tg = line/condition/branch/toggle coverage; F1/P1 = feature/test-plan F1; Ref = reference-model point accuracy; Kill = mutant kill rate. C/G and O are binary compile/golden and overall results. Tok = input plus output tokens in millions; Min = elapsed minutes. Overall uses exact model correctness and all four coverage thresholds. \par\normalsize
\begingroup
\fontsize{8.2}{10.2}\selectfont
\setlength{\tabcolsep}{3pt}\renewcommand{\arraystretch}{1.04}
\setlength{\LTleft}{0pt}\setlength{\LTright}{0pt}
\Needspace{10\baselineskip}
\captionsetup{hypcap=false}
\captionof{table}{Per-task measurements for C03. The companion CSV retains unrounded values, precision/recall, exact-model flags, input/output token splits, and source-cell coordinates.}\label{tab:app-detail-C03}
\addtocounter{table}{-1}
\begin{longtable}{>{\raggedright\arraybackslash}p{\dimexpr 0.1050000\textwidth-2\tabcolsep\relax}>{\raggedleft\arraybackslash}p{\dimexpr 0.0745833\textwidth-2\tabcolsep\relax}>{\raggedleft\arraybackslash}p{\dimexpr 0.0745833\textwidth-2\tabcolsep\relax}>{\raggedleft\arraybackslash}p{\dimexpr 0.0745833\textwidth-2\tabcolsep\relax}>{\raggedleft\arraybackslash}p{\dimexpr 0.0745833\textwidth-2\tabcolsep\relax}>{\raggedleft\arraybackslash}p{\dimexpr 0.0745833\textwidth-2\tabcolsep\relax}>{\raggedleft\arraybackslash}p{\dimexpr 0.0745833\textwidth-2\tabcolsep\relax}>{\raggedleft\arraybackslash}p{\dimexpr 0.0745833\textwidth-2\tabcolsep\relax}>{\raggedleft\arraybackslash}p{\dimexpr 0.0745833\textwidth-2\tabcolsep\relax}>{\raggedleft\arraybackslash}p{\dimexpr 0.0745833\textwidth-2\tabcolsep\relax}>{\raggedleft\arraybackslash}p{\dimexpr 0.0745833\textwidth-2\tabcolsep\relax}>{\raggedleft\arraybackslash}p{\dimexpr 0.0745833\textwidth-2\tabcolsep\relax}>{\raggedleft\arraybackslash}p{\dimexpr 0.0745833\textwidth-2\tabcolsep\relax}}
\toprule
\rowcolor{ReleasePale}\textbf{Task} & \textbf{Ln} & \textbf{Cn} & \textbf{Br} & \textbf{Tg} & \textbf{F1} & \textbf{P1} & \textbf{Ref} & \textbf{C/G} & \textbf{Kill} & \textbf{O} & \textbf{Tok} & \textbf{Min} \\
\midrule\endfirsthead
\multicolumn{13}{l}{\small\color{ReleaseMuted}Table~\ref{tab:app-detail-C03} (continued)}\\
\toprule
\rowcolor{ReleasePale}\textbf{Task} & \textbf{Ln} & \textbf{Cn} & \textbf{Br} & \textbf{Tg} & \textbf{F1} & \textbf{P1} & \textbf{Ref} & \textbf{C/G} & \textbf{Kill} & \textbf{O} & \textbf{Tok} & \textbf{Min} \\
\midrule\endhead
\midrule
\multicolumn{13}{r}{\footnotesize\color{ReleaseMuted}Continued on next page}\\\endfoot
\bottomrule\endlastfoot
dv-001 & 100.0 & 93.0 & 100.0 & 96.9 & 100.0 & 100.0 & 100.0 & 1/1 & 100.0 & \textbf{\color{ReleaseBlue}1} & 0.415 & 3.37 \\
dv-002 & 100.0 & 100.0 & 100.0 & 97.5 & 100.0 & 66.2 & 100.0 & 1/1 & 100.0 & \textbf{\color{ReleaseBlue}1} & 0.472 & 6.96 \\
dv-003 & 100.0 & 100.0 & 100.0 & 100.0 & 82.4 & 100.0 & 100.0 & 1/1 & 100.0 & \textbf{\color{ReleaseBlue}1} & 0.538 & 6.60 \\
dv-004 & 99.0 & 97.5 & 96.2 & 88.3 & 100.0 & 18.0 & 100.0 & 1/1 & 100.0 & 0 & 0.594 & 14.59 \\
dv-005 & 100.0 & 99.3 & 100.0 & 98.9 & 80.0 & 100.0 & 100.0 & 1/1 & 100.0 & \textbf{\color{ReleaseBlue}1} & 0.528 & 12.60 \\
dv-006 & 99.5 & 99.8 & 99.8 & 92.8 & 100.0 & 74.6 & 100.0 & 1/1 & 100.0 & \textbf{\color{ReleaseBlue}1} & 0.861 & 45.77 \\
dv-007 & 100.0 & 100.0 & 100.0 & 100.0 & 100.0 & 40.0 & 100.0 & 1/1 & 80.0 & 0 & 0.228 & 5.13 \\
dv-008 & 98.3 & 91.8 & 96.6 & 97.5 & 81.2 & 0.0 & 100.0 & 1/1 & 100.0 & \textbf{\color{ReleaseBlue}1} & 2.227 & 38.37 \\
dv-009 & 100.0 & 91.3 & 100.0 & 100.0 & 63.8 & 52.3 & 100.0 & 1/1 & 30.0 & 0 & 0.375 & 6.14 \\
dv-010 & 100.0 & 100.0 & 100.0 & 99.3 & 86.7 & 88.0 & 100.0 & 1/1 & 100.0 & \textbf{\color{ReleaseBlue}1} & 0.352 & 7.32 \\
dv-011 & 98.4 & 93.5 & 91.3 & 52.9 & 88.9 & 96.2 & 100.0 & 1/1 & 100.0 & 0 & 0.673 & 7.71 \\
dv-012 & 100.0 & 100.0 & 100.0 & 100.0 & 100.0 & 22.1 & 100.0 & 1/1 & 100.0 & \textbf{\color{ReleaseBlue}1} & 0.867 & 67.16 \\
dv-013 & 100.0 & 100.0 & 100.0 & 94.7 & 92.9 & 73.8 & 100.0 & 1/1 & 100.0 & \textbf{\color{ReleaseBlue}1} & 0.670 & 51.66 \\
dv-014 & 0.0 & 0.0 & 0.0 & 0.0 & 0.0 & 0.0 & 0.0 & 0/0 & 0.0 & 0 & 0.097 & 24.24 \\
dv-015 & 100.0 & 100.0 & 97.3 & 89.4 & 75.5 & 66.2 & 28.4 & 1/1 & 100.0 & 0 & 1.665 & 18.62 \\
dv-016 & 99.3 & 99.2 & 98.9 & 78.5 & 55.6 & 79.5 & 100.0 & 1/1 & 60.0 & 0 & 7.558 & 42.44 \\
dv-017 & 96.8 & 81.3 & 88.4 & 90.6 & 93.5 & 60.0 & 87.9 & 1/1 & 100.0 & 0 & 1.458 & 16.70 \\
dv-018 & 97.4 & 100.0 & 95.7 & 94.8 & 100.0 & 0.0 & 100.0 & 1/1 & 100.0 & \textbf{\color{ReleaseBlue}1} & 0.609 & 8.82 \\
dv-019 & 97.4 & 84.7 & 96.4 & 99.0 & 69.6 & 34.8 & 0.0 & 1/1 & 70.0 & 0 & 1.042 & 19.28 \\
dv-020 & 97.6 & 93.8 & 88.2 & 95.3 & 85.9 & 87.1 & 100.0 & 1/1 & 50.0 & 0 & 1.409 & 28.72 \\
dv-021 & 93.9 & 80.0 & 89.4 & 84.2 & 81.7 & 47.0 & 100.0 & 1/1 & 90.0 & 0 & 0.592 & 15.78 \\
dv-022 & 100.0 & 100.0 & 98.1 & 89.4 & 47.5 & 0.0 & 18.3 & 1/1 & 100.0 & 0 & 3.611 & 30.26 \\
dv-023 & 100.0 & 100.0 & 100.0 & 100.0 & 100.0 & 15.3 & 100.0 & 1/0 & 0.0 & 0 & 0.555 & 11.80 \\
dv-024 & 44.8 & 67.5 & 29.1 & 38.7 & 84.1 & 0.0 & 100.0 & 1/1 & 10.0 & 0 & 0.753 & 17.38 \\
dv-025 & 40.7 & 54.1 & 23.9 & 26.9 & 92.7 & 47.0 & 100.0 & 1/1 & 20.0 & 0 & 0.871 & 12.08 \\
dv-026 & 43.8 & 62.8 & 27.0 & 20.8 & 81.6 & 80.3 & 100.0 & 1/1 & 80.0 & 0 & 0.555 & 12.88 \\
dv-027 & 97.8 & 91.3 & 96.4 & 58.4 & 59.3 & 55.3 & 77.5 & 1/1 & 90.0 & 0 & 18.615 & 113.18 \\
dv-028 & 100.0 & 98.4 & 100.0 & 98.7 & 85.0 & 84.3 & 100.0 & 1/0 & 0.0 & 0 & 0.629 & 7.82 \\
dv-029 & 100.0 & 98.9 & 95.7 & 99.5 & 63.6 & 66.3 & 58.8 & 1/1 & 70.0 & 0 & 3.688 & 24.12 \\
dv-030 & 81.3 & 60.3 & 72.8 & 88.8 & 83.2 & 70.1 & 100.0 & 1/0 & 0.0 & 0 & 11.300 & 205.08 \\
dv-031 & 62.7 & 4.0 & 38.4 & 31.8 & 100.0 & 81.5 & 99.7 & 1/1 & 80.0 & 0 & 0.639 & 15.29 \\
dv-032 & 97.9 & 88.3 & 96.8 & 95.6 & 51.1 & 0.0 & 27.8 & 1/0 & 0.0 & 0 & 11.344 & 81.67 \\
dv-033 & 78.3 & 74.8 & 76.4 & 88.3 & 91.6 & 22.1 & 100.0 & 1/1 & 90.0 & 0 & 1.139 & 25.72 \\
dv-034 & 78.8 & 88.1 & 72.3 & 38.9 & 85.4 & 75.0 & 100.0 & 1/1 & 100.0 & 0 & 0.461 & 8.34 \\
dv-035 & 79.0 & 53.3 & 62.9 & 95.6 & 80.0 & 32.2 & 65.1 & 1/1 & 90.0 & 0 & 1.161 & 12.53 \\
dv-036 & 99.2 & 92.5 & 95.7 & 82.4 & 63.5 & 56.4 & 91.0 & 1/1 & 80.0 & 0 & 22.486 & 491.32 \\
dv-037 & 99.7 & 99.4 & 98.9 & 61.0 & 65.7 & 0.0 & 100.0 & 1/1 & 80.0 & 0 & 3.307 & 27.56 \\
dv-038 & 100.0 & 97.7 & 100.0 & 99.6 & 81.0 & 0.0 & 100.0 & 1/1 & 90.0 & 0 & 0.757 & 12.03 \\
dv-039 & 99.6 & 99.0 & 99.6 & 98.7 & 100.0 & 0.0 & 100.0 & 1/1 & 80.0 & 0 & 0.449 & 10.60 \\
dv-040 & 100.0 & 98.9 & 99.2 & 68.5 & 96.3 & 35.2 & 100.0 & 1/1 & 90.0 & 0 & 0.489 & 10.80 \\
dv-041 & 99.3 & 99.5 & 85.0 & 99.7 & 28.0 & 28.3 & 100.0 & 1/1 & 90.0 & 0 & 0.788 & 15.16 \\
dv-042 & 100.0 & 100.0 & 100.0 & 74.4 & 89.7 & 22.1 & 100.0 & 1/1 & 100.0 & 0 & 0.388 & 9.29 \\
dv-043 & 91.1 & 100.0 & 85.2 & 100.0 & 96.3 & 90.4 & 100.0 & 1/1 & 100.0 & 0 & 0.267 & 4.63 \\
dv-044 & 98.6 & 81.8 & 99.0 & 94.4 & 100.0 & 58.9 & 100.0 & 1/1 & 100.0 & 0 & 1.570 & 73.74 \\
dv-045 & 100.0 & 89.8 & 94.9 & 91.4 & 82.8 & 56.5 & 100.0 & 1/1 & 90.0 & 0 & 3.222 & 57.82 \\
dv-046 & 98.7 & 97.5 & 96.8 & 75.8 & 96.8 & 22.1 & 100.0 & 1/1 & 100.0 & 0 & 2.806 & 102.57 \\
dv-047 & 100.0 & 100.0 & 100.0 & 100.0 & 65.8 & 28.0 & 100.0 & 1/1 & 100.0 & \textbf{\color{ReleaseBlue}1} & 0.717 & 14.50 \\
dv-048 & 100.0 & 100.0 & 100.0 & 98.7 & 90.6 & 0.0 & 0.0 & 1/1 & 100.0 & 0 & 0.519 & 11.23 \\
dv-049 & 99.7 & 99.8 & 95.5 & 98.3 & 96.8 & 22.0 & 100.0 & 1/1 & 90.0 & 0 & 0.605 & 9.07 \\
dv-050 & 100.0 & 80.0 & 99.1 & 100.0 & 96.3 & 77.7 & 100.0 & 1/1 & 80.0 & 0 & 0.738 & 13.08 \\
dv-051 & 100.0 & 98.3 & 100.0 & 93.6 & 64.1 & 0.0 & 100.0 & 1/1 & 90.0 & 0 & 1.281 & 26.76 \\
dv-052 & 99.9 & 38.9 & 97.5 & 99.9 & 100.0 & 54.1 & 100.0 & 1/1 & 90.0 & 0 & 0.415 & 6.55 \\
dv-053 & 100.0 & 24.5 & 100.0 & 97.2 & 74.4 & 47.2 & 100.0 & 1/1 & 100.0 & 0 & 0.862 & 16.47 \\
dv-054 & 100.0 & 99.9 & 100.0 & 100.0 & 86.6 & 9.4 & 100.0 & 1/1 & 30.0 & 0 & 0.873 & 18.16 \\
dv-055 & 90.0 & 64.2 & 91.6 & 80.9 & 93.3 & 85.7 & 100.0 & 1/1 & 100.0 & 0 & 0.528 & 9.67 \\
dv-056 & 99.8 & 98.3 & 96.2 & 97.1 & 98.0 & 81.8 & 100.0 & 1/1 & 80.0 & 0 & 1.267 & 17.43 \\
dv-057 & 88.3 & 77.9 & 74.7 & 98.0 & 70.5 & 43.8 & 95.7 & 1/1 & 100.0 & 0 & 1.028 & 19.79 \\
dv-058 & 91.2 & 91.9 & 86.4 & 98.3 & 87.1 & 37.1 & 0.0 & 1/1 & 100.0 & 0 & 2.072 & 60.78 \\
dv-059 & 99.1 & 100.0 & 97.1 & 100.0 & 89.5 & 15.2 & 4.7 & 1/1 & 100.0 & 0 & 1.189 & 20.23 \\
dv-060 & 100.0 & 90.8 & 98.2 & 96.8 & 86.1 & 10.5 & 100.0 & 1/1 & 100.0 & \textbf{\color{ReleaseBlue}1} & 0.653 & 9.06 \\
dv-061 & 99.4 & 99.4 & 99.0 & 99.8 & 100.0 & 65.7 & 100.0 & 1/1 & 100.0 & \textbf{\color{ReleaseBlue}1} & 1.000 & 31.12 \\
dv-062 & 98.7 & 65.0 & 69.3 & 81.7 & 96.8 & 92.6 & 98.2 & 1/1 & 80.0 & 0 & 1.102 & 12.87 \\
dv-063 & 100.0 & 100.0 & 99.7 & 99.4 & 77.1 & 22.0 & 100.0 & 1/1 & 90.0 & 0 & 8.490 & 48.25 \\
dv-064 & 99.2 & 96.9 & 98.6 & 75.9 & 80.3 & 18.1 & 100.0 & 1/1 & 90.0 & 0 & 1.025 & 17.09 \\
dv-065 & 95.5 & 72.8 & 75.1 & 90.6 & 67.3 & 0.0 & 31.9 & 1/1 & 70.0 & 0 & 1.065 & 17.06 \\
dv-066 & 98.4 & 98.7 & 96.3 & 34.9 & 100.0 & 11.7 & 100.0 & 1/1 & 80.0 & 0 & 1.008 & 13.61 \\
dv-067 & 80.2 & 95.5 & 81.7 & 99.6 & 96.3 & 0.0 & 99.6 & 1/1 & 90.0 & 0 & 0.501 & 15.94 \\
dv-068 & 48.2 & 79.6 & 33.1 & 41.7 & 94.7 & 0.0 & 100.0 & 1/1 & 20.0 & 0 & 0.801 & 16.17 \\
dv-069 & 100.0 & 80.7 & 100.0 & 82.6 & 100.0 & 24.7 & 100.0 & 1/1 & 100.0 & 0 & 0.388 & 6.88 \\
dv-070 & 79.7 & 63.4 & 65.2 & 42.3 & 43.5 & 38.2 & 100.0 & 1/1 & 60.0 & 0 & 5.073 & 60.88 \\
dv-071 & 99.6 & 92.5 & 97.4 & 88.7 & 30.4 & 0.0 & 100.0 & 1/1 & 100.0 & 0 & 0.669 & 10.61 \\
dv-072 & 100.0 & 100.0 & 100.0 & 100.0 & 73.3 & 54.6 & 47.5 & 1/1 & 100.0 & 0 & 1.293 & 15.45 \\
dv-073 & 96.8 & 84.2 & 91.5 & 89.9 & 53.7 & 74.5 & 100.0 & 1/1 & 100.0 & 0 & 0.445 & 8.62 \\
dv-074 & 100.0 & 100.0 & 100.0 & 92.6 & 100.0 & 96.8 & 20.7 & 1/1 & 100.0 & 0 & 0.322 & 5.52 \\
dv-075 & 0.0 & 0.0 & 0.0 & 0.0 & 71.7 & 50.6 & 0.0 & 1/1 & 40.0 & 0 & 6.971 & 46.61 \\
dv-076 & 72.8 & 77.8 & 56.4 & 76.5 & 100.0 & 73.4 & 100.0 & 1/1 & 100.0 & 0 & 0.470 & 6.95 \\
dv-077 & 99.6 & 84.5 & 93.8 & 74.0 & 51.1 & 0.0 & 0.0 & 1/0 & 0.0 & 0 & 10.875 & 94.76 \\
dv-078 & 70.0 & 76.6 & 60.9 & 45.0 & 72.6 & 84.3 & 100.0 & 1/1 & 100.0 & 0 & 0.931 & 12.87 \\
dv-079 & 91.8 & 77.2 & 84.1 & 76.4 & 65.7 & 49.0 & 100.0 & 1/1 & 90.0 & 0 & 2.866 & 65.07 \\
dv-080 & 95.9 & 88.4 & 92.0 & 94.8 & 72.0 & 15.2 & 100.0 & 1/1 & 80.0 & 0 & 1.761 & 37.38 \\
dv-081 & 52.6 & 59.2 & 56.0 & 80.4 & 61.7 & 34.6 & 31.0 & 1/1 & 60.0 & 0 & 3.696 & 42.53 \\
dv-082 & 96.2 & 100.0 & 93.8 & 99.4 & 97.9 & 8.0 & 100.0 & 1/1 & 100.0 & \textbf{\color{ReleaseBlue}1} & 0.442 & 6.36 \\
dv-083 & 100.0 & 97.7 & 100.0 & 99.6 & 79.1 & 59.5 & 100.0 & 1/1 & 100.0 & \textbf{\color{ReleaseBlue}1} & 2.978 & 31.04 \\
dv-084 & 100.0 & 89.9 & 96.3 & 84.6 & 52.7 & 47.6 & 16.2 & 1/1 & 90.0 & 0 & 24.333 & 185.90 \\
dv-085 & 98.6 & 91.2 & 95.5 & 96.4 & 95.8 & 57.1 & 100.0 & 1/1 & 90.0 & 0 & 0.963 & 12.56 \\
dv-086 & 96.1 & 86.8 & 94.2 & 97.8 & 0.0 & 0.0 & 100.0 & 1/1 & 90.0 & 0 & 1.112 & 17.07 \\
dv-087 & 100.0 & 91.1 & 100.0 & 77.4 & 58.2 & 30.3 & 100.0 & 1/1 & 80.0 & 0 & 0.772 & 19.03 \\
dv-088 & 92.9 & 88.6 & 85.1 & 84.2 & 11.6 & 30.5 & 100.0 & 1/1 & 70.0 & 0 & 5.117 & 48.80 \\
dv-089 & 99.6 & 97.2 & 98.6 & 89.1 & 0.0 & 28.1 & 91.6 & 1/1 & 60.0 & 0 & 6.560 & 38.81 \\
dv-090 & 36.7 & 61.4 & 38.5 & 65.3 & 65.4 & 65.1 & 100.0 & 1/1 & 100.0 & 0 & 0.597 & 8.59 \\
dv-091 & 97.3 & 100.0 & 95.6 & 99.4 & 84.3 & 11.7 & 100.0 & 1/1 & 100.0 & \textbf{\color{ReleaseBlue}1} & 1.230 & 15.80 \\
dv-092 & 97.2 & 95.0 & 94.3 & 84.6 & 97.1 & 28.3 & 100.0 & 1/1 & 90.0 & 0 & 0.516 & 14.41 \\
dv-093 & 99.1 & 89.5 & 98.6 & 90.8 & 89.7 & 66.7 & 100.0 & 1/1 & 100.0 & 0 & 0.659 & 16.18 \\
dv-094 & 99.4 & 100.0 & 97.2 & 61.1 & 96.8 & 53.7 & 100.0 & 1/1 & 80.0 & 0 & 0.531 & 10.75 \\
dv-095 & 99.9 & 94.7 & 99.7 & 96.8 & 79.2 & 70.0 & 65.5 & 1/1 & 90.0 & 0 & 0.688 & 12.77 \\
\end{longtable}\endgroup

\clearpage
\subsection{C04: Claude Code + GPT-6.1 Sol}
\noindent\footnotesize Percentages: Ln/Cn/Br/Tg = line/condition/branch/toggle coverage; F1/P1 = feature/test-plan F1; Ref = reference-model point accuracy; Kill = mutant kill rate. C/G and O are binary compile/golden and overall results. Tok = input plus output tokens in millions; Min = elapsed minutes. Overall uses exact model correctness and all four coverage thresholds. \par\normalsize
\begingroup
\fontsize{8.2}{10.2}\selectfont
\setlength{\tabcolsep}{3pt}\renewcommand{\arraystretch}{1.04}
\setlength{\LTleft}{0pt}\setlength{\LTright}{0pt}
\Needspace{10\baselineskip}
\captionsetup{hypcap=false}
\captionof{table}{Per-task measurements for C04. The companion CSV retains unrounded values, precision/recall, exact-model flags, input/output token splits, and source-cell coordinates.}\label{tab:app-detail-C04}
\addtocounter{table}{-1}
\begin{longtable}{>{\raggedright\arraybackslash}p{\dimexpr 0.1050000\textwidth-2\tabcolsep\relax}>{\raggedleft\arraybackslash}p{\dimexpr 0.0745833\textwidth-2\tabcolsep\relax}>{\raggedleft\arraybackslash}p{\dimexpr 0.0745833\textwidth-2\tabcolsep\relax}>{\raggedleft\arraybackslash}p{\dimexpr 0.0745833\textwidth-2\tabcolsep\relax}>{\raggedleft\arraybackslash}p{\dimexpr 0.0745833\textwidth-2\tabcolsep\relax}>{\raggedleft\arraybackslash}p{\dimexpr 0.0745833\textwidth-2\tabcolsep\relax}>{\raggedleft\arraybackslash}p{\dimexpr 0.0745833\textwidth-2\tabcolsep\relax}>{\raggedleft\arraybackslash}p{\dimexpr 0.0745833\textwidth-2\tabcolsep\relax}>{\raggedleft\arraybackslash}p{\dimexpr 0.0745833\textwidth-2\tabcolsep\relax}>{\raggedleft\arraybackslash}p{\dimexpr 0.0745833\textwidth-2\tabcolsep\relax}>{\raggedleft\arraybackslash}p{\dimexpr 0.0745833\textwidth-2\tabcolsep\relax}>{\raggedleft\arraybackslash}p{\dimexpr 0.0745833\textwidth-2\tabcolsep\relax}>{\raggedleft\arraybackslash}p{\dimexpr 0.0745833\textwidth-2\tabcolsep\relax}}
\toprule
\rowcolor{ReleasePale}\textbf{Task} & \textbf{Ln} & \textbf{Cn} & \textbf{Br} & \textbf{Tg} & \textbf{F1} & \textbf{P1} & \textbf{Ref} & \textbf{C/G} & \textbf{Kill} & \textbf{O} & \textbf{Tok} & \textbf{Min} \\
\midrule\endfirsthead
\multicolumn{13}{l}{\small\color{ReleaseMuted}Table~\ref{tab:app-detail-C04} (continued)}\\
\toprule
\rowcolor{ReleasePale}\textbf{Task} & \textbf{Ln} & \textbf{Cn} & \textbf{Br} & \textbf{Tg} & \textbf{F1} & \textbf{P1} & \textbf{Ref} & \textbf{C/G} & \textbf{Kill} & \textbf{O} & \textbf{Tok} & \textbf{Min} \\
\midrule\endhead
\midrule
\multicolumn{13}{r}{\footnotesize\color{ReleaseMuted}Continued on next page}\\\endfoot
\bottomrule\endlastfoot
dv-001 & 100.0 & 93.3 & 100.0 & 96.9 & 97.4 & 100.0 & 100.0 & 1/1 & 100.0 & \textbf{\color{ReleaseBlue}1} & 0.358 & 7.94 \\
dv-002 & 100.0 & 100.0 & 100.0 & 97.5 & 97.6 & 65.1 & 100.0 & 1/1 & 100.0 & \textbf{\color{ReleaseBlue}1} & 0.144 & 5.65 \\
dv-003 & 100.0 & 100.0 & 100.0 & 100.0 & 100.0 & 100.0 & 100.0 & 1/1 & 100.0 & \textbf{\color{ReleaseBlue}1} & 0.285 & 7.98 \\
dv-004 & 99.0 & 96.0 & 96.2 & 89.0 & 100.0 & 56.8 & 100.0 & 1/1 & 100.0 & 0 & 1.052 & 15.89 \\
dv-005 & 100.0 & 99.3 & 100.0 & 98.9 & 100.0 & 100.0 & 100.0 & 1/1 & 100.0 & \textbf{\color{ReleaseBlue}1} & 0.633 & 23.01 \\
dv-006 & 99.5 & 99.8 & 99.8 & 92.8 & 100.0 & 100.0 & 100.0 & 1/1 & 100.0 & \textbf{\color{ReleaseBlue}1} & 0.959 & 13.22 \\
dv-007 & 100.0 & 100.0 & 100.0 & 100.0 & 100.0 & 66.7 & 100.0 & 1/1 & 90.0 & 0 & 0.219 & 5.33 \\
dv-008 & 98.3 & 91.8 & 96.6 & 97.5 & 75.7 & 65.9 & 98.2 & 1/1 & 100.0 & 0 & 4.536 & 50.74 \\
dv-009 & 100.0 & 91.3 & 100.0 & 100.0 & 100.0 & 66.7 & 63.3 & 1/0 & 0.0 & 0 & 0.352 & 7.30 \\
dv-010 & 100.0 & 100.0 & 100.0 & 99.3 & 87.0 & 100.0 & 100.0 & 1/1 & 100.0 & \textbf{\color{ReleaseBlue}1} & 0.436 & 10.69 \\
dv-011 & 98.4 & 93.5 & 91.3 & 52.9 & 82.4 & 94.1 & 100.0 & 1/1 & 100.0 & 0 & 0.273 & 5.98 \\
dv-012 & 100.0 & 100.0 & 100.0 & 100.0 & 100.0 & 100.0 & 100.0 & 1/1 & 100.0 & \textbf{\color{ReleaseBlue}1} & 0.910 & 14.10 \\
dv-013 & 100.0 & 100.0 & 100.0 & 94.7 & 100.0 & 98.7 & 100.0 & 1/1 & 100.0 & \textbf{\color{ReleaseBlue}1} & 0.672 & 24.57 \\
dv-014 & 86.8 & 77.6 & 84.0 & 68.2 & 92.4 & 75.9 & 86.8 & 1/1 & 100.0 & 0 & 1.696 & 31.44 \\
dv-015 & 100.0 & 100.0 & 97.3 & 89.4 & 83.9 & 65.4 & 28.4 & 1/1 & 100.0 & 0 & 0.870 & 18.35 \\
dv-016 & 99.3 & 96.9 & 98.9 & 57.6 & 89.1 & 60.8 & 100.0 & 1/1 & 60.0 & 0 & 1.434 & 21.60 \\
dv-017 & 96.8 & 81.3 & 88.4 & 90.6 & 93.3 & 100.0 & 89.9 & 1/1 & 100.0 & 0 & 0.814 & 18.79 \\
dv-018 & 97.4 & 100.0 & 95.7 & 94.8 & 100.0 & 99.2 & 100.0 & 1/1 & 100.0 & \textbf{\color{ReleaseBlue}1} & 0.711 & 16.07 \\
dv-019 & 98.3 & 88.7 & 97.0 & 99.0 & 96.3 & 66.3 & 0.0 & 1/1 & 80.0 & 0 & 1.369 & 20.66 \\
dv-020 & 97.6 & 93.8 & 88.2 & 95.3 & 99.2 & 92.7 & 100.0 & 1/1 & 100.0 & 0 & 3.931 & 37.32 \\
dv-021 & 93.9 & 88.0 & 90.6 & 93.8 & 85.4 & 89.7 & 100.0 & 1/0 & 0.0 & 0 & 0.642 & 24.25 \\
dv-022 & 100.0 & 97.5 & 98.1 & 97.7 & 98.0 & 28.6 & 15.0 & 1/1 & 100.0 & 0 & 1.038 & 14.55 \\
dv-023 & 100.0 & 100.0 & 100.0 & 100.0 & 100.0 & 28.6 & 100.0 & 1/1 & 100.0 & \textbf{\color{ReleaseBlue}1} & 0.640 & 11.71 \\
dv-024 & 43.3 & 59.7 & 27.1 & 35.3 & 100.0 & 0.0 & 100.0 & 1/1 & 80.0 & 0 & 1.704 & 20.94 \\
dv-025 & 40.7 & 54.3 & 23.9 & 30.0 & 95.0 & 65.8 & 100.0 & 1/1 & 20.0 & 0 & 0.196 & 7.91 \\
dv-026 & 42.1 & 58.9 & 24.9 & 20.8 & 95.9 & 89.7 & 100.0 & 1/1 & 80.0 & 0 & 1.051 & 14.38 \\
dv-027 & 99.7 & 91.3 & 97.8 & 58.2 & 100.0 & 43.3 & 69.4 & 1/1 & 70.0 & 0 & 4.485 & 93.56 \\
dv-028 & 100.0 & 98.4 & 100.0 & 98.7 & 98.9 & 100.0 & 100.0 & 1/1 & 100.0 & \textbf{\color{ReleaseBlue}1} & 1.279 & 12.64 \\
dv-029 & 99.7 & 98.9 & 95.0 & 96.6 & 89.3 & 100.0 & 0.0 & 1/1 & 80.0 & 0 & 1.360 & 18.15 \\
dv-030 & 81.3 & 57.7 & 73.8 & 86.2 & 98.9 & 65.6 & 100.0 & 1/0 & 0.0 & 0 & 1.545 & 93.59 \\
dv-031 & 62.2 & 3.9 & 38.0 & 31.8 & 99.1 & 89.1 & 99.7 & 1/1 & 60.0 & 0 & 0.742 & 12.62 \\
dv-032 & 97.0 & 84.1 & 95.3 & 95.4 & 100.0 & 71.4 & 53.5 & 1/1 & 80.0 & 0 & 2.105 & 22.96 \\
dv-033 & 78.3 & 73.5 & 76.4 & 73.3 & 96.6 & 39.8 & 96.7 & 1/1 & 90.0 & 0 & 1.591 & 25.01 \\
dv-034 & 78.3 & 86.9 & 71.4 & 38.4 & 91.7 & 100.0 & 100.0 & 1/0 & 0.0 & 0 & 0.408 & 9.26 \\
dv-035 & 79.0 & 53.3 & 62.9 & 95.6 & 79.1 & 99.1 & 65.1 & 1/1 & 90.0 & 0 & 1.093 & 14.59 \\
dv-036 & 97.6 & 91.4 & 95.7 & 83.6 & 82.6 & 71.8 & 59.5 & 1/1 & 80.0 & 0 & 3.075 & 155.10 \\
dv-037 & 99.7 & 96.7 & 98.9 & 76.2 & 80.5 & 60.0 & 100.0 & 1/1 & 80.0 & 0 & 1.392 & 18.71 \\
dv-038 & 100.0 & 97.7 & 100.0 & 99.6 & 100.0 & 0.0 & 100.0 & 1/1 & 90.0 & 0 & 0.438 & 8.62 \\
dv-039 & 99.6 & 99.0 & 99.6 & 98.7 & 100.0 & 83.3 & 100.0 & 1/1 & 80.0 & 0 & 0.400 & 9.92 \\
dv-040 & 100.0 & 98.9 & 99.2 & 68.5 & 100.0 & 100.0 & 100.0 & 1/1 & 100.0 & 0 & 0.242 & 7.79 \\
dv-041 & 99.3 & 99.5 & 85.0 & 99.7 & 98.7 & 65.7 & 100.0 & 1/1 & 80.0 & 0 & 0.660 & 68.34 \\
dv-042 & 100.0 & 100.0 & 100.0 & 74.4 & 65.1 & 54.5 & 100.0 & 1/1 & 100.0 & 0 & 0.685 & 71.06 \\
dv-043 & 91.1 & 100.0 & 85.2 & 100.0 & 100.0 & 78.3 & 99.6 & 1/1 & 100.0 & 0 & 0.231 & 6.24 \\
dv-044 & 98.6 & 81.8 & 99.0 & 94.4 & 100.0 & 100.0 & 100.0 & 1/1 & 100.0 & 0 & 1.532 & 32.63 \\
dv-045 & 100.0 & 91.2 & 95.9 & 93.2 & 0.0 & 94.3 & 100.0 & 1/1 & 100.0 & \textbf{\color{ReleaseBlue}1} & 7.108 & 45.80 \\
dv-046 & 98.1 & 97.5 & 96.4 & 75.8 & 95.3 & 54.2 & 100.0 & 1/1 & 100.0 & 0 & 1.224 & 15.25 \\
dv-047 & 100.0 & 100.0 & 100.0 & 100.0 & 100.0 & 28.4 & 100.0 & 1/1 & 100.0 & \textbf{\color{ReleaseBlue}1} & 0.545 & 13.51 \\
dv-048 & 100.0 & 100.0 & 100.0 & 98.7 & 97.7 & 44.4 & 0.0 & 1/1 & 100.0 & 0 & 0.677 & 13.77 \\
dv-049 & 99.7 & 99.8 & 95.5 & 98.3 & 100.0 & 81.5 & 100.0 & 1/1 & 90.0 & 0 & 0.402 & 11.81 \\
dv-050 & 100.0 & 80.0 & 99.1 & 100.0 & 100.0 & 82.8 & 100.0 & 1/1 & 90.0 & 0 & 0.567 & 11.13 \\
dv-051 & 98.9 & 98.3 & 97.5 & 93.6 & 100.0 & 76.2 & 100.0 & 1/1 & 100.0 & \textbf{\color{ReleaseBlue}1} & 1.156 & 16.50 \\
dv-052 & 99.9 & 38.9 & 99.9 & 99.9 & 100.0 & 65.3 & 100.0 & 1/1 & 90.0 & 0 & 0.351 & 7.37 \\
dv-053 & 100.0 & 24.5 & 100.0 & 97.2 & 100.0 & 60.9 & 100.0 & 1/1 & 100.0 & 0 & 0.989 & 13.90 \\
dv-054 & 100.0 & 99.8 & 100.0 & 100.0 & 99.1 & 33.1 & 100.0 & 1/1 & 90.0 & 0 & 1.185 & 14.92 \\
dv-055 & 90.0 & 66.0 & 91.6 & 80.8 & 100.0 & 93.3 & 100.0 & 1/1 & 100.0 & 0 & 0.726 & 11.42 \\
dv-056 & 99.8 & 98.3 & 94.0 & 94.8 & 98.3 & 66.2 & 100.0 & 1/1 & 100.0 & \textbf{\color{ReleaseBlue}1} & 1.044 & 12.59 \\
dv-057 & 88.4 & 77.9 & 74.7 & 97.4 & 100.0 & 100.0 & 95.7 & 1/1 & 100.0 & 0 & 0.679 & 12.12 \\
dv-058 & 91.3 & 94.6 & 86.7 & 98.5 & 93.4 & 89.4 & 0.0 & 1/0 & 0.0 & 0 & 1.079 & 34.90 \\
dv-059 & 99.1 & 100.0 & 97.1 & 100.0 & 100.0 & 0.0 & 4.7 & 1/1 & 100.0 & 0 & 2.704 & 23.47 \\
dv-060 & 100.0 & 90.3 & 98.2 & 96.8 & 97.3 & 50.0 & 100.0 & 1/1 & 100.0 & \textbf{\color{ReleaseBlue}1} & 0.708 & 10.26 \\
dv-061 & 99.9 & 91.1 & 99.8 & 99.4 & 100.0 & 95.6 & 100.0 & 1/1 & 100.0 & \textbf{\color{ReleaseBlue}1} & 4.398 & 31.04 \\
dv-062 & 98.7 & 65.3 & 69.3 & 81.7 & 97.3 & 88.8 & 97.7 & 1/1 & 90.0 & 0 & 1.044 & 14.17 \\
dv-063 & 100.0 & 100.0 & 99.9 & 99.4 & 97.1 & 66.7 & 100.0 & 1/1 & 70.0 & 0 & 4.748 & 41.30 \\
dv-064 & 99.2 & 96.7 & 98.6 & 76.4 & 100.0 & 55.6 & 100.0 & 1/1 & 90.0 & 0 & 0.842 & 13.70 \\
dv-065 & 95.5 & 72.7 & 75.1 & 90.6 & 99.0 & 0.0 & 100.0 & 1/1 & 80.0 & 0 & 2.269 & 21.07 \\
dv-066 & 98.4 & 98.7 & 96.3 & 48.2 & 100.0 & 54.5 & 100.0 & 1/1 & 90.0 & 0 & 0.460 & 8.61 \\
dv-067 & 78.9 & 95.5 & 79.8 & 99.6 & 98.0 & 59.7 & 99.6 & 1/1 & 60.0 & 0 & 0.887 & 13.46 \\
dv-068 & 48.2 & 78.3 & 33.1 & 40.2 & 100.0 & 60.0 & 100.0 & 1/1 & 40.0 & 0 & 0.499 & 11.81 \\
dv-069 & 100.0 & 80.7 & 100.0 & 82.6 & 100.0 & 66.7 & 100.0 & 1/1 & 100.0 & 0 & 0.559 & 10.19 \\
dv-070 & 79.4 & 61.5 & 64.1 & 44.0 & 59.4 & 75.9 & 100.0 & 1/1 & 60.0 & 0 & 3.042 & 31.51 \\
dv-071 & 99.6 & 91.9 & 97.4 & 88.7 & 53.9 & 92.5 & 100.0 & 1/1 & 100.0 & 0 & 0.685 & 9.62 \\
dv-072 & 100.0 & 100.0 & 100.0 & 100.0 & 100.0 & 100.0 & 47.5 & 1/1 & 100.0 & 0 & 0.613 & 8.98 \\
dv-073 & 96.8 & 84.2 & 92.2 & 89.7 & 97.8 & 88.2 & 100.0 & 1/1 & 90.0 & 0 & 0.420 & 7.32 \\
dv-074 & 100.0 & 100.0 & 100.0 & 92.6 & 100.0 & 92.3 & 20.7 & 1/1 & 100.0 & 0 & 0.422 & 9.14 \\
dv-075 & 98.6 & 93.8 & 90.4 & 87.1 & 100.0 & 75.0 & 0.0 & 1/1 & 80.0 & 0 & 1.723 & 20.73 \\
dv-076 & 72.8 & 80.2 & 56.4 & 76.5 & 100.0 & 0.0 & 100.0 & 1/1 & 100.0 & 0 & 0.407 & 7.45 \\
dv-077 & 100.0 & 91.6 & 99.7 & 52.7 & 77.5 & 11.7 & 0.0 & 1/1 & 70.0 & 0 & 5.959 & 43.34 \\
dv-078 & 70.0 & 76.6 & 60.9 & 45.0 & 96.8 & 94.4 & 100.0 & 1/1 & 100.0 & 0 & 0.463 & 10.40 \\
dv-079 & 93.2 & 78.5 & 86.5 & 76.7 & 85.9 & 85.1 & 100.0 & 1/0 & 0.0 & 0 & 1.489 & 38.67 \\
dv-080 & 95.9 & 88.4 & 92.0 & 90.1 & 98.3 & 65.4 & 0.0 & 1/1 & 80.0 & 0 & 1.100 & 19.87 \\
dv-081 & 37.3 & 34.3 & 31.0 & 13.3 & 91.8 & 100.0 & 0.0 & 1/1 & 100.0 & 0 & 2.060 & 24.35 \\
dv-082 & 94.8 & 100.0 & 91.1 & 99.4 & 100.0 & 58.8 & 100.0 & 1/1 & 90.0 & 0 & 0.177 & 6.93 \\
dv-083 & 0.0 & 0.0 & 0.0 & 0.0 & 91.6 & 96.8 & 100.0 & 0/0 & 0.0 & 0 & 3.694 & 23.84 \\
dv-084 & 100.0 & 89.1 & 95.0 & 94.5 & 98.2 & 65.6 & 13.4 & 1/1 & 100.0 & 0 & 3.691 & 30.84 \\
dv-085 & 98.6 & 91.2 & 95.5 & 96.4 & 93.0 & 75.0 & 100.0 & 1/1 & 80.0 & 0 & 1.199 & 17.40 \\
dv-086 & 96.1 & 86.6 & 94.2 & 97.8 & 100.0 & 0.0 & 100.0 & 1/1 & 80.0 & 0 & 0.493 & 11.00 \\
dv-087 & 99.6 & 91.1 & 97.7 & 77.4 & 100.0 & 93.3 & 100.0 & 1/1 & 80.0 & 0 & 0.728 & 15.60 \\
dv-088 & 92.7 & 88.3 & 84.9 & 76.0 & 96.8 & 66.7 & 100.0 & 1/1 & 80.0 & 0 & 4.246 & 45.71 \\
dv-089 & 99.6 & 95.6 & 98.6 & 77.8 & 93.7 & 79.2 & 60.7 & 1/1 & 50.0 & 0 & 3.128 & 35.42 \\
dv-090 & 36.8 & 61.9 & 38.9 & 65.6 & 84.1 & 90.9 & 100.0 & 1/1 & 90.0 & 0 & 0.327 & 7.75 \\
dv-091 & 97.3 & 100.0 & 95.6 & 99.4 & 100.0 & 60.9 & 100.0 & 1/1 & 70.0 & 0 & 2.820 & 29.15 \\
dv-092 & 97.2 & 95.0 & 94.3 & 84.6 & 100.0 & 87.5 & 100.0 & 1/1 & 100.0 & 0 & 0.313 & 8.67 \\
dv-093 & 99.1 & 89.2 & 98.6 & 89.3 & 100.0 & 93.3 & 100.0 & 1/1 & 100.0 & 0 & 1.002 & 13.24 \\
dv-094 & 99.4 & 95.6 & 97.2 & 61.1 & 100.0 & 89.7 & 100.0 & 1/0 & 0.0 & 0 & 0.681 & 10.79 \\
dv-095 & 100.0 & 97.9 & 100.0 & 96.8 & 100.0 & 81.5 & 7.2 & 1/1 & 80.0 & 0 & 0.953 & 13.74 \\
\end{longtable}\endgroup

\clearpage
\subsection{C05: Claude Code + GPT-6 Astra}
\noindent\footnotesize Percentages: Ln/Cn/Br/Tg = line/condition/branch/toggle coverage; F1/P1 = feature/test-plan F1; Ref = reference-model point accuracy; Kill = mutant kill rate. C/G and O are binary compile/golden and overall results. Tok = input plus output tokens in millions; Min = elapsed minutes. Overall uses exact model correctness and all four coverage thresholds. \par\normalsize
\begingroup
\fontsize{8.2}{10.2}\selectfont
\setlength{\tabcolsep}{3pt}\renewcommand{\arraystretch}{1.04}
\setlength{\LTleft}{0pt}\setlength{\LTright}{0pt}
\Needspace{10\baselineskip}
\captionsetup{hypcap=false}
\captionof{table}{Per-task measurements for C05. The companion CSV retains unrounded values, precision/recall, exact-model flags, input/output token splits, and source-cell coordinates.}\label{tab:app-detail-C05}
\addtocounter{table}{-1}
\begin{longtable}{>{\raggedright\arraybackslash}p{\dimexpr 0.1050000\textwidth-2\tabcolsep\relax}>{\raggedleft\arraybackslash}p{\dimexpr 0.0745833\textwidth-2\tabcolsep\relax}>{\raggedleft\arraybackslash}p{\dimexpr 0.0745833\textwidth-2\tabcolsep\relax}>{\raggedleft\arraybackslash}p{\dimexpr 0.0745833\textwidth-2\tabcolsep\relax}>{\raggedleft\arraybackslash}p{\dimexpr 0.0745833\textwidth-2\tabcolsep\relax}>{\raggedleft\arraybackslash}p{\dimexpr 0.0745833\textwidth-2\tabcolsep\relax}>{\raggedleft\arraybackslash}p{\dimexpr 0.0745833\textwidth-2\tabcolsep\relax}>{\raggedleft\arraybackslash}p{\dimexpr 0.0745833\textwidth-2\tabcolsep\relax}>{\raggedleft\arraybackslash}p{\dimexpr 0.0745833\textwidth-2\tabcolsep\relax}>{\raggedleft\arraybackslash}p{\dimexpr 0.0745833\textwidth-2\tabcolsep\relax}>{\raggedleft\arraybackslash}p{\dimexpr 0.0745833\textwidth-2\tabcolsep\relax}>{\raggedleft\arraybackslash}p{\dimexpr 0.0745833\textwidth-2\tabcolsep\relax}>{\raggedleft\arraybackslash}p{\dimexpr 0.0745833\textwidth-2\tabcolsep\relax}}
\toprule
\rowcolor{ReleasePale}\textbf{Task} & \textbf{Ln} & \textbf{Cn} & \textbf{Br} & \textbf{Tg} & \textbf{F1} & \textbf{P1} & \textbf{Ref} & \textbf{C/G} & \textbf{Kill} & \textbf{O} & \textbf{Tok} & \textbf{Min} \\
\midrule\endfirsthead
\multicolumn{13}{l}{\small\color{ReleaseMuted}Table~\ref{tab:app-detail-C05} (continued)}\\
\toprule
\rowcolor{ReleasePale}\textbf{Task} & \textbf{Ln} & \textbf{Cn} & \textbf{Br} & \textbf{Tg} & \textbf{F1} & \textbf{P1} & \textbf{Ref} & \textbf{C/G} & \textbf{Kill} & \textbf{O} & \textbf{Tok} & \textbf{Min} \\
\midrule\endhead
\midrule
\multicolumn{13}{r}{\footnotesize\color{ReleaseMuted}Continued on next page}\\\endfoot
\bottomrule\endlastfoot
dv-001 & 100.0 & 93.2 & 100.0 & 96.9 & 97.0 & 100.0 & 100.0 & 1/1 & 100.0 & \textbf{\color{ReleaseBlue}1} & 0.239 & 4.83 \\
dv-002 & 100.0 & 100.0 & 100.0 & 97.5 & 0.0 & 66.7 & 100.0 & 1/1 & 100.0 & \textbf{\color{ReleaseBlue}1} & 0.631 & 7.82 \\
dv-003 & 100.0 & 100.0 & 100.0 & 100.0 & 90.5 & 100.0 & 100.0 & 1/1 & 100.0 & \textbf{\color{ReleaseBlue}1} & 0.282 & 5.25 \\
dv-004 & 99.0 & 96.4 & 96.2 & 88.2 & 97.1 & 57.1 & 100.0 & 1/1 & 100.0 & 0 & 0.567 & 7.74 \\
dv-005 & 100.0 & 99.3 & 100.0 & 98.9 & 76.9 & 100.0 & 100.0 & 1/1 & 100.0 & \textbf{\color{ReleaseBlue}1} & 0.348 & 12.96 \\
dv-006 & 99.5 & 99.8 & 99.8 & 92.8 & 0.0 & 100.0 & 100.0 & 1/1 & 100.0 & \textbf{\color{ReleaseBlue}1} & 0.627 & 8.79 \\
dv-007 & 100.0 & 100.0 & 100.0 & 100.0 & 96.0 & 39.7 & 100.0 & 1/1 & 80.0 & 0 & 0.283 & 5.83 \\
dv-008 & 97.9 & 87.1 & 96.6 & 95.7 & 68.7 & 81.0 & 0.1 & 1/1 & 100.0 & 0 & 2.601 & 41.92 \\
dv-009 & 100.0 & 91.3 & 100.0 & 100.0 & 100.0 & 66.7 & 100.0 & 1/1 & 40.0 & 0 & 0.251 & 5.53 \\
dv-010 & 100.0 & 100.0 & 100.0 & 99.3 & 84.7 & 88.9 & 100.0 & 1/1 & 100.0 & \textbf{\color{ReleaseBlue}1} & 0.384 & 6.59 \\
dv-011 & 98.4 & 93.5 & 91.3 & 52.9 & 88.9 & 100.0 & 100.0 & 1/1 & 100.0 & 0 & 0.345 & 6.33 \\
dv-012 & 100.0 & 100.0 & 100.0 & 100.0 & 100.0 & 100.0 & 100.0 & 1/1 & 100.0 & \textbf{\color{ReleaseBlue}1} & 0.619 & 5.58 \\
dv-013 & 100.0 & 100.0 & 100.0 & 94.7 & 100.0 & 99.2 & 100.0 & 1/1 & 100.0 & \textbf{\color{ReleaseBlue}1} & 0.495 & 10.01 \\
dv-014 & 86.8 & 77.6 & 84.0 & 68.2 & 100.0 & 70.2 & 100.0 & 1/1 & 100.0 & 0 & 1.913 & 19.59 \\
dv-015 & 100.0 & 100.0 & 97.3 & 89.4 & 83.8 & 65.9 & 28.4 & 1/1 & 100.0 & 0 & 2.574 & 18.16 \\
dv-016 & 99.3 & 98.6 & 98.9 & 62.0 & 92.4 & 80.0 & 100.0 & 1/1 & 50.0 & 0 & 3.507 & 21.87 \\
dv-017 & 96.8 & 81.3 & 88.4 & 90.6 & 93.9 & 92.3 & 89.9 & 1/1 & 100.0 & 0 & 0.198 & 5.07 \\
dv-018 & 100.0 & 100.0 & 100.0 & 94.8 & 100.0 & 99.2 & 100.0 & 1/1 & 100.0 & \textbf{\color{ReleaseBlue}1} & 0.476 & 8.06 \\
dv-019 & 98.4 & 87.1 & 97.0 & 99.0 & 100.0 & 92.3 & 99.9 & 1/1 & 80.0 & 0 & 0.764 & 12.63 \\
dv-020 & 97.6 & 93.2 & 88.2 & 95.3 & 100.0 & 66.3 & 100.0 & 1/1 & 100.0 & 0 & 4.167 & 26.05 \\
dv-021 & 96.6 & 91.3 & 93.1 & 93.8 & 100.0 & 93.3 & 100.0 & 1/1 & 100.0 & \textbf{\color{ReleaseBlue}1} & 0.622 & 19.77 \\
dv-022 & 100.0 & 97.5 & 98.1 & 95.8 & 76.3 & 85.1 & 10.4 & 1/1 & 90.0 & 0 & 0.996 & 10.53 \\
dv-023 & 100.0 & 100.0 & 100.0 & 100.0 & 100.0 & 15.4 & 100.0 & 1/1 & 100.0 & \textbf{\color{ReleaseBlue}1} & 0.305 & 6.31 \\
dv-024 & 43.3 & 60.1 & 27.1 & 35.1 & 98.0 & 0.0 & 100.0 & 1/1 & 80.0 & 0 & 0.472 & 7.37 \\
dv-025 & 40.7 & 55.8 & 23.9 & 29.2 & 95.5 & 54.5 & 100.0 & 1/1 & 20.0 & 0 & 0.326 & 6.39 \\
dv-026 & 42.1 & 58.5 & 24.9 & 20.8 & 96.5 & 89.7 & 100.0 & 1/1 & 80.0 & 0 & 0.353 & 6.97 \\
dv-027 & 97.5 & 87.2 & 94.9 & 57.0 & 98.2 & 49.6 & 73.3 & 1/1 & 70.0 & 0 & 5.060 & 25.09 \\
dv-028 & 100.0 & 98.4 & 100.0 & 98.7 & 97.9 & 97.3 & 100.0 & 1/1 & 100.0 & \textbf{\color{ReleaseBlue}1} & 1.139 & 8.71 \\
dv-029 & 99.7 & 97.7 & 95.0 & 72.9 & 72.9 & 71.0 & 63.5 & 1/1 & 70.0 & 0 & 1.450 & 18.36 \\
dv-030 & 70.2 & 50.0 & 66.0 & 82.9 & 95.7 & 13.3 & 100.0 & 1/0 & 0.0 & 0 & 1.240 & 62.98 \\
dv-031 & 62.2 & 3.9 & 38.0 & 31.8 & 98.0 & 96.8 & 99.9 & 1/1 & 70.0 & 0 & 0.440 & 6.64 \\
dv-032 & 96.6 & 84.1 & 94.6 & 95.4 & 91.8 & 36.4 & 50.3 & 1/1 & 80.0 & 0 & 2.886 & 17.80 \\
dv-033 & 78.3 & 74.2 & 76.4 & 52.0 & 98.0 & 58.6 & 99.9 & 1/1 & 90.0 & 0 & 1.623 & 13.76 \\
dv-034 & 78.8 & 88.1 & 72.3 & 38.4 & 92.9 & 100.0 & 100.0 & 1/0 & 0.0 & 0 & 0.300 & 5.83 \\
dv-035 & 79.0 & 53.3 & 61.4 & 95.6 & 85.4 & 87.7 & 65.1 & 1/1 & 90.0 & 0 & 0.878 & 9.81 \\
dv-036 & 99.2 & 90.9 & 96.6 & 84.2 & 72.7 & 66.1 & 39.8 & 1/1 & 70.0 & 0 & 3.780 & 31.80 \\
dv-037 & 99.7 & 97.2 & 97.8 & 77.4 & 81.3 & 13.3 & 100.0 & 1/1 & 80.0 & 0 & 1.567 & 16.25 \\
dv-038 & 100.0 & 97.7 & 100.0 & 99.6 & 100.0 & 60.6 & 100.0 & 1/1 & 90.0 & 0 & 0.449 & 9.07 \\
dv-039 & 99.6 & 99.0 & 99.6 & 98.7 & 100.0 & 83.3 & 100.0 & 1/1 & 90.0 & 0 & 0.235 & 5.11 \\
dv-040 & 100.0 & 98.9 & 99.2 & 68.5 & 100.0 & 92.3 & 100.0 & 1/1 & 90.0 & 0 & 0.507 & 8.85 \\
dv-041 & 99.3 & 99.5 & 85.0 & 99.7 & 100.0 & 90.9 & 100.0 & 1/1 & 80.0 & 0 & 0.425 & 6.68 \\
dv-042 & 100.0 & 100.0 & 100.0 & 74.4 & 97.7 & 66.7 & 100.0 & 1/1 & 100.0 & 0 & 0.902 & 10.22 \\
dv-043 & 91.1 & 100.0 & 85.2 & 100.0 & 100.0 & 100.0 & 71.4 & 1/1 & 100.0 & 0 & 0.287 & 6.30 \\
dv-044 & 98.6 & 81.8 & 99.0 & 94.4 & 100.0 & 99.2 & 100.0 & 1/1 & 100.0 & 0 & 0.731 & 9.77 \\
dv-045 & 100.0 & 91.2 & 95.8 & 93.1 & 88.8 & 95.7 & 100.0 & 1/1 & 90.0 & 0 & 2.311 & 18.09 \\
dv-046 & 98.1 & 97.5 & 96.4 & 75.8 & 93.8 & 22.2 & 100.0 & 1/1 & 100.0 & 0 & 0.473 & 7.07 \\
dv-047 & 100.0 & 100.0 & 100.0 & 100.0 & 78.7 & 100.0 & 100.0 & 1/1 & 100.0 & \textbf{\color{ReleaseBlue}1} & 0.309 & 8.38 \\
dv-048 & 100.0 & 100.0 & 100.0 & 98.7 & 97.9 & 52.6 & 0.0 & 1/1 & 100.0 & 0 & 0.376 & 7.25 \\
dv-049 & 99.7 & 99.8 & 95.5 & 98.3 & 100.0 & 69.2 & 100.0 & 1/1 & 100.0 & \textbf{\color{ReleaseBlue}1} & 0.259 & 6.23 \\
dv-050 & 100.0 & 80.0 & 99.1 & 100.0 & 100.0 & 88.0 & 100.0 & 1/1 & 100.0 & 0 & 0.524 & 9.21 \\
dv-051 & 99.6 & 98.3 & 99.2 & 93.6 & 100.0 & 89.7 & 100.0 & 1/1 & 90.0 & 0 & 1.049 & 11.01 \\
dv-052 & 99.9 & 38.9 & 99.9 & 99.9 & 100.0 & 100.0 & 100.0 & 1/1 & 90.0 & 0 & 0.329 & 9.21 \\
dv-053 & 100.0 & 24.5 & 100.0 & 97.2 & 100.0 & 0.0 & 100.0 & 1/1 & 100.0 & 0 & 0.416 & 7.92 \\
dv-054 & 100.0 & 99.8 & 100.0 & 100.0 & 100.0 & 0.0 & 100.0 & 1/1 & 90.0 & 0 & 0.364 & 6.98 \\
dv-055 & 90.0 & 66.0 & 91.6 & 80.7 & 100.0 & 100.0 & 100.0 & 1/1 & 100.0 & 0 & 0.298 & 6.17 \\
dv-056 & 99.8 & 98.3 & 94.0 & 93.9 & 96.9 & 82.4 & 100.0 & 1/1 & 100.0 & \textbf{\color{ReleaseBlue}1} & 0.638 & 10.04 \\
dv-057 & 88.4 & 77.9 & 74.7 & 97.4 & 100.0 & 82.7 & 95.7 & 1/1 & 100.0 & 0 & 0.445 & 7.63 \\
dv-058 & 91.3 & 93.2 & 86.7 & 98.5 & 100.0 & 79.1 & 100.0 & 1/1 & 100.0 & 0 & 0.852 & 23.06 \\
dv-059 & 99.1 & 100.0 & 97.1 & 100.0 & 100.0 & 0.0 & 99.9 & 1/1 & 100.0 & 0 & 1.615 & 13.55 \\
dv-060 & 100.0 & 90.3 & 98.2 & 96.8 & 100.0 & 100.0 & 100.0 & 1/1 & 100.0 & \textbf{\color{ReleaseBlue}1} & 0.396 & 7.08 \\
dv-061 & 99.9 & 88.9 & 99.8 & 98.6 & 100.0 & 0.0 & 100.0 & 1/1 & 80.0 & 0 & 1.550 & 12.78 \\
dv-062 & 98.7 & 65.5 & 69.3 & 81.7 & 100.0 & 93.3 & 98.2 & 1/1 & 100.0 & 0 & 0.569 & 8.13 \\
dv-063 & 100.0 & 100.0 & 99.9 & 99.4 & 97.3 & 66.7 & 100.0 & 1/1 & 90.0 & 0 & 3.290 & 23.99 \\
dv-064 & 99.2 & 96.4 & 98.6 & 76.4 & 100.0 & 33.3 & 100.0 & 1/1 & 90.0 & 0 & 0.459 & 7.19 \\
dv-065 & 100.0 & 76.4 & 82.1 & 90.6 & 91.6 & 0.0 & 100.0 & 1/1 & 80.0 & 0 & 1.918 & 17.27 \\
dv-066 & 98.4 & 98.7 & 96.3 & 43.7 & 100.0 & 11.8 & 100.0 & 1/1 & 90.0 & 0 & 0.411 & 6.55 \\
dv-067 & 80.2 & 95.5 & 81.7 & 99.6 & 98.1 & 44.2 & 100.0 & 1/1 & 90.0 & 0 & 0.494 & 9.65 \\
dv-068 & 48.2 & 78.3 & 33.1 & 42.5 & 100.0 & 52.6 & 100.0 & 1/1 & 30.0 & 0 & 0.367 & 6.90 \\
dv-069 & 100.0 & 80.7 & 100.0 & 82.6 & 100.0 & 66.7 & 100.0 & 1/1 & 100.0 & 0 & 0.384 & 6.90 \\
dv-070 & 90.1 & 69.0 & 79.3 & 57.4 & 72.0 & 88.7 & 72.9 & 1/0 & 0.0 & 0 & 3.010 & 57.76 \\
dv-071 & 99.6 & 95.9 & 97.4 & 88.7 & 76.9 & 100.0 & 100.0 & 1/1 & 100.0 & 0 & 0.609 & 9.35 \\
dv-072 & 100.0 & 100.0 & 100.0 & 100.0 & 94.7 & 100.0 & 47.5 & 1/1 & 100.0 & 0 & 0.397 & 6.99 \\
dv-073 & 96.8 & 84.2 & 92.2 & 89.9 & 100.0 & 88.9 & 100.0 & 1/1 & 100.0 & 0 & 0.511 & 8.35 \\
dv-074 & 100.0 & 100.0 & 100.0 & 92.6 & 100.0 & 95.9 & 24.3 & 1/1 & 100.0 & 0 & 0.266 & 5.47 \\
dv-075 & 98.6 & 94.2 & 90.4 & 84.7 & 100.0 & 71.0 & 0.0 & 1/1 & 80.0 & 0 & 1.819 & 13.12 \\
dv-076 & 72.8 & 80.2 & 56.4 & 76.5 & 100.0 & 60.9 & 100.0 & 1/1 & 100.0 & 0 & 0.258 & 5.91 \\
dv-077 & 99.8 & 92.3 & 99.3 & 70.7 & 93.5 & 0.0 & 0.0 & 1/0 & 0.0 & 0 & 1.192 & 12.92 \\
dv-078 & 70.0 & 76.6 & 60.9 & 45.0 & 88.1 & 93.3 & 100.0 & 1/1 & 100.0 & 0 & 0.460 & 7.86 \\
dv-079 & 92.1 & 75.9 & 84.7 & 75.5 & 100.0 & 71.0 & 100.0 & 1/0 & 0.0 & 0 & 0.671 & 26.86 \\
dv-080 & 95.9 & 87.4 & 92.0 & 85.2 & 84.2 & 54.0 & 0.0 & 1/1 & 70.0 & 0 & 1.736 & 16.22 \\
dv-081 & 45.3 & 41.2 & 39.1 & 24.1 & 100.0 & 99.2 & 100.0 & 1/1 & 100.0 & 0 & 2.418 & 18.60 \\
dv-082 & 94.8 & 100.0 & 91.1 & 99.4 & 100.0 & 34.4 & 100.0 & 1/1 & 100.0 & \textbf{\color{ReleaseBlue}1} & 0.496 & 8.44 \\
dv-083 & 100.0 & 97.7 & 100.0 & 99.6 & 98.3 & 84.4 & 100.0 & 1/1 & 100.0 & \textbf{\color{ReleaseBlue}1} & 0.342 & 6.20 \\
dv-084 & 100.0 & 87.4 & 95.0 & 94.4 & 93.5 & 65.4 & 16.9 & 1/1 & 100.0 & 0 & 2.168 & 17.19 \\
dv-085 & 98.6 & 91.2 & 95.5 & 96.4 & 83.0 & 87.9 & 100.0 & 1/1 & 80.0 & 0 & 1.448 & 12.32 \\
dv-086 & 96.1 & 85.8 & 94.2 & 90.7 & 100.0 & 93.3 & 100.0 & 1/1 & 90.0 & 0 & 0.419 & 7.91 \\
dv-087 & 99.6 & 90.2 & 97.6 & 75.8 & 96.9 & 80.9 & 100.0 & 1/1 & 100.0 & 0 & 0.560 & 9.95 \\
dv-088 & 92.8 & 87.5 & 85.1 & 73.4 & 93.3 & 54.5 & 98.3 & 1/1 & 80.0 & 0 & 10.406 & 38.23 \\
dv-089 & 99.6 & 93.5 & 98.6 & 74.0 & 92.8 & 65.8 & 88.1 & 1/0 & 0.0 & 0 & 2.677 & 16.23 \\
dv-090 & 36.8 & 61.9 & 38.9 & 66.0 & 89.3 & 90.9 & 100.0 & 1/1 & 90.0 & 0 & 0.278 & 5.58 \\
dv-091 & 97.3 & 100.0 & 95.6 & 99.4 & 85.7 & 80.5 & 100.0 & 1/1 & 80.0 & 0 & 0.386 & 6.55 \\
dv-092 & 97.2 & 95.0 & 94.3 & 84.6 & 100.0 & 87.5 & 100.0 & 1/1 & 100.0 & 0 & 0.482 & 9.07 \\
dv-093 & 99.1 & 89.1 & 98.6 & 88.5 & 93.3 & 66.7 & 100.0 & 1/1 & 100.0 & 0 & 0.622 & 8.15 \\
dv-094 & 98.9 & 100.0 & 95.8 & 61.1 & 100.0 & 93.3 & 100.0 & 1/1 & 80.0 & 0 & 0.602 & 9.55 \\
dv-095 & 99.9 & 94.7 & 99.7 & 96.8 & 89.7 & 60.3 & 7.2 & 1/1 & 60.0 & 0 & 0.446 & 7.21 \\
\end{longtable}\endgroup

\clearpage
\subsection{C06: Codex + GLM 5.3}
\noindent\footnotesize Percentages: Ln/Cn/Br/Tg = line/condition/branch/toggle coverage; F1/P1 = feature/test-plan F1; Ref = reference-model point accuracy; Kill = mutant kill rate. C/G and O are binary compile/golden and overall results. Tok = input plus output tokens in millions; Min = elapsed minutes. Overall uses exact model correctness and all four coverage thresholds. \par\normalsize
\begingroup
\fontsize{8.2}{10.2}\selectfont
\setlength{\tabcolsep}{3pt}\renewcommand{\arraystretch}{1.04}
\setlength{\LTleft}{0pt}\setlength{\LTright}{0pt}
\Needspace{10\baselineskip}
\captionsetup{hypcap=false}
\captionof{table}{Per-task measurements for C06. The companion CSV retains unrounded values, precision/recall, exact-model flags, input/output token splits, and source-cell coordinates.}\label{tab:app-detail-C06}
\addtocounter{table}{-1}
\begin{longtable}{>{\raggedright\arraybackslash}p{\dimexpr 0.1050000\textwidth-2\tabcolsep\relax}>{\raggedleft\arraybackslash}p{\dimexpr 0.0745833\textwidth-2\tabcolsep\relax}>{\raggedleft\arraybackslash}p{\dimexpr 0.0745833\textwidth-2\tabcolsep\relax}>{\raggedleft\arraybackslash}p{\dimexpr 0.0745833\textwidth-2\tabcolsep\relax}>{\raggedleft\arraybackslash}p{\dimexpr 0.0745833\textwidth-2\tabcolsep\relax}>{\raggedleft\arraybackslash}p{\dimexpr 0.0745833\textwidth-2\tabcolsep\relax}>{\raggedleft\arraybackslash}p{\dimexpr 0.0745833\textwidth-2\tabcolsep\relax}>{\raggedleft\arraybackslash}p{\dimexpr 0.0745833\textwidth-2\tabcolsep\relax}>{\raggedleft\arraybackslash}p{\dimexpr 0.0745833\textwidth-2\tabcolsep\relax}>{\raggedleft\arraybackslash}p{\dimexpr 0.0745833\textwidth-2\tabcolsep\relax}>{\raggedleft\arraybackslash}p{\dimexpr 0.0745833\textwidth-2\tabcolsep\relax}>{\raggedleft\arraybackslash}p{\dimexpr 0.0745833\textwidth-2\tabcolsep\relax}>{\raggedleft\arraybackslash}p{\dimexpr 0.0745833\textwidth-2\tabcolsep\relax}}
\toprule
\rowcolor{ReleasePale}\textbf{Task} & \textbf{Ln} & \textbf{Cn} & \textbf{Br} & \textbf{Tg} & \textbf{F1} & \textbf{P1} & \textbf{Ref} & \textbf{C/G} & \textbf{Kill} & \textbf{O} & \textbf{Tok} & \textbf{Min} \\
\midrule\endfirsthead
\multicolumn{13}{l}{\small\color{ReleaseMuted}Table~\ref{tab:app-detail-C06} (continued)}\\
\toprule
\rowcolor{ReleasePale}\textbf{Task} & \textbf{Ln} & \textbf{Cn} & \textbf{Br} & \textbf{Tg} & \textbf{F1} & \textbf{P1} & \textbf{Ref} & \textbf{C/G} & \textbf{Kill} & \textbf{O} & \textbf{Tok} & \textbf{Min} \\
\midrule\endhead
\midrule
\multicolumn{13}{r}{\footnotesize\color{ReleaseMuted}Continued on next page}\\\endfoot
\bottomrule\endlastfoot
dv-001 & 100.0 & 92.8 & 100.0 & 96.0 & 92.9 & 95.7 & 100.0 & 1/1 & 100.0 & \textbf{\color{ReleaseBlue}1} & 1.820 & 21.03 \\
dv-002 & 100.0 & 100.0 & 100.0 & 97.5 & 85.7 & 57.7 & 99.4 & 1/0 & 0.0 & 0 & 7.646 & 85.28 \\
dv-003 & 100.0 & 40.0 & 50.0 & 0.6 & 96.8 & 47.0 & 100.0 & 1/0 & 0.0 & 0 & 9.061 & 43.93 \\
dv-004 & 0.0 & 0.0 & 0.0 & 0.0 & 94.7 & 0.0 & 100.0 & 0/0 & 0.0 & 0 & 1.939 & 38.19 \\
dv-005 & 0.0 & 0.0 & 0.0 & 0.0 & 75.0 & 0.0 & 69.6 & 1/0 & 0.0 & 0 & 2.767 & 58.04 \\
dv-006 & 0.0 & 0.0 & 0.0 & 0.0 & 64.8 & 63.3 & 98.8 & 1/1 & 40.0 & 0 & 7.108 & 63.22 \\
dv-007 & 100.0 & 100.0 & 100.0 & 100.0 & 85.7 & 0.0 & 100.0 & 1/0 & 0.0 & 0 & 5.121 & 49.53 \\
dv-008 & 84.1 & 87.6 & 61.3 & 94.2 & 73.9 & 46.3 & 0.0 & 1/0 & 0.0 & 0 & 28.411 & 185.46 \\
dv-009 & 59.3 & 60.0 & 47.8 & 29.5 & 65.3 & 0.0 & 97.2 & 1/0 & 0.0 & 0 & 6.432 & 48.36 \\
dv-010 & 100.0 & 100.0 & 100.0 & 99.3 & 72.9 & 57.1 & 100.0 & 1/1 & 100.0 & \textbf{\color{ReleaseBlue}1} & 2.126 & 35.98 \\
dv-011 & 98.4 & 93.5 & 91.3 & 52.9 & 75.0 & 54.4 & 100.0 & 1/1 & 100.0 & 0 & 0.925 & 17.98 \\
dv-012 & 6.2 & 1.4 & 50.0 & 0.2 & 96.8 & 66.7 & 27.3 & 1/1 & 0.0 & 0 & 1.472 & 22.28 \\
dv-013 & 100.0 & 95.5 & 100.0 & 93.1 & 72.5 & 38.7 & 45.1 & 1/0 & 0.0 & 0 & 18.698 & 108.75 \\
dv-014 & 0.0 & 0.0 & 0.0 & 0.0 & 90.9 & 19.8 & 0.1 & 0/0 & 0.0 & 0 & 13.300 & 127.49 \\
dv-015 & 0.0 & 0.0 & 0.0 & 0.0 & 85.1 & 15.3 & 3.2 & 1/0 & 0.0 & 0 & 10.620 & 82.48 \\
dv-016 & 0.0 & 0.0 & 0.0 & 0.0 & 0.0 & 0.0 & 0.6 & 1/0 & 0.0 & 0 & 2.640 & 58.09 \\
dv-017 & 0.0 & 0.0 & 0.0 & 0.0 & 86.3 & 98.3 & 89.9 & 0/0 & 0.0 & 0 & 25.718 & 155.64 \\
dv-018 & 92.1 & 87.5 & 87.2 & 78.0 & 0.0 & 0.0 & 100.0 & 1/0 & 0.0 & 0 & 6.302 & 71.05 \\
dv-019 & 54.1 & 41.1 & 32.9 & 42.9 & 0.0 & 0.0 & 1.3 & 1/0 & 0.0 & 0 & 2.903 & 47.97 \\
dv-020 & 93.5 & 91.4 & 83.6 & 84.1 & 58.7 & 44.3 & 70.5 & 1/0 & 0.0 & 0 & 16.544 & 151.32 \\
dv-021 & 0.0 & 0.0 & 0.0 & 0.0 & 0.0 & 0.0 & 0.0 & 0/0 & 0.0 & 0 & 0.423 & 38.26 \\
dv-022 & 0.0 & 0.0 & 0.0 & 0.0 & 84.3 & 0.0 & 0.0 & 0/0 & 0.0 & 0 & 10.838 & 71.87 \\
dv-023 & 0.0 & 0.0 & 0.0 & 0.0 & 65.5 & 0.0 & 61.0 & 1/0 & 0.0 & 0 & 3.881 & 65.66 \\
dv-024 & 0.0 & 0.0 & 0.0 & 0.0 & 89.7 & 0.0 & 100.0 & 1/0 & 0.0 & 0 & 2.049 & 51.85 \\
dv-025 & 41.1 & 58.1 & 24.3 & 6.9 & 87.2 & 11.8 & 100.0 & 1/1 & 20.0 & 0 & 1.691 & 31.53 \\
dv-026 & 0.0 & 0.0 & 0.0 & 0.0 & 96.8 & 21.9 & 100.0 & 1/1 & 60.0 & 0 & 10.561 & 85.36 \\
dv-027 & 66.8 & 37.9 & 51.8 & 9.6 & 66.5 & 10.5 & 0.0 & 1/0 & 0.0 & 0 & 25.630 & 239.48 \\
dv-028 & 0.0 & 0.0 & 0.0 & 0.0 & 96.8 & 53.3 & 100.0 & 1/1 & 100.0 & 0 & 4.554 & 39.91 \\
dv-029 & 27.1 & 19.3 & 15.0 & 0.1 & 0.0 & 10.4 & 0.0 & 1/1 & 0.0 & 0 & 9.361 & 58.31 \\
dv-030 & 0.0 & 0.0 & 0.0 & 0.0 & 0.0 & 0.0 & 15.7 & 1/0 & 0.0 & 0 & 16.125 & 134.01 \\
dv-031 & 62.2 & 3.8 & 38.0 & 31.8 & 0.0 & 0.0 & 99.7 & 1/1 & 70.0 & 0 & 5.153 & 57.28 \\
dv-032 & 0.0 & 0.0 & 0.0 & 0.0 & 87.5 & 0.0 & 50.6 & 1/1 & 30.0 & 0 & 17.862 & 148.55 \\
dv-033 & 73.5 & 67.1 & 66.9 & 30.2 & 0.0 & 0.0 & 0.0 & 1/0 & 0.0 & 0 & 2.293 & 48.44 \\
dv-034 & 0.0 & 0.0 & 0.0 & 0.0 & 79.4 & 17.9 & 15.8 & 1/0 & 0.0 & 0 & 21.971 & 121.68 \\
dv-035 & 0.0 & 0.0 & 0.0 & 0.0 & 78.9 & 56.5 & 0.0 & 0/0 & 0.0 & 0 & 0.987 & 28.59 \\
dv-036 & 0.0 & 0.0 & 0.0 & 0.0 & 0.0 & 0.0 & 0.0 & 0/0 & 0.0 & 0 & 6.833 & 123.55 \\
dv-037 & 91.8 & 84.0 & 85.4 & 24.8 & 0.0 & 0.0 & 51.0 & 1/0 & 0.0 & 0 & 18.920 & 139.18 \\
dv-038 & 100.0 & 97.7 & 100.0 & 98.7 & 82.8 & 0.0 & 100.0 & 1/1 & 90.0 & 0 & 6.315 & 51.55 \\
dv-039 & 0.0 & 0.0 & 0.0 & 0.0 & 76.8 & 13.3 & 14.5 & 1/0 & 0.0 & 0 & 1.117 & 27.21 \\
dv-040 & 0.0 & 0.0 & 0.0 & 0.0 & 0.0 & 0.0 & 100.0 & 1/0 & 0.0 & 0 & 14.743 & 93.70 \\
dv-041 & 0.0 & 0.0 & 0.0 & 0.0 & 0.0 & 0.0 & 69.4 & 1/0 & 0.0 & 0 & 15.882 & 84.55 \\
dv-042 & 89.4 & 84.3 & 85.4 & 45.2 & 78.5 & 0.0 & 37.0 & 1/1 & 80.0 & 0 & 2.874 & 40.35 \\
dv-043 & 91.1 & 100.0 & 85.2 & 100.0 & 76.8 & 66.7 & 89.4 & 1/0 & 0.0 & 0 & 20.110 & 108.84 \\
dv-044 & 73.7 & 30.5 & 58.9 & 39.0 & 0.0 & 0.0 & 0.0 & 1/0 & 0.0 & 0 & 9.622 & 81.53 \\
dv-045 & 100.0 & 64.8 & 82.9 & 69.1 & 87.3 & 84.7 & 0.0 & 1/1 & 10.0 & 0 & 12.515 & 77.11 \\
dv-046 & 0.0 & 0.0 & 0.0 & 0.0 & 70.8 & 0.0 & 0.0 & 0/0 & 0.0 & 0 & 3.923 & 67.04 \\
dv-047 & 100.0 & 100.0 & 100.0 & 100.0 & 39.3 & 0.0 & 100.0 & 1/1 & 90.0 & 0 & 21.225 & 124.40 \\
dv-048 & 100.0 & 100.0 & 100.0 & 98.6 & 67.8 & 0.0 & 0.0 & 1/1 & 90.0 & 0 & 17.445 & 120.38 \\
dv-049 & 99.7 & 99.8 & 95.5 & 98.3 & 100.0 & 31.1 & 100.0 & 1/1 & 90.0 & 0 & 25.798 & 237.00 \\
dv-050 & 100.0 & 80.0 & 99.1 & 100.0 & 92.3 & 59.7 & 100.0 & 1/1 & 80.0 & 0 & 1.808 & 58.37 \\
dv-051 & 0.0 & 0.0 & 0.0 & 0.0 & 0.0 & 0.0 & 56.0 & 1/0 & 0.0 & 0 & 10.273 & 73.75 \\
dv-052 & 97.1 & 38.9 & 94.7 & 99.9 & 74.2 & 0.0 & 1.7 & 1/0 & 0.0 & 0 & 26.498 & 250.18 \\
dv-053 & 100.0 & 24.0 & 98.9 & 92.4 & 0.0 & 0.0 & 66.1 & 1/0 & 0.0 & 0 & 27.164 & 214.11 \\
dv-054 & 100.0 & 90.3 & 100.0 & 95.0 & 48.2 & 0.0 & 100.0 & 1/1 & 90.0 & 0 & 4.184 & 58.89 \\
dv-055 & 0.0 & 0.0 & 0.0 & 0.0 & 85.7 & 76.1 & 66.8 & 1/1 & 70.0 & 0 & 5.583 & 42.44 \\
dv-056 & 99.7 & 93.2 & 93.2 & 57.1 & 0.0 & 0.0 & 41.4 & 1/0 & 0.0 & 0 & 13.429 & 110.62 \\
dv-057 & 0.0 & 0.0 & 0.0 & 0.0 & 66.7 & 51.9 & 95.7 & 1/0 & 0.0 & 0 & 1.321 & 34.01 \\
dv-058 & 0.0 & 0.0 & 0.0 & 0.0 & 41.1 & 7.3 & 99.9 & 1/0 & 0.0 & 0 & 27.818 & 304.32 \\
dv-059 & 99.1 & 100.0 & 97.1 & 99.5 & 65.9 & 0.0 & 2.7 & 1/1 & 100.0 & 0 & 18.492 & 126.71 \\
dv-060 & 87.0 & 89.4 & 83.9 & 96.7 & 82.8 & 0.0 & 100.0 & 1/1 & 90.0 & 0 & 7.223 & 75.25 \\
dv-061 & 60.7 & 54.3 & 44.2 & 24.3 & 21.3 & 60.5 & 2.9 & 1/0 & 0.0 & 0 & 18.875 & 105.48 \\
dv-062 & 0.0 & 0.0 & 0.0 & 0.0 & 73.4 & 30.9 & 0.0 & 0/0 & 0.0 & 0 & 1.725 & 77.52 \\
dv-063 & 34.8 & 100.0 & 33.8 & 99.0 & 0.0 & 0.0 & 21.3 & 1/0 & 0.0 & 0 & 7.029 & 65.76 \\
dv-064 & 64.7 & 40.3 & 57.9 & 30.2 & 75.2 & 0.0 & 95.0 & 1/1 & 30.0 & 0 & 8.344 & 98.93 \\
dv-065 & 95.4 & 53.8 & 66.9 & 16.3 & 0.0 & 0.0 & 76.9 & 1/0 & 0.0 & 0 & 13.352 & 81.55 \\
dv-066 & 88.8 & 80.1 & 76.4 & 14.7 & 0.0 & 0.0 & 100.0 & 1/1 & 60.0 & 0 & 2.834 & 40.39 \\
dv-067 & 74.3 & 84.1 & 69.7 & 55.9 & 53.6 & 0.0 & 47.8 & 1/1 & 40.0 & 0 & 5.745 & 89.48 \\
dv-068 & 0.0 & 0.0 & 0.0 & 0.0 & 49.5 & 0.0 & 0.0 & 0/0 & 0.0 & 0 & 6.998 & 75.63 \\
dv-069 & 100.0 & 80.6 & 100.0 & 82.5 & 71.5 & 51.9 & 100.0 & 1/1 & 100.0 & 0 & 1.957 & 40.93 \\
dv-070 & 0.0 & 0.0 & 0.0 & 0.0 & 0.0 & 0.0 & 0.0 & 0/0 & 0.0 & 0 & 19.270 & 158.31 \\
dv-071 & 99.6 & 56.2 & 95.6 & 88.3 & 53.8 & 0.0 & 16.4 & 1/0 & 0.0 & 0 & 7.274 & 70.84 \\
dv-072 & 0.0 & 0.0 & 0.0 & 0.0 & 0.0 & 0.0 & 0.3 & 1/0 & 0.0 & 0 & 8.511 & 58.83 \\
dv-073 & 0.0 & 0.0 & 0.0 & 0.0 & 0.0 & 0.0 & 90.2 & 1/0 & 0.0 & 0 & 1.726 & 40.12 \\
dv-074 & 100.0 & 100.0 & 100.0 & 92.6 & 96.8 & 88.9 & 2.0 & 1/1 & 100.0 & 0 & 25.246 & 222.65 \\
dv-075 & 0.0 & 0.0 & 0.0 & 0.0 & 0.0 & 0.0 & 0.0 & 0/0 & 0.0 & 0 & 0.958 & 45.27 \\
dv-076 & 72.8 & 77.8 & 55.4 & 75.2 & 0.0 & 11.8 & 100.0 & 1/1 & 80.0 & 0 & 2.465 & 56.81 \\
dv-077 & 0.0 & 0.0 & 0.0 & 0.0 & 0.0 & 0.0 & 0.0 & 1/0 & 0.0 & 0 & 18.036 & 231.92 \\
dv-078 & 66.4 & 76.6 & 54.7 & 42.9 & 67.0 & 66.2 & 100.0 & 1/1 & 80.0 & 0 & 11.335 & 131.67 \\
dv-079 & 0.0 & 0.0 & 0.0 & 0.0 & 0.0 & 0.0 & 99.7 & 1/0 & 0.0 & 0 & 1.610 & 108.07 \\
dv-080 & 0.0 & 0.0 & 0.0 & 0.0 & 73.5 & 0.0 & 0.0 & 0/0 & 0.0 & 0 & 23.567 & 238.46 \\
dv-081 & 0.0 & 0.0 & 0.0 & 0.0 & 0.0 & 0.0 & 0.0 & 0/0 & 0.0 & 0 & 0.249 & 35.32 \\
dv-082 & 96.2 & 93.9 & 93.8 & 99.4 & 97.9 & 0.0 & 100.0 & 1/1 & 90.0 & 0 & 24.876 & 282.34 \\
dv-083 & 0.0 & 0.0 & 0.0 & 0.0 & 0.0 & 0.0 & 28.6 & 1/0 & 0.0 & 0 & 6.694 & 114.03 \\
dv-084 & 93.0 & 76.0 & 82.0 & 64.7 & 62.0 & 62.0 & 1.4 & 1/0 & 0.0 & 0 & 24.011 & 188.88 \\
dv-085 & 0.0 & 0.0 & 0.0 & 0.0 & 0.0 & 0.0 & 98.0 & 1/1 & 0.0 & 0 & 2.341 & 77.19 \\
dv-086 & 95.9 & 84.2 & 92.8 & 96.2 & 64.9 & 22.2 & 69.3 & 1/0 & 0.0 & 0 & 27.598 & 324.43 \\
dv-087 & 81.9 & 53.3 & 76.1 & 50.7 & 0.0 & 46.7 & 99.9 & 1/1 & 30.0 & 0 & 24.545 & 262.69 \\
dv-088 & 73.1 & 36.0 & 51.4 & 0.7 & 44.5 & 0.0 & 1.6 & 1/1 & 0.0 & 0 & 5.158 & 111.54 \\
dv-089 & 0.0 & 0.0 & 0.0 & 0.0 & 0.0 & 0.0 & 0.4 & 1/0 & 0.0 & 0 & 3.849 & 121.37 \\
dv-090 & 0.0 & 0.0 & 0.0 & 0.0 & 80.0 & 73.7 & 100.0 & 1/1 & 80.0 & 0 & 1.853 & 81.69 \\
dv-091 & 97.3 & 100.0 & 95.6 & 41.5 & 95.9 & 0.0 & 13.7 & 1/0 & 0.0 & 0 & 23.014 & 195.75 \\
dv-092 & 0.0 & 0.0 & 0.0 & 0.0 & 0.0 & 0.0 & 0.0 & 0/0 & 0.0 & 0 & 0.414 & 24.31 \\
dv-093 & 0.0 & 0.0 & 0.0 & 0.0 & 80.6 & 54.2 & 57.2 & 1/1 & 70.0 & 0 & 2.488 & 40.81 \\
dv-094 & 0.0 & 0.0 & 0.0 & 0.0 & 68.8 & 53.8 & 72.1 & 0/0 & 0.0 & 0 & 2.256 & 51.05 \\
dv-095 & 0.0 & 0.0 & 0.0 & 0.0 & 0.0 & 0.0 & 7.2 & 1/0 & 0.0 & 0 & 5.746 & 54.21 \\
\end{longtable}\endgroup

\clearpage
\subsection{C07: Codex + DeepSeek V4.1 Flash 1017}
\noindent\footnotesize Percentages: Ln/Cn/Br/Tg = line/condition/branch/toggle coverage; F1/P1 = feature/test-plan F1; Ref = reference-model point accuracy; Kill = mutant kill rate. C/G and O are binary compile/golden and overall results. Tok = input plus output tokens in millions; Min = elapsed minutes. Overall uses exact model correctness and all four coverage thresholds. \par\normalsize
\begingroup
\fontsize{8.2}{10.2}\selectfont
\setlength{\tabcolsep}{3pt}\renewcommand{\arraystretch}{1.04}
\setlength{\LTleft}{0pt}\setlength{\LTright}{0pt}
\Needspace{10\baselineskip}
\captionsetup{hypcap=false}
\captionof{table}{Per-task measurements for C07. The companion CSV retains unrounded values, precision/recall, exact-model flags, input/output token splits, and source-cell coordinates.}\label{tab:app-detail-C07}
\addtocounter{table}{-1}
\begin{longtable}{>{\raggedright\arraybackslash}p{\dimexpr 0.1050000\textwidth-2\tabcolsep\relax}>{\raggedleft\arraybackslash}p{\dimexpr 0.0745833\textwidth-2\tabcolsep\relax}>{\raggedleft\arraybackslash}p{\dimexpr 0.0745833\textwidth-2\tabcolsep\relax}>{\raggedleft\arraybackslash}p{\dimexpr 0.0745833\textwidth-2\tabcolsep\relax}>{\raggedleft\arraybackslash}p{\dimexpr 0.0745833\textwidth-2\tabcolsep\relax}>{\raggedleft\arraybackslash}p{\dimexpr 0.0745833\textwidth-2\tabcolsep\relax}>{\raggedleft\arraybackslash}p{\dimexpr 0.0745833\textwidth-2\tabcolsep\relax}>{\raggedleft\arraybackslash}p{\dimexpr 0.0745833\textwidth-2\tabcolsep\relax}>{\raggedleft\arraybackslash}p{\dimexpr 0.0745833\textwidth-2\tabcolsep\relax}>{\raggedleft\arraybackslash}p{\dimexpr 0.0745833\textwidth-2\tabcolsep\relax}>{\raggedleft\arraybackslash}p{\dimexpr 0.0745833\textwidth-2\tabcolsep\relax}>{\raggedleft\arraybackslash}p{\dimexpr 0.0745833\textwidth-2\tabcolsep\relax}>{\raggedleft\arraybackslash}p{\dimexpr 0.0745833\textwidth-2\tabcolsep\relax}}
\toprule
\rowcolor{ReleasePale}\textbf{Task} & \textbf{Ln} & \textbf{Cn} & \textbf{Br} & \textbf{Tg} & \textbf{F1} & \textbf{P1} & \textbf{Ref} & \textbf{C/G} & \textbf{Kill} & \textbf{O} & \textbf{Tok} & \textbf{Min} \\
\midrule\endfirsthead
\multicolumn{13}{l}{\small\color{ReleaseMuted}Table~\ref{tab:app-detail-C07} (continued)}\\
\toprule
\rowcolor{ReleasePale}\textbf{Task} & \textbf{Ln} & \textbf{Cn} & \textbf{Br} & \textbf{Tg} & \textbf{F1} & \textbf{P1} & \textbf{Ref} & \textbf{C/G} & \textbf{Kill} & \textbf{O} & \textbf{Tok} & \textbf{Min} \\
\midrule\endhead
\midrule
\multicolumn{13}{r}{\footnotesize\color{ReleaseMuted}Continued on next page}\\\endfoot
\bottomrule\endlastfoot
dv-001 & 100.0 & 92.8 & 100.0 & 96.0 & 0.0 & 94.5 & 100.0 & 1/1 & 100.0 & \textbf{\color{ReleaseBlue}1} & 1.551 & 4.73 \\
dv-002 & 100.0 & 100.0 & 100.0 & 97.5 & 95.7 & 66.5 & 100.0 & 1/1 & 100.0 & \textbf{\color{ReleaseBlue}1} & 3.017 & 7.95 \\
dv-003 & 100.0 & 100.0 & 100.0 & 98.8 & 100.0 & 0.0 & 100.0 & 1/1 & 100.0 & \textbf{\color{ReleaseBlue}1} & 3.596 & 7.40 \\
dv-004 & 99.0 & 92.9 & 96.2 & 76.0 & 100.0 & 0.0 & 100.0 & 1/1 & 100.0 & 0 & 7.984 & 11.25 \\
dv-005 & 100.0 & 99.1 & 100.0 & 98.9 & 87.1 & 50.0 & 100.0 & 1/1 & 100.0 & \textbf{\color{ReleaseBlue}1} & 2.876 & 18.91 \\
dv-006 & 0.0 & 0.0 & 0.0 & 0.0 & 85.2 & 60.4 & 100.0 & 1/1 & 50.0 & 0 & 14.944 & 16.09 \\
dv-007 & 100.0 & 100.0 & 100.0 & 100.0 & 100.0 & 0.0 & 100.0 & 1/1 & 80.0 & 0 & 0.878 & 4.71 \\
dv-008 & 98.3 & 92.9 & 96.6 & 97.4 & 60.0 & 0.0 & 82.1 & 1/1 & 100.0 & 0 & 66.299 & 60.72 \\
dv-009 & 100.0 & 91.3 & 100.0 & 100.0 & 81.1 & 40.0 & 100.0 & 1/1 & 30.0 & 0 & 1.880 & 6.66 \\
dv-010 & 100.0 & 100.0 & 100.0 & 99.3 & 86.3 & 66.7 & 100.0 & 1/1 & 100.0 & \textbf{\color{ReleaseBlue}1} & 1.617 & 6.01 \\
dv-011 & 98.4 & 93.5 & 91.3 & 52.9 & 75.0 & 82.3 & 100.0 & 1/1 & 100.0 & 0 & 1.152 & 3.97 \\
dv-012 & 100.0 & 100.0 & 100.0 & 100.0 & 100.0 & 63.9 & 100.0 & 1/1 & 100.0 & \textbf{\color{ReleaseBlue}1} & 3.462 & 7.57 \\
dv-013 & 100.0 & 93.9 & 100.0 & 94.7 & 100.0 & 51.1 & 100.0 & 1/1 & 90.0 & 0 & 1.619 & 6.31 \\
dv-014 & 85.8 & 77.6 & 82.3 & 64.3 & 81.4 & 42.8 & 100.0 & 1/1 & 90.0 & 0 & 22.340 & 29.81 \\
dv-015 & 100.0 & 100.0 & 97.3 & 81.0 & 64.3 & 0.0 & 100.0 & 1/1 & 90.0 & 0 & 14.313 & 26.41 \\
dv-016 & 98.9 & 98.0 & 98.4 & 52.6 & 82.0 & 0.0 & 100.0 & 1/1 & 60.0 & 0 & 19.095 & 19.27 \\
dv-017 & 96.8 & 81.3 & 88.4 & 90.6 & 85.1 & 100.0 & 87.9 & 1/1 & 100.0 & 0 & 2.140 & 6.19 \\
dv-018 & 100.0 & 98.2 & 100.0 & 94.8 & 53.6 & 0.0 & 100.0 & 1/1 & 100.0 & \textbf{\color{ReleaseBlue}1} & 1.523 & 5.50 \\
dv-019 & 95.7 & 76.6 & 93.4 & 98.3 & 89.4 & 34.4 & 29.1 & 1/1 & 70.0 & 0 & 20.469 & 21.26 \\
dv-020 & 97.6 & 91.4 & 86.4 & 93.6 & 59.4 & 54.3 & 100.0 & 1/1 & 60.0 & 0 & 36.534 & 28.17 \\
dv-021 & 85.5 & 66.0 & 78.8 & 75.3 & 84.3 & 66.7 & 78.8 & 1/1 & 90.0 & 0 & 25.504 & 31.86 \\
dv-022 & 100.0 & 95.0 & 98.1 & 59.6 & 73.7 & 28.5 & 6.2 & 1/1 & 30.0 & 0 & 7.020 & 12.81 \\
dv-023 & 100.0 & 100.0 & 100.0 & 100.0 & 76.0 & 89.7 & 100.0 & 1/1 & 100.0 & \textbf{\color{ReleaseBlue}1} & 1.321 & 6.56 \\
dv-024 & 43.3 & 60.3 & 27.1 & 36.5 & 95.1 & 0.0 & 100.0 & 1/1 & 20.0 & 0 & 12.397 & 18.93 \\
dv-025 & 40.7 & 57.1 & 23.9 & 8.3 & 30.6 & 22.1 & 100.0 & 1/1 & 20.0 & 0 & 4.903 & 22.70 \\
dv-026 & 42.1 & 62.2 & 24.9 & 14.2 & 100.0 & 92.6 & 100.0 & 1/1 & 80.0 & 0 & 5.184 & 11.95 \\
dv-027 & 95.7 & 83.6 & 92.0 & 41.8 & 78.2 & 0.0 & 0.0 & 1/1 & 40.0 & 0 & 48.378 & 26.22 \\
dv-028 & 100.0 & 98.0 & 100.0 & 98.7 & 100.0 & 96.8 & 100.0 & 1/1 & 100.0 & \textbf{\color{ReleaseBlue}1} & 14.309 & 13.30 \\
dv-029 & 99.7 & 98.9 & 95.0 & 45.3 & 59.9 & 28.0 & 56.4 & 1/1 & 70.0 & 0 & 30.720 & 23.05 \\
dv-030 & 54.9 & 44.9 & 57.3 & 75.3 & 94.4 & 13.3 & 100.0 & 1/1 & 60.0 & 0 & 33.719 & 91.59 \\
dv-031 & 62.8 & 4.0 & 38.8 & 31.8 & 100.0 & 31.2 & 99.9 & 1/1 & 60.0 & 0 & 4.213 & 11.45 \\
dv-032 & 96.6 & 83.5 & 94.6 & 92.7 & 62.2 & 0.0 & 55.5 & 1/0 & 0.0 & 0 & 43.321 & 45.05 \\
dv-033 & 77.4 & 74.2 & 70.1 & 30.4 & 74.9 & 15.3 & 94.1 & 1/1 & 80.0 & 0 & 20.035 & 30.19 \\
dv-034 & 75.3 & 89.3 & 69.7 & 39.1 & 93.9 & 56.2 & 100.0 & 1/1 & 100.0 & 0 & 2.670 & 9.18 \\
dv-035 & 79.0 & 52.0 & 62.9 & 81.8 & 83.9 & 0.0 & 65.1 & 1/1 & 90.0 & 0 & 6.746 & 13.81 \\
dv-036 & 90.5 & 75.4 & 80.3 & 61.8 & 92.3 & 0.0 & 28.1 & 1/1 & 50.0 & 0 & 48.713 & 89.75 \\
dv-037 & 99.7 & 97.2 & 95.5 & 41.7 & 43.5 & 0.0 & 99.8 & 1/1 & 70.0 & 0 & 46.600 & 53.72 \\
dv-038 & 100.0 & 97.7 & 100.0 & 99.6 & 71.2 & 0.0 & 100.0 & 1/1 & 90.0 & 0 & 5.412 & 12.50 \\
dv-039 & 99.6 & 97.1 & 99.6 & 91.9 & 92.3 & 13.3 & 100.0 & 1/1 & 80.0 & 0 & 2.544 & 9.33 \\
dv-040 & 100.0 & 91.7 & 99.2 & 67.6 & 80.4 & 100.0 & 100.0 & 1/1 & 70.0 & 0 & 5.478 & 15.75 \\
dv-041 & 99.3 & 78.4 & 85.0 & 98.0 & 90.0 & 0.0 & 100.0 & 1/1 & 80.0 & 0 & 12.014 & 16.79 \\
dv-042 & 89.0 & 85.8 & 83.7 & 61.7 & 80.4 & 0.0 & 100.0 & 1/1 & 90.0 & 0 & 12.597 & 27.99 \\
dv-043 & 91.1 & 100.0 & 85.2 & 100.0 & 92.3 & 57.6 & 97.5 & 1/1 & 100.0 & 0 & 3.439 & 11.51 \\
dv-044 & 0.0 & 0.0 & 0.0 & 0.0 & 78.6 & 60.1 & 100.0 & 1/1 & 50.0 & 0 & 31.640 & 56.55 \\
dv-045 & 100.0 & 84.6 & 92.8 & 87.7 & 56.0 & 66.7 & 100.0 & 1/1 & 80.0 & 0 & 13.446 & 28.05 \\
dv-046 & 97.8 & 96.8 & 95.9 & 75.8 & 76.9 & 0.0 & 37.1 & 1/1 & 100.0 & 0 & 17.693 & 33.38 \\
dv-047 & 100.0 & 100.0 & 100.0 & 100.0 & 54.7 & 0.0 & 100.0 & 1/1 & 90.0 & 0 & 12.672 & 23.30 \\
dv-048 & 100.0 & 100.0 & 100.0 & 98.7 & 100.0 & 0.0 & 0.0 & 1/1 & 100.0 & 0 & 4.557 & 16.33 \\
dv-049 & 99.7 & 99.8 & 95.5 & 98.3 & 83.8 & 39.7 & 100.0 & 1/1 & 90.0 & 0 & 9.403 & 18.25 \\
dv-050 & 100.0 & 80.0 & 99.1 & 100.0 & 92.3 & 66.3 & 100.0 & 1/1 & 90.0 & 0 & 2.234 & 12.86 \\
dv-051 & 89.1 & 97.9 & 78.0 & 93.6 & 73.9 & 0.0 & 92.4 & 1/1 & 80.0 & 0 & 10.185 & 20.64 \\
dv-052 & 99.3 & 38.9 & 96.9 & 99.9 & 93.3 & 39.4 & 100.0 & 1/1 & 90.0 & 0 & 2.271 & 11.34 \\
dv-053 & 100.0 & 23.7 & 100.0 & 89.9 & 91.6 & 31.4 & 95.2 & 1/1 & 100.0 & 0 & 24.789 & 40.09 \\
dv-054 & 100.0 & 99.2 & 100.0 & 96.4 & 88.9 & 0.0 & 100.0 & 1/1 & 90.0 & 0 & 4.192 & 17.45 \\
dv-055 & 89.6 & 65.6 & 91.3 & 76.0 & 74.9 & 54.0 & 100.0 & 1/1 & 90.0 & 0 & 3.266 & 13.69 \\
dv-056 & 99.8 & 88.1 & 95.7 & 87.5 & 84.0 & 25.6 & 92.3 & 1/1 & 90.0 & 0 & 12.780 & 24.31 \\
dv-057 & 88.4 & 77.9 & 74.7 & 97.2 & 77.0 & 69.9 & 95.7 & 1/1 & 100.0 & 0 & 4.223 & 16.68 \\
dv-058 & 84.6 & 87.2 & 73.1 & 98.0 & 89.4 & 26.6 & 0.0 & 1/1 & 100.0 & 0 & 20.487 & 53.51 \\
dv-059 & 98.6 & 97.5 & 95.7 & 95.2 & 78.2 & 0.0 & 4.8 & 1/1 & 100.0 & 0 & 18.765 & 35.14 \\
dv-060 & 98.7 & 89.9 & 97.3 & 96.8 & 49.1 & 0.0 & 100.0 & 1/1 & 100.0 & 0 & 31.431 & 34.87 \\
dv-061 & 94.2 & 77.2 & 91.5 & 99.1 & 22.0 & 31.1 & 91.3 & 1/1 & 90.0 & 0 & 70.514 & 66.59 \\
dv-062 & 98.7 & 63.4 & 69.3 & 81.6 & 100.0 & 75.4 & 95.2 & 1/1 & 80.0 & 0 & 4.369 & 14.71 \\
dv-063 & 79.9 & 100.0 & 79.4 & 99.4 & 98.7 & 0.0 & 100.0 & 1/1 & 80.0 & 0 & 19.956 & 35.59 \\
dv-064 & 97.3 & 92.6 & 95.0 & 75.4 & 83.6 & 18.1 & 100.0 & 1/1 & 90.0 & 0 & 9.783 & 20.26 \\
dv-065 & 95.5 & 72.0 & 75.1 & 90.2 & 83.9 & 0.0 & 95.9 & 1/1 & 70.0 & 0 & 21.652 & 35.44 \\
dv-066 & 98.4 & 98.7 & 96.3 & 44.3 & 70.4 & 11.6 & 100.0 & 1/1 & 60.0 & 0 & 8.169 & 24.89 \\
dv-067 & 74.3 & 93.9 & 73.4 & 99.6 & 76.5 & 0.0 & 100.0 & 1/1 & 80.0 & 0 & 9.223 & 24.30 \\
dv-068 & 47.0 & 62.5 & 31.5 & 13.8 & 57.4 & 0.0 & 100.0 & 1/1 & 40.0 & 0 & 24.519 & 32.28 \\
dv-069 & 100.0 & 80.7 & 100.0 & 82.6 & 100.0 & 59.0 & 100.0 & 1/1 & 100.0 & 0 & 2.687 & 10.37 \\
dv-070 & 82.3 & 55.9 & 67.4 & 42.5 & 43.3 & 21.8 & 72.5 & 1/0 & 0.0 & 0 & 55.296 & 136.03 \\
dv-071 & 99.6 & 58.2 & 95.6 & 88.7 & 39.3 & 38.5 & 100.0 & 1/1 & 90.0 & 0 & 5.722 & 16.01 \\
dv-072 & 100.0 & 100.0 & 100.0 & 100.0 & 65.0 & 56.8 & 47.5 & 1/1 & 100.0 & 0 & 2.261 & 12.23 \\
dv-073 & 96.8 & 84.2 & 90.7 & 88.3 & 80.5 & 66.3 & 100.0 & 1/1 & 80.0 & 0 & 2.963 & 12.52 \\
dv-074 & 100.0 & 100.0 & 100.0 & 92.6 & 95.2 & 68.9 & 9.3 & 1/1 & 100.0 & 0 & 1.683 & 8.45 \\
dv-075 & 94.0 & 85.4 & 83.0 & 80.7 & 59.4 & 66.3 & 0.0 & 1/1 & 70.0 & 0 & 26.029 & 40.52 \\
dv-076 & 72.8 & 77.8 & 55.4 & 76.5 & 96.8 & 66.7 & 100.0 & 1/1 & 90.0 & 0 & 2.118 & 10.79 \\
dv-077 & 100.0 & 90.8 & 97.6 & 25.3 & 84.4 & 0.0 & 0.0 & 1/0 & 0.0 & 0 & 16.204 & 23.42 \\
dv-078 & 66.8 & 76.6 & 55.5 & 44.9 & 65.2 & 57.5 & 100.0 & 1/1 & 90.0 & 0 & 4.040 & 12.60 \\
dv-079 & 89.9 & 70.9 & 80.0 & 71.1 & 86.5 & 56.4 & 99.8 & 1/1 & 50.0 & 0 & 16.697 & 41.88 \\
dv-080 & 95.9 & 80.0 & 92.0 & 65.7 & 78.8 & 15.3 & 0.0 & 1/1 & 70.0 & 0 & 18.054 & 26.91 \\
dv-081 & 35.2 & 30.8 & 27.5 & 3.8 & 85.0 & 54.3 & 100.0 & 1/1 & 80.0 & 0 & 8.902 & 19.55 \\
dv-082 & 96.2 & 100.0 & 93.8 & 96.4 & 80.0 & 0.0 & 100.0 & 1/1 & 90.0 & 0 & 15.991 & 33.62 \\
dv-083 & 100.0 & 95.8 & 97.7 & 99.6 & 81.5 & 60.6 & 100.0 & 1/1 & 100.0 & \textbf{\color{ReleaseBlue}1} & 9.774 & 23.27 \\
dv-084 & 98.0 & 81.3 & 87.6 & 77.6 & 72.8 & 38.9 & 16.8 & 1/1 & 50.0 & 0 & 12.106 & 40.48 \\
dv-085 & 98.6 & 91.2 & 95.5 & 96.4 & 55.1 & 32.8 & 98.8 & 1/1 & 80.0 & 0 & 7.381 & 23.39 \\
dv-086 & 96.1 & 81.9 & 90.9 & 97.3 & 100.0 & 0.0 & 100.0 & 1/1 & 80.0 & 0 & 12.728 & 26.05 \\
dv-087 & 99.6 & 89.6 & 97.7 & 74.2 & 82.2 & 78.9 & 100.0 & 1/1 & 90.0 & 0 & 11.265 & 31.05 \\
dv-088 & 88.5 & 70.6 & 74.1 & 17.7 & 30.4 & 0.0 & 32.3 & 1/1 & 80.0 & 0 & 59.478 & 47.29 \\
dv-089 & 99.1 & 88.0 & 96.7 & 70.6 & 68.5 & 44.4 & 86.5 & 1/1 & 40.0 & 0 & 30.532 & 43.41 \\
dv-090 & 36.6 & 60.9 & 38.1 & 63.6 & 68.9 & 80.0 & 100.0 & 1/1 & 90.0 & 0 & 2.444 & 12.00 \\
dv-091 & 97.3 & 100.0 & 95.6 & 99.4 & 71.5 & 0.0 & 100.0 & 1/1 & 40.0 & 0 & 34.252 & 39.60 \\
dv-092 & 97.2 & 95.0 & 94.3 & 84.6 & 100.0 & 0.0 & 100.0 & 1/1 & 100.0 & 0 & 3.013 & 12.17 \\
dv-093 & 99.1 & 85.5 & 97.3 & 48.2 & 88.0 & 65.5 & 38.4 & 1/1 & 100.0 & 0 & 7.025 & 22.42 \\
dv-094 & 98.9 & 94.1 & 94.4 & 61.1 & 81.5 & 53.3 & 100.0 & 1/1 & 80.0 & 0 & 10.022 & 13.30 \\
dv-095 & 99.9 & 91.6 & 99.5 & 96.6 & 69.0 & 60.0 & 83.5 & 1/1 & 60.0 & 0 & 8.158 & 27.05 \\
\end{longtable}\endgroup

\clearpage
\subsection{C08: Codex + GPT-6.1 Sol}
\noindent\footnotesize Percentages: Ln/Cn/Br/Tg = line/condition/branch/toggle coverage; F1/P1 = feature/test-plan F1; Ref = reference-model point accuracy; Kill = mutant kill rate. C/G and O are binary compile/golden and overall results. Tok = input plus output tokens in millions; Min = elapsed minutes. Overall uses exact model correctness and all four coverage thresholds. \par\normalsize
\begingroup
\fontsize{8.2}{10.2}\selectfont
\setlength{\tabcolsep}{3pt}\renewcommand{\arraystretch}{1.04}
\setlength{\LTleft}{0pt}\setlength{\LTright}{0pt}
\Needspace{10\baselineskip}
\captionsetup{hypcap=false}
\captionof{table}{Per-task measurements for C08. The companion CSV retains unrounded values, precision/recall, exact-model flags, input/output token splits, and source-cell coordinates.}\label{tab:app-detail-C08}
\addtocounter{table}{-1}
\begin{longtable}{>{\raggedright\arraybackslash}p{\dimexpr 0.1050000\textwidth-2\tabcolsep\relax}>{\raggedleft\arraybackslash}p{\dimexpr 0.0745833\textwidth-2\tabcolsep\relax}>{\raggedleft\arraybackslash}p{\dimexpr 0.0745833\textwidth-2\tabcolsep\relax}>{\raggedleft\arraybackslash}p{\dimexpr 0.0745833\textwidth-2\tabcolsep\relax}>{\raggedleft\arraybackslash}p{\dimexpr 0.0745833\textwidth-2\tabcolsep\relax}>{\raggedleft\arraybackslash}p{\dimexpr 0.0745833\textwidth-2\tabcolsep\relax}>{\raggedleft\arraybackslash}p{\dimexpr 0.0745833\textwidth-2\tabcolsep\relax}>{\raggedleft\arraybackslash}p{\dimexpr 0.0745833\textwidth-2\tabcolsep\relax}>{\raggedleft\arraybackslash}p{\dimexpr 0.0745833\textwidth-2\tabcolsep\relax}>{\raggedleft\arraybackslash}p{\dimexpr 0.0745833\textwidth-2\tabcolsep\relax}>{\raggedleft\arraybackslash}p{\dimexpr 0.0745833\textwidth-2\tabcolsep\relax}>{\raggedleft\arraybackslash}p{\dimexpr 0.0745833\textwidth-2\tabcolsep\relax}>{\raggedleft\arraybackslash}p{\dimexpr 0.0745833\textwidth-2\tabcolsep\relax}}
\toprule
\rowcolor{ReleasePale}\textbf{Task} & \textbf{Ln} & \textbf{Cn} & \textbf{Br} & \textbf{Tg} & \textbf{F1} & \textbf{P1} & \textbf{Ref} & \textbf{C/G} & \textbf{Kill} & \textbf{O} & \textbf{Tok} & \textbf{Min} \\
\midrule\endfirsthead
\multicolumn{13}{l}{\small\color{ReleaseMuted}Table~\ref{tab:app-detail-C08} (continued)}\\
\toprule
\rowcolor{ReleasePale}\textbf{Task} & \textbf{Ln} & \textbf{Cn} & \textbf{Br} & \textbf{Tg} & \textbf{F1} & \textbf{P1} & \textbf{Ref} & \textbf{C/G} & \textbf{Kill} & \textbf{O} & \textbf{Tok} & \textbf{Min} \\
\midrule\endhead
\midrule
\multicolumn{13}{r}{\footnotesize\color{ReleaseMuted}Continued on next page}\\\endfoot
\bottomrule\endlastfoot
dv-001 & 100.0 & 93.2 & 100.0 & 96.9 & 97.0 & 100.0 & 100.0 & 1/1 & 100.0 & \textbf{\color{ReleaseBlue}1} & 0.421 & 7.61 \\
dv-002 & 100.0 & 100.0 & 100.0 & 97.5 & 97.4 & 73.7 & 100.0 & 1/1 & 100.0 & \textbf{\color{ReleaseBlue}1} & 0.737 & 8.85 \\
dv-003 & 100.0 & 100.0 & 100.0 & 100.0 & 93.3 & 100.0 & 100.0 & 1/1 & 100.0 & \textbf{\color{ReleaseBlue}1} & 0.618 & 8.46 \\
dv-004 & 99.0 & 96.3 & 96.2 & 89.0 & 100.0 & 33.2 & 100.0 & 1/1 & 100.0 & 0 & 1.583 & 17.82 \\
dv-005 & 100.0 & 99.3 & 100.0 & 98.9 & 77.3 & 100.0 & 100.0 & 1/1 & 100.0 & \textbf{\color{ReleaseBlue}1} & 1.834 & 17.17 \\
dv-006 & 99.5 & 99.8 & 99.8 & 92.8 & 100.0 & 100.0 & 100.0 & 1/1 & 100.0 & \textbf{\color{ReleaseBlue}1} & 0.969 & 11.77 \\
dv-007 & 100.0 & 100.0 & 100.0 & 100.0 & 96.3 & 66.3 & 100.0 & 1/1 & 100.0 & \textbf{\color{ReleaseBlue}1} & 0.357 & 6.81 \\
dv-008 & 98.3 & 92.4 & 96.6 & 97.5 & 90.8 & 81.1 & 100.0 & 1/1 & 100.0 & \textbf{\color{ReleaseBlue}1} & 6.385 & 45.87 \\
dv-009 & 100.0 & 91.3 & 100.0 & 100.0 & 100.0 & 93.3 & 100.0 & 1/1 & 40.0 & 0 & 0.645 & 11.16 \\
dv-010 & 100.0 & 100.0 & 100.0 & 99.3 & 87.2 & 100.0 & 100.0 & 1/1 & 100.0 & \textbf{\color{ReleaseBlue}1} & 0.708 & 9.28 \\
dv-011 & 98.4 & 93.5 & 91.3 & 52.9 & 94.7 & 99.0 & 100.0 & 1/1 & 90.0 & 0 & 0.562 & 8.07 \\
dv-012 & 100.0 & 100.0 & 100.0 & 100.0 & 100.0 & 98.8 & 100.0 & 1/1 & 100.0 & \textbf{\color{ReleaseBlue}1} & 7.333 & 52.87 \\
dv-013 & 100.0 & 100.0 & 100.0 & 94.7 & 100.0 & 100.0 & 100.0 & 1/1 & 100.0 & \textbf{\color{ReleaseBlue}1} & 0.881 & 15.96 \\
dv-014 & 87.0 & 79.6 & 84.3 & 68.2 & 98.2 & 70.8 & 100.0 & 1/1 & 100.0 & 0 & 3.459 & 21.48 \\
dv-015 & 100.0 & 100.0 & 97.3 & 89.4 & 84.2 & 95.7 & 28.4 & 1/1 & 100.0 & 0 & 2.750 & 20.59 \\
dv-016 & 99.3 & 97.8 & 98.9 & 63.5 & 100.0 & 86.7 & 100.0 & 1/1 & 80.0 & 0 & 5.022 & 26.02 \\
dv-017 & 96.8 & 81.3 & 88.4 & 90.6 & 89.3 & 99.2 & 89.9 & 1/1 & 100.0 & 0 & 1.418 & 16.55 \\
dv-018 & 100.0 & 100.0 & 100.0 & 94.8 & 100.0 & 99.2 & 100.0 & 1/1 & 100.0 & \textbf{\color{ReleaseBlue}1} & 2.372 & 18.50 \\
dv-019 & 98.0 & 86.3 & 97.0 & 99.0 & 100.0 & 66.4 & 0.0 & 1/1 & 80.0 & 0 & 1.277 & 12.56 \\
dv-020 & 97.6 & 93.8 & 88.2 & 95.3 & 96.8 & 99.1 & 100.0 & 1/1 & 100.0 & 0 & 14.533 & 42.36 \\
dv-021 & 0.0 & 0.0 & 0.0 & 0.0 & 99.2 & 96.8 & 100.0 & 1/1 & 100.0 & 0 & 11.294 & 62.74 \\
dv-022 & 100.0 & 100.0 & 98.1 & 94.3 & 96.4 & 40.0 & 15.9 & 1/1 & 90.0 & 0 & 1.539 & 13.02 \\
dv-023 & 100.0 & 100.0 & 100.0 & 100.0 & 100.0 & 50.0 & 100.0 & 1/1 & 100.0 & \textbf{\color{ReleaseBlue}1} & 0.987 & 11.61 \\
dv-024 & 43.3 & 59.8 & 27.1 & 34.5 & 90.4 & 39.8 & 100.0 & 1/1 & 90.0 & 0 & 0.925 & 12.95 \\
dv-025 & 40.7 & 54.3 & 23.9 & 30.0 & 100.0 & 100.0 & 100.0 & 1/1 & 20.0 & 0 & 0.908 & 12.75 \\
dv-026 & 42.1 & 58.6 & 24.9 & 20.8 & 93.3 & 93.3 & 100.0 & 1/1 & 80.0 & 0 & 0.561 & 13.11 \\
dv-027 & 0.0 & 0.0 & 0.0 & 0.0 & 100.0 & 56.0 & 69.4 & 1/1 & 80.0 & 0 & 6.707 & 34.68 \\
dv-028 & 100.0 & 98.4 & 100.0 & 98.7 & 100.0 & 100.0 & 100.0 & 1/1 & 100.0 & \textbf{\color{ReleaseBlue}1} & 1.204 & 12.63 \\
dv-029 & 0.0 & 0.0 & 0.0 & 0.0 & 100.0 & 97.1 & 0.0 & 1/1 & 80.0 & 0 & 3.642 & 30.15 \\
dv-030 & 79.4 & 57.7 & 70.9 & 87.8 & 86.2 & 86.3 & 100.0 & 1/0 & 0.0 & 0 & 20.590 & 128.68 \\
dv-031 & 62.2 & 3.9 & 38.0 & 31.8 & 100.0 & 100.0 & 99.7 & 1/1 & 70.0 & 0 & 1.451 & 13.61 \\
dv-032 & 97.0 & 84.5 & 95.3 & 94.6 & 100.0 & 100.0 & 1.3 & 1/0 & 0.0 & 0 & 2.195 & 44.22 \\
dv-033 & 78.3 & 74.8 & 76.4 & 89.2 & 98.4 & 80.0 & 100.0 & 1/1 & 100.0 & 0 & 7.038 & 34.27 \\
dv-034 & 0.0 & 0.0 & 0.0 & 0.0 & 92.9 & 73.9 & 100.0 & 1/1 & 100.0 & 0 & 0.556 & 9.07 \\
dv-035 & 0.0 & 0.0 & 0.0 & 0.0 & 84.3 & 87.5 & 65.1 & 1/0 & 0.0 & 0 & 1.277 & 14.10 \\
dv-036 & 99.2 & 92.0 & 97.4 & 82.2 & 94.5 & 59.0 & 72.3 & 1/1 & 80.0 & 0 & 5.710 & 87.53 \\
dv-037 & 99.7 & 98.3 & 98.9 & 68.2 & 92.9 & 52.5 & 100.0 & 1/1 & 80.0 & 0 & 2.354 & 18.87 \\
dv-038 & 100.0 & 97.7 & 100.0 & 99.6 & 100.0 & 65.9 & 100.0 & 1/1 & 90.0 & 0 & 1.221 & 12.09 \\
dv-039 & 99.6 & 99.0 & 99.6 & 98.7 & 100.0 & 83.3 & 100.0 & 1/1 & 80.0 & 0 & 0.426 & 8.20 \\
dv-040 & 100.0 & 98.9 & 99.2 & 68.5 & 100.0 & 92.3 & 100.0 & 1/1 & 90.0 & 0 & 0.715 & 10.20 \\
dv-041 & 0.0 & 0.0 & 0.0 & 0.0 & 100.0 & 80.0 & 100.0 & 1/1 & 90.0 & 0 & 1.325 & 9.57 \\
dv-042 & 100.0 & 100.0 & 100.0 & 74.4 & 99.0 & 100.0 & 100.0 & 1/1 & 100.0 & 0 & 1.074 & 12.77 \\
dv-043 & 91.1 & 100.0 & 85.2 & 100.0 & 100.0 & 100.0 & 99.6 & 1/1 & 100.0 & 0 & 0.570 & 7.54 \\
dv-044 & 98.6 & 81.8 & 99.0 & 94.4 & 97.1 & 99.2 & 100.0 & 1/1 & 100.0 & 0 & 2.396 & 22.30 \\
dv-045 & 100.0 & 90.5 & 95.5 & 92.4 & 32.2 & 94.5 & 100.0 & 1/1 & 100.0 & \textbf{\color{ReleaseBlue}1} & 3.895 & 23.13 \\
dv-046 & 98.1 & 97.5 & 96.4 & 75.8 & 89.4 & 84.1 & 100.0 & 1/1 & 100.0 & 0 & 1.166 & 15.22 \\
dv-047 & 100.0 & 100.0 & 100.0 & 100.0 & 89.1 & 100.0 & 100.0 & 1/1 & 100.0 & \textbf{\color{ReleaseBlue}1} & 1.282 & 14.87 \\
dv-048 & 100.0 & 100.0 & 100.0 & 98.7 & 98.0 & 44.4 & 0.0 & 1/0 & 0.0 & 0 & 1.440 & 15.13 \\
dv-049 & 99.7 & 99.8 & 95.5 & 98.3 & 98.0 & 89.7 & 100.0 & 1/1 & 100.0 & \textbf{\color{ReleaseBlue}1} & 1.160 & 12.33 \\
dv-050 & 100.0 & 80.0 & 99.1 & 100.0 & 100.0 & 88.0 & 100.0 & 1/1 & 90.0 & 0 & 1.215 & 17.27 \\
dv-051 & 99.6 & 98.3 & 99.2 & 93.6 & 100.0 & 100.0 & 100.0 & 1/1 & 90.0 & 0 & 1.203 & 11.86 \\
dv-052 & 99.9 & 38.9 & 97.6 & 99.9 & 97.7 & 99.3 & 100.0 & 1/1 & 100.0 & 0 & 0.600 & 9.12 \\
dv-053 & 100.0 & 24.8 & 100.0 & 97.2 & 90.0 & 54.2 & 100.0 & 1/1 & 100.0 & 0 & 1.293 & 12.96 \\
dv-054 & 100.0 & 99.9 & 100.0 & 100.0 & 93.7 & 33.3 & 100.0 & 1/1 & 90.0 & 0 & 2.553 & 18.55 \\
dv-055 & 90.0 & 70.8 & 91.6 & 80.7 & 100.0 & 100.0 & 100.0 & 1/1 & 100.0 & 0 & 1.064 & 13.65 \\
dv-056 & 99.8 & 98.3 & 96.2 & 88.8 & 98.9 & 88.9 & 100.0 & 1/1 & 100.0 & 0 & 1.313 & 14.84 \\
dv-057 & 88.4 & 77.9 & 74.7 & 97.4 & 96.3 & 95.5 & 95.7 & 1/1 & 100.0 & 0 & 1.076 & 12.43 \\
dv-058 & 91.3 & 94.6 & 86.7 & 98.5 & 100.0 & 83.6 & 0.0 & 1/1 & 100.0 & 0 & 6.746 & 51.85 \\
dv-059 & 99.1 & 100.0 & 97.1 & 100.0 & 89.3 & 49.8 & 99.9 & 1/1 & 100.0 & 0 & 2.381 & 17.60 \\
dv-060 & 100.0 & 89.9 & 98.2 & 96.8 & 100.0 & 71.4 & 100.0 & 1/1 & 100.0 & 0 & 2.397 & 16.75 \\
dv-061 & 99.9 & 99.0 & 99.8 & 99.7 & 100.0 & 92.8 & 100.0 & 1/1 & 100.0 & \textbf{\color{ReleaseBlue}1} & 4.188 & 21.87 \\
dv-062 & 98.7 & 65.4 & 69.3 & 81.7 & 94.4 & 96.8 & 97.3 & 1/1 & 100.0 & 0 & 1.529 & 16.10 \\
dv-063 & 100.0 & 100.0 & 99.9 & 99.4 & 100.0 & 66.7 & 100.0 & 1/1 & 90.0 & 0 & 6.847 & 34.49 \\
dv-064 & 99.2 & 96.4 & 98.6 & 76.4 & 100.0 & 62.1 & 100.0 & 1/1 & 90.0 & 0 & 2.700 & 26.75 \\
dv-065 & 98.4 & 74.9 & 79.6 & 90.6 & 100.0 & 49.7 & 100.0 & 1/1 & 80.0 & 0 & 3.250 & 18.77 \\
dv-066 & 98.4 & 98.7 & 96.3 & 45.1 & 100.0 & 40.0 & 100.0 & 1/1 & 100.0 & 0 & 1.027 & 12.56 \\
dv-067 & 80.2 & 95.5 & 81.7 & 97.9 & 96.8 & 72.0 & 99.6 & 1/1 & 90.0 & 0 & 0.772 & 12.13 \\
dv-068 & 0.0 & 0.0 & 0.0 & 0.0 & 98.3 & 91.8 & 100.0 & 1/1 & 20.0 & 0 & 1.353 & 16.67 \\
dv-069 & 100.0 & 80.7 & 100.0 & 82.6 & 100.0 & 66.7 & 100.0 & 1/1 & 100.0 & 0 & 1.155 & 67.96 \\
dv-070 & 83.0 & 68.1 & 66.3 & 53.8 & 62.0 & 39.0 & 86.5 & 1/0 & 0.0 & 0 & 11.618 & 157.16 \\
dv-071 & 99.6 & 99.6 & 97.4 & 88.7 & 38.7 & 40.0 & 100.0 & 1/1 & 90.0 & 0 & 1.905 & 13.22 \\
dv-072 & 100.0 & 100.0 & 100.0 & 100.0 & 100.0 & 100.0 & 47.5 & 1/1 & 100.0 & 0 & 0.932 & 11.40 \\
dv-073 & 96.8 & 84.2 & 92.2 & 89.9 & 97.7 & 88.9 & 100.0 & 1/1 & 100.0 & 0 & 0.775 & 11.18 \\
dv-074 & 100.0 & 100.0 & 100.0 & 92.6 & 98.5 & 100.0 & 24.3 & 1/1 & 100.0 & 0 & 1.033 & 9.52 \\
dv-075 & 98.6 & 94.2 & 90.4 & 91.5 & 98.1 & 78.8 & 0.0 & 1/1 & 90.0 & 0 & 8.934 & 49.27 \\
dv-076 & 72.8 & 80.2 & 56.4 & 76.5 & 100.0 & 60.9 & 100.0 & 1/1 & 100.0 & 0 & 0.872 & 11.37 \\
dv-077 & 100.0 & 92.7 & 99.7 & 85.6 & 85.9 & 39.7 & 0.0 & 1/1 & 80.0 & 0 & 8.882 & 50.21 \\
dv-078 & 70.0 & 76.6 & 60.9 & 45.0 & 90.2 & 100.0 & 100.0 & 1/1 & 100.0 & 0 & 0.793 & 11.40 \\
dv-079 & 92.3 & 77.2 & 84.7 & 76.2 & 79.7 & 81.4 & 100.0 & 1/1 & 80.0 & 0 & 8.536 & 100.76 \\
dv-080 & 95.9 & 87.4 & 92.0 & 91.7 & 100.0 & 85.7 & 19.8 & 1/1 & 80.0 & 0 & 5.545 & 34.44 \\
dv-081 & 35.2 & 30.8 & 27.5 & 3.8 & 80.3 & 98.1 & 31.0 & 1/1 & 100.0 & 0 & 6.631 & 36.08 \\
dv-082 & 94.8 & 100.0 & 91.1 & 99.4 & 100.0 & 100.0 & 100.0 & 1/1 & 100.0 & \textbf{\color{ReleaseBlue}1} & 1.586 & 19.65 \\
dv-083 & 100.0 & 97.7 & 100.0 & 90.1 & 100.0 & 100.0 & 100.0 & 1/1 & 100.0 & \textbf{\color{ReleaseBlue}1} & 4.431 & 73.27 \\
dv-084 & 100.0 & 88.1 & 95.0 & 93.2 & 97.9 & 66.2 & 1.7 & 1/1 & 100.0 & 0 & 3.715 & 63.82 \\
dv-085 & 98.6 & 91.2 & 95.5 & 96.4 & 97.9 & 74.6 & 100.0 & 1/1 & 70.0 & 0 & 3.489 & 61.70 \\
dv-086 & 96.1 & 86.8 & 94.2 & 97.8 & 100.0 & 0.0 & 100.0 & 1/1 & 80.0 & 0 & 1.131 & 75.82 \\
dv-087 & 100.0 & 91.1 & 100.0 & 77.4 & 95.1 & 96.8 & 100.0 & 1/1 & 80.0 & 0 & 6.554 & 237.53 \\
dv-088 & 92.8 & 88.2 & 85.1 & 78.2 & 96.8 & 39.4 & 100.0 & 1/1 & 80.0 & 0 & 7.908 & 96.61 \\
dv-089 & 99.6 & 93.1 & 98.1 & 73.8 & 94.3 & 54.5 & 59.9 & 1/1 & 60.0 & 0 & 2.178 & 62.04 \\
dv-090 & 36.8 & 62.3 & 38.9 & 66.8 & 70.0 & 90.9 & 100.0 & 1/1 & 90.0 & 0 & 0.677 & 49.58 \\
dv-091 & 97.3 & 100.0 & 95.6 & 99.4 & 100.0 & 93.3 & 100.0 & 1/1 & 70.0 & 0 & 1.556 & 56.36 \\
dv-092 & 97.2 & 95.0 & 94.3 & 84.6 & 100.0 & 94.1 & 100.0 & 1/1 & 90.0 & 0 & 0.778 & 15.65 \\
dv-093 & 99.1 & 89.5 & 98.6 & 90.1 & 100.0 & 100.0 & 100.0 & 1/1 & 100.0 & 0 & 2.337 & 21.31 \\
dv-094 & 100.0 & 100.0 & 98.6 & 61.1 & 100.0 & 100.0 & 100.0 & 1/1 & 90.0 & 0 & 1.391 & 73.56 \\
dv-095 & 100.0 & 92.6 & 100.0 & 96.8 & 95.8 & 96.8 & 91.4 & 1/1 & 60.0 & 0 & 1.406 & 17.46 \\
\end{longtable}\endgroup

\clearpage
\subsection{C09: Codex + GPT-6 Astra}
\noindent\footnotesize Percentages: Ln/Cn/Br/Tg = line/condition/branch/toggle coverage; F1/P1 = feature/test-plan F1; Ref = reference-model point accuracy; Kill = mutant kill rate. C/G and O are binary compile/golden and overall results. Tok = input plus output tokens in millions; Min = elapsed minutes. Overall uses exact model correctness and all four coverage thresholds. \par\normalsize
\begingroup
\fontsize{8.2}{10.2}\selectfont
\setlength{\tabcolsep}{3pt}\renewcommand{\arraystretch}{1.04}
\setlength{\LTleft}{0pt}\setlength{\LTright}{0pt}
\Needspace{10\baselineskip}
\captionsetup{hypcap=false}
\captionof{table}{Per-task measurements for C09. The companion CSV retains unrounded values, precision/recall, exact-model flags, input/output token splits, and source-cell coordinates.}\label{tab:app-detail-C09}
\addtocounter{table}{-1}
\begin{longtable}{>{\raggedright\arraybackslash}p{\dimexpr 0.1050000\textwidth-2\tabcolsep\relax}>{\raggedleft\arraybackslash}p{\dimexpr 0.0745833\textwidth-2\tabcolsep\relax}>{\raggedleft\arraybackslash}p{\dimexpr 0.0745833\textwidth-2\tabcolsep\relax}>{\raggedleft\arraybackslash}p{\dimexpr 0.0745833\textwidth-2\tabcolsep\relax}>{\raggedleft\arraybackslash}p{\dimexpr 0.0745833\textwidth-2\tabcolsep\relax}>{\raggedleft\arraybackslash}p{\dimexpr 0.0745833\textwidth-2\tabcolsep\relax}>{\raggedleft\arraybackslash}p{\dimexpr 0.0745833\textwidth-2\tabcolsep\relax}>{\raggedleft\arraybackslash}p{\dimexpr 0.0745833\textwidth-2\tabcolsep\relax}>{\raggedleft\arraybackslash}p{\dimexpr 0.0745833\textwidth-2\tabcolsep\relax}>{\raggedleft\arraybackslash}p{\dimexpr 0.0745833\textwidth-2\tabcolsep\relax}>{\raggedleft\arraybackslash}p{\dimexpr 0.0745833\textwidth-2\tabcolsep\relax}>{\raggedleft\arraybackslash}p{\dimexpr 0.0745833\textwidth-2\tabcolsep\relax}>{\raggedleft\arraybackslash}p{\dimexpr 0.0745833\textwidth-2\tabcolsep\relax}}
\toprule
\rowcolor{ReleasePale}\textbf{Task} & \textbf{Ln} & \textbf{Cn} & \textbf{Br} & \textbf{Tg} & \textbf{F1} & \textbf{P1} & \textbf{Ref} & \textbf{C/G} & \textbf{Kill} & \textbf{O} & \textbf{Tok} & \textbf{Min} \\
\midrule\endfirsthead
\multicolumn{13}{l}{\small\color{ReleaseMuted}Table~\ref{tab:app-detail-C09} (continued)}\\
\toprule
\rowcolor{ReleasePale}\textbf{Task} & \textbf{Ln} & \textbf{Cn} & \textbf{Br} & \textbf{Tg} & \textbf{F1} & \textbf{P1} & \textbf{Ref} & \textbf{C/G} & \textbf{Kill} & \textbf{O} & \textbf{Tok} & \textbf{Min} \\
\midrule\endhead
\midrule
\multicolumn{13}{r}{\footnotesize\color{ReleaseMuted}Continued on next page}\\\endfoot
\bottomrule\endlastfoot
dv-001 & 100.0 & 93.0 & 100.0 & 96.9 & 86.0 & 98.9 & 100.0 & 1/1 & 100.0 & \textbf{\color{ReleaseBlue}1} & 0.406 & 6.14 \\
dv-002 & 100.0 & 100.0 & 100.0 & 97.5 & 85.9 & 66.7 & 100.0 & 1/1 & 100.0 & \textbf{\color{ReleaseBlue}1} & 1.213 & 10.17 \\
dv-003 & 100.0 & 100.0 & 100.0 & 100.0 & 97.6 & 100.0 & 100.0 & 1/1 & 100.0 & \textbf{\color{ReleaseBlue}1} & 0.951 & 9.33 \\
dv-004 & 99.0 & 96.3 & 96.2 & 89.8 & 97.2 & 75.0 & 100.0 & 1/1 & 100.0 & 0 & 0.507 & 10.09 \\
dv-005 & 100.0 & 99.3 & 100.0 & 98.9 & 100.0 & 100.0 & 100.0 & 1/1 & 100.0 & \textbf{\color{ReleaseBlue}1} & 1.293 & 15.15 \\
dv-006 & 99.5 & 99.8 & 99.8 & 92.8 & 100.0 & 100.0 & 100.0 & 1/1 & 100.0 & \textbf{\color{ReleaseBlue}1} & 0.822 & 16.63 \\
dv-007 & 100.0 & 100.0 & 100.0 & 100.0 & 96.0 & 40.0 & 100.0 & 1/1 & 90.0 & 0 & 0.311 & 6.60 \\
dv-008 & 98.3 & 92.4 & 96.6 & 97.5 & 99.2 & 86.3 & 100.0 & 1/1 & 100.0 & \textbf{\color{ReleaseBlue}1} & 7.305 & 47.27 \\
dv-009 & 100.0 & 91.3 & 100.0 & 100.0 & 100.0 & 85.7 & 100.0 & 1/1 & 40.0 & 0 & 0.506 & 6.72 \\
dv-010 & 100.0 & 100.0 & 100.0 & 99.3 & 91.2 & 100.0 & 100.0 & 1/1 & 100.0 & \textbf{\color{ReleaseBlue}1} & 0.618 & 8.91 \\
dv-011 & 98.4 & 93.5 & 91.3 & 52.9 & 88.9 & 100.0 & 100.0 & 1/1 & 100.0 & 0 & 0.375 & 5.40 \\
dv-012 & 100.0 & 100.0 & 100.0 & 100.0 & 100.0 & 100.0 & 100.0 & 1/1 & 100.0 & \textbf{\color{ReleaseBlue}1} & 0.621 & 6.51 \\
dv-013 & 100.0 & 100.0 & 100.0 & 94.7 & 97.7 & 100.0 & 100.0 & 1/1 & 100.0 & \textbf{\color{ReleaseBlue}1} & 0.393 & 9.71 \\
dv-014 & 86.8 & 77.6 & 84.0 & 68.2 & 83.0 & 71.4 & 68.7 & 1/1 & 100.0 & 0 & 1.751 & 19.67 \\
dv-015 & 100.0 & 100.0 & 97.3 & 89.4 & 84.3 & 89.0 & 28.4 & 1/1 & 100.0 & 0 & 2.204 & 21.00 \\
dv-016 & 99.3 & 98.6 & 98.9 & 61.4 & 98.0 & 70.3 & 100.0 & 1/1 & 60.0 & 0 & 1.551 & 15.54 \\
dv-017 & 96.8 & 81.3 & 88.4 & 90.6 & 97.9 & 100.0 & 89.9 & 1/1 & 100.0 & 0 & 0.396 & 8.62 \\
dv-018 & 100.0 & 100.0 & 100.0 & 94.8 & 100.0 & 100.0 & 100.0 & 1/1 & 100.0 & \textbf{\color{ReleaseBlue}1} & 0.313 & 7.32 \\
dv-019 & 97.7 & 85.5 & 96.4 & 99.0 & 96.3 & 96.3 & 99.8 & 1/1 & 90.0 & 0 & 0.818 & 28.50 \\
dv-020 & 99.6 & 93.2 & 90.0 & 95.3 & 80.5 & 99.2 & 100.0 & 1/1 & 100.0 & \textbf{\color{ReleaseBlue}1} & 5.641 & 37.99 \\
dv-021 & 96.8 & 92.0 & 93.9 & 89.5 & 100.0 & 93.3 & 100.0 & 1/1 & 100.0 & 0 & 2.451 & 20.08 \\
dv-022 & 100.0 & 100.0 & 98.1 & 98.9 & 97.4 & 89.6 & 10.4 & 1/1 & 100.0 & 0 & 1.440 & 13.48 \\
dv-023 & 100.0 & 100.0 & 100.0 & 100.0 & 100.0 & 28.6 & 100.0 & 1/1 & 100.0 & \textbf{\color{ReleaseBlue}1} & 0.525 & 8.44 \\
dv-024 & 43.8 & 59.9 & 27.8 & 34.8 & 98.1 & 0.0 & 100.0 & 1/1 & 40.0 & 0 & 0.690 & 10.21 \\
dv-025 & 40.7 & 54.8 & 23.9 & 30.0 & 95.2 & 60.9 & 100.0 & 1/1 & 20.0 & 0 & 0.340 & 9.82 \\
dv-026 & 43.8 & 60.5 & 27.0 & 20.8 & 82.4 & 89.7 & 100.0 & 1/1 & 90.0 & 0 & 0.458 & 11.44 \\
dv-027 & 97.8 & 89.2 & 95.6 & 60.9 & 94.1 & 50.0 & 73.3 & 1/1 & 70.0 & 0 & 5.343 & 28.26 \\
dv-028 & 100.0 & 98.4 & 100.0 & 98.7 & 93.3 & 100.0 & 100.0 & 1/1 & 100.0 & \textbf{\color{ReleaseBlue}1} & 0.486 & 6.85 \\
dv-029 & 99.7 & 98.9 & 95.0 & 89.1 & 88.5 & 89.8 & 9.6 & 1/1 & 80.0 & 0 & 2.955 & 22.10 \\
dv-030 & 71.6 & 51.3 & 65.0 & 83.6 & 98.9 & 44.4 & 100.0 & 1/0 & 0.0 & 0 & 9.165 & 83.08 \\
dv-031 & 62.2 & 3.9 & 38.0 & 31.8 & 97.0 & 81.5 & 99.9 & 1/1 & 80.0 & 0 & 0.482 & 8.99 \\
dv-032 & 96.8 & 84.1 & 94.9 & 95.4 & 95.8 & 66.7 & 43.5 & 1/1 & 80.0 & 0 & 2.872 & 22.66 \\
dv-033 & 78.3 & 74.8 & 76.4 & 68.7 & 92.5 & 58.4 & 99.9 & 1/1 & 90.0 & 0 & 2.403 & 24.61 \\
dv-034 & 79.8 & 94.0 & 74.8 & 39.6 & 92.3 & 100.0 & 100.0 & 1/1 & 100.0 & 0 & 0.673 & 35.85 \\
dv-035 & 79.0 & 53.3 & 62.9 & 95.6 & 82.4 & 73.2 & 65.1 & 1/1 & 90.0 & 0 & 1.947 & 18.36 \\
dv-036 & 99.2 & 91.4 & 96.6 & 84.2 & 90.6 & 86.1 & 100.0 & 1/1 & 80.0 & 0 & 3.049 & 26.71 \\
dv-037 & 99.7 & 98.9 & 98.9 & 92.8 & 97.7 & 100.0 & 100.0 & 1/1 & 80.0 & 0 & 1.398 & 21.37 \\
dv-038 & 100.0 & 97.7 & 100.0 & 99.6 & 96.8 & 60.9 & 100.0 & 1/1 & 90.0 & 0 & 1.581 & 48.28 \\
dv-039 & 99.6 & 99.0 & 99.6 & 98.7 & 100.0 & 82.8 & 100.0 & 1/1 & 80.0 & 0 & 0.310 & 6.47 \\
dv-040 & 100.0 & 98.9 & 99.2 & 68.5 & 100.0 & 96.3 & 100.0 & 1/1 & 90.0 & 0 & 0.419 & 9.51 \\
dv-041 & 99.3 & 99.5 & 85.0 & 99.7 & 90.9 & 79.0 & 100.0 & 1/1 & 90.0 & 0 & 1.838 & 21.43 \\
dv-042 & 100.0 & 100.0 & 100.0 & 74.4 & 94.7 & 66.7 & 100.0 & 1/1 & 100.0 & 0 & 0.547 & 10.87 \\
dv-043 & 91.1 & 100.0 & 85.2 & 100.0 & 100.0 & 100.0 & 99.6 & 1/1 & 100.0 & 0 & 0.487 & 9.09 \\
dv-044 & 98.6 & 81.8 & 99.0 & 94.4 & 98.6 & 97.1 & 100.0 & 1/1 & 100.0 & 0 & 0.666 & 10.32 \\
dv-045 & 0.0 & 0.0 & 0.0 & 0.0 & 28.0 & 95.7 & 100.0 & 1/0 & 0.0 & 0 & 1.652 & 13.21 \\
dv-046 & 98.1 & 97.5 & 96.4 & 75.8 & 96.0 & 22.2 & 100.0 & 1/1 & 100.0 & 0 & 0.638 & 13.48 \\
dv-047 & 100.0 & 100.0 & 100.0 & 100.0 & 100.0 & 100.0 & 100.0 & 1/1 & 100.0 & \textbf{\color{ReleaseBlue}1} & 0.412 & 8.21 \\
dv-048 & 100.0 & 100.0 & 100.0 & 98.7 & 97.9 & 44.4 & 0.0 & 1/1 & 100.0 & 0 & 0.704 & 12.64 \\
dv-049 & 99.7 & 99.8 & 95.5 & 98.3 & 100.0 & 81.5 & 100.0 & 1/1 & 100.0 & \textbf{\color{ReleaseBlue}1} & 0.420 & 9.82 \\
dv-050 & 100.0 & 80.0 & 99.1 & 100.0 & 100.0 & 100.0 & 100.0 & 1/1 & 100.0 & 0 & 0.718 & 10.33 \\
dv-051 & 100.0 & 98.3 & 100.0 & 93.6 & 99.2 & 100.0 & 100.0 & 1/1 & 100.0 & \textbf{\color{ReleaseBlue}1} & 2.787 & 26.49 \\
dv-052 & 99.9 & 38.9 & 97.6 & 99.9 & 100.0 & 100.0 & 100.0 & 1/1 & 100.0 & 0 & 0.303 & 6.05 \\
dv-053 & 100.0 & 24.7 & 100.0 & 97.2 & 99.0 & 0.0 & 100.0 & 1/1 & 100.0 & 0 & 1.303 & 17.53 \\
dv-054 & 100.0 & 99.8 & 100.0 & 100.0 & 100.0 & 40.0 & 100.0 & 1/1 & 90.0 & 0 & 1.002 & 13.47 \\
dv-055 & 90.0 & 70.8 & 91.6 & 80.7 & 100.0 & 100.0 & 100.0 & 1/1 & 100.0 & 0 & 0.462 & 7.60 \\
dv-056 & 99.8 & 98.3 & 94.0 & 94.9 & 98.0 & 88.9 & 100.0 & 1/1 & 100.0 & \textbf{\color{ReleaseBlue}1} & 0.998 & 14.43 \\
dv-057 & 88.4 & 77.9 & 74.7 & 97.4 & 96.3 & 94.9 & 95.7 & 1/1 & 100.0 & 0 & 1.757 & 19.03 \\
dv-058 & 91.3 & 91.9 & 86.2 & 98.5 & 100.0 & 76.2 & 100.0 & 1/1 & 100.0 & 0 & 1.884 & 24.61 \\
dv-059 & 99.1 & 100.0 & 97.1 & 100.0 & 100.0 & 28.6 & 99.9 & 1/1 & 100.0 & 0 & 0.718 & 10.70 \\
dv-060 & 100.0 & 89.9 & 98.2 & 96.8 & 94.1 & 61.5 & 100.0 & 1/1 & 100.0 & 0 & 1.419 & 14.22 \\
dv-061 & 99.9 & 99.4 & 99.8 & 99.8 & 90.7 & 100.0 & 100.0 & 1/1 & 100.0 & \textbf{\color{ReleaseBlue}1} & 2.414 & 19.67 \\
dv-062 & 98.7 & 65.4 & 69.3 & 81.7 & 100.0 & 81.5 & 98.2 & 1/1 & 80.0 & 0 & 0.704 & 9.26 \\
dv-063 & 100.0 & 100.0 & 99.9 & 99.4 & 100.0 & 66.7 & 100.0 & 1/1 & 90.0 & 0 & 3.375 & 34.85 \\
dv-064 & 99.2 & 96.9 & 98.6 & 76.4 & 84.8 & 0.0 & 100.0 & 1/1 & 90.0 & 0 & 0.997 & 10.23 \\
dv-065 & 95.5 & 73.0 & 75.1 & 90.6 & 100.0 & 49.4 & 100.0 & 1/1 & 80.0 & 0 & 3.164 & 18.70 \\
dv-066 & 98.4 & 98.7 & 96.3 & 59.8 & 86.8 & 66.1 & 100.0 & 1/1 & 80.0 & 0 & 1.094 & 16.54 \\
dv-067 & 80.2 & 95.5 & 81.7 & 99.6 & 95.5 & 72.7 & 100.0 & 1/1 & 90.0 & 0 & 0.881 & 13.14 \\
dv-068 & 48.2 & 78.5 & 33.1 & 42.9 & 100.0 & 72.7 & 100.0 & 1/1 & 30.0 & 0 & 0.696 & 10.15 \\
dv-069 & 100.0 & 80.7 & 100.0 & 82.6 & 100.0 & 65.7 & 100.0 & 1/1 & 100.0 & 0 & 0.569 & 8.75 \\
dv-070 & 75.9 & 57.7 & 56.5 & 38.5 & 66.7 & 76.2 & 45.7 & 1/0 & 0.0 & 0 & 3.372 & 27.39 \\
dv-071 & 99.6 & 95.2 & 97.4 & 88.7 & 59.4 & 100.0 & 100.0 & 1/1 & 100.0 & 0 & 1.079 & 13.09 \\
dv-072 & 100.0 & 100.0 & 100.0 & 100.0 & 100.0 & 100.0 & 47.5 & 1/1 & 100.0 & 0 & 0.626 & 7.79 \\
dv-073 & 96.8 & 84.2 & 92.2 & 89.9 & 100.0 & 94.1 & 100.0 & 1/1 & 100.0 & 0 & 0.598 & 9.74 \\
dv-074 & 100.0 & 100.0 & 100.0 & 92.6 & 98.3 & 100.0 & 24.3 & 1/1 & 100.0 & 0 & 0.675 & 8.61 \\
dv-075 & 98.6 & 94.2 & 90.4 & 87.5 & 100.0 & 82.4 & 0.0 & 1/1 & 80.0 & 0 & 1.284 & 13.84 \\
dv-076 & 72.8 & 80.2 & 56.4 & 76.5 & 100.0 & 66.7 & 100.0 & 1/1 & 100.0 & 0 & 0.327 & 6.02 \\
dv-077 & 100.0 & 91.9 & 99.7 & 86.0 & 96.8 & 0.0 & 0.0 & 1/0 & 0.0 & 0 & 1.432 & 16.68 \\
dv-078 & 70.0 & 76.6 & 60.9 & 45.0 & 93.1 & 93.3 & 100.0 & 1/1 & 100.0 & 0 & 0.387 & 8.24 \\
dv-079 & 92.9 & 77.8 & 85.9 & 76.4 & 86.2 & 82.4 & 100.0 & 1/0 & 0.0 & 0 & 3.630 & 34.46 \\
dv-080 & 95.9 & 88.4 & 92.0 & 88.5 & 100.0 & 73.7 & 0.0 & 1/1 & 80.0 & 0 & 1.739 & 14.56 \\
dv-081 & 37.3 & 34.3 & 31.0 & 11.5 & 100.0 & 99.1 & 100.0 & 1/1 & 100.0 & 0 & 1.693 & 15.45 \\
dv-082 & 94.8 & 100.0 & 91.1 & 99.4 & 98.5 & 58.6 & 100.0 & 1/1 & 100.0 & \textbf{\color{ReleaseBlue}1} & 0.632 & 9.30 \\
dv-083 & 100.0 & 97.0 & 100.0 & 99.6 & 97.5 & 89.7 & 100.0 & 1/1 & 100.0 & \textbf{\color{ReleaseBlue}1} & 0.614 & 11.74 \\
dv-084 & 100.0 & 87.6 & 95.0 & 94.8 & 80.8 & 57.1 & 5.9 & 1/1 & 100.0 & 0 & 3.489 & 27.95 \\
dv-085 & 98.6 & 91.2 & 95.5 & 96.4 & 96.8 & 57.1 & 100.0 & 1/1 & 90.0 & 0 & 0.720 & 10.62 \\
dv-086 & 96.1 & 86.7 & 94.2 & 97.8 & 100.0 & 85.7 & 100.0 & 1/1 & 90.0 & 0 & 0.801 & 10.52 \\
dv-087 & 99.6 & 90.2 & 97.6 & 75.8 & 98.4 & 99.3 & 100.0 & 1/1 & 100.0 & 0 & 1.558 & 30.59 \\
dv-088 & 92.8 & 87.6 & 85.1 & 75.4 & 78.5 & 76.4 & 100.0 & 1/1 & 80.0 & 0 & 7.029 & 44.26 \\
dv-089 & 99.6 & 96.8 & 98.6 & 85.0 & 100.0 & 80.0 & 88.1 & 1/1 & 50.0 & 0 & 2.359 & 24.65 \\
dv-090 & 36.7 & 61.4 & 38.5 & 63.5 & 78.0 & 66.7 & 100.0 & 1/1 & 80.0 & 0 & 1.345 & 7.77 \\
dv-091 & 97.3 & 100.0 & 95.6 & 99.4 & 95.8 & 89.7 & 100.0 & 1/1 & 80.0 & 0 & 0.638 & 10.89 \\
dv-092 & 97.2 & 95.0 & 94.3 & 84.6 & 100.0 & 87.5 & 100.0 & 1/1 & 100.0 & 0 & 0.871 & 12.50 \\
dv-093 & 99.1 & 89.1 & 98.6 & 90.7 & 88.4 & 66.7 & 100.0 & 1/1 & 100.0 & 0 & 0.748 & 13.60 \\
dv-094 & 99.4 & 100.0 & 97.2 & 61.1 & 98.3 & 85.7 & 100.0 & 1/1 & 90.0 & 0 & 0.984 & 13.69 \\
dv-095 & 99.9 & 94.7 & 99.7 & 96.8 & 90.2 & 59.5 & 7.2 & 1/1 & 70.0 & 0 & 0.822 & 12.84 \\
\end{longtable}\endgroup

\clearpage
\subsection{C10: OpenCode + GLM 5.3}
\noindent\footnotesize Percentages: Ln/Cn/Br/Tg = line/condition/branch/toggle coverage; F1/P1 = feature/test-plan F1; Ref = reference-model point accuracy; Kill = mutant kill rate. C/G and O are binary compile/golden and overall results. Tok = input plus output tokens in millions; Min = elapsed minutes. Overall uses exact model correctness and all four coverage thresholds. \par\normalsize
\begingroup
\fontsize{8.2}{10.2}\selectfont
\setlength{\tabcolsep}{3pt}\renewcommand{\arraystretch}{1.04}
\setlength{\LTleft}{0pt}\setlength{\LTright}{0pt}
\Needspace{10\baselineskip}
\captionsetup{hypcap=false}
\captionof{table}{Per-task measurements for C10. The companion CSV retains unrounded values, precision/recall, exact-model flags, input/output token splits, and source-cell coordinates.}\label{tab:app-detail-C10}
\addtocounter{table}{-1}
\begin{longtable}{>{\raggedright\arraybackslash}p{\dimexpr 0.1050000\textwidth-2\tabcolsep\relax}>{\raggedleft\arraybackslash}p{\dimexpr 0.0745833\textwidth-2\tabcolsep\relax}>{\raggedleft\arraybackslash}p{\dimexpr 0.0745833\textwidth-2\tabcolsep\relax}>{\raggedleft\arraybackslash}p{\dimexpr 0.0745833\textwidth-2\tabcolsep\relax}>{\raggedleft\arraybackslash}p{\dimexpr 0.0745833\textwidth-2\tabcolsep\relax}>{\raggedleft\arraybackslash}p{\dimexpr 0.0745833\textwidth-2\tabcolsep\relax}>{\raggedleft\arraybackslash}p{\dimexpr 0.0745833\textwidth-2\tabcolsep\relax}>{\raggedleft\arraybackslash}p{\dimexpr 0.0745833\textwidth-2\tabcolsep\relax}>{\raggedleft\arraybackslash}p{\dimexpr 0.0745833\textwidth-2\tabcolsep\relax}>{\raggedleft\arraybackslash}p{\dimexpr 0.0745833\textwidth-2\tabcolsep\relax}>{\raggedleft\arraybackslash}p{\dimexpr 0.0745833\textwidth-2\tabcolsep\relax}>{\raggedleft\arraybackslash}p{\dimexpr 0.0745833\textwidth-2\tabcolsep\relax}>{\raggedleft\arraybackslash}p{\dimexpr 0.0745833\textwidth-2\tabcolsep\relax}}
\toprule
\rowcolor{ReleasePale}\textbf{Task} & \textbf{Ln} & \textbf{Cn} & \textbf{Br} & \textbf{Tg} & \textbf{F1} & \textbf{P1} & \textbf{Ref} & \textbf{C/G} & \textbf{Kill} & \textbf{O} & \textbf{Tok} & \textbf{Min} \\
\midrule\endfirsthead
\multicolumn{13}{l}{\small\color{ReleaseMuted}Table~\ref{tab:app-detail-C10} (continued)}\\
\toprule
\rowcolor{ReleasePale}\textbf{Task} & \textbf{Ln} & \textbf{Cn} & \textbf{Br} & \textbf{Tg} & \textbf{F1} & \textbf{P1} & \textbf{Ref} & \textbf{C/G} & \textbf{Kill} & \textbf{O} & \textbf{Tok} & \textbf{Min} \\
\midrule\endhead
\midrule
\multicolumn{13}{r}{\footnotesize\color{ReleaseMuted}Continued on next page}\\\endfoot
\bottomrule\endlastfoot
dv-001 & 100.0 & 91.6 & 100.0 & 96.2 & 86.5 & 80.0 & 100.0 & 1/1 & 100.0 & \textbf{\color{ReleaseBlue}1} & 3.646 & 140.15 \\
dv-002 & 100.0 & 100.0 & 100.0 & 97.5 & 95.7 & 58.6 & 100.0 & 1/1 & 100.0 & \textbf{\color{ReleaseBlue}1} & 1.950 & 287.71 \\
dv-003 & 100.0 & 100.0 & 100.0 & 100.0 & 0.0 & 76.0 & 100.0 & 1/1 & 100.0 & \textbf{\color{ReleaseBlue}1} & 3.175 & 130.58 \\
dv-004 & 0.0 & 0.0 & 0.0 & 0.0 & 80.0 & 0.0 & 100.0 & 0/0 & 0.0 & 0 & 8.544 & 563.88 \\
dv-005 & 100.0 & 99.0 & 100.0 & 98.9 & 72.9 & 0.0 & 100.0 & 1/1 & 100.0 & \textbf{\color{ReleaseBlue}1} & 1.431 & 124.77 \\
dv-006 & 0.0 & 0.0 & 0.0 & 0.0 & 0.0 & 0.0 & 0.0 & 0/0 & 0.0 & 0 & 0.022 & 6.47 \\
dv-007 & 100.0 & 100.0 & 100.0 & 100.0 & 100.0 & 22.1 & 100.0 & 1/1 & 80.0 & 0 & 7.269 & 276.37 \\
dv-008 & 0.0 & 0.0 & 0.0 & 0.0 & 0.0 & 0.0 & 0.0 & 0/0 & 0.0 & 0 & 0.640 & 815.88 \\
dv-009 & 100.0 & 91.3 & 100.0 & 100.0 & 100.0 & 0.0 & 100.0 & 1/1 & 30.0 & 0 & 1.623 & 192.14 \\
dv-010 & 100.0 & 100.0 & 100.0 & 99.3 & 86.3 & 65.4 & 100.0 & 1/1 & 100.0 & \textbf{\color{ReleaseBlue}1} & 1.935 & 123.42 \\
dv-011 & 98.4 & 93.5 & 91.3 & 52.9 & 73.7 & 87.5 & 100.0 & 1/1 & 100.0 & 0 & 1.541 & 194.35 \\
dv-012 & 0.0 & 0.0 & 0.0 & 0.0 & 0.0 & 0.0 & 0.0 & 0/0 & 0.0 & 0 & 0.014 & 2.07 \\
dv-013 & 0.0 & 0.0 & 0.0 & 0.0 & 0.0 & 0.0 & 0.0 & 0/0 & 0.0 & 0 & 0.006 & 1.81 \\
dv-014 & 0.0 & 0.0 & 0.0 & 0.0 & 0.0 & 0.0 & 0.0 & 0/0 & 0.0 & 0 & 0.172 & 716.00 \\
dv-015 & 0.0 & 0.0 & 0.0 & 0.0 & 0.0 & 0.0 & 0.0 & 0/0 & 0.0 & 0 & 2.031 & 1269.92 \\
dv-016 & 0.0 & 0.0 & 0.0 & 0.0 & 0.0 & 0.0 & 0.0 & 0/0 & 0.0 & 0 & 0.467 & 768.22 \\
dv-017 & 0.0 & 0.0 & 0.0 & 0.0 & 0.0 & 0.0 & 0.0 & 0/0 & 0.0 & 0 & 0.596 & 841.68 \\
dv-018 & 0.0 & 0.0 & 0.0 & 0.0 & 73.7 & 0.0 & 100.0 & 0/0 & 0.0 & 0 & 4.641 & 408.91 \\
dv-019 & 0.0 & 0.0 & 0.0 & 0.0 & 0.0 & 0.0 & 0.0 & 0/0 & 0.0 & 0 & 0.190 & 715.10 \\
dv-020 & 0.0 & 0.0 & 0.0 & 0.0 & 0.0 & 0.0 & 0.0 & 0/0 & 0.0 & 0 & 0.186 & 714.21 \\
dv-021 & 0.0 & 0.0 & 0.0 & 0.0 & 0.0 & 0.0 & 0.0 & 0/0 & 0.0 & 0 & 0.990 & 778.19 \\
dv-022 & 0.0 & 0.0 & 0.0 & 0.0 & 0.0 & 0.0 & 0.0 & 0/0 & 0.0 & 0 & 0.591 & 813.65 \\
dv-023 & 0.0 & 0.0 & 0.0 & 0.0 & 0.0 & 0.0 & 0.0 & 0/0 & 0.0 & 0 & 0.255 & 719.57 \\
dv-024 & 0.0 & 0.0 & 0.0 & 0.0 & 0.0 & 0.0 & 0.0 & 0/0 & 0.0 & 0 & 0.464 & 785.86 \\
dv-025 & 0.0 & 0.0 & 0.0 & 0.0 & 0.0 & 0.0 & 100.0 & 0/0 & 0.0 & 0 & 0.699 & 809.95 \\
dv-026 & 0.0 & 0.0 & 0.0 & 0.0 & 0.0 & 0.0 & 0.0 & 0/0 & 0.0 & 0 & 0.230 & 661.77 \\
dv-027 & 0.0 & 0.0 & 0.0 & 0.0 & 0.0 & 0.0 & 0.0 & 0/0 & 0.0 & 0 & 0.284 & 665.43 \\
dv-028 & 0.0 & 0.0 & 0.0 & 0.0 & 0.0 & 0.0 & 0.0 & 0/0 & 0.0 & 0 & 1.745 & 910.43 \\
dv-029 & 0.0 & 0.0 & 0.0 & 0.0 & 0.0 & 0.0 & 0.0 & 0/0 & 0.0 & 0 & 0.530 & 719.81 \\
dv-030 & 0.0 & 0.0 & 0.0 & 0.0 & 0.0 & 0.0 & 0.0 & 0/0 & 0.0 & 0 & 0.584 & 756.96 \\
dv-031 & 0.0 & 0.0 & 0.0 & 0.0 & 100.0 & 31.4 & 99.7 & 0/0 & 0.0 & 0 & 1.458 & 899.57 \\
dv-032 & 0.0 & 0.0 & 0.0 & 0.0 & 0.0 & 0.0 & 0.0 & 0/0 & 0.0 & 0 & 0.301 & 637.59 \\
dv-033 & 0.0 & 0.0 & 0.0 & 0.0 & 0.0 & 0.0 & 0.0 & 0/0 & 0.0 & 0 & 0.877 & 797.92 \\
dv-034 & 0.0 & 0.0 & 0.0 & 0.0 & 0.0 & 0.0 & 0.0 & 0/0 & 0.0 & 0 & 0.709 & 635.98 \\
dv-035 & 0.0 & 0.0 & 0.0 & 0.0 & 0.0 & 0.0 & 0.0 & 0/0 & 0.0 & 0 & 0.157 & 621.00 \\
dv-036 & 70.6 & 72.7 & 70.1 & 57.8 & 85.0 & 0.0 & 0.0 & 1/0 & 0.0 & 0 & 11.284 & 1838.45 \\
dv-037 & 0.0 & 0.0 & 0.0 & 0.0 & 0.0 & 0.0 & 0.0 & 0/0 & 0.0 & 0 & 0.238 & 617.53 \\
dv-038 & 100.0 & 97.7 & 100.0 & 99.6 & 91.6 & 0.0 & 100.0 & 1/1 & 90.0 & 0 & 6.415 & 131.67 \\
dv-039 & 0.0 & 0.0 & 0.0 & 0.0 & 0.0 & 0.0 & 21.4 & 0/0 & 0.0 & 0 & 0.438 & 693.23 \\
dv-040 & 0.0 & 0.0 & 0.0 & 0.0 & 0.0 & 0.0 & 0.0 & 0/0 & 0.0 & 0 & 0.181 & 611.51 \\
dv-041 & 99.3 & 95.1 & 85.0 & 99.7 & 95.7 & 0.0 & 100.0 & 1/1 & 90.0 & 0 & 3.260 & 489.06 \\
dv-042 & 0.0 & 0.0 & 0.0 & 0.0 & 0.0 & 0.0 & 0.0 & 0/0 & 0.0 & 0 & 0.467 & 719.85 \\
dv-043 & 0.0 & 0.0 & 0.0 & 0.0 & 0.0 & 0.0 & 0.0 & 0/0 & 0.0 & 0 & 0.440 & 600.01 \\
dv-044 & 0.0 & 0.0 & 0.0 & 0.0 & 0.0 & 0.0 & 0.0 & 0/0 & 0.0 & 0 & 0.202 & 55.66 \\
dv-045 & 0.0 & 0.0 & 0.0 & 0.0 & 0.0 & 0.0 & 0.0 & 0/0 & 0.0 & 0 & 2.260 & 132.25 \\
dv-046 & 0.0 & 0.0 & 0.0 & 0.0 & 0.0 & 0.0 & 0.0 & 0/0 & 0.0 & 0 & 0.428 & 679.20 \\
dv-047 & 100.0 & 100.0 & 100.0 & 100.0 & 57.5 & 15.3 & 0.0 & 1/1 & 100.0 & 0 & 5.874 & 744.52 \\
dv-048 & 0.0 & 0.0 & 0.0 & 0.0 & 0.0 & 0.0 & 0.0 & 0/0 & 0.0 & 0 & 0.199 & 592.96 \\
dv-049 & 0.0 & 0.0 & 0.0 & 0.0 & 0.0 & 0.0 & 0.0 & 0/0 & 0.0 & 0 & 0.175 & 586.15 \\
dv-050 & 0.0 & 0.0 & 0.0 & 0.0 & 0.0 & 0.0 & 0.0 & 0/0 & 0.0 & 0 & 1.059 & 399.89 \\
dv-051 & 0.0 & 0.0 & 0.0 & 0.0 & 0.0 & 0.0 & 0.0 & 0/0 & 0.0 & 0 & 0.597 & 712.41 \\
dv-052 & 0.0 & 0.0 & 0.0 & 0.0 & 0.0 & 0.0 & 0.9 & 0/0 & 0.0 & 0 & 10.236 & 323.02 \\
dv-053 & 0.0 & 0.0 & 0.0 & 0.0 & 0.0 & 0.0 & 0.0 & 0/0 & 0.0 & 0 & 0.380 & 638.44 \\
dv-054 & 0.0 & 0.0 & 0.0 & 0.0 & 0.0 & 0.0 & 0.0 & 0/0 & 0.0 & 0 & 1.112 & 637.51 \\
dv-055 & 0.0 & 0.0 & 0.0 & 0.0 & 0.0 & 0.0 & 0.0 & 0/0 & 0.0 & 0 & 0.829 & 667.86 \\
dv-056 & 0.0 & 0.0 & 0.0 & 0.0 & 0.0 & 0.0 & 0.0 & 0/0 & 0.0 & 0 & 9.080 & 929.70 \\
dv-057 & 0.0 & 0.0 & 0.0 & 0.0 & 0.0 & 0.0 & 0.0 & 0/0 & 0.0 & 0 & 0.748 & 762.28 \\
dv-058 & 0.0 & 0.0 & 0.0 & 0.0 & 0.0 & 0.0 & 0.0 & 0/0 & 0.0 & 0 & 0.581 & 667.60 \\
dv-059 & 0.0 & 0.0 & 0.0 & 0.0 & 0.0 & 0.0 & 0.0 & 0/0 & 0.0 & 0 & 0.910 & 709.08 \\
dv-060 & 0.0 & 0.0 & 0.0 & 0.0 & 0.0 & 0.0 & 0.0 & 0/0 & 0.0 & 0 & 0.767 & 611.20 \\
dv-061 & 0.0 & 0.0 & 0.0 & 0.0 & 0.0 & 0.0 & 0.0 & 0/0 & 0.0 & 0 & 0.371 & 590.96 \\
dv-062 & 0.0 & 0.0 & 0.0 & 0.0 & 0.0 & 0.0 & 0.0 & 0/0 & 0.0 & 0 & 0.980 & 632.89 \\
dv-063 & 0.0 & 0.0 & 0.0 & 0.0 & 0.0 & 0.0 & 0.0 & 0/0 & 0.0 & 0 & 3.095 & 612.20 \\
dv-064 & 0.0 & 0.0 & 0.0 & 0.0 & 0.0 & 0.0 & 0.0 & 0/0 & 0.0 & 0 & 0.193 & 534.36 \\
dv-065 & 0.0 & 0.0 & 0.0 & 0.0 & 0.0 & 0.0 & 0.0 & 0/0 & 0.0 & 0 & 0.194 & 532.72 \\
dv-066 & 0.0 & 0.0 & 0.0 & 0.0 & 0.0 & 0.0 & 0.0 & 0/0 & 0.0 & 0 & 0.319 & 532.86 \\
dv-067 & 0.0 & 0.0 & 0.0 & 0.0 & 0.0 & 0.0 & 0.0 & 0/0 & 0.0 & 0 & 0.184 & 517.58 \\
dv-068 & 0.0 & 0.0 & 0.0 & 0.0 & 0.0 & 0.0 & 0.0 & 0/0 & 0.0 & 0 & 0.255 & 525.66 \\
dv-069 & 0.0 & 0.0 & 0.0 & 0.0 & 0.0 & 0.0 & 0.0 & 0/0 & 0.0 & 0 & 0.234 & 552.49 \\
dv-070 & 0.0 & 0.0 & 0.0 & 0.0 & 0.0 & 0.0 & 0.0 & 0/0 & 0.0 & 0 & 0.193 & 501.82 \\
dv-071 & 0.0 & 0.0 & 0.0 & 0.0 & 0.0 & 0.0 & 0.0 & 0/0 & 0.0 & 0 & 0.808 & 653.54 \\
dv-072 & 100.0 & 100.0 & 100.0 & 100.0 & 94.7 & 18.1 & 41.4 & 1/1 & 100.0 & 0 & 3.789 & 302.67 \\
dv-073 & 0.0 & 0.0 & 0.0 & 0.0 & 31.7 & 0.0 & 90.2 & 0/0 & 0.0 & 0 & 0.859 & 268.25 \\
dv-074 & 0.0 & 0.0 & 0.0 & 0.0 & 0.0 & 0.0 & 0.0 & 0/0 & 0.0 & 0 & 0.848 & 675.95 \\
dv-075 & 0.0 & 0.0 & 0.0 & 0.0 & 0.0 & 0.0 & 0.0 & 0/0 & 0.0 & 0 & 0.789 & 645.88 \\
dv-076 & 0.0 & 0.0 & 0.0 & 0.0 & 0.0 & 0.0 & 0.0 & 0/0 & 0.0 & 0 & 0.764 & 723.01 \\
dv-077 & 0.0 & 0.0 & 0.0 & 0.0 & 0.0 & 0.0 & 0.0 & 0/0 & 0.0 & 0 & 0.226 & 573.92 \\
dv-078 & 0.0 & 0.0 & 0.0 & 0.0 & 48.3 & 21.4 & 0.0 & 0/0 & 0.0 & 0 & 3.184 & 72.12 \\
dv-079 & 0.0 & 0.0 & 0.0 & 0.0 & 0.0 & 0.0 & 28.1 & 0/0 & 0.0 & 0 & 0.783 & 846.44 \\
dv-080 & 0.0 & 0.0 & 0.0 & 0.0 & 0.0 & 0.0 & 0.0 & 0/0 & 0.0 & 0 & 0.744 & 715.59 \\
dv-081 & 0.0 & 0.0 & 0.0 & 0.0 & 0.0 & 0.0 & 21.4 & 1/0 & 0.0 & 0 & 5.230 & 242.06 \\
dv-082 & 0.0 & 0.0 & 0.0 & 0.0 & 0.0 & 0.0 & 0.0 & 0/0 & 0.0 & 0 & 0.346 & 584.04 \\
dv-083 & 0.0 & 0.0 & 0.0 & 0.0 & 0.0 & 0.0 & 0.0 & 0/0 & 0.0 & 0 & 0.376 & 581.59 \\
dv-084 & 0.0 & 0.0 & 0.0 & 0.0 & 0.0 & 0.0 & 0.0 & 0/0 & 0.0 & 0 & 0.415 & 539.91 \\
dv-085 & 0.0 & 0.0 & 0.0 & 0.0 & 0.0 & 0.0 & 0.0 & 0/0 & 0.0 & 0 & 0.191 & 515.18 \\
dv-086 & 0.0 & 0.0 & 0.0 & 0.0 & 0.0 & 0.0 & 0.0 & 0/0 & 0.0 & 0 & 0.239 & 515.97 \\
dv-087 & 0.0 & 0.0 & 0.0 & 0.0 & 0.0 & 0.0 & 0.0 & 0/0 & 0.0 & 0 & 0.386 & 571.43 \\
dv-088 & 0.0 & 0.0 & 0.0 & 0.0 & 0.0 & 0.0 & 0.0 & 0/0 & 0.0 & 0 & 0.210 & 509.93 \\
dv-089 & 0.0 & 0.0 & 0.0 & 0.0 & 0.0 & 0.0 & 0.0 & 0/0 & 0.0 & 0 & 0.361 & 557.41 \\
dv-090 & 0.0 & 0.0 & 0.0 & 0.0 & 0.0 & 0.0 & 100.0 & 0/0 & 0.0 & 0 & 0.470 & 156.83 \\
dv-091 & 0.0 & 0.0 & 0.0 & 0.0 & 0.0 & 0.0 & 0.0 & 0/0 & 0.0 & 0 & 0.214 & 503.42 \\
dv-092 & 0.0 & 0.0 & 0.0 & 0.0 & 84.3 & 0.0 & 100.0 & 0/0 & 0.0 & 0 & 1.951 & 932.31 \\
dv-093 & 0.0 & 0.0 & 0.0 & 0.0 & 0.0 & 0.0 & 88.7 & 0/0 & 0.0 & 0 & 0.618 & 503.30 \\
dv-094 & 0.0 & 0.0 & 0.0 & 0.0 & 0.0 & 0.0 & 0.0 & 0/0 & 0.0 & 0 & 0.290 & 504.47 \\
dv-095 & 0.0 & 0.0 & 0.0 & 0.0 & 0.0 & 0.0 & 0.0 & 0/0 & 0.0 & 0 & 0.326 & 541.02 \\
\end{longtable}\endgroup

\clearpage
\subsection{C11: OpenCode + DeepSeek V4.1 Flash 1017}
\noindent\footnotesize Percentages: Ln/Cn/Br/Tg = line/condition/branch/toggle coverage; F1/P1 = feature/test-plan F1; Ref = reference-model point accuracy; Kill = mutant kill rate. C/G and O are binary compile/golden and overall results. Tok = input plus output tokens in millions; Min = elapsed minutes. Overall uses exact model correctness and all four coverage thresholds. \par\normalsize
\begingroup
\fontsize{8.2}{10.2}\selectfont
\setlength{\tabcolsep}{3pt}\renewcommand{\arraystretch}{1.04}
\setlength{\LTleft}{0pt}\setlength{\LTright}{0pt}
\Needspace{10\baselineskip}
\captionsetup{hypcap=false}
\captionof{table}{Per-task measurements for C11. The companion CSV retains unrounded values, precision/recall, exact-model flags, input/output token splits, and source-cell coordinates.}\label{tab:app-detail-C11}
\addtocounter{table}{-1}
\begin{longtable}{>{\raggedright\arraybackslash}p{\dimexpr 0.1050000\textwidth-2\tabcolsep\relax}>{\raggedleft\arraybackslash}p{\dimexpr 0.0745833\textwidth-2\tabcolsep\relax}>{\raggedleft\arraybackslash}p{\dimexpr 0.0745833\textwidth-2\tabcolsep\relax}>{\raggedleft\arraybackslash}p{\dimexpr 0.0745833\textwidth-2\tabcolsep\relax}>{\raggedleft\arraybackslash}p{\dimexpr 0.0745833\textwidth-2\tabcolsep\relax}>{\raggedleft\arraybackslash}p{\dimexpr 0.0745833\textwidth-2\tabcolsep\relax}>{\raggedleft\arraybackslash}p{\dimexpr 0.0745833\textwidth-2\tabcolsep\relax}>{\raggedleft\arraybackslash}p{\dimexpr 0.0745833\textwidth-2\tabcolsep\relax}>{\raggedleft\arraybackslash}p{\dimexpr 0.0745833\textwidth-2\tabcolsep\relax}>{\raggedleft\arraybackslash}p{\dimexpr 0.0745833\textwidth-2\tabcolsep\relax}>{\raggedleft\arraybackslash}p{\dimexpr 0.0745833\textwidth-2\tabcolsep\relax}>{\raggedleft\arraybackslash}p{\dimexpr 0.0745833\textwidth-2\tabcolsep\relax}>{\raggedleft\arraybackslash}p{\dimexpr 0.0745833\textwidth-2\tabcolsep\relax}}
\toprule
\rowcolor{ReleasePale}\textbf{Task} & \textbf{Ln} & \textbf{Cn} & \textbf{Br} & \textbf{Tg} & \textbf{F1} & \textbf{P1} & \textbf{Ref} & \textbf{C/G} & \textbf{Kill} & \textbf{O} & \textbf{Tok} & \textbf{Min} \\
\midrule\endfirsthead
\multicolumn{13}{l}{\small\color{ReleaseMuted}Table~\ref{tab:app-detail-C11} (continued)}\\
\toprule
\rowcolor{ReleasePale}\textbf{Task} & \textbf{Ln} & \textbf{Cn} & \textbf{Br} & \textbf{Tg} & \textbf{F1} & \textbf{P1} & \textbf{Ref} & \textbf{C/G} & \textbf{Kill} & \textbf{O} & \textbf{Tok} & \textbf{Min} \\
\midrule\endhead
\midrule
\multicolumn{13}{r}{\footnotesize\color{ReleaseMuted}Continued on next page}\\\endfoot
\bottomrule\endlastfoot
dv-001 & 100.0 & 92.8 & 100.0 & 96.0 & 100.0 & 84.0 & 100.0 & 1/1 & 100.0 & \textbf{\color{ReleaseBlue}1} & 0.657 & 4.72 \\
dv-002 & 100.0 & 100.0 & 100.0 & 97.5 & 100.0 & 64.7 & 100.0 & 1/1 & 100.0 & \textbf{\color{ReleaseBlue}1} & 1.689 & 7.72 \\
dv-003 & 100.0 & 100.0 & 100.0 & 100.0 & 77.3 & 88.7 & 100.0 & 1/1 & 100.0 & \textbf{\color{ReleaseBlue}1} & 5.898 & 13.00 \\
dv-004 & 98.6 & 92.0 & 94.7 & 79.6 & 100.0 & 0.0 & 100.0 & 1/1 & 100.0 & 0 & 3.945 & 11.95 \\
dv-005 & 100.0 & 99.3 & 100.0 & 98.9 & 82.4 & 39.0 & 100.0 & 1/1 & 100.0 & \textbf{\color{ReleaseBlue}1} & 3.174 & 26.33 \\
dv-006 & 99.5 & 99.8 & 99.8 & 92.8 & 94.1 & 77.1 & 100.0 & 1/1 & 100.0 & \textbf{\color{ReleaseBlue}1} & 11.024 & 17.46 \\
dv-007 & 100.0 & 100.0 & 100.0 & 100.0 & 74.3 & 0.0 & 100.0 & 1/1 & 90.0 & 0 & 4.490 & 14.05 \\
dv-008 & 98.3 & 91.8 & 96.6 & 97.5 & 67.9 & 22.1 & 94.1 & 1/1 & 100.0 & 0 & 19.963 & 56.99 \\
dv-009 & 100.0 & 91.3 & 100.0 & 100.0 & 93.8 & 53.6 & 100.0 & 1/1 & 30.0 & 0 & 3.423 & 11.79 \\
dv-010 & 100.0 & 100.0 & 100.0 & 99.3 & 86.2 & 80.5 & 100.0 & 1/1 & 100.0 & \textbf{\color{ReleaseBlue}1} & 2.372 & 8.96 \\
dv-011 & 98.4 & 93.5 & 91.3 & 52.9 & 100.0 & 85.1 & 100.0 & 1/1 & 100.0 & 0 & 1.766 & 7.09 \\
dv-012 & 100.0 & 100.0 & 100.0 & 100.0 & 90.8 & 71.1 & 100.0 & 1/1 & 100.0 & \textbf{\color{ReleaseBlue}1} & 5.518 & 12.11 \\
dv-013 & 100.0 & 100.0 & 100.0 & 94.3 & 100.0 & 64.3 & 100.0 & 1/1 & 100.0 & \textbf{\color{ReleaseBlue}1} & 15.026 & 25.34 \\
dv-014 & 87.0 & 79.6 & 84.3 & 66.7 & 84.5 & 19.7 & 100.0 & 1/1 & 100.0 & 0 & 22.815 & 52.20 \\
dv-015 & 100.0 & 96.6 & 97.3 & 89.4 & 66.7 & 39.3 & 99.9 & 1/1 & 90.0 & 0 & 14.535 & 26.52 \\
dv-016 & 99.3 & 97.8 & 98.9 & 55.7 & 100.0 & 93.7 & 100.0 & 1/1 & 70.0 & 0 & 8.278 & 18.41 \\
dv-017 & 96.8 & 79.5 & 88.4 & 90.6 & 90.0 & 98.2 & 91.9 & 1/1 & 100.0 & 0 & 4.602 & 13.96 \\
dv-018 & 97.4 & 100.0 & 95.7 & 94.8 & 94.7 & 56.5 & 100.0 & 1/1 & 100.0 & \textbf{\color{ReleaseBlue}1} & 7.063 & 21.47 \\
dv-019 & 97.0 & 85.5 & 95.8 & 99.0 & 86.5 & 0.0 & 100.0 & 1/1 & 80.0 & 0 & 8.719 & 22.15 \\
dv-020 & 97.6 & 93.8 & 88.2 & 95.3 & 70.4 & 65.0 & 20.1 & 1/1 & 90.0 & 0 & 17.265 & 40.29 \\
dv-021 & 93.5 & 80.0 & 88.6 & 88.4 & 83.1 & 21.2 & 100.0 & 1/1 & 100.0 & 0 & 12.617 & 78.97 \\
dv-022 & 100.0 & 97.5 & 98.1 & 69.9 & 63.4 & 15.3 & 6.2 & 1/1 & 90.0 & 0 & 18.476 & 41.82 \\
dv-023 & 100.0 & 100.0 & 100.0 & 100.0 & 93.5 & 28.4 & 100.0 & 1/1 & 100.0 & \textbf{\color{ReleaseBlue}1} & 5.152 & 19.43 \\
dv-024 & 43.6 & 59.8 & 27.4 & 35.1 & 72.7 & 83.7 & 99.8 & 1/1 & 60.0 & 0 & 20.120 & 37.29 \\
dv-025 & 40.7 & 54.3 & 23.9 & 10.9 & 86.9 & 31.6 & 100.0 & 1/1 & 20.0 & 0 & 4.909 & 17.84 \\
dv-026 & 42.1 & 58.5 & 24.9 & 18.0 & 96.8 & 71.6 & 100.0 & 1/1 & 80.0 & 0 & 5.556 & 18.06 \\
dv-027 & 97.5 & 87.2 & 95.6 & 57.2 & 86.4 & 0.0 & 80.1 & 1/1 & 50.0 & 0 & 48.384 & 60.94 \\
dv-028 & 100.0 & 98.4 & 100.0 & 98.7 & 87.5 & 65.6 & 100.0 & 1/1 & 100.0 & \textbf{\color{ReleaseBlue}1} & 4.204 & 15.18 \\
dv-029 & 99.4 & 95.5 & 94.3 & 51.5 & 41.6 & 42.8 & 72.8 & 1/1 & 70.0 & 0 & 38.464 & 45.26 \\
dv-030 & 68.7 & 48.7 & 62.1 & 82.2 & 67.9 & 24.5 & 100.0 & 1/0 & 0.0 & 0 & 12.452 & 229.47 \\
dv-031 & 62.8 & 4.0 & 38.8 & 31.8 & 100.0 & 46.9 & 99.9 & 1/1 & 70.0 & 0 & 13.704 & 25.27 \\
dv-032 & 97.1 & 84.1 & 95.6 & 95.0 & 76.9 & 65.8 & 55.8 & 1/1 & 60.0 & 0 & 16.228 & 39.20 \\
dv-033 & 78.3 & 74.8 & 71.7 & 49.2 & 79.7 & 15.3 & 0.3 & 1/1 & 90.0 & 0 & 18.967 & 34.24 \\
dv-034 & 79.8 & 94.0 & 74.8 & 39.8 & 95.7 & 57.1 & 100.0 & 1/1 & 100.0 & 0 & 2.981 & 13.66 \\
dv-035 & 79.0 & 53.3 & 62.9 & 95.6 & 82.0 & 0.0 & 65.1 & 1/1 & 90.0 & 0 & 7.991 & 22.04 \\
dv-036 & 98.8 & 89.3 & 94.9 & 70.1 & 56.6 & 0.0 & 20.2 & 1/1 & 70.0 & 0 & 18.405 & 82.78 \\
dv-037 & 99.7 & 98.9 & 98.9 & 78.1 & 57.3 & 0.0 & 79.2 & 1/1 & 70.0 & 0 & 40.106 & 63.62 \\
dv-038 & 100.0 & 97.7 & 100.0 & 99.6 & 100.0 & 0.0 & 100.0 & 1/1 & 90.0 & 0 & 6.069 & 21.01 \\
dv-039 & 99.6 & 97.1 & 99.6 & 96.2 & 92.3 & 44.4 & 100.0 & 1/1 & 80.0 & 0 & 3.054 & 13.69 \\
dv-040 & 100.0 & 97.8 & 99.2 & 68.5 & 100.0 & 0.0 & 100.0 & 1/1 & 80.0 & 0 & 1.275 & 11.76 \\
dv-041 & 99.3 & 96.8 & 85.0 & 99.7 & 100.0 & 0.0 & 99.8 & 1/1 & 90.0 & 0 & 6.484 & 18.13 \\
dv-042 & 100.0 & 100.0 & 100.0 & 72.3 & 100.0 & 0.0 & 100.0 & 1/1 & 90.0 & 0 & 5.194 & 22.67 \\
dv-043 & 91.1 & 100.0 & 85.2 & 100.0 & 100.0 & 0.0 & 97.5 & 1/1 & 100.0 & 0 & 2.084 & 10.42 \\
dv-044 & 98.6 & 81.6 & 98.7 & 94.4 & 94.1 & 60.1 & 100.0 & 1/1 & 100.0 & 0 & 9.022 & 30.69 \\
dv-045 & 100.0 & 89.3 & 95.1 & 91.7 & 90.9 & 64.4 & 100.0 & 1/1 & 80.0 & 0 & 4.967 & 23.66 \\
dv-046 & 98.1 & 97.5 & 96.4 & 75.8 & 91.9 & 54.5 & 100.0 & 1/1 & 100.0 & 0 & 16.274 & 46.69 \\
dv-047 & 100.0 & 100.0 & 100.0 & 100.0 & 72.6 & 15.3 & 93.7 & 1/1 & 100.0 & 0 & 4.009 & 19.28 \\
dv-048 & 100.0 & 100.0 & 100.0 & 98.7 & 93.8 & 0.0 & 0.0 & 1/1 & 100.0 & 0 & 12.048 & 30.60 \\
dv-049 & 99.7 & 99.8 & 95.5 & 98.3 & 91.1 & 46.5 & 100.0 & 1/1 & 90.0 & 0 & 5.835 & 19.54 \\
dv-050 & 100.0 & 80.0 & 99.1 & 100.0 & 96.3 & 43.9 & 100.0 & 1/1 & 70.0 & 0 & 1.569 & 14.93 \\
dv-051 & 96.3 & 98.3 & 91.5 & 93.6 & 91.5 & 30.8 & 96.6 & 1/1 & 100.0 & 0 & 16.863 & 33.43 \\
dv-052 & 99.8 & 38.9 & 97.5 & 99.9 & 64.6 & 22.2 & 100.0 & 1/1 & 100.0 & 0 & 4.015 & 16.60 \\
dv-053 & 100.0 & 24.7 & 100.0 & 82.6 & 100.0 & 0.0 & 94.8 & 1/1 & 100.0 & 0 & 8.572 & 29.85 \\
dv-054 & 100.0 & 99.4 & 100.0 & 100.0 & 100.0 & 0.0 & 100.0 & 1/1 & 90.0 & 0 & 9.400 & 35.50 \\
dv-055 & 89.6 & 63.9 & 91.3 & 77.5 & 85.7 & 66.7 & 100.0 & 1/1 & 100.0 & 0 & 2.194 & 12.61 \\
dv-056 & 99.8 & 93.8 & 93.5 & 90.6 & 78.6 & 63.0 & 92.3 & 1/1 & 100.0 & 0 & 19.896 & 51.18 \\
dv-057 & 88.4 & 77.9 & 74.7 & 97.4 & 100.0 & 77.6 & 95.7 & 1/1 & 100.0 & 0 & 4.486 & 30.35 \\
dv-058 & 89.6 & 90.5 & 82.7 & 98.2 & 78.4 & 46.2 & 100.0 & 1/1 & 100.0 & 0 & 18.041 & 66.31 \\
dv-059 & 99.1 & 100.0 & 97.1 & 100.0 & 95.7 & 0.0 & 4.8 & 1/1 & 100.0 & 0 & 15.233 & 33.42 \\
dv-060 & 100.0 & 89.9 & 98.2 & 96.1 & 94.1 & 36.0 & 100.0 & 1/1 & 100.0 & 0 & 4.300 & 16.56 \\
dv-061 & 86.1 & 96.1 & 79.9 & 99.6 & 11.7 & 59.3 & 84.1 & 1/1 & 90.0 & 0 & 52.323 & 82.88 \\
dv-062 & 98.7 & 64.1 & 69.3 & 81.6 & 100.0 & 64.1 & 98.2 & 1/1 & 80.0 & 0 & 6.681 & 22.29 \\
dv-063 & 76.4 & 100.0 & 76.0 & 99.4 & 95.0 & 0.0 & 100.0 & 1/1 & 70.0 & 0 & 15.112 & 44.83 \\
dv-064 & 98.8 & 94.3 & 97.9 & 76.0 & 80.2 & 9.4 & 100.0 & 1/1 & 90.0 & 0 & 5.527 & 28.16 \\
dv-065 & 95.5 & 72.1 & 75.1 & 90.4 & 92.2 & 0.0 & 100.0 & 1/1 & 70.0 & 0 & 9.557 & 23.96 \\
dv-066 & 95.8 & 95.3 & 93.2 & 91.3 & 91.6 & 0.0 & 99.7 & 1/1 & 90.0 & 0 & 4.393 & 15.45 \\
dv-067 & 71.5 & 95.5 & 73.4 & 99.5 & 58.3 & 0.0 & 100.0 & 1/1 & 80.0 & 0 & 8.660 & 41.75 \\
dv-068 & 48.2 & 74.5 & 33.1 & 37.0 & 75.9 & 13.3 & 100.0 & 1/1 & 20.0 & 0 & 6.691 & 23.87 \\
dv-069 & 100.0 & 80.7 & 100.0 & 82.6 & 96.3 & 66.7 & 100.0 & 1/1 & 100.0 & 0 & 2.865 & 18.18 \\
dv-070 & 79.2 & 56.3 & 59.8 & 38.6 & 59.7 & 21.8 & 63.6 & 1/1 & 40.0 & 0 & 24.347 & 157.86 \\
dv-071 & 99.6 & 80.2 & 97.4 & 88.7 & 46.8 & 0.0 & 100.0 & 1/1 & 100.0 & 0 & 10.647 & 27.77 \\
dv-072 & 100.0 & 100.0 & 100.0 & 100.0 & 80.4 & 57.1 & 41.4 & 1/1 & 100.0 & 0 & 1.054 & 11.21 \\
dv-073 & 96.2 & 84.2 & 90.7 & 83.1 & 100.0 & 0.0 & 100.0 & 1/1 & 80.0 & 0 & 10.089 & 25.12 \\
dv-074 & 100.0 & 100.0 & 100.0 & 92.6 & 90.3 & 71.8 & 24.3 & 1/1 & 100.0 & 0 & 2.543 & 12.30 \\
dv-075 & 98.1 & 85.4 & 88.5 & 83.4 & 74.4 & 49.9 & 0.0 & 1/1 & 80.0 & 0 & 34.051 & 58.09 \\
dv-076 & 72.8 & 77.8 & 55.4 & 76.5 & 96.8 & 72.0 & 100.0 & 1/1 & 90.0 & 0 & 1.547 & 10.44 \\
dv-077 & 0.0 & 0.0 & 0.0 & 0.0 & 44.5 & 0.0 & 0.0 & 0/0 & 0.0 & 0 & 10.020 & 29.55 \\
dv-078 & 66.8 & 76.6 & 55.5 & 44.9 & 79.5 & 52.0 & 100.0 & 1/1 & 90.0 & 0 & 3.877 & 15.00 \\
dv-079 & 90.7 & 74.1 & 81.8 & 73.5 & 17.6 & 64.0 & 100.0 & 1/1 & 80.0 & 0 & 15.734 & 72.40 \\
dv-080 & 92.8 & 69.5 & 85.3 & 73.8 & 93.5 & 0.0 & 0.0 & 1/1 & 80.0 & 0 & 21.802 & 42.73 \\
dv-081 & 38.6 & 36.3 & 33.2 & 18.7 & 56.7 & 81.9 & 100.0 & 1/1 & 90.0 & 0 & 5.416 & 21.20 \\
dv-082 & 96.2 & 100.0 & 93.8 & 93.6 & 96.1 & 0.0 & 100.0 & 1/1 & 90.0 & 0 & 15.821 & 33.49 \\
dv-083 & 100.0 & 97.0 & 100.0 & 68.2 & 76.0 & 46.6 & 100.0 & 1/1 & 100.0 & 0 & 3.958 & 40.14 \\
dv-084 & 98.0 & 85.1 & 93.2 & 78.3 & 45.9 & 0.0 & 11.1 & 1/1 & 80.0 & 0 & 20.648 & 49.36 \\
dv-085 & 98.6 & 91.2 & 95.5 & 96.4 & 83.6 & 0.0 & 100.0 & 1/1 & 80.0 & 0 & 11.600 & 34.59 \\
dv-086 & 96.1 & 86.6 & 94.2 & 97.8 & 92.1 & 0.0 & 100.0 & 1/1 & 80.0 & 0 & 7.614 & 24.40 \\
dv-087 & 100.0 & 90.1 & 99.6 & 75.1 & 84.3 & 46.3 & 100.0 & 1/1 & 90.0 & 0 & 4.673 & 29.61 \\
dv-088 & 88.6 & 70.6 & 73.2 & 37.0 & 52.8 & 0.0 & 0.0 & 1/1 & 80.0 & 0 & 12.738 & 38.66 \\
dv-089 & 97.3 & 81.6 & 94.8 & 74.3 & 90.6 & 33.9 & 86.3 & 1/1 & 50.0 & 0 & 29.010 & 48.25 \\
dv-090 & 36.6 & 60.9 & 38.1 & 64.5 & 54.8 & 58.0 & 100.0 & 1/1 & 90.0 & 0 & 2.571 & 11.51 \\
dv-091 & 97.3 & 100.0 & 95.6 & 99.4 & 84.5 & 22.1 & 100.0 & 1/1 & 80.0 & 0 & 9.834 & 32.67 \\
dv-092 & 97.2 & 95.0 & 94.3 & 84.6 & 100.0 & 36.0 & 100.0 & 1/1 & 100.0 & 0 & 11.087 & 28.35 \\
dv-093 & 99.1 & 87.6 & 98.0 & 74.1 & 95.0 & 30.8 & 100.0 & 1/1 & 100.0 & 0 & 4.579 & 22.42 \\
dv-094 & 98.3 & 98.5 & 93.0 & 61.1 & 64.2 & 72.0 & 100.0 & 1/1 & 90.0 & 0 & 7.926 & 20.09 \\
dv-095 & 99.9 & 92.6 & 99.5 & 96.7 & 88.2 & 47.1 & 69.1 & 1/1 & 30.0 & 0 & 20.835 & 34.65 \\
\end{longtable}\endgroup

\clearpage
\subsection{C12: OpenCode + GPT-6.1 Sol}
\noindent\footnotesize Percentages: Ln/Cn/Br/Tg = line/condition/branch/toggle coverage; F1/P1 = feature/test-plan F1; Ref = reference-model point accuracy; Kill = mutant kill rate. C/G and O are binary compile/golden and overall results. Tok = input plus output tokens in millions; Min = elapsed minutes. Overall uses exact model correctness and all four coverage thresholds. \par\normalsize
\begingroup
\fontsize{8.2}{10.2}\selectfont
\setlength{\tabcolsep}{3pt}\renewcommand{\arraystretch}{1.04}
\setlength{\LTleft}{0pt}\setlength{\LTright}{0pt}
\Needspace{10\baselineskip}
\captionsetup{hypcap=false}
\captionof{table}{Per-task measurements for C12. The companion CSV retains unrounded values, precision/recall, exact-model flags, input/output token splits, and source-cell coordinates.}\label{tab:app-detail-C12}
\addtocounter{table}{-1}
\begin{longtable}{>{\raggedright\arraybackslash}p{\dimexpr 0.1050000\textwidth-2\tabcolsep\relax}>{\raggedleft\arraybackslash}p{\dimexpr 0.0745833\textwidth-2\tabcolsep\relax}>{\raggedleft\arraybackslash}p{\dimexpr 0.0745833\textwidth-2\tabcolsep\relax}>{\raggedleft\arraybackslash}p{\dimexpr 0.0745833\textwidth-2\tabcolsep\relax}>{\raggedleft\arraybackslash}p{\dimexpr 0.0745833\textwidth-2\tabcolsep\relax}>{\raggedleft\arraybackslash}p{\dimexpr 0.0745833\textwidth-2\tabcolsep\relax}>{\raggedleft\arraybackslash}p{\dimexpr 0.0745833\textwidth-2\tabcolsep\relax}>{\raggedleft\arraybackslash}p{\dimexpr 0.0745833\textwidth-2\tabcolsep\relax}>{\raggedleft\arraybackslash}p{\dimexpr 0.0745833\textwidth-2\tabcolsep\relax}>{\raggedleft\arraybackslash}p{\dimexpr 0.0745833\textwidth-2\tabcolsep\relax}>{\raggedleft\arraybackslash}p{\dimexpr 0.0745833\textwidth-2\tabcolsep\relax}>{\raggedleft\arraybackslash}p{\dimexpr 0.0745833\textwidth-2\tabcolsep\relax}>{\raggedleft\arraybackslash}p{\dimexpr 0.0745833\textwidth-2\tabcolsep\relax}}
\toprule
\rowcolor{ReleasePale}\textbf{Task} & \textbf{Ln} & \textbf{Cn} & \textbf{Br} & \textbf{Tg} & \textbf{F1} & \textbf{P1} & \textbf{Ref} & \textbf{C/G} & \textbf{Kill} & \textbf{O} & \textbf{Tok} & \textbf{Min} \\
\midrule\endfirsthead
\multicolumn{13}{l}{\small\color{ReleaseMuted}Table~\ref{tab:app-detail-C12} (continued)}\\
\toprule
\rowcolor{ReleasePale}\textbf{Task} & \textbf{Ln} & \textbf{Cn} & \textbf{Br} & \textbf{Tg} & \textbf{F1} & \textbf{P1} & \textbf{Ref} & \textbf{C/G} & \textbf{Kill} & \textbf{O} & \textbf{Tok} & \textbf{Min} \\
\midrule\endhead
\midrule
\multicolumn{13}{r}{\footnotesize\color{ReleaseMuted}Continued on next page}\\\endfoot
\bottomrule\endlastfoot
dv-001 & 100.0 & 93.3 & 100.0 & 96.9 & 97.1 & 100.0 & 100.0 & 1/1 & 100.0 & \textbf{\color{ReleaseBlue}1} & 0.291 & 6.75 \\
dv-002 & 100.0 & 100.0 & 100.0 & 97.5 & 97.4 & 66.7 & 100.0 & 1/1 & 100.0 & \textbf{\color{ReleaseBlue}1} & 0.484 & 9.16 \\
dv-003 & 100.0 & 100.0 & 100.0 & 100.0 & 88.0 & 100.0 & 100.0 & 1/1 & 100.0 & \textbf{\color{ReleaseBlue}1} & 1.420 & 21.14 \\
dv-004 & 99.0 & 96.5 & 96.2 & 88.6 & 97.9 & 88.9 & 100.0 & 1/1 & 100.0 & 0 & 1.372 & 20.22 \\
dv-005 & 100.0 & 99.3 & 100.0 & 98.9 & 100.0 & 100.0 & 100.0 & 1/1 & 100.0 & \textbf{\color{ReleaseBlue}1} & 0.340 & 12.19 \\
dv-006 & 99.5 & 99.8 & 99.8 & 92.8 & 100.0 & 92.1 & 100.0 & 1/1 & 100.0 & \textbf{\color{ReleaseBlue}1} & 1.305 & 36.31 \\
dv-007 & 100.0 & 100.0 & 100.0 & 100.0 & 96.0 & 76.9 & 100.0 & 1/1 & 80.0 & 0 & 0.497 & 9.58 \\
dv-008 & 98.3 & 92.4 & 96.6 & 97.5 & 89.8 & 89.1 & 99.7 & 1/1 & 100.0 & 0 & 3.491 & 49.20 \\
dv-009 & 100.0 & 91.3 & 100.0 & 100.0 & 100.0 & 93.3 & 100.0 & 1/1 & 30.0 & 0 & 0.919 & 13.23 \\
dv-010 & 100.0 & 100.0 & 100.0 & 99.3 & 94.1 & 98.9 & 100.0 & 1/1 & 100.0 & \textbf{\color{ReleaseBlue}1} & 1.090 & 14.99 \\
dv-011 & 98.4 & 93.5 & 91.3 & 52.9 & 88.9 & 100.0 & 100.0 & 1/1 & 100.0 & 0 & 0.593 & 11.21 \\
dv-012 & 100.0 & 100.0 & 100.0 & 100.0 & 100.0 & 100.0 & 100.0 & 1/1 & 100.0 & \textbf{\color{ReleaseBlue}1} & 4.419 & 80.07 \\
dv-013 & 100.0 & 100.0 & 100.0 & 94.7 & 100.0 & 100.0 & 100.0 & 1/1 & 100.0 & \textbf{\color{ReleaseBlue}1} & 1.061 & 30.15 \\
dv-014 & 87.0 & 79.6 & 84.3 & 68.2 & 100.0 & 71.4 & 100.0 & 1/1 & 100.0 & 0 & 2.413 & 27.09 \\
dv-015 & 100.0 & 100.0 & 97.3 & 89.4 & 89.3 & 49.3 & 28.4 & 1/1 & 90.0 & 0 & 2.482 & 23.48 \\
dv-016 & 99.3 & 97.5 & 98.9 & 66.6 & 100.0 & 79.2 & 100.0 & 1/1 & 50.0 & 0 & 2.643 & 24.76 \\
dv-017 & 96.8 & 81.3 & 88.4 & 90.6 & 96.6 & 100.0 & 89.9 & 1/1 & 100.0 & 0 & 3.927 & 31.73 \\
dv-018 & 100.0 & 100.0 & 100.0 & 94.8 & 100.0 & 100.0 & 100.0 & 1/1 & 100.0 & \textbf{\color{ReleaseBlue}1} & 2.269 & 22.20 \\
dv-019 & 97.7 & 85.5 & 96.4 & 99.0 & 92.3 & 77.4 & 0.0 & 1/1 & 80.0 & 0 & 1.056 & 15.82 \\
dv-020 & 97.6 & 93.2 & 88.2 & 95.3 & 97.0 & 99.3 & 99.8 & 1/1 & 100.0 & 0 & 6.990 & 30.64 \\
dv-021 & 92.7 & 85.3 & 89.4 & 93.8 & 98.2 & 89.7 & 100.0 & 1/0 & 0.0 & 0 & 1.296 & 41.76 \\
dv-022 & 100.0 & 100.0 & 98.1 & 95.8 & 98.0 & 66.3 & 15.0 & 1/1 & 80.0 & 0 & 1.551 & 16.94 \\
dv-023 & 100.0 & 100.0 & 100.0 & 100.0 & 100.0 & 28.6 & 100.0 & 1/1 & 100.0 & \textbf{\color{ReleaseBlue}1} & 0.812 & 14.85 \\
dv-024 & 43.3 & 60.8 & 27.1 & 34.2 & 89.1 & 66.7 & 100.0 & 1/1 & 80.0 & 0 & 1.162 & 13.16 \\
dv-025 & 40.7 & 54.9 & 23.9 & 29.9 & 100.0 & 0.0 & 100.0 & 1/1 & 20.0 & 0 & 0.716 & 13.33 \\
dv-026 & 43.8 & 64.4 & 27.0 & 24.7 & 96.3 & 93.3 & 100.0 & 1/1 & 80.0 & 0 & 1.003 & 23.62 \\
dv-027 & 97.5 & 89.7 & 95.6 & 58.6 & 100.0 & 66.7 & 76.9 & 1/0 & 0.0 & 0 & 7.755 & 47.95 \\
dv-028 & 100.0 & 98.4 & 100.0 & 98.7 & 89.4 & 100.0 & 100.0 & 1/1 & 100.0 & \textbf{\color{ReleaseBlue}1} & 2.980 & 29.11 \\
dv-029 & 100.0 & 97.7 & 95.7 & 98.6 & 78.8 & 83.9 & 26.3 & 1/1 & 80.0 & 0 & 3.482 & 25.50 \\
dv-030 & 64.6 & 48.7 & 63.1 & 82.6 & 94.1 & 97.8 & 100.0 & 1/0 & 0.0 & 0 & 3.177 & 109.70 \\
dv-031 & 62.2 & 3.9 & 38.0 & 31.8 & 100.0 & 71.1 & 99.7 & 1/1 & 60.0 & 0 & 0.748 & 15.91 \\
dv-032 & 97.5 & 84.5 & 96.2 & 95.3 & 100.0 & 98.9 & 54.5 & 1/1 & 80.0 & 0 & 6.923 & 38.27 \\
dv-033 & 78.3 & 74.8 & 76.4 & 84.4 & 98.4 & 58.8 & 98.9 & 1/1 & 90.0 & 0 & 2.797 & 23.17 \\
dv-034 & 79.3 & 92.9 & 73.9 & 39.6 & 87.0 & 100.0 & 100.0 & 1/1 & 100.0 & 0 & 0.684 & 12.06 \\
dv-035 & 79.0 & 53.3 & 61.4 & 95.6 & 80.0 & 87.7 & 65.1 & 1/1 & 90.0 & 0 & 1.672 & 19.60 \\
dv-036 & 99.2 & 91.4 & 97.4 & 84.2 & 89.6 & 71.2 & 91.0 & 1/1 & 80.0 & 0 & 8.083 & 267.76 \\
dv-037 & 99.7 & 98.9 & 98.9 & 72.8 & 76.8 & 72.1 & 100.0 & 1/1 & 70.0 & 0 & 2.818 & 30.61 \\
dv-038 & 100.0 & 97.7 & 100.0 & 99.6 & 98.2 & 0.0 & 100.0 & 1/1 & 90.0 & 0 & 1.301 & 19.97 \\
dv-039 & 99.6 & 99.0 & 99.6 & 98.7 & 100.0 & 72.7 & 100.0 & 1/1 & 80.0 & 0 & 0.827 & 21.94 \\
dv-040 & 100.0 & 98.9 & 99.2 & 68.5 & 100.0 & 81.8 & 100.0 & 1/1 & 90.0 & 0 & 0.583 & 13.79 \\
dv-041 & 99.3 & 99.5 & 85.0 & 99.7 & 100.0 & 50.0 & 100.0 & 1/1 & 90.0 & 0 & 3.556 & 32.58 \\
dv-042 & 100.0 & 100.0 & 100.0 & 74.4 & 100.0 & 76.9 & 100.0 & 1/1 & 100.0 & 0 & 0.791 & 15.85 \\
dv-043 & 91.1 & 100.0 & 85.2 & 100.0 & 100.0 & 100.0 & 99.6 & 1/1 & 100.0 & 0 & 0.576 & 10.29 \\
dv-044 & 98.6 & 81.8 & 99.0 & 94.4 & 97.1 & 100.0 & 100.0 & 1/1 & 100.0 & 0 & 2.359 & 48.43 \\
dv-045 & 100.0 & 90.8 & 95.5 & 92.6 & 100.0 & 66.2 & 100.0 & 1/1 & 80.0 & 0 & 4.629 & 41.12 \\
dv-046 & 98.1 & 97.5 & 96.4 & 75.8 & 99.3 & 40.0 & 100.0 & 1/1 & 100.0 & 0 & 2.483 & 32.89 \\
dv-047 & 100.0 & 100.0 & 100.0 & 100.0 & 100.0 & 79.0 & 100.0 & 1/1 & 90.0 & 0 & 1.189 & 19.02 \\
dv-048 & 100.0 & 100.0 & 100.0 & 98.7 & 98.2 & 44.4 & 0.0 & 1/1 & 100.0 & 0 & 1.612 & 23.28 \\
dv-049 & 99.7 & 99.8 & 95.5 & 98.3 & 100.0 & 76.0 & 100.0 & 1/1 & 90.0 & 0 & 1.457 & 21.00 \\
dv-050 & 100.0 & 80.0 & 99.1 & 100.0 & 100.0 & 92.3 & 100.0 & 1/1 & 90.0 & 0 & 1.405 & 18.76 \\
dv-051 & 99.6 & 98.3 & 99.2 & 93.6 & 100.0 & 100.0 & 100.0 & 1/1 & 90.0 & 0 & 1.768 & 21.25 \\
dv-052 & 99.9 & 38.9 & 99.9 & 99.9 & 100.0 & 100.0 & 100.0 & 1/1 & 90.0 & 0 & 0.533 & 14.84 \\
dv-053 & 100.0 & 24.6 & 100.0 & 97.2 & 83.1 & 0.0 & 92.0 & 1/1 & 100.0 & 0 & 1.399 & 22.13 \\
dv-054 & 100.0 & 99.8 & 100.0 & 100.0 & 92.2 & 88.9 & 100.0 & 1/1 & 90.0 & 0 & 2.329 & 28.41 \\
dv-055 & 90.0 & 66.0 & 91.6 & 80.9 & 100.0 & 100.0 & 100.0 & 1/1 & 100.0 & 0 & 1.346 & 16.56 \\
dv-056 & 99.8 & 98.3 & 94.0 & 96.1 & 99.0 & 96.3 & 100.0 & 1/1 & 100.0 & \textbf{\color{ReleaseBlue}1} & 1.530 & 21.98 \\
dv-057 & 88.4 & 77.9 & 74.7 & 97.4 & 98.0 & 100.0 & 95.7 & 1/1 & 100.0 & 0 & 3.678 & 36.97 \\
dv-058 & 91.3 & 93.9 & 86.7 & 98.5 & 100.0 & 76.2 & 0.0 & 1/0 & 0.0 & 0 & 1.626 & 36.08 \\
dv-059 & 99.1 & 100.0 & 97.1 & 100.0 & 98.2 & 28.6 & 99.9 & 1/0 & 0.0 & 0 & 2.657 & 18.31 \\
dv-060 & 100.0 & 90.3 & 98.2 & 96.8 & 100.0 & 71.4 & 100.0 & 1/1 & 100.0 & \textbf{\color{ReleaseBlue}1} & 1.987 & 23.28 \\
dv-061 & 99.9 & 99.0 & 99.8 & 99.8 & 97.9 & 96.8 & 100.0 & 1/1 & 100.0 & \textbf{\color{ReleaseBlue}1} & 2.818 & 27.67 \\
dv-062 & 98.7 & 65.2 & 69.3 & 81.7 & 97.1 & 96.8 & 97.7 & 1/1 & 80.0 & 0 & 2.030 & 22.30 \\
dv-063 & 100.0 & 100.0 & 99.9 & 99.4 & 100.0 & 66.7 & 100.0 & 1/1 & 90.0 & 0 & 7.550 & 35.62 \\
dv-064 & 99.2 & 96.7 & 98.6 & 76.4 & 100.0 & 82.0 & 100.0 & 1/1 & 90.0 & 0 & 1.390 & 16.72 \\
dv-065 & 95.5 & 72.5 & 75.1 & 90.6 & 100.0 & 71.4 & 100.0 & 1/1 & 80.0 & 0 & 1.577 & 20.74 \\
dv-066 & 98.4 & 98.7 & 96.3 & 62.7 & 100.0 & 54.5 & 100.0 & 1/1 & 80.0 & 0 & 0.866 & 13.66 \\
dv-067 & 79.9 & 95.5 & 81.2 & 99.6 & 100.0 & 44.1 & 100.0 & 1/1 & 90.0 & 0 & 1.212 & 16.94 \\
dv-068 & 48.2 & 78.3 & 33.1 & 44.2 & 100.0 & 35.3 & 100.0 & 1/1 & 20.0 & 0 & 1.834 & 19.06 \\
dv-069 & 100.0 & 80.7 & 100.0 & 82.6 & 100.0 & 100.0 & 100.0 & 1/1 & 100.0 & 0 & 1.059 & 14.79 \\
dv-070 & 78.0 & 59.2 & 62.0 & 49.8 & 60.1 & 99.2 & 100.0 & 1/1 & 60.0 & 0 & 3.871 & 32.63 \\
dv-071 & 99.6 & 97.7 & 97.4 & 88.7 & 53.8 & 100.0 & 100.0 & 1/1 & 100.0 & 0 & 0.964 & 9.91 \\
dv-072 & 100.0 & 100.0 & 100.0 & 100.0 & 100.0 & 100.0 & 47.5 & 1/1 & 100.0 & 0 & 0.659 & 10.31 \\
dv-073 & 96.8 & 84.2 & 92.2 & 89.9 & 87.0 & 82.4 & 100.0 & 1/1 & 100.0 & 0 & 0.632 & 10.48 \\
dv-074 & 100.0 & 100.0 & 100.0 & 92.6 & 97.0 & 100.0 & 20.7 & 1/1 & 100.0 & 0 & 0.685 & 12.17 \\
dv-075 & 0.0 & 0.0 & 0.0 & 0.0 & 60.7 & 65.6 & 2.0 & 0/0 & 0.0 & 0 & 0.483 & 11.05 \\
dv-076 & 72.8 & 80.2 & 56.4 & 76.5 & 100.0 & 60.9 & 100.0 & 1/1 & 90.0 & 0 & 1.077 & 15.51 \\
dv-077 & 100.0 & 92.5 & 99.7 & 67.0 & 96.4 & 0.0 & 0.0 & 1/1 & 80.0 & 0 & 7.386 & 34.21 \\
dv-078 & 70.0 & 76.6 & 60.9 & 45.0 & 91.1 & 100.0 & 100.0 & 1/1 & 100.0 & 0 & 0.823 & 10.50 \\
dv-079 & 93.2 & 79.1 & 86.5 & 76.6 & 100.0 & 78.3 & 100.0 & 1/0 & 0.0 & 0 & 2.129 & 42.62 \\
dv-080 & 95.9 & 86.3 & 92.0 & 92.9 & 100.0 & 79.6 & 19.6 & 1/1 & 70.0 & 0 & 3.879 & 29.14 \\
dv-081 & 45.3 & 41.1 & 39.1 & 24.4 & 92.8 & 99.3 & 100.0 & 1/1 & 100.0 & 0 & 6.700 & 38.75 \\
dv-082 & 94.8 & 100.0 & 91.1 & 99.4 & 100.0 & 85.7 & 100.0 & 1/1 & 100.0 & \textbf{\color{ReleaseBlue}1} & 1.007 & 16.75 \\
dv-083 & 0.0 & 0.0 & 0.0 & 0.0 & 96.7 & 100.0 & 100.0 & 0/0 & 0.0 & 0 & 4.429 & 29.72 \\
dv-084 & 100.0 & 88.1 & 95.0 & 94.0 & 91.7 & 73.1 & 1.8 & 1/1 & 100.0 & 0 & 2.082 & 20.60 \\
dv-085 & 98.6 & 91.2 & 95.5 & 96.4 & 98.0 & 75.0 & 100.0 & 1/1 & 70.0 & 0 & 2.163 & 18.86 \\
dv-086 & 96.1 & 86.6 & 94.2 & 97.8 & 100.0 & 72.0 & 100.0 & 1/1 & 80.0 & 0 & 1.299 & 20.42 \\
dv-087 & 99.6 & 91.1 & 97.7 & 77.4 & 96.1 & 93.3 & 100.0 & 1/1 & 80.0 & 0 & 2.820 & 29.05 \\
dv-088 & 93.0 & 87.2 & 85.2 & 75.8 & 100.0 & 54.4 & 100.0 & 1/1 & 70.0 & 0 & 4.419 & 33.33 \\
dv-089 & 99.6 & 96.8 & 98.6 & 76.6 & 97.3 & 90.9 & 91.2 & 1/1 & 60.0 & 0 & 3.857 & 28.89 \\
dv-090 & 36.8 & 61.9 & 38.9 & 66.4 & 82.1 & 90.9 & 100.0 & 1/1 & 90.0 & 0 & 0.647 & 11.32 \\
dv-091 & 97.3 & 100.0 & 95.6 & 99.4 & 98.0 & 71.3 & 100.0 & 1/1 & 80.0 & 0 & 2.142 & 19.75 \\
dv-092 & 97.2 & 95.0 & 94.3 & 84.6 & 97.9 & 87.5 & 100.0 & 1/1 & 90.0 & 0 & 1.013 & 16.59 \\
dv-093 & 99.1 & 89.3 & 98.6 & 90.8 & 96.8 & 0.0 & 100.0 & 1/1 & 100.0 & 0 & 0.877 & 15.53 \\
dv-094 & 99.4 & 100.0 & 97.2 & 61.1 & 100.0 & 100.0 & 100.0 & 1/0 & 0.0 & 0 & 1.443 & 15.68 \\
dv-095 & 99.9 & 94.7 & 99.7 & 96.8 & 89.4 & 98.9 & 41.6 & 1/1 & 60.0 & 0 & 1.923 & 17.85 \\
\end{longtable}\endgroup

\clearpage
\subsection{C13: OpenCode + GPT-6 Astra}
\noindent\footnotesize Percentages: Ln/Cn/Br/Tg = line/condition/branch/toggle coverage; F1/P1 = feature/test-plan F1; Ref = reference-model point accuracy; Kill = mutant kill rate. C/G and O are binary compile/golden and overall results. Tok = input plus output tokens in millions; Min = elapsed minutes. Overall uses exact model correctness and all four coverage thresholds. \par\normalsize
\begingroup
\fontsize{8.2}{10.2}\selectfont
\setlength{\tabcolsep}{3pt}\renewcommand{\arraystretch}{1.04}
\setlength{\LTleft}{0pt}\setlength{\LTright}{0pt}
\Needspace{10\baselineskip}
\captionsetup{hypcap=false}
\captionof{table}{Per-task measurements for C13. The companion CSV retains unrounded values, precision/recall, exact-model flags, input/output token splits, and source-cell coordinates.}\label{tab:app-detail-C13}
\addtocounter{table}{-1}
\begin{longtable}{>{\raggedright\arraybackslash}p{\dimexpr 0.1050000\textwidth-2\tabcolsep\relax}>{\raggedleft\arraybackslash}p{\dimexpr 0.0745833\textwidth-2\tabcolsep\relax}>{\raggedleft\arraybackslash}p{\dimexpr 0.0745833\textwidth-2\tabcolsep\relax}>{\raggedleft\arraybackslash}p{\dimexpr 0.0745833\textwidth-2\tabcolsep\relax}>{\raggedleft\arraybackslash}p{\dimexpr 0.0745833\textwidth-2\tabcolsep\relax}>{\raggedleft\arraybackslash}p{\dimexpr 0.0745833\textwidth-2\tabcolsep\relax}>{\raggedleft\arraybackslash}p{\dimexpr 0.0745833\textwidth-2\tabcolsep\relax}>{\raggedleft\arraybackslash}p{\dimexpr 0.0745833\textwidth-2\tabcolsep\relax}>{\raggedleft\arraybackslash}p{\dimexpr 0.0745833\textwidth-2\tabcolsep\relax}>{\raggedleft\arraybackslash}p{\dimexpr 0.0745833\textwidth-2\tabcolsep\relax}>{\raggedleft\arraybackslash}p{\dimexpr 0.0745833\textwidth-2\tabcolsep\relax}>{\raggedleft\arraybackslash}p{\dimexpr 0.0745833\textwidth-2\tabcolsep\relax}>{\raggedleft\arraybackslash}p{\dimexpr 0.0745833\textwidth-2\tabcolsep\relax}}
\toprule
\rowcolor{ReleasePale}\textbf{Task} & \textbf{Ln} & \textbf{Cn} & \textbf{Br} & \textbf{Tg} & \textbf{F1} & \textbf{P1} & \textbf{Ref} & \textbf{C/G} & \textbf{Kill} & \textbf{O} & \textbf{Tok} & \textbf{Min} \\
\midrule\endfirsthead
\multicolumn{13}{l}{\small\color{ReleaseMuted}Table~\ref{tab:app-detail-C13} (continued)}\\
\toprule
\rowcolor{ReleasePale}\textbf{Task} & \textbf{Ln} & \textbf{Cn} & \textbf{Br} & \textbf{Tg} & \textbf{F1} & \textbf{P1} & \textbf{Ref} & \textbf{C/G} & \textbf{Kill} & \textbf{O} & \textbf{Tok} & \textbf{Min} \\
\midrule\endhead
\midrule
\multicolumn{13}{r}{\footnotesize\color{ReleaseMuted}Continued on next page}\\\endfoot
\bottomrule\endlastfoot
dv-001 & 100.0 & 93.3 & 100.0 & 96.9 & 97.1 & 100.0 & 100.0 & 1/1 & 100.0 & \textbf{\color{ReleaseBlue}1} & 0.296 & 7.65 \\
dv-002 & 100.0 & 100.0 & 100.0 & 97.5 & 100.0 & 66.7 & 100.0 & 1/1 & 100.0 & \textbf{\color{ReleaseBlue}1} & 1.186 & 12.54 \\
dv-003 & 100.0 & 100.0 & 100.0 & 100.0 & 87.2 & 100.0 & 100.0 & 1/1 & 100.0 & \textbf{\color{ReleaseBlue}1} & 1.044 & 18.35 \\
dv-004 & 99.0 & 96.9 & 96.2 & 89.1 & 96.7 & 74.6 & 100.0 & 1/1 & 100.0 & 0 & 0.941 & 17.21 \\
dv-005 & 100.0 & 99.3 & 100.0 & 98.9 & 97.4 & 97.9 & 100.0 & 1/1 & 100.0 & \textbf{\color{ReleaseBlue}1} & 0.484 & 11.96 \\
dv-006 & 99.5 & 99.8 & 99.8 & 92.8 & 100.0 & 100.0 & 100.0 & 1/1 & 100.0 & \textbf{\color{ReleaseBlue}1} & 0.926 & 13.61 \\
dv-007 & 100.0 & 100.0 & 100.0 & 100.0 & 96.0 & 66.7 & 100.0 & 1/1 & 90.0 & 0 & 0.353 & 8.24 \\
dv-008 & 98.3 & 91.8 & 96.6 & 97.5 & 88.4 & 93.6 & 100.0 & 1/1 & 100.0 & \textbf{\color{ReleaseBlue}1} & 3.424 & 43.69 \\
dv-009 & 100.0 & 91.3 & 100.0 & 100.0 & 100.0 & 93.3 & 100.0 & 1/1 & 30.0 & 0 & 0.455 & 11.18 \\
dv-010 & 100.0 & 100.0 & 100.0 & 99.3 & 90.5 & 100.0 & 100.0 & 1/1 & 100.0 & \textbf{\color{ReleaseBlue}1} & 0.697 & 9.79 \\
dv-011 & 98.4 & 93.5 & 91.3 & 52.9 & 92.6 & 100.0 & 100.0 & 1/1 & 100.0 & 0 & 0.320 & 7.99 \\
dv-012 & 100.0 & 100.0 & 100.0 & 100.0 & 100.0 & 100.0 & 100.0 & 1/1 & 100.0 & \textbf{\color{ReleaseBlue}1} & 1.946 & 14.19 \\
dv-013 & 100.0 & 100.0 & 100.0 & 94.7 & 96.3 & 100.0 & 100.0 & 1/1 & 100.0 & \textbf{\color{ReleaseBlue}1} & 0.639 & 16.46 \\
dv-014 & 87.0 & 79.6 & 84.3 & 68.2 & 92.2 & 69.9 & 100.0 & 1/1 & 100.0 & 0 & 1.599 & 30.48 \\
dv-015 & 100.0 & 100.0 & 97.3 & 89.4 & 85.1 & 99.2 & 28.4 & 1/1 & 100.0 & 0 & 1.861 & 24.71 \\
dv-016 & 99.3 & 99.2 & 98.9 & 72.2 & 100.0 & 90.4 & 100.0 & 1/1 & 70.0 & 0 & 3.133 & 25.92 \\
dv-017 & 96.8 & 81.3 & 88.4 & 90.6 & 95.5 & 99.2 & 89.9 & 1/1 & 100.0 & 0 & 1.341 & 16.63 \\
dv-018 & 100.0 & 100.0 & 100.0 & 94.8 & 100.0 & 99.0 & 100.0 & 1/1 & 100.0 & \textbf{\color{ReleaseBlue}1} & 0.570 & 14.41 \\
dv-019 & 98.4 & 87.1 & 97.0 & 99.0 & 96.3 & 91.7 & 99.9 & 1/1 & 90.0 & 0 & 0.909 & 17.14 \\
dv-020 & 97.6 & 93.2 & 88.2 & 95.3 & 100.0 & 99.1 & 100.0 & 1/1 & 100.0 & 0 & 5.646 & 26.00 \\
dv-021 & 96.8 & 91.3 & 93.5 & 93.8 & 100.0 & 93.3 & 100.0 & 1/1 & 100.0 & \textbf{\color{ReleaseBlue}1} & 0.638 & 27.09 \\
dv-022 & 100.0 & 97.5 & 98.1 & 98.8 & 85.7 & 95.7 & 9.8 & 1/1 & 100.0 & 0 & 1.922 & 19.23 \\
dv-023 & 100.0 & 100.0 & 100.0 & 100.0 & 100.0 & 28.6 & 100.0 & 1/1 & 100.0 & \textbf{\color{ReleaseBlue}1} & 0.454 & 11.27 \\
dv-024 & 43.3 & 61.6 & 27.1 & 38.1 & 97.3 & 0.0 & 100.0 & 1/1 & 80.0 & 0 & 0.583 & 12.67 \\
dv-025 & 40.7 & 54.8 & 23.9 & 30.0 & 88.5 & 100.0 & 100.0 & 1/1 & 20.0 & 0 & 0.677 & 15.13 \\
dv-026 & 42.1 & 58.6 & 24.9 & 20.8 & 94.7 & 89.7 & 100.0 & 1/1 & 80.0 & 0 & 0.861 & 15.73 \\
dv-027 & 97.5 & 87.7 & 94.9 & 58.0 & 100.0 & 36.2 & 80.0 & 1/1 & 70.0 & 0 & 2.069 & 21.85 \\
dv-028 & 0.0 & 0.0 & 0.0 & 0.0 & 0.0 & 0.0 & 0.0 & 0/0 & 0.0 & 0 & 0.223 & 2.25 \\
dv-029 & 99.7 & 97.7 & 95.0 & 84.8 & 72.5 & 92.9 & 0.0 & 1/1 & 90.0 & 0 & 2.977 & 19.44 \\
dv-030 & 80.9 & 55.1 & 70.9 & 86.8 & 98.1 & 59.7 & 100.0 & 1/0 & 0.0 & 0 & 0.904 & 63.99 \\
dv-031 & 62.2 & 3.9 & 38.0 & 31.8 & 98.2 & 89.1 & 99.9 & 1/1 & 80.0 & 0 & 0.640 & 13.39 \\
dv-032 & 96.8 & 84.5 & 94.9 & 95.4 & 98.5 & 83.9 & 58.0 & 1/1 & 70.0 & 0 & 2.247 & 23.76 \\
dv-033 & 78.3 & 74.8 & 76.4 & 83.3 & 98.6 & 66.7 & 100.0 & 1/1 & 90.0 & 0 & 2.392 & 22.38 \\
dv-034 & 79.8 & 94.0 & 74.8 & 39.6 & 87.0 & 100.0 & 100.0 & 1/1 & 100.0 & 0 & 0.344 & 9.25 \\
dv-035 & 79.0 & 53.3 & 62.9 & 95.6 & 89.6 & 99.1 & 65.1 & 1/1 & 90.0 & 0 & 1.643 & 19.54 \\
dv-036 & 99.2 & 92.0 & 97.4 & 84.1 & 89.3 & 90.7 & 99.7 & 1/1 & 80.0 & 0 & 1.746 & 26.88 \\
dv-037 & 99.7 & 97.2 & 98.9 & 83.7 & 96.2 & 82.6 & 100.0 & 1/1 & 80.0 & 0 & 3.809 & 25.98 \\
dv-038 & 100.0 & 97.7 & 100.0 & 99.6 & 98.7 & 99.4 & 100.0 & 1/1 & 90.0 & 0 & 2.123 & 25.20 \\
dv-039 & 99.6 & 99.0 & 99.6 & 98.7 & 100.0 & 44.0 & 100.0 & 1/1 & 90.0 & 0 & 0.365 & 9.26 \\
dv-040 & 100.0 & 98.9 & 99.2 & 68.5 & 100.0 & 95.5 & 100.0 & 1/1 & 90.0 & 0 & 0.750 & 15.24 \\
dv-041 & 99.3 & 99.5 & 85.0 & 99.7 & 100.0 & 100.0 & 100.0 & 1/1 & 90.0 & 0 & 2.013 & 29.56 \\
dv-042 & 100.0 & 100.0 & 100.0 & 74.4 & 87.3 & 66.7 & 100.0 & 1/1 & 100.0 & 0 & 1.376 & 18.58 \\
dv-043 & 91.1 & 100.0 & 85.2 & 100.0 & 100.0 & 100.0 & 99.6 & 1/1 & 100.0 & 0 & 0.380 & 9.92 \\
dv-044 & 98.6 & 81.8 & 99.0 & 94.4 & 97.1 & 97.1 & 100.0 & 1/1 & 100.0 & 0 & 0.731 & 13.05 \\
dv-045 & 100.0 & 90.3 & 95.3 & 92.2 & 100.0 & 88.0 & 100.0 & 1/0 & 0.0 & 0 & 2.028 & 12.65 \\
dv-046 & 0.0 & 0.0 & 0.0 & 0.0 & 95.4 & 40.0 & 100.0 & 1/1 & 100.0 & 0 & 1.542 & 17.48 \\
dv-047 & 100.0 & 100.0 & 100.0 & 100.0 & 100.0 & 80.0 & 100.0 & 1/1 & 100.0 & \textbf{\color{ReleaseBlue}1} & 0.497 & 11.36 \\
dv-048 & 100.0 & 100.0 & 100.0 & 98.7 & 100.0 & 60.0 & 0.0 & 1/1 & 100.0 & 0 & 0.991 & 13.21 \\
dv-049 & 99.7 & 99.8 & 95.5 & 98.3 & 100.0 & 80.5 & 100.0 & 1/1 & 90.0 & 0 & 0.462 & 12.48 \\
dv-050 & 100.0 & 80.0 & 99.1 & 100.0 & 100.0 & 96.3 & 100.0 & 1/1 & 100.0 & 0 & 0.780 & 15.62 \\
dv-051 & 100.0 & 98.3 & 100.0 & 93.6 & 100.0 & 76.1 & 100.0 & 1/1 & 90.0 & 0 & 1.078 & 19.94 \\
dv-052 & 99.9 & 38.9 & 99.9 & 99.9 & 100.0 & 100.0 & 100.0 & 1/1 & 90.0 & 0 & 0.353 & 9.55 \\
dv-053 & 100.0 & 24.8 & 100.0 & 97.1 & 100.0 & 85.7 & 92.0 & 1/1 & 100.0 & 0 & 1.323 & 21.88 \\
dv-054 & 100.0 & 99.8 & 100.0 & 100.0 & 98.2 & 40.0 & 100.0 & 1/1 & 90.0 & 0 & 0.995 & 15.98 \\
dv-055 & 90.0 & 70.8 & 91.6 & 80.9 & 100.0 & 100.0 & 100.0 & 1/1 & 100.0 & 0 & 0.556 & 15.96 \\
dv-056 & 99.8 & 98.3 & 96.2 & 95.5 & 98.0 & 88.9 & 100.0 & 1/1 & 100.0 & \textbf{\color{ReleaseBlue}1} & 1.862 & 20.52 \\
dv-057 & 88.4 & 77.9 & 74.7 & 97.4 & 100.0 & 83.3 & 95.7 & 1/1 & 100.0 & 0 & 1.114 & 15.50 \\
dv-058 & 91.3 & 93.9 & 86.4 & 98.5 & 100.0 & 79.1 & 100.0 & 1/1 & 100.0 & 0 & 0.514 & 19.53 \\
dv-059 & 99.1 & 100.0 & 97.1 & 100.0 & 100.0 & 0.0 & 99.9 & 1/1 & 100.0 & 0 & 1.863 & 22.44 \\
dv-060 & 100.0 & 89.9 & 98.2 & 96.8 & 100.0 & 87.5 & 100.0 & 1/1 & 100.0 & 0 & 1.113 & 19.25 \\
dv-061 & 99.9 & 97.4 & 99.8 & 99.8 & 89.7 & 76.9 & 100.0 & 1/1 & 100.0 & \textbf{\color{ReleaseBlue}1} & 3.202 & 26.27 \\
dv-062 & 98.7 & 65.6 & 69.3 & 81.7 & 96.3 & 93.3 & 98.2 & 1/1 & 90.0 & 0 & 0.563 & 11.16 \\
dv-063 & 100.0 & 100.0 & 99.9 & 99.4 & 97.6 & 66.3 & 100.0 & 1/1 & 80.0 & 0 & 12.384 & 49.23 \\
dv-064 & 99.2 & 96.9 & 98.6 & 76.4 & 100.0 & 100.0 & 100.0 & 1/1 & 90.0 & 0 & 0.418 & 10.90 \\
dv-065 & 95.5 & 72.3 & 75.1 & 90.6 & 100.0 & 61.0 & 100.0 & 1/1 & 80.0 & 0 & 2.152 & 27.74 \\
dv-066 & 98.4 & 98.7 & 96.3 & 46.2 & 85.5 & 40.0 & 100.0 & 1/1 & 80.0 & 0 & 0.643 & 12.76 \\
dv-067 & 79.3 & 93.9 & 80.3 & 99.6 & 96.4 & 72.7 & 100.0 & 1/1 & 90.0 & 0 & 0.495 & 13.50 \\
dv-068 & 48.2 & 79.4 & 33.1 & 44.4 & 100.0 & 60.0 & 100.0 & 1/1 & 30.0 & 0 & 0.686 & 11.95 \\
dv-069 & 100.0 & 80.7 & 100.0 & 82.6 & 98.0 & 66.7 & 100.0 & 1/1 & 100.0 & 0 & 0.388 & 11.01 \\
dv-070 & 90.8 & 76.1 & 80.4 & 57.2 & 72.0 & 83.6 & 100.0 & 1/0 & 0.0 & 0 & 4.079 & 104.31 \\
dv-071 & 99.6 & 97.5 & 97.4 & 88.7 & 47.0 & 100.0 & 100.0 & 1/1 & 100.0 & 0 & 1.192 & 14.98 \\
dv-072 & 100.0 & 100.0 & 100.0 & 100.0 & 94.4 & 87.3 & 47.5 & 1/1 & 100.0 & 0 & 0.504 & 8.59 \\
dv-073 & 97.0 & 84.2 & 93.0 & 89.9 & 100.0 & 94.7 & 100.0 & 1/1 & 100.0 & 0 & 0.566 & 12.21 \\
dv-074 & 100.0 & 100.0 & 100.0 & 92.6 & 100.0 & 100.0 & 24.3 & 1/1 & 100.0 & 0 & 0.559 & 10.71 \\
dv-075 & 98.6 & 94.2 & 90.4 & 89.2 & 100.0 & 66.1 & 0.0 & 1/1 & 90.0 & 0 & 1.824 & 17.15 \\
dv-076 & 72.8 & 80.2 & 56.4 & 76.5 & 100.0 & 66.7 & 100.0 & 1/1 & 100.0 & 0 & 0.621 & 14.23 \\
dv-077 & 99.8 & 94.1 & 99.3 & 85.4 & 93.3 & 0.0 & 0.0 & 1/0 & 0.0 & 0 & 1.336 & 19.74 \\
dv-078 & 70.0 & 76.6 & 60.9 & 45.0 & 91.3 & 96.8 & 100.0 & 1/1 & 100.0 & 0 & 0.526 & 12.34 \\
dv-079 & 93.2 & 77.8 & 86.5 & 76.7 & 100.0 & 71.0 & 100.0 & 1/0 & 0.0 & 0 & 1.743 & 59.60 \\
dv-080 & 95.9 & 87.4 & 92.0 & 91.2 & 100.0 & 58.8 & 0.0 & 1/1 & 80.0 & 0 & 3.048 & 24.75 \\
dv-081 & 37.3 & 34.3 & 31.0 & 13.0 & 92.5 & 100.0 & 100.0 & 1/1 & 100.0 & 0 & 1.501 & 19.58 \\
dv-082 & 94.8 & 100.0 & 91.1 & 98.6 & 80.0 & 0.0 & 100.0 & 1/1 & 100.0 & \textbf{\color{ReleaseBlue}1} & 0.762 & 4.14 \\
dv-083 & 100.0 & 97.0 & 100.0 & 99.2 & 95.2 & 100.0 & 100.0 & 1/1 & 100.0 & \textbf{\color{ReleaseBlue}1} & 0.852 & 20.48 \\
dv-084 & 100.0 & 88.9 & 95.0 & 94.8 & 94.4 & 73.1 & 5.8 & 1/1 & 100.0 & 0 & 2.455 & 22.72 \\
dv-085 & 98.6 & 91.2 & 95.5 & 96.4 & 98.2 & 88.9 & 100.0 & 1/1 & 90.0 & 0 & 1.371 & 18.48 \\
dv-086 & 96.1 & 86.5 & 94.2 & 97.2 & 100.0 & 0.0 & 100.0 & 1/1 & 80.0 & 0 & 0.739 & 14.53 \\
dv-087 & 99.6 & 90.2 & 97.6 & 76.0 & 95.2 & 96.8 & 100.0 & 1/1 & 100.0 & 0 & 0.913 & 18.33 \\
dv-088 & 92.6 & 85.2 & 84.4 & 76.5 & 66.1 & 0.0 & 98.3 & 1/1 & 80.0 & 0 & 4.459 & 27.89 \\
dv-089 & 99.6 & 96.5 & 98.6 & 82.8 & 98.9 & 75.9 & 88.0 & 1/1 & 60.0 & 0 & 3.268 & 29.45 \\
dv-090 & 36.8 & 62.3 & 38.9 & 66.8 & 89.4 & 90.9 & 100.0 & 1/1 & 90.0 & 0 & 0.397 & 10.43 \\
dv-091 & 97.3 & 100.0 & 95.6 & 99.4 & 93.3 & 92.6 & 100.0 & 1/1 & 80.0 & 0 & 0.913 & 13.03 \\
dv-092 & 97.2 & 95.0 & 94.3 & 84.6 & 100.0 & 87.5 & 100.0 & 1/1 & 100.0 & 0 & 1.093 & 15.85 \\
dv-093 & 99.1 & 89.0 & 98.6 & 88.7 & 93.3 & 96.8 & 100.0 & 1/1 & 100.0 & 0 & 0.510 & 13.32 \\
dv-094 & 99.4 & 98.5 & 97.2 & 61.1 & 99.0 & 100.0 & 100.0 & 1/1 & 90.0 & 0 & 0.854 & 15.11 \\
dv-095 & 99.9 & 92.6 & 99.7 & 96.8 & 100.0 & 100.0 & 91.3 & 1/1 & 60.0 & 0 & 0.750 & 12.31 \\
\end{longtable}\endgroup

\end{document}